\documentclass[phd,10pt,upper,appendixpart,oneside,bold]{ubcthesis}
\newcommand{\chap}{\chapter}

\newcommand{\bie}{\begin{itemize}}
\newcommand{\eie}{\end{itemize}}
\newcommand{\ben}{\begin{enumerate}}
\newcommand{\een}{\end{enumerate}}

\newcommand{\scr}{\scriptstyle}
\newcommand{\noi}{\noindent}
\newcommand{\nn}{\nonumber}
\newcommand{\mbf}{\mathbf}

\newcommand{\lab}{\label}

\newcommand{\beq}{\begin{equation}}
\newcommand{\eeq}{\end{equation}}
\newcommand{\bea}{\begin{eqnarray}}
\newcommand{\eea}{\end{eqnarray}}
\newcommand{\ba}{\begin{array}}
\newcommand{\ea}{\end{array}}

\newcommand{\na}{\nabla}
\newcommand{\pa}{\partial}
\newcommand{\eq}{\equiv}
\newcommand{\fr}{\frac}
\newcommand{\sr}{\sqrt}
\newcommand{\ha}{\fr{1}{2}}

\newcommand{\al}{\alpha}
\newcommand{\bt}{\beta}
\newcommand{\ga}{\gamma}
\newcommand{\de}{\delta}
\newcommand{\ep}{\epsilon}

\newcommand{\ze}{\zeta}
\newcommand{\et}{\eta}
\newcommand{\te}{\theta}

\newcommand{\ka}{\kappa}
\newcommand{\la}{\lambda}

\newcommand{\rh}{\rho}

\newcommand{\si}{\sigma}

\newcommand{\ta}{\tau}

\newcommand{\ph}{\phi}
\newcommand{\vp}{\varphi}
\newcommand{\ch}{\chi}
\newcommand{\ps}{\psi}
\newcommand{\om}{\omega}
\newcommand{\Ga}{\Gamma}
\newcommand{\De}{\Delta}

\newcommand{\La}{\Lambda}
\newcommand{\Si}{\Sigma}

\newcommand{\Ph}{\Phi}
\newcommand{\Ps}{\Psi}
\newcommand{\Om}{\Omega}

\usepackage{setspace}
\usepackage{graphicx}
\usepackage{amsmath,amssymb,latexsym} 
\usepackage{ifthen} 

\newcommand{\resetcounters}{ 
	\setcounter{equation}{0} 
	\setcounter{figure}{0} 
	\setcounter{table}{0}   }       
\newcommand{\graxi}{{\tt graxi}}
\newcommand{\highQ}{false}

\newcommand{\krr}{K^r{}_r}
\newcommand{\ktt}{K^\te{}_\te}
\newcommand{\kpp}{K^\vp{}_\vp}

\newcommand{\omss}{(1-s^2)}
\newcommand{\br}{\ze}
\newcommand{\bomr}{1-\br}
\newcommand{\bomror}{\fr{\bomr}{\br}}
\newcommand{\bomrors}{\left(\fr{\bomr}{\br}\right)^2}

\newcommand{\bal}{\bar{\al}}
\newcommand{\bbt}{\bar{\bt}}
\newcommand{\bga}{\bar{\ga}}
\newcommand{\bh}{\bar{h}}
\newcommand{\Oeps}{O(\ep^2)}

\newcommand{\rtoi}{r\to\infty}
\newcommand{\Rtoi}{R\to\infty}

\newcommand{\nj}{^{^{\scr n}}_{_{\scr j}}}
\newcommand{\nNr}{^{^{\scr n}}_{_{\scr N_r}}}
\newcommand{\no}{^{^{\scr n}}_{_{\scr 1}}}

\newcommand{\npoj}{^{^{\scr n+1}}_{_{\scr j}}}

\newcommand{\njpo}{^{^{\scr n}}_{_{\scr j+1}}}
\newcommand{\njmo}{^{^{\scr n}}_{_{\scr j-1}}}
\newcommand{\njpt}{^{^{\scr n}}_{_{\scr j+2}}}
\newcommand{\njmt}{^{^{\scr n}}_{_{\scr j-2}}}
\newcommand{\njph}{^{^{\scr n}}_{_{\scr j+\ha}}}
\newcommand{\njmh}{^{^{\scr n}}_{_{\scr j-\ha}}}
\newcommand{\npojpo}{^{^{\scr n+1}}_{_{\scr j+1}}}
\newcommand{\npojmo}{^{^{\scr n+1}}_{_{\scr j-1}}}

\newcommand{\Dtp}{\De^{^{\scr t}}_{_{\scr +}}}
\newcommand{\Ddiss}{\De^{^{\scr t}}_{_{\scr +KO}}}

\newcommand{\Drz}{\De^{^{\scr r}}_{_{\scr 0}}}
\newcommand{\Drp}{\De^{^{\scr r}}_{_{\scr +}}}
\newcommand{\Drm}{\De^{^{\scr r}}_{_{\scr -}}}
\newcommand{\Drzb}{\De^{^{\scr r}}_{_{\scr 0 b}}}
\newcommand{\Dhrz}{\De^{^{\scr \fr{r}{2}}}_{_{\scr 0}}}
\newcommand{\Drsz}{\De^{^{\scr r^2}}_{_{\scr 0}}}
\newcommand{\Drcz}{\De^{^{\scr r^3}}_{_{\scr 0}}}
\newcommand{\Dhrcz}{\De^{^{\scr \fr{r^3}{2}}}_{_{\scr 0}}}

\newcommand{\mutp}{\mu^{^{\scr t}}_{_{\scr +}}}

\newcommand{\murm}{\mu^{^{\scr r}}_{_{\scr -}}}
\newcommand{\bmurp}{\bar{\mu}^{^{\scr r}}_{_{\scr +}}}
\newcommand{\bmurm}{\bar{\mu}^{^{\scr r}}_{_{\scr -}}}

\newcommand{\omegabar}{\bar{\Omega}}
\newcommand{\sigmabar}{\bar{\sigma}}
\newcommand{\brho}{\beta^\rho}
\newcommand{\bz}{\beta^z}

\newcommand{\hx}{{\mathcal H}}
\input epsf  
\epsfverbosetrue
\usepackage{psfrag}

\institution{The University Of British Columbia}
\institutionaddress{Vancouver, Canada}
\department{Department of Physics and Astronomy}
\numberofsignatures{4}          
 
\previousdegree{B.Sc., The Chinese University of Hong Kong, 1996}
\previousdegree{M.Phil., The Chinese University of Hong Kong, 1998}

\title{A Numerical Study of Boson Stars}
\author{Chi Wai (Kevin) Lai}
\copyrightyear{2004}
\submitdate{September 15, 2004}    

\renewcommand\thepart         {\Roman{part}}
\renewcommand\thechapter      {\arabic{chapter}}
\renewcommand\thesection      {\thechapter.\arabic{section}}
\renewcommand\thesubsection   {\thesection.\arabic{subsection}}
\renewcommand\thesubsubsection{\thesubsection.\arabic{subsubsection}}
\renewcommand\theparagraph    {\thesubsubsection.\arabic{paragraph}}

\frontmatter

\begin{document}
 \onehalfspacing
 \bibliographystyle{unsrt}
  \maketitle
\authorizationform
\begin{abstract}
In this thesis we present a numerical study of general relativistic
boson stars in both spherical symmetry and axisymmetry.  We consider both time-independent problems, 
involving the solution of equilibrium equations for rotating boson stars, and time-dependent
problems, focusing on black hole critical behaviour associated with boson stars.

Boson stars are localized solutions of the equations governing 
a massive complex scalar field coupled to the gravitational field.
They can be simulated using more straightforward numerical techniques than
are required for fluid stars.  In particular, the evolution of smooth initial
data for a scalar field tends to stay smooth, in sharp contrast to 
hydrodynamical evolution, which tends to develop discontinuities, even
from smooth initial conditions.  At the same time, relativistic boson stars
share many of the same features with respect to the strong-field gravitational
interaction  as their fermionic counterparts.  A detailed study of their 
dynamics can thus potentially lead to a better understanding of the dynamics 
of compact fermionic stars (such as neutron stars), while the relative ease 
with which they can be treated numerically makes them ideal for use in 
theoretical studies of strong gravity.

In this last vein, the study of the critical phenomena that arise at
the threshold of black hole formation has been a subject of intense 
interest among relativists and applied mathematicians over the past decade.
Type I critical phenomena, in which the black hole mass jumps discontinuously
at threshold, were previously observed in the dynamics of spherically 
symmetric boson stars by Hawley and 
Choptuik~\cite{shawley:phd,scott_matt:2000}.
We extend this work and show that, contrary to previous claims, 
the subcritical end-state is well described by a stable boson star 
executing a large amplitude oscillation with a frequency in
good agreement with that predicted 
for the fundamental normal mode of the end-state star
from linear perturbation theory.

We then extend our studies of critical phenomena to the axisymmetric case,
studying two distinct classes of parametrized families of initial data whose 
evolution generates families of spacetimes that ``interpolate'' between 
those than contain a black hole and those that do not.  In both cases we 
find strong evidence for a Type I transition at threshold, and are able 
to demonstrate scaling of the lifetime for near-critical configurations
of the type expected for such a transition. This is the first time that Type I critical solutions have 
been simulated in axisymmetry (all previous general relativistic 
calculations of this sort imposed spherical symmetry).

In addition, we develop an efficient algorithm for 
constructing equilibrium configurations of rotating boson stars, which 
are characterized by discrete values of an angular momentum parameter, $k$ (an
azimuthal quantum number).  We construct families of solutions for $k=1$ and 
$k=2$, and demonstrate the existence of a maximum mass in each case. 
\end{abstract}
 
\tableofcontents
\listoftables
\listoffigures
 
\acknowledgements
I would like to express my sincere gratitude to my thesis supervisor,
Matthew Choptuik, for his guidance, help and encouragement throughout the years.  
I would also like to thank my collaborator, Dale Choi, for many useful and
valuable discussions on boson stars.
Thanks to Frans Pretorius, for his help on the usage of {\tt graxi}. 
Thanks to Scott Hawley, for
generously providing the code for perturbation analysis in the study of
spherically symmetric boson stars.  
Thanks to my supervisory committee---Kristin Schleich, Douglas Scott and William
Unruh, in addition to Matthew Choptuik---for their comments and suggestions.
Thanks to all members of the UBC numerical relativity group, for 
creating a pleasant and stimulating research environment.  
Additional thanks to Huang Hai, for teaching me condensed matter physics.
Special thanks to 
Eric Kwan, Chi Yui Chan, Jonathan Chau, Samuel Wai and Hottman Sin, for their 
friendship and support.  Finally I would like to thank my family, for their 
endless love in the past and the future.
 
\newpage

  \mainmatter
  \resetcounters

\chapter{Introduction} \lab{intro}

This thesis is concerned with the numerical simulation of boson stars
within the framework of Einstein's theory of general relativity.
Boson stars are self-gravitating compact objects~\footnote
{
	By compact, we mean {\em gravitationally compact}, so that the size, $R$,
	of the star is comparable to its Schwarzschild radius, $R_S$.  The
	Schwarzschild radius associated with a mass $M$ is $R_S = (2G/c^2) M$, where
	$G$ is Newton's gravitational constant, and $c$ is the speed of light.
}
composed of scalar particles~\cite{kaup:1968},
and are the bosonic
counterparts of the more well-known compact fermionic stars, which include 
white dwarfs and especially neutron stars.  
In contrast to fermionic stars, there
is no observational evidence that boson stars exist in nature,
nor has the existence of any {\em fundamental} scalar particle yet been verified by experiments.
It has been proposed, however, that 
boson stars could be a candidate for, or at least make up a considerable fraction
of, the dark matter in our universe~\cite{jetzer}.  Were this true, it is quite probable
that studies of their properties would lead to a better understanding of 
astrophysical phenomena.  However, at the current time boson stars remain purely
theoretical entities.

Their current hypothetical nature notwithstanding, boson stars might indeed exist
in our universe, and, more importantly for this thesis, they  
are excellent matter models for the numerical study of compact objects in strong 
gravitational fields, 
a subject that continues to be a key concern of numerical
relativity~\cite{luis:2001} ({\em numerical relativity:} the numerical simulation of 
the Einstein field equations, as well as the field equations for
any matter fields being modeled).
The boson stars we consider can be viewed as zero-temperature, ground-state, Bose-Einstein
condensates, with enormous occupation numbers, so that the stellar material
is described by a single complex scalar field, $\phi(t,{\bf x})$, that satisfies 
a simple, classical Klein-Gordon equation.
Many of the nice modeling properties associated with boson stars 
derives from the fact that the dynamics is governed by a partial differential
equation (PDE) that does {\em not} tend to develop discontinuities from
smooth initial data.  This is not the case for 
fermionic stars, which are usually treated as perfect
fluids, and which thus satisfy hydrodynamical equations with phenomenologically-determined
equations of state.  Relativistic fluid evolution will generically produce shock waves and
other discontinuities, even from smooth initial data.
Therefore, today's state-of-the-art relativistic hydrodynamics codes use sophisticated 
{\em high resolution shock capturing} (HRSC) schemes in order to more accurately and efficiently simulate the 
fluid physics.  The fact that the solutions of the Klein-Gordon equation stay smooth
is a tremendous boon when it comes to discretizing the equation in a stable and 
accurate manner.

In addition, the effective integration of HRSC schemes with other advanced
numerical techniques such as adaptive mesh refinement~\cite{BO}  (AMR) is still 
in its infancy, at least for the case of general relativistic applications.  
For bosonic matter, on the other hand, it has proven relatively easy to 
use rather generic AMR algorithms (such as that of Berger and Oliger~\cite{BO}) in
conjunction with straightforward second-order finite-difference discretization to 
achieve essentially unbounded dynamical range in an efficient
fashion~\cite{choptuik,fransp:phd,graxi:2003}.

Of course, we do not primarily study boson stars because they are 
easy to simulate.  As mentioned already, a good fraction of the work 
in numerical relativity is concerned with the strong-field 
dynamics of gravitationally compact
objects~\cite{jetzer,schunck:2003,baumgarte:2003b}.
Despite the large amount of 
effort that has been devoted to this subject, it is fair to say that much 
remains to be learned, and much of what remains to be discovered is 
likely to be found through simulation.  Although the strong-field
gravitational physics of boson stars may not compare in detail with that of 
fermionic stars in all respects,
there are clearly some key features of the usual
stars that are shared by their bosonic counterparts.  For example,
in analogy with relativistic fermionic stars, spherically symmetric 
boson stars typically come in one-parameter families, where the 
parameter can be viewed as the central density of the star (or 
an analogue thereof).  Moreover, in relativistic cases
the boson
star families share with the fermionic sequences the surprising 
property that there comes a point when increasing the density of 
the star at the center actually {\em decreases} the total gravitating 
mass of the star.  In both cases this leads to maximal masses for any given
family of stars.  In addition, in both instances, stars near that limit are 
naturally very strongly self-gravitating.

Suffice it to say, then, that a careful study of general relativistic boson 
stars is likely to lead to insights into strong-field gravitational
physics, even if there are no immediate astrophysical applications, and that
this is the primary motivation for the calculations described below. 
In particular, and again as with the fermionic case, black 
hole formation is a crucial process that can, and will, generically 
occur in the strong-field dynamics of one or more boson stars.
For the most part, this is simply because (1) as gravitationally compact
objects, boson stars are, by definition, close to the point of collapse, and 
(2) the process of gravitational collapse involving matter with positive 
energy tends to be very unstable in many senses, including the fact 
that black hole areas can only increase. 

Over the past decade or so, the careful study of gravitational collapse 
and black hole formation has lead to the discovery of black hole 
critical phenomena~\cite{choptuik}, wherein the process of black hole formation,
studied in solution space, takes on many of the features of 
a phase transition in a statistical mechanical system.
A primary goal of the work presented in this thesis is to study
so-called Type I critical phenomena of boson stars in both spherical
symmetry and axisymmetry, and this will be explained in more detail
below.  The simulations are carried out via finite difference solution
of the governing PDEs. 
We also develop and apply a new algorithm for constructing solutions 
of the time-independent form of the PDEs in axisymmetry; these solutions 
represent relativistic {\em rotating} boson stars.

\section{An Overview of Boson Stars}

The study of boson stars can be traced back to the work of Wheeler.
Wheeler studied self-gravitating objects whose constituent element is the 
electromagnetic field and named the resulting ``photonic'' configurations geons~\cite{wheeler:1955}.
Wheeler's original intent was to construct a self-consistent, classical and field-theoretical notion of body, 
thus providing a divergence-free model for the Newtonian concept of body.  In the late 1960s,
Kaup \cite{kaup:1968} adopted the geon idea, but coupled a massive complex scalar field, rather than
the electromagnetic field, to general-relativistic gravity. 
Assuming time-independence and spherical symmetry he found solutions of the coupled equations which he called Klein-Gordon 
geons.  Subsequently, Ruffini \& Bonazzola \cite{ruffini:1969} 
studied field quantization of a real scalar field and considered the ground state
configurations of a system of such particles.  The expectation value of the field
operators gives the same energy-momentum tensor as those given by Kaup, and hence the
different approaches followed in the two studies give essentially the same macroscopic results.

Later, these Klein-Gordon geons were given the name boson stars, and the 
nomenclature {\em boson star} now
generally refers to compact self-gravitating objects that are
regular everywhere and that are made up of scalar fields.  Variations of the 
original model studied by Kaup and Ruffini \& Bonazzola include
self-interacting boson stars (described by a Klein-Gordon field with one or more self-interaction terms), charged boson stars
(boson stars coupled to the electromagnetic field) and rotating boson stars (boson stars possessing
angular momentum), to name a few.
When stable, all of these objects are held together by the balance between the
attractive gravitational force and a pressure that can be viewed as
arising from 
Heisenberg's uncertainty principle, as well as any explicit repulsive self-interaction
between the bosons that is incorporated in the model.
Depending on the mass of the constituent particles, and on
the value of the self-interaction coupling constant(s), the size of the stars can in 
principle vary from the atomic scale to an astrophysically-relevant scale~\cite{schunck:2003}. 
As mentioned previously, any given
boson star is typically only one member of a continuous family of equilibrium solutions. 
In spherical symmetry, the family can be conveniently parametrized using 
the central value of the modulus of the scalar field, in analogy
to the central pressure of perfect fluid stars.  Moreover, the equilibrium configurations
generally have an exponentially decaying tail at large distances from the stellar core, in contrast
to fluid-models stars which tend to have sharp, well-defined edges.

\subsection{Stability of Boson Stars} \lab{stability_BS}

The dynamical stability of equilibrium, compact fermionic (fluid) stars
against gravitational collapse can be studied
using the linear perturbation analysis of infinitesimal radial oscillations
that conserve the total particle number, $N$,  and mass/energy, $M$. One important theorem in this 
regard
concerns the transition between stable and unstable equilibrium
\cite{HTW:grav_theory}\cite[pp.305]{weinberg:grav_cos}.  
The theorem 
states that a perfect fluid star with constant chemical composition and constant entropy per nucleon 
becomes unstable with respect to some radial mode only at central densities $\rho(0)$ 
such that 
\bea
  \fr{\pa M(\rh(0),s,\cdots)}{\pa \rh(0)} &=& 0\,, \\
  \fr{\pa N(\rh(0),s,\cdots)}{\pa \rh(0)} &=& 0\,.
\eea
\noi
In other words, a change in stability can only occur at those points in a curve 
of total mass $M(\rho(0))$ {\em vs} $\rho(0)$, where an extrema is attained.  
This results directly
from the fact that the eigenvalues, $\si^2$, of the associated pulsation equation change
sign at those points.  

Similar results hold
for various boson star models.  
As just mentioned, in the case of boson stars we use the
central value of the scalar field modulus, denoted $\ph_0(0)$, as the parameter for the family
of solutions, and it has been shown that the pulsation equation has a zero mode
at the stationary points in the $M(\ph_0(0))$ plot~~\cite{lee_pang:1989,glesier_watkins:1989}.  
Numerical verification of the instability of configurations past the 
mass maximum using the full dynamical equations in spherical symmetry was studied in
\cite{seidel_suen:1990,Balakrishna:1998} and \cite{scott_matt:2000}.

\subsection{Maximum Mass of Boson Stars}
As already mentioned, 
another important feature of stellar structure that is largely 
due to relativistic effects is the existence of
a maximum allowable mass for a particular species of stars.  Depending on
the mechanism of stellar pressure (degenerate electron pressure for white
dwarfs, degenerate neutron pressure for neutron stars, Heisenberg
uncertainty principle for boson stars) these values are different.  
White dwarfs and neutron stars share the same dependence on the mass of the
constituent particles ($\sim M_{\rm pl}^3/m^2$), while the dependence for 
boson stars is quite
different ($\sim M_{\rm pl}^2/m$).  The origin of
the difference can be understood via the following 
heuristic argument.
The ground state stationary boson stars are macroscopic quantum states of
cold, degenerate bosons, whose existence is the result of the balance
between the attractive gravitational force and the dispersive nature of
the wave function.  By the uncertainty principle, if the bosons are confined in a region of
size $R$, we have $p R \sim \hbar$, where $p$ is the typical momentum of the
bosons.  For a moderately relativistic boson we have $p \sim mc$ 
and hence $R \sim \hbar/(mc)$.  Equating this with the Schwarzschild radius
we have $\hbar/(mc) \sim 2GM_{\mathrm{max}}/c^2$.  Hence
$M_{\mathrm{max}}\sim (\hbar c)/(2Gm) = 0.5 M_{\rm pl}^2/m$.  Numerical
calculation shows that $M_{\mathrm{max}} \approx 0.633 M_{\rm pl}^2/m$ (see
Sec.~\ref{family_statsol}),
surprisingly close to the above estimate.

The maximum mass of neutron stars can be estimated in a similar way.~\footnote{
\cite{shapiro_teukolsky:BHWD} gives another heuristic argument due to Landau
(1932), which applies to both neutron stars and white dwarfs.  The current argument applies to neutron stars only; for white
dwarfs the mass of the star is dominated by baryons, while the pressure is
provided by electrons.}
The existence of these stars is the result of the balance
between the attractive gravitational force and the pressure due to
degenerate neutrons (fermions).  Suppose there
are $N$ fermions confined in a region of size $R$. Then by Pauli's exclusion
principle, each particle occupies a volume $1/n$, where $n \eq N/R^3$ is the
number density.  Effectively, each particle has a size of $R/N^{1/3}$.
Again, by the uncertainty principle we have $ pR/N^{1/3} \sim \hbar$.  Following
the same argument as for the boson star case, we have $R \sim \hbar N^{1/3}/(mc)$, and  hence
$2GM_{\mathrm{max}}/c^2 \sim \hbar N^{1/3}/(mc) \sim \hbar
{M_{\mathrm{max}}}^{1/3}/(m^{4/3}c)$.  Thus we have $M_{\mathrm{max}} \sim
0.35 {M_{\rm pl}}^3/m^2$.  
In contrast to the bosonic case, then, the maximum mass of fermionic stars scales as
${M_{\mathrm{max}}} \sim {M_{\rm pl}}^3/m^2$.

\subsection{Rotating Boson Stars}

Although spherically symmetric boson star solutions were found 
as early as 1968,
for many years it was unclear whether solutions describing time-independent boson 
stars with angular momentum existed or not.
The first attempt to construct such stars was due to Kobayashi, Kasai \&
Futamase in 1994~\cite{kobayashi:1994}.  These authors followed the same approach as 
Hartle and others~\cite{hartle:1967,hartle:1968}, in which 
the {\em slow} rotation of a general relativistic star is treated 
via perturbation of a 
spherically symmetric equilibrium configuration. 
In their study they found that slowly
rotating boson stars solutions coupled with a $U(1)$ gauge field (charged
boson stars) do {\em not} exist, at least perturbatively.  
Shortly thereafter, however, it was demonstrated that rotating boson stars
{\em could} be constructed on the basis of an ansatz which leads to a quantized
(or perhaps more properly, discretized) angular momentum.
It was later understood that rotating boson stars have quantized angular
momenta and that the concomitantly discrete nature of the on-axis regularity 
conditions prohibit a continuous (perturbative) change from 
non-rotating to rotating configurations. 

A year later, Silveira \& de~Sousa~\cite{sousa_silveira:1995}, following the
approach of Ferrell \& Gleiser~\cite{ferrell_gleiser:1989}, succeeded in
obtaining equilibrium solutions of rotating boson stars within the framework of
Newtonian gravity.  Specifically, they adopted the ansatz
\footnote{Note that (\ref{2dansatz}) is clearly not the most general ansatz we could make
for stationary solutions of the Einstein-Klein-Gordon system. For example, we could consider
$p$ independent bosonic fields $\ph_i(t,{\bf x}), i = 1,2,\cdots, p$, each satisfying ansatz
(\ref{2dansatz}) for specific values of $(k,\om) = (k_i, \om_i)$, i.e.\ $\ph_i(t,{\bf x}) =
\ph_i(r,\te) e^{i \left( \om_i t + k_i \vp\right)}$, with a total stress energy tensor
given by $T_{\mu \nu} = \sum_{i=1}^p T_{\mu \nu}^{\,i}$, with each of the $T_{\mu \nu}^{\,i}$ given
by  (\ref{Ctmunu}).  The stationary solutions $\ph_i(t,{\bf x}; k_i, \om_i(k_j))$ would then
be labelled by $p$ quantum numbers ($k_i$) with $p$ associated eigenvalues $\left(
\om_i(k_j)\right)$, and the spectrum would still be discrete.}

\beq \lab{2dansatz}
 \ph(t,r,\te,\vp) = \ph_0(r,\te) e^{i \left( \om t + k \vp\right)}\,,
\eeq
\noi
where $k$ is an integer (we use the symbol $k$, instead of the symbol $m$,
that is commonly used in quantum mechanics, to avoid confusion with the
particle mass, $m$), so that the stress-energy tensor $T^\phi_{\mu \nu}$
is independent of both time $t$ and the azimuthal angle $\vp$.
In the non-relativistic limit~\footnote{
By which we mean that {\em both} the gravitational and matter fields are treated using 
non-relativistic equations of motion.
} 
the governing field equations constitute a 
coupled Poisson-Schr\"{o}dinger (PS) system 
(for simplicity, we have dropped the subscript ``0'' so that $\ph_0(r,\te) \to \ph(r,\te)$)

\beq \lab{V_eq}
 \na^2 V = 8 \pi m^2 \ph \ph^{\ast}\,,
\eeq
\beq \lab{ph_eq}
-\fr{1}{2m} \na^2 \ph + m V \ph = E \ph\,,
\eeq
\noi
where $V$ is the Newtonian gravitational potential, and $E$ is an
energy eigenvalue.  
The scalar field $\ph_0(r,\te)$ is then expanded in associated Legendre functions,
$P_l^k(\te)$:
\beq \lab{ph_exp}
 \ph_0(r,\te) = \fr{1}{\sr{4 \pi}} \sum^{\infty}_{l=k} R_l(r) P^k_l(\te)\,,
\eeq 
Similarly, the potential $V(r,\te)$ is assumed to have no $\vp$-dependence, and thus can be expanded
as
\beq\lab{V_exp}
 V(r,\te) = \sum_{l=0}^{\infty} V_l(r) P_l(\te)\,.
\eeq
Eqs.~(\ref{ph_exp}) and (\ref{V_exp}) are then substituted into (\ref{V_eq})
and (\ref{ph_eq}), and multiplied by $P_{l_0}(\te)$ to obtain a system of
equations for any particular value of $l_0$
(using the orthogonality of the associated Legendre functions).
Using this strategy, Silveira and de~Sousa were 
able to obtain solutions for each specific combination of $l_0$ and $k$.  
A main difference of these solutions relative to the
spherically symmetric ones, is that the scalar field vanishes at
the origin, and hence the rotating star solutions have toroidal level
surfaces of the matter field, 
rather than spheroidal level surfaces as in the spherical case.

In 1996 Schunck \& 
Mielke~\cite{schunck_mielke:1996} used the ansatz (\ref{2dansatz}) to 
construct rotating boson star solutions.
Specifically, 
they chose some particular values of $k$, specifically $k=0,1\cdots,10$ and $k=500$
and showed that solutions to the 
fully general relativistic equations {\em did} exist for those cases.
They also showed that 
the angular momentum, $J$, and the total particle number, $N$, of the stars are related by
\beq
 J = k N\,,
\eeq
\noi
where $k$ is the integer defined in (\ref{2dansatz}).  An important implication of
the above equation is that if we consider equilibrium configurations with
the same total particle number $N$, then the total angular momentum
has to be quantized.  This property is in clear contrast with that of a
perfect fluid star.

Later, in the work most relevant to the study of rotating boson stars described 
in this thesis,
Yoshida \& Eriguchi~\cite{yoshida_eriguchi:1997b} 
used a self-consistent-field method~\cite{yoshida_eriguchi:1997} 
to obtain the whole family of solutions
for $k=1$, as well as part of the family for $k=2$.  The maximum mass they found
for the $k=1$ case was $1.314M_{\rm pl}^2/m$.  However, 
their code broke down before they could compute the star with maximum mass 
for the $k=2$ case.

In Table~\ref{fermionboson} we summarize
some similarities and differences between rotating fermion stars and
rotating boson stars.  For further background information on boson 
stars we suggest that readers consult the review by Jetzer~\cite{jetzer}, or
the more up-to-date survey by Schunck \& Mielke~\cite{schunck:2003}.

\vspace{0.5cm}
\begin{table}[htbp]
\begin{center}
\begin{tabular}[l]{cc}
\hline
 {\bf Relativistic Rotating Fermion Stars }& {\bf Relativistic Rotating Boson Stars }\\ \hline
\hline
\multicolumn{2}{c}{{\em  Similarities }}  \\
\multicolumn{2}{c}{\small Come in families of solutions parametrized by a
single value} \\
             \multicolumn{2}{c}{\small Each family has a maximum possible mass} \\
\hline
\multicolumn{2}{c}{{\em  Differences}}  \\
{\small Parametrized by $p(0)$} & {\small Parametrized by 
$|\partial^{(k)}_r\phi(0)|$} \\
  {\small Spheroidal level surfaces of rotating matter field} &   {\small Toroidal level surfaces
of matter field for $k=1,2,\cdots$}\\
 {\small Finite size, abrupt change in $\partial_r p, \partial_r \rh$ at surface} & {\small
Exponential decay to infinity} \\
 {\small Angular momentum can vary continuously} & {\small Angular momentum
quantized: $k=1,2,\cdots$} \\
\hline
\end{tabular}
\caption[Similarities and differences between relativistic rotating fermion
stars and relativistic rotating boson stars]
{Similarities and differences between relativistic rotating fermion
stars and relativistic rotating boson stars.}
\end{center}
\label{fermionboson}
\end{table}

\section{Critical Phenomena in Gravitational Collapse} 

Over the past decade, intricate and unexpected phenomena related to 
black holes have been discovered through the detailed numerical 
study of various models for gravitational collapse,
starting with Choptuik's investigation of the spherically symmetric 
collapse of a massless scalar field~\cite{choptuik}.
These studies generally concern the {\em threshold} of black hole 
formation (a concept described below), and the phenomena observed
near threshold are collectively called (black hole) critical phenomena,
since they share many of the features associated with critical phenomena
in statistical mechanical systems.  The study of critical phenomena 
continues to be an active area of research in numerical relativity, 
and we refer the interested reader to the recent review article
by Gundlach~\cite{gundlach:2003} for full details on the subject. 
Here we will simply summarize some key points that are most germane 
to the work in this thesis.

To understand black hole critical phenomena, one must understand
the notion of the ``threshold of black hole formation".  
The basic idea is to consider {\em families} of solutions of
the coupled dynamical equations for the gravitational field 
and the matter field that is undergoing collapse (the complex 
scalar field, $\phi$, in our case).  Since we are considering a dynamical problem,
and since we assume that the overall dynamics is uniquely determined 
by the initial conditions, we can view the families as being 
parametrized by the initial conditions---variations in one or more 
of the parameters that fix the initial values will then generate 
various solution families.  We also emphasize that we are considering 
{\em collapse} problems.  This means that we will generically 
be studying the dynamics of systems that have length scales 
comparable to their Schwarzschild radii, {\em for at least some period of time 
during the dynamical evolution}.  We also note that we will often
take advantage of the complete freedom we have as numerical
experimentalists to choose initial conditions that lead to collapse, but which may 
be highly unlikely to occur in an astrophysical setting.

We now focus attention on {\em single parameter} families of 
data, so that the specification of the initial data is fixed 
up to the value of {\em the} family parameter, $p$.  We will generally 
view $p$ as a non-linear {\em control parameter} that will
be used to govern how strong the gravitational field becomes
in the subsequent evolution of the initial data, and in particular,
whether a black hole forms or not. Specifically, we will always 
demand that any one-parameter family of solutions has the 
following properties:
\begin{enumerate}  
\item For sufficiently small values of $p$ the dynamics remains regular 
      for all time, and no black hole forms.  
\item For sufficiently large values of $p$, complete gravitational collapse 
      sets in at some point during the dynamical development of the initial
      data, and a black hole forms.
\end{enumerate}
From the point of view of simulation, it turns out to be a relatively 
easy task for many models of collapse to construct such families,
and then to identify 2 specific parameter values, $p^-$ ($p^+$) which do not (do)
lead to black hole formation.  Once such a  ``bracket'' $[p^-,p^+]$ has been
found, it is straightforward in principle to use a technique such as 
binary search to hone in on a {\em critical parameter value}, $p^\star$, such 
that all solutions with $p<p^\star$ ($p>p^\star$) do not (do) contain
black holes. A solution corresponding to $p=p^\star$ thus sits at the 
threshold of black hole formation, and is known as a {\em critical solution}.
It should be emphasized that underlying the existence of critical solutions 
are the facts that (1) the end states (infinite-time behaviour)  corresponding to
properties 1.~and 2.~above are {\em distinct} (a spacetime containing a black hole 
{\em vs} a spacetime not containing a black hole) and (2) the process 
characterizing the black hole threshold (i.e.\ gravitational collapse) 
is {\em unstable}. We also note that we will term evolutions with $p<p^\star$
{\em subcritical}, while those with $p>p^\star$ will be called {\em supercritical}.

Having discussed the basic concepts underlying black hole critical phenomena,
we now briefly describe the features of critical collapse that are most 
relevant to the work in this thesis.  

First, critical solutions {\em do}
exist for all matter models that have been studied to date, and for 
any given matter model, almost certainly constitute discrete sets.  In
fact, for some models, there may be only {\em one} critical solution,
and we therefore have a form of universality.  

Second, critical solutions 
tend to have additional symmetry beyond that which has been adopted in the 
specification of the model (e.g. we will impose spherical and axial 
symmetry in our calculations).  

Third, the critical solutions known
thus far, and the black hole thresholds associated with them, come 
in two broad classes.  The first, dubbed Type I, is characterized 
by static or periodic critical solutions (i.e.\ the additional symmetry 
is a continuous or discrete time-translational symmetry), and by
the fact that the black hole mass just above threshold is {\em finite}
(i.e.\ so that there is a minimum black hole mass that can be 
formed from the collapse). 
The second class, called Type II, is characterized by continuously or 
discretely self-similar critical solutions (i.e.\ the additional
symmetry is a continuous or discrete scaling symmetry), and by the 
fact that the black hole mass just above threshold is {\em infinitesimal}
(i.e.\ so that there is {\em no} minimum for the black hole mass that 
can be formed).    The nomenclature Type I and Type II is by analogy 
with first and second order phase transitions in statistical mechanics,
and where the black hole mass is viewed as an order parameter.

Fourth, solutions close to criticality exhibit various scaling laws.
For example, in the case of Type I collapse, where the critical solution is an
unstable, time-independent (or periodic) compact object, the amount 
of time, $\tau$, that the dynamically evolved configuration is well 
approximated by the critical solution {\em per se} satisfies a scaling law of 
the form
\beq
	\label{tau-scaling}
	\tau(p) \sim -\gamma \ln | p - p^\star | \,,
\eeq
where $\gamma$ is a {\em universal} exponent in the sense of not 
depending on which particular family of initial data is used to 
generate the critical solution, and $\sim$ indicates that the relation 
(\ref{tau-scaling}) is expected to hold in the limit $p \to p^\star$.

Fifth, and finally, much insight into critical phenomena comes 
from the observation that although unstable, critical solutions 
tend to be {\em minimally} unstable, in the sense that they 
tend to have only a few, and perhaps only one, unstable modes 
in perturbation theory.  In fact, if one assumes that a Type I
solution, for example, has only a single unstable mode, then 
the growth factor (Lyapunov exponent) associated with that mode can 
be immediately related to the scaling exponent~$\gamma$ defined by
(\ref{tau-scaling}).

In this thesis we will be exclusively concerned with Type I critical 
phenomena, where the threshold solutions will generally turn
out to be unstable boson stars.  Previous work relevant to ours
includes studies by (1) Hawley \cite{shawley:phd} and Hawley \& Choptuik~\cite{scott_matt:2000} of 
boson stars in spherically symmetry, (2) Noble~\cite{scn:phd} of 
fluid stars in spherical symmetry and (3) Rousseau~\cite{rousseau:master} of 
axisymmetric boson stars within the context of the  conformally flat approximation
to general relativity.  Evidence for Type I transitions have been found in 
all three cases.

\section{Layout}

The remaining chapters of this thesis are organized as follows.  In
Chap.~\ref{MathForm} we summarize the mathematical formalism used in the
work of this thesis.  This includes a brief summary of the mathematical
model of spacetime, in which the key ingredient to be used is the Einstein
field equation.  We then summarize the ADM (3+1) formalism, which will be used in
the study of boson stars in spherical symmetry, as well as  the (2+1)+1 formalism,
which is used in the study of boson stars in axisymmetry.  We also describe
the Einstein-Klein-Gordon system, which is the fundamental set of PDEs 
underlying all of our studies.

In Chap.~\ref{num_method} we summarize the numerical methods used in the
thesis.  This includes finite differencing techniques which are central to
all the calculations shown;  the multigrid method, which is used in the construction
of rotating boson stars, as well as in the solution of the 
constraint equations in the
dynamical study of boson stars in axisymmetry;  adaptive mesh refinement,
which is essential in the study of critical phenomena in axisymmetry;
excision techniques which are used in the study of boson stars in spherical
symmetry; and the technique of spatial compactification, which is used in the 
construction of rotating boson stars.

In Chap.~\ref{bs1d} we study Type I critical phenomena of boson stars in spherical symmetry.
This research can be viewed as an extension of the work reported by Hawley \&
Choptuik in~\cite{scott_matt:2000}.  Our principal new result is 
compelling numerical evidence for the existence of oscillatory final states of
subcritical evolutions.  We also perform perturbation analyses and show 
that the simulation results agree very well with those obtained from perturbation
theory.  We then present a rudimentary, but stable and convergent implementation of the black hole excision
method for the model.  Supercritical simulations using excision show that the spacetimes 
approach static black holes at late time, so there is no impediment to very long run
times (in physical time).

In Chap.~\ref{bs2d} we describe a study of boson stars in axisymmetry.
We first present an algorithm to construct the equilibrium configurations of rotating
boson stars that is based on the multigrid technique.  We argue that our method
is more computationally efficient than methods previously used and reported 
in the literature.  
More importantly, we obtain 
numerical solutions in the highly relativistic regime for an angular momentum parameter
$k=2$.  We then discuss studies of the
dynamics of boson stars in axisymmetry.  
Following Choi's~\cite{dale} work in the Newtonian limit, we show that 
solitonic behaviour occurs in the head-on collision of boson stars with 
sufficiently large relative initial velocities.
We also present a study of Type I critical phenomena in the model.
The two classes of simulations (collisions of boson stars, and boson stars gravitationally
perturbed by a massless real scalar field) provide evidence for Type I 
black hole transitions, as well as scaling laws for the lifetime of near-critical 
configurations of the form~(\ref{tau-scaling}).  This represents the first 
time that Type I behaviour has been 
observed in the context of {\em axisymmetric} collapse.

The final chapter summarizes the results, gives overall conclusions and points to some 
directions of future work.  Several appendices providing various technical details 
are also included.

\section{Conventions, Notation and Units}\lab{notation}
Throughout this thesis the signature of 
metric is taken to be ($-$ + + +).  Spacetime indices of four dimensional (1 temporal +
3 spatial) tensors are
labeled by lower case Greek letters ($\al, \bt, \ga, \cdots$). 
Spatial indices of three dimensional (3 spatial) tensors are labeled by lower case Latin 
letters starting from $i$ ($i, j, k, \cdots$).
Spacetime indices of three dimensional (1 temporal + 2 spatial) tensors are
labeled by 
the first few Latin indices ($a, b, c,\cdots, h$), and
spatial indices of two dimensional (2 spatial) tensors
are labelled by upper case Latin letter ($A, B, C,\cdots$).
The Einstein summation convention is implied for all types of indices. 
That is, repeated (one upper and one
lower) indices are automatically summed over the appropriate range.
Covariant derivatives are denoted by $\na$ or by a semi-colon ``;".
Ordinary partial derivatives are denoted by $\pa$ or by a comma ``,". 
Conventions
for the Riemann and Ricci tensors are $\na_{[\al} \na_{\bt]} V_{\ga} = \ha R_{\al
\bt \ga \de} V^{\de}, R_{\mu \nu} = R^{\al}{}_{\mu \al \nu}$.  

We adopt a system of ``natural" units 
in which $G=c=\hbar=1$.  
In this system the unit time, unit length and unit mass are known
as the Planck time, Planck length and Planck mass, respectively.
Specifically, we have

\bea
 T_{\rm{pl}} &=& \sr{\fr{\hbar G}{c^5}} = 5.39\times 10^{-44}\mbox{s}
\,,\\ 
 L_{\rm{pl}}&=&\sr{\fr{\hbar G}{c^3}} = 1.62\times10^{-35}\mbox{m}\,, \\
 M_{\rm{pl}} &=& \sr{\fr{\hbar c}{G}} = 2.18 \times 10^{-8}\mbox{kg}\,.
\eea

\noi
The corresponding energy scale is $M_{\rm{pl}}\,c^2 \approx
10^{19}\mbox{GeV}$.  
For the scalar field, the action has dimension $[\hbar]$.  Hence
$\left[ d^4x (\pa \ph)^2 \right] = [\hbar]$. Therefore, $[\ph] =
[\sr{\hbar}]/L=\sr{M/T}$.  Thus, $\ph$ is measured in units of
$\sr{M_{\rm{pl}}/T_{\rm{pl}}} = 6.36\times 10^{17} \sr{\mbox{kg/s}}$.

Finally, we adopt a nomenclature commonly used in numerical work, whereby 
we refer to calculations requiring the solution of 
partial differential equations in $n$ {\em spatial} dimensions
as ``$n$D''.  In particular, we will often refer to spherically symmetric 
computations (whether time-dependent or not) as ``1D'', and axisymmetric
calculations (again, irrespective of any time-dependence) as ``2D''.

  \resetcounters
\def\calH{{\mathcal H}}
\def\calT{{\mathcal T}}
\def\calR{{\mathcal R}}
\def\calG{{\mathcal G}}
\def\H#1#2{{\mathcal H}^{#1}{}_{#2}}

\chap{Mathematical Formalism} \lab{MathForm}

In this chapter we briefly summarize the mathematical description of spacetime,
as well as the ADM (or 3+1) and (2+1)+1 decompositions of spacetime.  
These decompositions---which form the bases for the numerical calculations described in the 
remainder of this thesis---are well described in the literature, and we therefore restrict 
our discussion here to a summary of the physical picture, and statements of the key equations 
that will be used in subsequent chapters.

\section{Mathematical Model of Spacetime}
In general relativity the elementary entities are events, each of which is characterized by
a time, $t$, and location, $x^i$, at which the particular event occurs.  According to 
Einstein's geometric view of the gravitational interaction, gravitational physics
is concerned with the relationship between these events: in other words, it describes the
``structure" of a collection of events $\mathcal{M}=\{(t,x^i)\}$.
Specifically, we are to view the
collection of events as a surface with a certain structure imposed on it.  In 
mathematical language,
the collection of events forms a manifold, where manifold is simply a generalization
of the concept of surface, and the structure that describes the gravitational physics turns out to
be a metric on the manifold.  The pair of objects, the manifold, $\mathcal{M}$, and the
metric, $g_{\mu \nu}$, comprise {\em spacetime}, which is of primary interest
in general relativity.  
The study of spacetime thus becomes the study of the explicit
form of the metric, or, equivalently, of the geometry of spacetime. 

The physics that we are looking for should also tell us how the distribution of
matter affects the structure of spacetime.  In other words, the theory should have a way
of prescribing matter fields, and the relationship between these matter fields and the
geometry of spacetime.  Mathematically, we describe the matter as some tensor fields on the manifold,
and the relationship between the matter and the geometry is written in the form of
field equations.  We will make some further assumptions about the spacetime so that 
some general physical principles---of whose validity we are confident---will be satisfied.
Following Hawking and Ellis
\cite{hawking_ellis:LSSST}, these assumptions are:

\begin{enumerate}
\item {\em Local causality}: special relativity posits that no signal can travel faster than the
speed of light. Therefore, equations for any matter fields should respect this
property.  Causality effectively partitions the neighborhood of any point $p$ into two
sets: those
points which are causally connected to $p$ and those which are not.  This assumption
allow us to determine the metric up to a conformal factor (i.e.\ up to an overall location-dependent
scale).

\item \emph{Local conservation of energy and momentum}: 
conservation laws generally reflect symmetries in physical laws.
Energy and momentum conservation---related to invariance of physical laws under temporal and
spatial translations---is a cornerstone of physics, with an extremely solid empirical 
basis.
Conservation of energy and momentum can be expressed
as a condition on the energy-momentum tensor, $T_{\mu \nu}$, which describes the matter
content of the spacetime. Knowledge of this tensor determines the conformal factor.
In general the conservation equations would not be satisfied for a connection derived
from a metric $\hat{g} = \Om^2 g$.  Observing two paths $\ga(t)$ and $\ga'(t)$ of small test
particles would allow us to determine $\Om$ up to a constant factor, which simply reflects
one's units of measurement. 

\item \emph{Field equations}: this is the most stringent assumption that we make, but
it also yields the key equation that we wish to solve, namely the Einstein field equation:
\beq
\label{einstein-eqn}
 G_{\mu \nu} \eq R_{\mu \nu} - \ha g_{\mu \nu} R = 8 \pi T_{\mu \nu}\,.
\eeq
Here $G_{\mu \nu}$, $R_{\mu \nu}$ and $R$ are, respectively, the Einstein tensor, 
the Ricci tensor and the Ricci scalar.  (We note that, in contrast to~\cite{hawking_ellis:LSSST}, 
we further assume that the cosmological constant vanishes.)
We observe that 
predictions made on the basis of this 
equation agree, within experimental/observational uncertainties, with all experiments and 
observations that have been performed to date.  
(A survey of experimental tests of general relativity can be found in \cite{will:TEGP}, and 
an up-to-date review of the state of the art in experimental verification 
of the theory is given in
\cite{will:2001,will:2003}).
The Einstein equation is the mathematical expression of the statement that ``matter tells 
spacetime how to curve''.  In turn, the curvature of spacetime ``tells matter how to move''
through the appearance of the metric, and gradients of the metric in the equations of 
motion for the matter fields.
\footnote{We note that Einstein gravity can be formulated in a coordinate-free (i.e.\
geometrical fashion); this in fact is a necessary consequence of the general covariance of
the theory.  Nonetheless to {\em  construct} spacetimes (i.e.\ to solve Einstein's
equations, particularly via numerical means, we must adopt specific coordinate systems (also
known as charts or an atlas). Although this can be a subtle issue, we will assume in the
following that the coordinate systems we choose cover the region of spacetime of interest
(i.e.\ we adopt global coordinate systems).}

\end{enumerate}

\section{The ADM (3+1) Formalism}\lab{ADMform}
One of the great conceptual revolutions of relativity is the unification of the concepts of
space and time into a single entity. This unification is partly motivated by
the fact that, under general
coordinate transformations, time and space ``mix" together, and it is thus no longer
meaningful to talk about absolute space or absolute time.  In particular, the field 
equation~(\ref{einstein-eqn}), which is invariant under general coordinate transformations, 
relates the entire spacetime geometry to the distribution of matter-energy in space and time.
The value of physics, on the
other hand, is often rooted in its ability to predict what happens in the future, given
certain information that is known at some initial time (where time must generally be defined 
with respect to a given family of observers).
In other words, one traditional way of solving a physics problem is through the study 
of the {\em dynamics} of the system under consideration; i.e.\ given 
the state of a physical system at a specific instant of time, we use dynamical 
equations of motion to evolve the state to yield 
information about the system at future (or past) times.  This approach of viewing
the solution of~(\ref{einstein-eqn}) as the time development of some initial data is known as the 
Cauchy problem for relativity. 
Note that realization of this strategy will in some sense undo Einstein's 
feat of unification, i.e.\ it will split up spacetime into 3 ``space" dimensions plus
1 ``time" dimension---hence the terminology ``3+1'' split.  The split will involve the 
introduction of some specific coordinate system, including a definition of time,
and a prescription of initial data at the specified initial time.  The initial
data are to be chosen so that, as much as possible, the resulting evolution will
accurately model the physical system of interest. 
We will then march the data forward in time using the dynamical form of 
the field equation. 
Our task then, is to rewrite 
equation~(\ref{einstein-eqn}) as a set of partial differential equations (PDEs) suitable 
for solution as a dynamical problem.  As it turns out, the PDEs themselves naturally 
decompose into two classes: one which involves only quantities defined at a
given instant of time (the constraints), and another
that describe how the dynamical variables evolve in time (evolution equations). 

The split of spacetime into space and time as sketched above, was first introduced
by Arnowitt, Deser and Misner \cite{ADM:1962,york:1978} (hence the name ADM formalism).  
The original motivation for the development of the formalism was, in fact, the desire
to cast the field equations in a form suitable for quantization. 
Although this so-called canonical quantization of the gravitational field remains a largely unsolved 
problem, the ADM approach has proven to be a very useful basis for {\em classical} computations 
in numerical relativity. 

The explicit 3+1 decomposition proceeds
as follows.  We suppose that spacetime is orientable, which,  roughly speaking means
that we can time order the events in spacetime.  We then slice up the
spacetime into different slices of constant time, and
label each slice with a value of the time parameter, $t$.  We demand that the slicing is such that
any two events on any given
slice are spacelike-separated; i.e.\ that the slices are 
{\em spacelike hypersurfaces}~\footnote{Roughly speaking, a hypersurface is a
$d-1$ surface in a $d$-dimensional manifold.  More precisely, it is the
image of an imbedding of a $d-1$ dimensional manifold.}.  
The collection of these slices
$\{\Si_t\}$ becomes a particular ``foliation" of spacetime.  
Mathematically this foliation is described, at least locally, 
by a closed one form, $\Om =
dt$ (the closed nature of the form enables us to use $t$ as a unique label for the slice).
The norm (length) of this form 
quantifies how far (proper distance) it is from one slice, $\Sigma_t$, to a 
nearby slice, $\Sigma_{t+dt}$:
\beq
\|\Omega\|^2 \equiv g^{\mu\nu} \Omega_{\mu} \Omega_{\nu} \equiv -
\alpha^{-2} \,\,.
\eeq
\noi
Here, the positive function, $\al$, is called the lapse function and encapsulates one of the four 
degrees of coordinate freedom in the theory. 
The unit-normalized one form, $n_\mu$
\beq \lab{unit_norm}
 n_{\mu} = - \al \Om_{\mu}  \,,
\eeq
\noi
and the associated unit-norm vector, $n^{\mu} = g^{\mu \nu} n_{\nu}$, then allow us to
define a projection 
operator, $\perp^{\mu}{}_{\nu}$
\beq
 \perp^{\mu}{}_{\nu} \eq \de^{\mu}{}_{\nu} + n^{\mu}n_{\nu}\,.
\eeq
\noi
Note that since $n^\mu$ is timelike, we have $g_{\mu\nu}n^\mu n^\nu = g^{\mu\nu}n_\mu n_\nu = 
n^\mu n_\mu = -1$. 
Using the projection operator, we can construct tensors that ``live in" the hypersurfaces 
by projecting all of the indices of generic 4-dimensional (spacetime) tensors.  For example, for a 
rank-3 tensor with one contravariant and two covariant indices, we can define

\beq
 \perp T^{\al}{}_{\bt \ga} \eq \perp^{\al}{}_{\ka} \perp^{\mu}{}_{\bt}
\perp^{\nu}{}_{\ga} T^{\ka}{}_{\mu \nu}\,.
\eeq   
\noi
By construction, any tensor defined in this fashion is orthogonal to the
unit normal, so that, for example, we have $n_\alpha \perp T^{\al}{}_{\bt \ga} =
n^\beta \perp T^{\al}{}_{\bt \ga} = n^\gamma \perp T^{\al}{}_{\bt \ga} = 0$.
Hence such a tensor naturally 
lives in the 3-dimensional hypersurfaces, and is known as a {\em spatial} tensor.  
In particular, we construct the induced (or spatial) metric, 
$\ga_{\mu \nu}$, via

\beq
 \ga_{\mu \nu} \eq  \perp g_{\mu \nu} = g_{\mu \nu} + n_{\mu} n_{\nu}  \, ,
\eeq

\noi
and note that $\gamma_{\mu\nu}$ describes the {\em intrinsic} geometry of the hypersurfaces. 
Similarly, we can use projection to define a 3-covariant derivative, $D_\alpha$, that is 
compatible with the 3-metric, $\gamma_{\mu\nu}$ (i.e.\ so that 
$D_\alpha \gamma_{\mu \nu} = 0$).  For example, for a spatial vector, $V^\mu$
($V^\mu n_\mu = 0$), we have
\beq
	D_\alpha V^\mu \equiv \perp \nabla_\alpha V^\mu = \perp^\beta{}_\alpha \perp^\mu{}_\nu
\nabla_\beta V^\nu\,,
\eeq
where $\nabla_\alpha$ is the spacetime covariant derivative. 
Using $D_\alpha$ and standard formulae of tensor calculus, we can then compute the various 3-dimensional 
curvature tensors (Riemann tensor, Ricci tensor, Ricci scalar etc.) that characterize the 
geometry of the hypersurfaces.

In the 3+1 approach, the components of the 3-metric, $\gamma_{\mu\nu}$, are to be viewed as 
dynamical variables of the theory; the geometry of spacetime then becomes the time history of the 
geometry of an initial hypersurface.  In order to characterize how the geometries of two nearby 
slices differ, we must consider the manner in which the slices are embedded in the enveloping 
spacetime.  This information is encoded in the (symmetric) {\em extrinsic curvature} tensor,
which, up to a sign, is simply the projected gradient of the unit normal to the hypersurfaces:
\beq \lab{def_Kij}
K_{\mu \nu} = K_{\nu\mu} \eq -\perp\na_{\mu} n_{\nu} \,.
\eeq
\noi
From the above definition, it is clear that $K_{\mu\nu}$ is a spatial tensor, and 
thus can also be viewed as living in the hypersurfaces.
Roughly speaking, the components of the extrinsic curvature can be 
thought of as the velocities (time 
derivatives) of the 3-metric 
components (see equation~(\ref{metric_evolution_eq})), so 
that $\gamma_{\mu\nu}$ and $K_{\mu\nu}$ are, loosely, dynamical conjugates. 
With the above definitions of the 
spatial metric and the extrinsic curvature (also known as the first and second
fundamental forms, respectively), one can derive the 
\emph{Gauss-Codazzi equations}~\footnote{See, for instance, \cite{hawking_ellis:LSSST}.}
\bea
 \perp R_{\al \bt \ga \de } &=& {}^{(3)}R_{\al \bt \ga \de } + K_{\de \bt }K_{\ga \al } -
K_{\ga \bt }
K_{\de \al }\,, \\
\perp R_{\al \bt \ga \hat{n}} &=& D_{\bt } K_{\al \ga } - D_\al  K_{\bt \ga }\,,
\eea
\noi
where ${}^{(3)}R_{\al\bt \ga \de }$ denotes the Riemann tensor constructed from the
spatial metric, $\ga_{\mu \nu}$, and where, following York~\cite{york:1978},
a $\hat{n}$ in a covariant index position
denotes that the index has been 
contracted with $n^{\mu}$---for instance, $W_{\hat{n}} \eq W_\mu n^{\mu}$.
(However, and again following York, a $\hat{n}$ in a contravariant index position
denotes contraction 
with $-n_{\mu}$.)

The Einstein equation, together with the following definitions:
\bea \lab{energydensity}
 \rh &\eq& T_{\hat{n}\hat{n}} = T_{\mu \nu} n^\mu n^\nu\,,\\
\lab{momentumdensity}
 j^{\mu} &\eq& \perp T^{\mu \hat{n}} = - \perp \left( T^{\mu \nu} n_{\nu}
\right) \,,
\eea
\noi
where $T^{\mu \nu}$ is again the stress energy tensor, allow us to rewrite the
Gauss-Codazzi equations as
\beq \lab{hamiltonian_constraint}
  {}^{(3)}R+ K^2 -K_{ij}\,K^{ij} =16 \pi \rh \,\,,
\eeq
\beq \lab{momentum_constraint}
 D_j K^{ij}-D^i K = 8 \pi j^i \,\,.
\eeq
\noi
Here, $^{(3)}R$ is the 3-dimensional Ricci scalar, $K \eq \ga^{ij} K_{ij}$ is the 
trace of $K_{ij}$ and we remind the 
reader that 
Latin indices range over the spatial values $1,2,3$. 
We also note that spatial indices on spatial tensors are raised and lowered using $\gamma^{ij}$ and 
$\gamma_{ij}$, respectively, and that $\gamma_{ij}\gamma^{jk} = \delta^i{}_k$.

(\ref{hamiltonian_constraint}) and (\ref{momentum_constraint}) involve only
spatial tensors and spatial derivatives of such tensors, and hence, from the point of view of dynamics,
represent {\em constraints} that must be satisfied on each slice, including the initial slice. 
They are commonly called the \emph{Hamiltonian constraint} and \emph{momentum constraint}
respectively, and are a direct consequence of the four-fold coordinate freedom of the 
theory.

\begin{figure}[h]
\begin{center}
\includegraphics[width=12cm,clip=true]{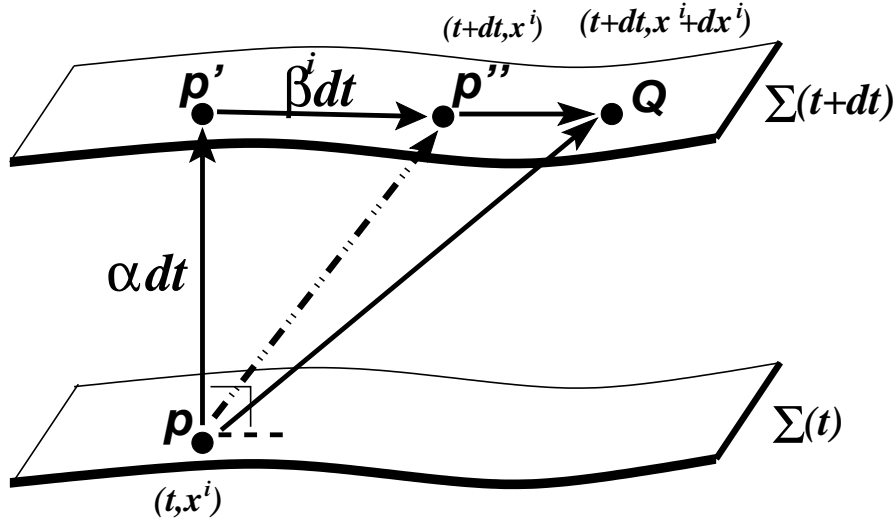}
\end{center}
\caption
[A schematic representation of the ADM (or 3+1) decomposition]{This figure illustrates a schematic
representation of the ADM, or 3+1, decomposition. $\al$ is the lapse function and
encodes the proper distance, $\alpha dt$, between two infinitesimally separated 
hypersurfaces, $\Si(t)$ and $\Si(t+dt)$.
$\bt^i$ is the shift vector and describes the shifting of spatial
coordinates, $x^i$, relative to normal propagation. 
$PP''$ is the vector $t^{\mu}=\al n^{\mu} + \bt^{\mu}$, where $n^\mu$ is the unit timelike normal 
to the hypersurfaces, and is
proportional to the 4-velocity of an observer with fixed spatial coordinates.
}
\label{3p1}
\end{figure}

So far we have used only one coordinate degree of freedom, which was in our 
specification of the manner in which we foliated the spacetime into 
a family of spacelike hypersurfaces---or, equivalently, in the specification
of the lapse function, $\alpha$, at all events of the spacetime.
For any fixed time slicing, however, we are allowed to relabel the spatial 
coordinates of events intrinsic to any of the slices as illustrated in Fig.~\ref{3p1}.
As shown in that figure, it is convenient to describe the {\em general} labeling of events 
with spatial coordinates relative to the {\em specific} labelling one would get by 
propagating the spatial coordinates along the unit normal field, $n^\mu$.  In general,
the spatial coordinates, will be ``shifted'' relative to normal propagation, and this 
shift will be quantified by a spatial vector, $\beta^\mu$, known, naturally enough, 
as the {\em shift vector}.  An observer who is at rest with respect to the coordinate
system illustrated in Fig.~\ref{3p1} will thus move along a worldline with a tangent vector
given by
\beq \lab{def_t}
t^{\mu} = \al n^{\mu} + \bt^{\mu}\,,  
\eeq
\noi
where we again emphasize that $\beta^\mu n_\mu = 0$.  

The specification of the three independent 
components of the shift vector, along with the lapse function exhausts the coordinate freedom of 
general relativity, and defines a coordinate system $(t,x^i)$ (sometimes called a 3+1 coordinate 
system) that we will assume covers the 
region of spacetime in which we are interested.   In such a coordinate system, and using a four-dimensional
generalization of the Pythagorean theorem, we can then write the spacetime displacement in 3+1 form
\beq \lab{metric}
  ds^2 = ( -\al^2+\bt_i \bt^i)\, dt^2+2 \bt_i\, dt\, dx^i+\gamma_{ij}
\, dx^i\, dx^j \,\,.
\eeq
\noi

\vspace{0.5cm}

\noi
From the definitions of the vector field, $t^{\mu}$~(\ref{def_t}), and the
extrinsic curvature~(\ref{def_Kij}), one can show that
\beq \lab{metric_evolution_eq}
  \pounds_t \ga_{ij} =-2 \al K_{ij}+ D_i \bt_j + D_j \bt_i \,\,,
\eeq
where $\pounds_t$ denotes Lie differentiation along $t^\mu$  (in a 3+1 coordinate system
$\pounds_t$ reduces to partial differentiation $\partial_t$).
This last expression can be viewed as a set 
of evolution equations for the 3-metric components, and makes the interpretation of the 
extrinsic curvature components as ``velocities'' of the $\gamma_{ij}$ manifest.

Making the further definitions
\bea \lab{Sij_def}
 S_{\mu \nu} &\eq& \bot T_{\mu \nu}\,, \\ \lab{S_def}
 S &\eq & \ga^{ij} S_{ij}\,,
\eea

\noi
the remaining components of the Einstein equation can be written as: 
\bea \lab{extrinsic_curvature_evolution_eq}
\pounds_t\,K_{ij}&=&-D_i D_j\, \al+\al
\left\{  {}^{(3)}R_{ij}-2K_{ik}K^k{}_j + K_{ij} K - 
8 \pi \left[ S_{ij}
-\ha \ga_{ij}\, \left( S-\rh\right)\right]
\right\} \nn\\
&& + \bt^k{D}_k K_{ij}+K_{ik}{D}_j\bt^k+
K_{kj}{D}_i\beta^k \,,
\eea
\noi
where ${}^{(3)}R_{ij}$ is the 3-dimensional Ricci tensor.
These last equations are to be viewed as evolution equations for 
the extrinsic curvature components.

In summary, the 3+1 decomposition applied to the Einstein field 
equation~(\ref{einstein-eqn}) yields a set of 4 constraint equations, (\ref{hamiltonian_constraint}),
(\ref{momentum_constraint}), and 12 (first order in time) evolution equations,
(\ref{metric_evolution_eq}) and 
(\ref{extrinsic_curvature_evolution_eq}).   We note that, as a consequence of the contracted 
Bianchi identity (which itself follows from the coordinate invariance of the theory), the evolution
equations guarantee that data that satisfies the constraints at some instant in time (at the initial
time, for example), is evolved to data that satisfies the constraints at any future or past times.

We conclude this section by observing that a key idea in the 3+1 approach is to use a timelike vector 
field, $n^{\mu}$ (whose existence is guaranteed by the Lorentzian signature of the spacetime metric), 
to construct a projection operator that is then used to define quantities on surfaces orthogonal 
to $n^{\mu}$ (surfaces of simultaneity), and to derive equations governing those quantities from
the covariant field equations.  This same idea is used
in the (2+1)+1 formalism described in the next section, the only difference being that we will 
now decompose with respect to a spacelike vector field, $\xi^{\mu}$, which is a
Killing vector field.

\section{The (2+1)+1 Formalism}
\begin{figure}[h]
\begin{center}
\includegraphics[width=11cm,clip=true]{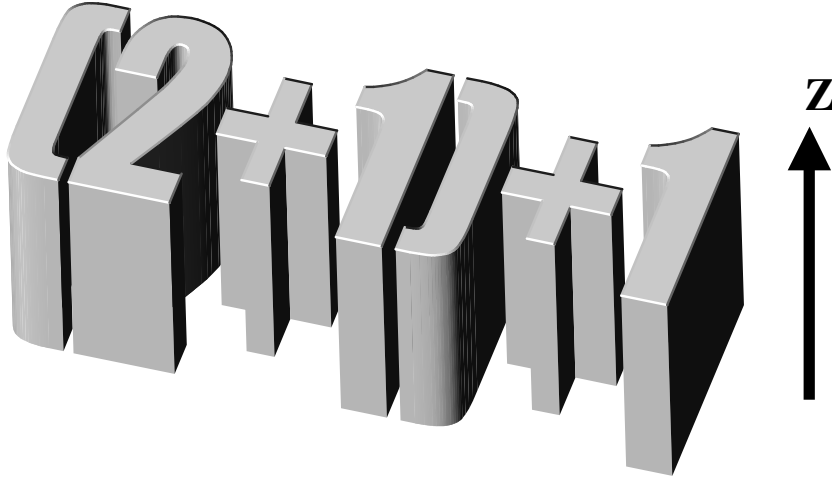}
\end{center}
\caption
[A 3-dimensional object with continuous symmetry]{A 3-dimensional object with continuous symmetry, which can be
described as having a cross section that is the planar text string ``(2+1)+1", and which has a symmetry
along the $z$ direction.}
\label{fig8}
\end{figure}

Suppose we stack a collection of identical planar figures---such as the 
text string ``(2+1)+1"---to create a 3-dimensional structure as shown in 
Fig.~\ref{fig8}.   For the purposes of discussion, we will 
think of the 3-dimensional object as being of infinite extent in the $z$-direction.  To describe this
structure it suffices to state that its $z={\rm const.}$ cross section is the planar text ``(2+1)+1",
and that it has a symmetry in the $z$ direction.  Similarly if there is a
symmetry in spacetime, we can state what the symmetry is, then focus on the
detailed structure of a ``cross section'' orthogonal to the symmetry direction.  This
technique of ``dividing out'' the symmetry and then restricting attention to the 
cross section, or projected space, (or, in mathematical parlance, the quotient 
space~\footnote{The equivalent relation can be defined as $p\sim q$ iff $p$ and $q$ lie on the same integral curve
of the Killing vector field $\xi$.}),
underlies the (2+1)+1 formalism.  In particular, if the spacetime
in which we are interested is axisymmetric, we can first project quantities and equations onto a 
``plane" perpendicular
to this symmetry, and then perform the equivalent of an ADM space-plus-time
decomposition on the dimensionally reduced spacetime.

\vspace{7mm}

The (2+1)+1 formalism described here is based on the work of
Maeda and his collaborators \cite{maeda}, and, as just stated, combines (1) the method
of dimensional reduction (dividing out the symmetry, or projection along
the symmetry) 
of a spacetime possessing a Killing vector field, as originally developed by 
Geroch \cite{geroch:1971}, and 
(2) a 3-dimensional analogue of the ADM approach introduced in the
previous section (Sec.~\ref{ADMform}).  Other references for the application 
of this formalism to problems in  numerical relativity include
\cite{liebling:phd, fransp:phd}. 

As previously mentioned, we are interested in the case of axisymmetry, and thus 
assume that the spacetime under consideration contains a spacelike Killing vector field, $\xi^\mu$,
with closed orbits.  Choosing an azimuthal coordinate, $\vp$, adapted to the symmetry,
the Killing vector can be written as: 

\beq
 \xi^\mu \eq \left(\fr{\pa}{\pa \vp}\right)^\mu \,.
\eeq

\noi

To perform the projection, we first construct a projection operator, $\H\al\bt$, similar
to that previously used in the 3+1 decomposition~\footnote{This definition of projection
operator, which is constructed from a vector ($\xi$) instead of a
1-form ($\Om$) makes a significant difference for the relation between 4-dimensional 
objects and 3-dimensional ones.  It is straightforward to get the 
contravariant objects in the former situation, while it is straightforward
to get the covariant objects in the latter case.  The main reason is that
$\xi$ has only 1 non-zero component in the contravariant form, but $\Om$
has only 1 non-zero component in the covariant form. For instance,
$^{(3)}V^a \simeq \perp^a_{\mu}{}^{(4)}V^{\mu} = (\de^a_{\mu} - \xi_{\mu}
\xi^a/s^2) {}^{(4)}V^{\mu} = {}^{(4)}V^a$, since $\xi^a = 0$.}:

\beq
 \mathcal{H}^{\al}{}_{\bt} \eq  \de^{\al}{}_{\bt} -
\fr{\xi^\al\xi_\bt}{\xi^{\ga}\xi_{\ga}}\,.
\eeq

\noi
Using this operator, we can construct tensors intrinsic to the 
cross sections via projections of spacetime tensors.  In particular, we can
project the spacetime metric itself (or equivalently, lower the contravariant 
index of $\H\al\bt$) to yield the metric, $\calH_{\al\bt}$, on the 
dimensionally reduced space:

\beq
 \mathcal{H}_{\al \bt} \eq  g_{\al \bt} -
\fr{\xi_\al\xi_\bt}{\xi^{\ga}\xi_{\ga}} \,.
\eeq

Again paralleling the 3+1 development, we can also use projection of the spacetime covariant
derivative, $\nabla_\al$, to define an induced covariant derivative operator,
$D_\al$, in the cross sections that satisfies $D_\al\calH_{\mu\nu}=0$. 

Since it is a (Lorentzian) metric on a 3-dimensional manifold, $\calH_{\al\bt}$ 
clearly has fewer degrees of freedom than the spacetime metric, $g_{\al\bt}$.  The 
``missing'' degrees of freedom can be conveniently encoded in the norm, $s$, and 
the twist, $\omega^\alpha$, of the Killing vector, $\xi^\mu$:

\bea \lab{s2_def}
 s^2 &\eq& \xi^\al \xi_\al \,, \\ \lab{om_def}
 \om^\al &\eq& \ep^\al{}_{\bt \ga \de}\, \xi^\bt \xi^{\de;\ga}\,.
\eea

\noi
where $\ep^\al{}_{\bt \ga \de}$ is the usual Levi-Civita symbol.
Note that since $\om_\al \xi^\al=\mathcal{L}_{\xi} \om_\al = \mathcal{L}_\xi s^2 = 0$,
$s^2$ and $\om_\al$ are tensors intrinsic to the cross sections.
For notational convenience, we further define
\bea
 \om^2 &\eq& \om^\al \om_\al \, ,\\
  \la &\eq & s^2 \, .
\eea
\noi
The basic equations for a spacetime with a Killing vector field $\xi^\al$
can now be written as~\footnote{See appendix of  \cite{geroch:1971} for a detailed
derivation.}:
\bea \lab{2p1p1bea}
D_{[\al}\om_{\bt]} &=& - \ep_{\al\bt \ga \de} \,\xi^\ga R^\de{}_\ep \,\xi^\ep \,, \\
D^\al \om_\al &=& \fr{3}{2 \la} \om_{\bt} D^\bt\la \,, \\
D^2 \la &=&  \fr{1}{2\la} D^\al{\la}D_\al{\la} - \fr{\om^2}{\la} - 2R_{\al
\bt}\xi^\al \xi^\bt \,,\\ \lab{2p1p1bed}
\calR_{\al \bt} &=& \fr{1}{2\la^2} \left( \om_\al \om_\bt - \hx_{\al \bt} \om^2\right) + \fr{1}{s} D_{\al}
D_{\bt} s + \H\mu\al \H\nu\bt R_{\mu\nu}\,,
\eea

\noi
where $\calR_{\al \bt}$ denotes the Ricci tensor associated with the induced metric
$\hx_{\al\bt}$.
Note that (\ref{2p1p1bea})--(\ref{2p1p1bed}) completely describe the
dimensional reduction piece of the (2+1)+1 formalism, and are written in 
purely geometrical terms.  To incorporate matter fields, we
can manipulate (\ref{2p1p1bea})--(\ref{2p1p1bed}) and use the Einstein equation to get:
\bea
D_{[a}\omega_{b]} &=& 8\pi s\,\epsilon_{abc}\tau^c, \label{curl_w}\\
D^a\left[\frac{\omega_a}{s^3}\right] &=& 0\,, \label{div_w}\\
\frac{1}{s}D^2s &=& -\frac{\omega^2}{2 s^4} -
4\pi\left(\frac{T_{\phi\phi}}{s^2}
-\tau\right)\label{div_s}, \\
\calG_{ab} \equiv \calR_{ab} - \frac12 \calH_{ab} \calR &=& 8\pi{\cal T}_{ab} \label{G_3D}\,,
\eea
where $\calR = \calH^{ab} \calR_{ab}$.
\noi
Here we note that indices $a, b, \cdots$ range over the temporal and (two) spatial dimensions of the 
reduced space, and that we have defined the following quantities:
\bea
 \tau_{ab} &\eq& \hx^{\gamma}{}_a \hx^{\delta}{}_b\,T_{\gamma\delta}\,, \\
\tau^c&\eq& \hx_{\alpha}{}^{c}\,T^\alpha{}_{\phi} \,,\\
\tau &\eq& T-\frac{T_{\phi\phi}}{s^2}\,, \\
\calT_{ab} &\eq&  \tau_{ab}
-\frac{1}{4}\hx_{ab}\left(\tau-\frac{T_{\phi\phi}}{s^2}\right)
+\frac{1}{8\pi}\left\{\frac{\omega_a\omega_b}{2 s^4}
+\frac{1}{s}\left[D_a D_b s -\frac{1}{2}\hx_{ab}
D^2s\right]\right\}.
\eea
\noi
Note that the 4-dimensional Einstein equations have been effectively replaced by 
(\ref{curl_w})--(\ref{G_3D}), where the first two equations determine
the twist $\om^a$~\footnote{Helmholtz's decomposition theorem tells us that
a vector field can be determined by its curl and divergence.},
the third equation determines the norm, $s$, of the Killing vector field,
and the fourth governs the reduced metric, $\calH_{\al\bt}$. 

The last stage in the derivation of the (2+1)+1 equations involves 
a space-plus-time split (completely analogous to the 3+1 split) of 
the reduced field equations.  
Again, the interested reader is referred 
to~\cite{maeda} for full details.

\section{Boson Stars---the Einstein-Klein-Gordon System}
In this section we briefly describe the theoretical framework for boson 
stars, which are equilibrium configurations of a massive, complex 
Klein-Gordon field coupled
to gravity---we will thus refer to the model as the Einstein-Klein-Gordon system.  

We derive the equations for the Einstein-Klein-Gordon system using a variational approach.
The model is described by an action
\beq
  S =  \int d^4 x \sr{ -g }\, \left(\mathcal{L}_g + \mathcal{L}_\ph
\right)\,, 
\eeq

\noi
where
\beq
\mathcal{L}_g \eq  \fr{ 1}{16 \pi } R \,,
\eeq
\noi
and
\beq
  \mathcal{L}_\ph \eq - \fr{ 1}{2} \left( \na^\mu \ph \na_\mu \ph^\ast + m^2 |\ph|^2 \right)\,.
\eeq

\noi
Here $R$ is the spacetime Ricci scalar and $\ph$ is a massive 
complex scalar field with bare (particle) mass $m$.  
Variation of the action with respect to the metric $g_{\mu \nu}$ yields the Einstein
equation 
\beq \lab{einstein_eq}
 R_{\mu \nu} - \ha g_{\mu \nu} R = 8 \pi T_{\mu \nu}\,,
\eeq
\noi
where 
\beq \lab{Ctmunu}
T_{\mu \nu} \equiv T^\phi_{\mu \nu} \eq 
\ha \left[  \left( \na_{\mu} \ph \na_{\nu} \ph^\ast + \na_{\nu} \ph
\na_{\mu} \ph^\ast \right)
- g_{\mu \nu} \left( \ \na^\al \ph \na_\al \ph^\ast + m^2 |\ph|^2
\right)\right] \,.
\eeq

\noi
Similarly, variation with respect to the field $\ph$ gives the Klein-Gordon equation
\beq \lab{klein_gordon_eq}
\na^{\mu}\na_\mu \ph - m^2 \ph   = 0\,.
\eeq
\noi
The coupled equations (\ref{einstein_eq}) and (\ref{klein_gordon_eq}) constitute the 
equations of motion for the Einstein-Klein-Gordon system, and, in particular, are the fundamental equations 
used in our subsequent study of boson stars.
When we (gravitationally) perturb boson stars using an additional real, massless scalar field, $\phi_3$, we simply add 
an extra term
$\mathcal{L}_{\phi_3}$ to the total Lagrangian density
\beq
 \mathcal{L} = \mathcal{L}_g + \mathcal{L}_{\ph} + \mathcal{L}_{\phi_3} \,,
\eeq
\noi
with
\beq
 \mathcal{L}_{\phi_3} \eq -\ha \na^{\mu} \na_{\mu} \phi_3 \, .
\eeq
\noi
This yields an additional contribution, $T_{\mu \nu}^{\phi_3}$, to the total stress energy
tensor
\beq
T_{\mu \nu} = T_{\mu \nu}^\phi + T_{\mu \nu}^{\phi_3}\,,
\eeq
\noi
where
\beq
T_{\mu \nu}^{\phi_3} = 
  \na_{\mu} \phi_3 \na_{\nu} \phi_3
- \ha g_{\mu \nu}  \ \na^\al \phi_3 \na_\al \phi_3 \,,
\eeq
\noi
as well as a field equation  (the massless Klein-Gordon equation), for $\phi_3$
\beq
\na^{\mu}\na_\mu \phi_3 = 0\,.
\eeq

\noi
In the following chapters we will consider more explicit forms of the above equations of motion, once 
we have imposed symmetries and fixed specific coordinate systems.

  \resetcounters

\chapter{Numerical Methods} \lab{num_method}

In this chapter we briefly summarize the numerical methods used in this thesis, as well 
as some continuum strategies central to our numerical calculations.  As with 
the previous chapter, since most of the topics discussed here are 
well documented in the literature, we refer the interested reader to 
appropriate references for additional details.
We also note that we use the common subscript notation for partial differentiation,
e.g. $u_x \equiv \partial u / \partial x$, in this chapter. 

\section{Finite Difference Techniques}

The basic technique that we use in our computational solution of model
physical problems is finite
differencing.  Finite differencing is a very commonly used method in
the numerical solution of partial differential equations (PDEs), and continues to 
dominate work in numerical relativity.  
The basic idea of the approach is to replace
partial derivatives by suitable algebraic difference quotients, and the fundamental 
assumption underlying the efficacy of the technique is that the
functions to be approximated are smooth so that they can be Taylor-expanded about any point 
in the computational domain. 
The first step in the development of a finite difference approximation (FDA) to a PDE (or more 
generally, to a system of PDEs)
is the introduction of a discrete grid (or lattice, or mesh) of points, which replaces
the continuum physical domain.
We then approximate continuum field quantities
${\bf u} = \left( u, v, w,\cdots \right)$ by a set of grid functions
${{\bf u^h}} = (u^h,v^h,w^h,\cdots )$ which will constitute a solution of our FDA.
For simplicity of presentation, we will now assume that our vector of fields, ${\bf u}$, has 
a single component, $u(t,x)$, which is a function of time and one spatial dimension, but the
approach and techniques discussed in the following are easily generalized to the cases of 
multiple fields and/or multiple spatial dimensions.

We will generally restrict attention to  so-called uniform grids wherein the spacing 
between adjacent grid points is constant in each of the coordinate directions. 
Denoting these regular mesh spacings in space and time as $h$  and $k$, respectively, and 
assuming that we are solving our PDE on the continuum domain, $x_{\rm min} \le x \le x_{\rm max}$,
$t \ge 0$, our finite difference grid is given by $(x_j , t^n)$, where 
\beq
  x_j = x_{\rm min} + \left( j-1\right) h\,,\,\,j = 1,\cdots,N_x\, ,
\eeq
with
\beq
h = \frac{x_{\rm max} - x_{\rm min}}{N_x -1},
\eeq
\noi
and 
\beq
t^n = n k\,,\,\, n = 0, 1, \cdots\,.
\eeq
\noi
For any grid function $u^h \in {\bf u^h}$, the value at $(x_j ,
t^n)$ is denoted by $u^n_j$ and is an approximation
of the continuum value $u(x_j , t^n)$.
Convergence of the finite difference approximation is then the statement that 
$\lim_{h\to0,k\to0} u^n_j = u(x_j , t^n)$.  We remark at this point that, operationally, 
investigation of the behaviour of a finite difference solution, $u^h$, as a function 
of the mesh spacings $h,k$ (holding other problem parameters fixed) provides a very 
powerful and general methodology for assessing the accuracy of the solution.   We have
used this strategy of {\em convergence testing} extensively in 
assessing the correctness and accuracy of the results 
described in later chapters. 
\footnote{Our determination of rotating boson star data in Chap.~\ref{bs2d} is an exception.
We typically compute there on a $N_r\times N_\te = 100\times 15$ grid.  Calculations with,
e.g.\ $N_r\times N_{\te} = 200\times 30$ do not converge.  We feel this non-convergence is
due to the regularity problems that we report in that chapter.}
In addition, for all of our time-dependent calculations, when we vary 
the spatial mesh spacing(s) during a convergence test, we also vary the time step, $k$, so
that (for the one dimensional case), the ratio $\lambda = k / h$ is held constant ($\lambda$
is often called the {\em Courant factor}, or {\em Courant number}).  Thus, our FDAs tend 
to be characterized by a {\em single} overall discretization scale. 

\begin{figure}[h]
\begin{center}
\epsfxsize=5.0cm
\includegraphics[width=5.0cm,clip=true]{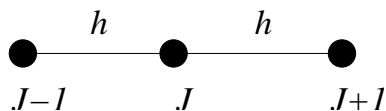}
\caption
[A 3-point finite difference stencil suitable for computation of a second-order 
approximation to a first derivative]
{A 3-point finite difference stencil suitable for computation of a second-order 
approximation to a first derivative}
\label{3pt}
\end{center}
\end{figure}

We now sketch one general technique that can be used to derive finite difference formulae.
As mentioned above, we want to use difference quotients (algebraic combinations of grid-function values) 
to approximate derivatives.  For instance, suppose we wish to approximate
the first derivative $u_x(x)$ of our function $u(x)$.  Suppressing the time index, $n$, we 
Taylor series expand about $x=x_j$
\beq
 u_{j+ih} = [u]_j + i h [u_x]_j + \ha (ih)^2 [u_{xx}]_j + O(h^3)\, ,
\eeq
\noi
for $i = -I, \cdots, -2, -1, 0, 1, 2, \cdots, \tilde{I}$.
Now suppose we want to use a three-point formula (or a three-point {\em stencil}) to 
calculate our approximation of the first derivative (see Fig.~\ref{3pt}).

\vspace{0.5cm}

\noi
We consider a linear combination, $\sum_{i=-1}^{1} c_i u_{j+ih}$, of the grid function values at the 
grid points $x_{i-1}, x_i, x_{i+1}$,
and equate it to the first derivative, $u_x$, evaluated at $x=x_j$:
\beq
 \sum_{i=-1}^{1} c_i u_{j+ih} \sim [u_x]_j \,. 
\eeq
Solving for the coefficients $c_i$, we find 
\bea  \nn
 c_{-1} &=& -\fr{1}{2h} \,,\\
 c_{0} &=& 0 \,,\\ \nn
 c_{1} &=& +\fr{1}{2h} \,,
\eea
\noi
and hence our FDA is 
\beq
 u_x(x_j) = \fr{u_{j+1} - u_{j-1}}{2h} + O(h^2)\,.
\eeq
\noi
Since the truncation error of this formula is $O(h^2)$, we call the approximation {\em second order}.
Similarly, we can derive expressions for other derivatives (with varying degrees of accuracy)
using appropriate stencils.  For example, we have \cite{CFD}
\bea
 u_x(x_j) &=& \fr{u_{j+1} - u_{j}}{h} + O(h)\,,\\
 u_x(x_j) &=& \fr{u_{j} - u_{j-1}}{h} + O(h)\,,\\
 u_{xx}(x_j) &=& \fr{u_{j+1} - 2 u_j + u_{j-1}}{h^2} + O(h^2)\,,\\
 u_{xx}(x_j) &=& \fr{-u_{j+2}+16 u_{j+1} - 30 u_j + 16 u_{j-1} - u_{j-2}}{ 12 h^2} + O(h^4)\, .
\eea

\noi
In discretizing time-dependent PDEs we make exclusive use of Crank-Nicholson schemes,
with second order spatial differences. 
For the case of one space dimension (1+1 problems), such a scheme
uses the stencil shown in~Fig.~\ref{cn}:

\vspace{0.5cm}
\begin{figure}[htp]
\begin{center}
\epsfxsize=7.0cm
\includegraphics[width=7.0cm,clip=true]{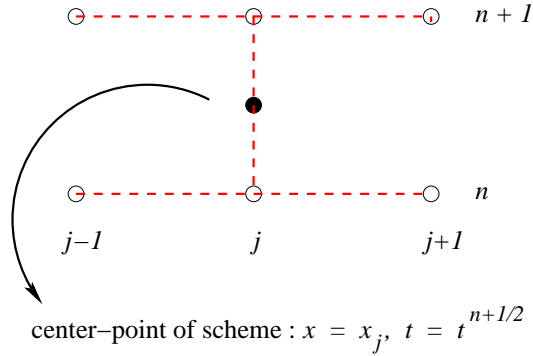}
\caption
[Stencil for $O(h^2)$ Crank-Nicholson Scheme]
{Stencil for an $O(h^2)$ Crank-Nicholson scheme for a PDE in one space dimension and time.}
\label{cn}
\end{center}
\end{figure}

\noi
The key idea of a Crank-Nicholson method is to keep the differencing {\em centered} in time 
as well as in space.  As an example, we consider the simple advection equation
\beq
 u_t(t,x) = a u_x \, ,
\eeq
\noi
where $a$ is a constant.  The Crank-Nicholson scheme for the equation is then
\beq
 \fr{u\npoj - u\nj}{\De t} = \fr{a}{2}\, \left[  \fr{u\njpo - u\njmo}{2 \De x} + 
\fr{u\npojpo - u\npojmo}{2 \De x}  \right]\,,
\eeq
and is $O(\Delta t^2,\Delta x^2)$ accurate as can be verified via Taylor series
expansion about the center-point of the stencil,
$(x_j,t^{n+1/2})$ (see Fig.~\ref{cn}).
Note that the scheme involves the FDA of the spatial derivative applied at both the 
retarded ($n$) and advanced ($n+1$) time levels.  This leads to coupling of the advanced-time
unknowns, $u^{n+1}_j$ (i.e.\ the scheme is {\em implicit}), and is crucial for 
both the accuracy and stability of the scheme.

\noi 
The Crank-Nicholson approach can be applied in a straightforward manner to any set 
of time-dependent PDEs that have been cast in first-order-in-time form, and for the 
case of the evolution equations treated in this thesis, the discrete equations 
that result can be solved iteratively in an efficient manner. 

\noi
For time-dependent FDAs, the notions of convergence and {\em stability} are intimately 
related.  We have generally {\em not} encountered serious stability problems with the 
difference schemes used below, but nonetheless refer the interested reader to~\cite{KO} for 
discussion of this important subject.

\section{Multigrid Method}

In this section we introduce the basic concepts and techniques of the multigrid method for 
the solution of FDAs arising from the discretization of elliptic boundary value problems.
This method (including a modification for eigenvalue problems) is used extensively 
in our construction of stationary solutions describing rotating boson stars, and in the 
solution of the elliptic constraint equations and coordinate conditions that arise 
in our study of the dynamical evolution of axisymmetric boson stars.
A more detailed introductory discussion of the method can be found in~\cite{matt:master,numrec}, 
while~\cite{MGM} provides an extensive reference source.

Traditional methods for solving elliptic PDEs, such as successive over-relaxation (SOR),
are easy to implement, but suffer from slow convergence rates (particularly in the limit 
of small mesh sizes), and have contributed to the general impression that, relative
to equations of evolutionary type, the finite difference solution of elliptic
equations is computationally expensive.  However, the multigrid method, first introduced by Brandt 
in the 1970s \cite{brandt}, is capable of solving discrete versions of elliptic PDEs 
(even systems of nonlinear PDEs) in a time that is proportional to the total number, $N$, of 
grid points in the discretization---i.e.\ in $O(N)$ time. 
The basic idea underlying 
multigrid is to use a {\em hierarchy} of grids with different resolutions (mesh spacings) to solve a 
particular problem instead of using
a single grid.  By an intelligent transfer of the problem back and forth between the various
grids, one can speed up the convergence rate of the solution process on the finest grid.  
Central to the efficient operation of most multigrid solvers is the observation that
straightforward relaxation (in particular, Gauss-Seidel relaxation, \cite{numrec,ueberhuber:NC2})
is very good at removing high frequency error components, but
very bad at removing low frequency ones.  At the same time, once the solution error has been smoothed
by relaxation, we can sensibly pose a coarse grid version of the problem at hand, the solution of which 
can be computed in a fraction of the time required to solve the fine grid problem. 

Perhaps the best way of illustrating the multigrid method is by consideration of a simple
example.  Suppose we want to solve the following linear elliptic equation
\beq
 L u = f\,,
\eeq
\noi
where $L$ is some linear elliptic operator, $u$ is the continuum solution, and $f$ is a source function
(we assume $L$ is 
linear strictly to keep the presentation simple; as already mentioned, the multigrid technique
can also be applied to nonlinear elliptic equations).  As in the
previous section we replace the PDE by a finite difference approximation, defined on a uniform
grid characterized by a mesh spacing, $h$
\beq
\label{lh}
 L^h u^h = f^h\, .
\eeq
\noi
Here $L^h$ is our FDA of $L$ and $u^h$ and $f^h$ are the grid versions of the solution and source 
function, respectively.
Suppose we have an
approximate solution (or guess) $\tilde{u}^h$ to the above equation.  Then we consider the
difference, $v^h$, between the exact discrete solution, $u^h$, and the (current) approximation,
$\tilde{u}^h$, to  $u^h$:
\beq
 v^h = u^h - \tilde{u}^h\,.
\eeq
We call $v^h$ the error or {\em correction}, since it is the quantity by which we must correct $\tilde{u}^h$,
so that we get the solution of the finite difference approximation.
\noi
Solving this last expression for $u^h$, and substituting the result in~(\ref{lh}) we have
\beq
 L^h \left( \tilde{u}^h+v^h\right) = f^h\,,
\eeq
\noi
or
\beq \lab{resideq}
 L^h v^h =  -\left( L^h \tilde{u}^h -f^h\right) \eq -r^h\,,
\eeq
\noi
where we have defined the {\em residual} or {\em defect}, $r^h$.  
Focus is now shifted to the solution of 
(\ref{resideq}) for the correction, $v^h$.   Once $v^h$---or more generally, some 
approximate solution, ${\tilde v}^h$, of~(\ref{resideq})---is in hand, we can update
$\tilde{u}^h$ via
\beq
 \tilde{u}^h := \tilde{u}^h + v^h\,.
\eeq
\noi
Of course, (\ref{resideq}) is fundamentally no easier to solve than~(\ref{lh}).  
However, if we apply a few relaxation sweeps to~(\ref{resideq}), thus smoothing the 
residual, $r^h$, as well as the (unknown) correction, $v^h$, 
we can transfer the problem to a coarser grid with mesh size $H$, where
$H$ is a multiple of $h$ (typically $H = 2h$).  That is, we consider the coarse grid problem
\beq
 L^H v^H = -r^H = -I_h^H r^h\,,
\eeq
\noi
where the coarse grid source function, $r^H$, is obtained by the application
of a suitable {\em restriction} operator $I_h^H$ to the fine grid residual, $r^h$ (i.e.\ $I_h^H$ 
transfers a fine grid function to a coarse grid).

Particularly for 2D and 3D cases, the coarse grid problem will be significantly 
less expensive to solve than the fine grid problem.
Once the solution $v^H$ is obtained we can approximate the desired 
correction $v^h$ by using an {\em prolongation} operator, $I^h_H$, which transfers 
a coarse grid function to a fine grid:
\beq
 \tilde{v}^h \eq I^h_H v^H \,,
\eeq
\noi
Once the fine grid function has been updated using the prolonged correction, a few more relaxation
sweeps are applied to smooth out any high frequency error components that are introduced in 
the coarse-to-fine transfer of the correction.

We note that the choice of suitable restriction and prolongation operators will generally 
depend on the discretization that is used, as well as on the specific type of relaxation
strategy that is employed to smooth the residuals/corrections. 
More importantly, the above sketched process of a two level correction can be easily
generalized to a multi-level correction, wherein one uses an entire 
hierarchy of mesh scales (for example $h, 2h, 4h, \cdots$) and where the cost of {\em solving}
the discrete equations on the coarsest grid can be made computationally negligible.
The algorithmic process by which the solution is manipulated first on the finest grid, then 
on the auxiliary coarse grids, and finally on the finest grid again is called a multigrid {\em
cycle}.  Although it is sometimes advantageous to use other types of cycles, we use 
$V$-cycles exclusively in our work, wherein the problem solution sequence is $l_f, l_f-1,\cdots,l_c+1,l_c,l_c+1,\cdots l_f-1,l_f$,
and where the integers $l_c$ and $l_f$ ($l_f > l_c$) label the coarsest and finest levels, respectively, of refinement used.

\section{Adaptive Mesh Refinement}

In this section we introduce the method of adaptive mesh refinement (AMR).  This method
is important in the finite difference solution of problems that exhibit significant 
variation in the length and time scales that must be resolved (dynamic range), 
particularly when the specific resolution requirements (e.g.\ in which regions of the 
computational domain the smallest scale features will develop) are not known {\em a priori}.

In numerical relativity, the use of AMR has played a crucial role in the study of 
black hole critical phenomena, one class of which is characterized by self-similar 
solutions, which thus involve features on arbitrarily small scales (see~\cite{choptuik}
for example).  Other examples where AMR is likely to ultimately play an important role
include the calculations aimed at predicting the gravitational waves generated from 
the inspiral and merger of a black hole binary.  Here one basic length scale is set 
by the mass, $M$, of one of the holes (we assume roughly equal mass black holes)---clearly,
in the vicinity of the black holes our finite difference mesh must be sufficiently 
fine to resolve that length scale.  On the other hand, the gravitational waves that are generated 
in the late stages of the binary evolution will have wavelengths of order $10M$, and 
the computational grid may have to extend to a distance of order $100M$ from the sources in
order that the wave train that would register in a terrestrial detector can be accurately 
read off.  Use of a single uniform grid with $h$ chosen small enough to resolve the smallest scale 
features would not only be extremely inefficient in such a case, it would likely be 
prohibitively expensive (both in terms of CPU time and memory), even with the largest 
computers currently available.

One can attempt to deal with the multi-scale nature of such a problem through the use of 
a non-uniform grid (or several component grids, each of which is
uniform) so that the local mesh spacing at least roughly matches the local resolution 
requirement.  Here we are envisioning the use of {\em a priori} information concerning the 
nature of the solution, and would further assume that the resolution demands are 
{\em static}, so that, for example, the regions requiring highest resolution would not 
move around during the course of the solution.  Such a refinement strategy is often known
as {\em fixed mesh refinement} (FMR) and has been used successfully in some recent calculations 
in numerical relativity (see, for instance, \cite{schnetter:2004,Brugmann:2004}).
However, in many
situations---such as the modeling of objects that propagate through the computational domain---we will 
{\em not} have detailed {\em a priori} information concerning the 
resolution requirements, and thus algorithms that {\em adaptively} respond to the local  
development of solution features become attractive.   Note that such algorithms need
to have the capability both to introduce additional resolution when and where it is needed and 
to coarsen the local discretization scale when and where resolution demands become less
stringent.  

Here we briefly describe the version of AMR introduced by Berger and Oliger in
1984 \cite{BO}, which has been previously extended to accommodate the 
treatment of elliptic equations and then incorporated into the 
axisymmetric code, {\tt graxi}~\cite{fransp:phd,graxi:2003,graxi:2003b,graxi:2004}, used to generate the results discussed in Chap.~\ref{bs2d}.   
We note that the {\tt graxi} implementation of AMR introduces several simplifying 
restrictions relative to the original Berger and Oliger work---the most important 
of these include the requirements that (1) the computational domain be covered by 
a {\em single} base-level grid, (2) all component grids be aligned with the 
coordinate axes, (3) each level, $l$, of refinement be characterized by a unique spatial 
discretization scale, $h_l$, (4) the refinement factor between any two successive 
discretization levels be a unique integer $\rho$, and  (5) distinct grids at the same
level are not allowed to overlap.

The Berger and Oliger algorithm was designed for the finite difference solution 
of general systems of hyperbolic partial differential equations.  Adaptivity is achieved via 
the use of a hierarchy of component grids, where each component has uniform mesh 
spacings in each of the coordinate directions (and in our case, the same mesh spacing in 
{\em all} directions).  Specifically, the hierarchy is 
characterized by a sequence of discretization {\em levels}, $l=0,1,2,\cdots l_{\rm max}$,
where $l=0$ represents the coarsest level of grid, and $l_{\rm max}$ the finest. 
Associated with each level is a spatial discretization scale, $h_l$, and the scales 
at different levels satisfy 
\beq
 h_l \eq \rh h_{l+1}\,,
\eeq
where $\rho \ge 2$ is an integer (typically 2 in our calculations).  The grid 
hierarchy can be denoted $G_{l,j}$, where the index $l$ ranges over the discretization
levels, while the index $j$, for each level, ranges over the number of component 
grids at that level.  At the coarsest level, $l=0$, there is only a single 
grid, $G_{0,0}$, which covers the entire computational domain.  At finer 
levels, $l\ge1$,  there can be an arbitrary number of grids each of which is required to (1) have 
extrema ($x_{\rm min}, x_{\rm max}, y_{\rm min}$ etc.) that coincide with mesh points of {\em some}
level $l-1$ grid, and (2) be completely contained within that grid.  The level $l-1$ grid,
$G_{l-1,j'}$ that contains $G_{l,j}$ is called the parent grid, while $G_{l,j}$ is known
as a child of $G_{l-1,j'}$.  
As mentioned previously, distinct grids at the same level are not allowed to overlap. 

\begin{figure}[htp]
\begin{center}
\epsfxsize=8.0cm
\includegraphics[width=8.0cm,clip=true]{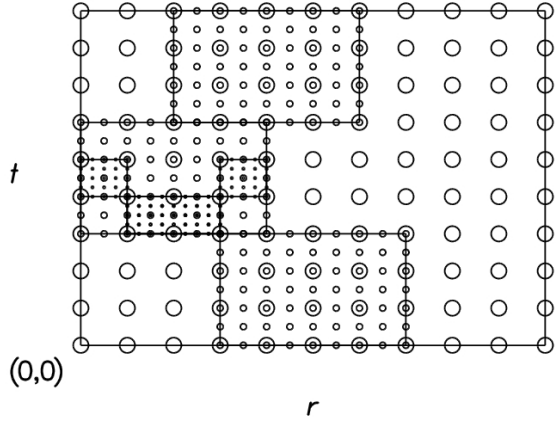}
\caption
[Schematic representation of the grid structure for a 1+1 AMR example]
{Schematic representation of a typical grid structure for a 1+1 calculation
using Berger and Oliger adaptive mesh refinement (AMR).  In this figure
we have $l_{\mathrm{max}}=2$ and $\rh =2$.}
\label{agrid}
\end{center}
\end{figure}

Fig.~\ref{agrid} shows a schematic representation of a typical grid hierarchy for a 1+1 dimensional problem 
with $l_{\rm max} = 2$ and a refinement ratio $\rho=2$. Note that the mesh refinement occurs 
in time as well as in space, and that the discrete time steps, $k_l$, used in the various levels of the 
hierarchy satisfy 
\beq
 k_l \eq \rh k_{l+1}\,,
\eeq
(i.e.\ a constant Courant factor is maintained across the hierarchy). Also notice that the structure 
of the hierarchy changes in time---resolution is increased where needed through the introduction of 
new grids, and decreased where it is not needed through the deletion of one or more existing grids.

The core of the Berger and Oliger algorithm is a recursive time stepping algorithm whereby 
equations on coarse grids are advanced before those on fine grids.  This allows the fine 
grid values at so-called refinement boundaries (i.e.\ boundaries of grids $G_{l,j}, l > 0$ 
that do {\em not} coincide with boundaries of the global computational domain) to be 
computed via {\em interpolation} of parental values.  Otherwise the same set of finite 
difference equations (including discretized versions of the physical boundary conditions) 
are used to update discrete unknowns on all component grids.  Once a single time step has 
been completed at level $l$, and if {\em any} level $l+1$ grids exist, the recursive 
time stepping algorithm is invoked to take $\rho$ steps at level $l+1$, so that the fine 
grid equations are integrated to the same physical time as the coarse ones.

Another key component of the AMR algorithm is the regridding procedure in which resolution 
demands are periodically assessed, and the grid hierarchy is correspondingly reconfigured.
The crucial task here is to decide when and where a particular solution region requires 
refinement or coarsening.  Although other approaches are possible, {\tt graxi} implements 
the strategy described in~\cite{BO} that ties regridding to local (truncation) error estimates 
based on Richardson-type procedures.  To illustrate this method, assume that our FDA 
can be written as 
\beq
 u^{n+1} = Q^h u^n\,,
\eeq 
\noi
where $u^{n+1}$ and $u^{n}$ are the advanced ($t=t^{n+1}$) and retarded ($t=t^n$) finite difference unknowns, respectively
(spatial grid indices are suppressed),
and $Q^h$ is the finite difference update operator that advances the solution from one 
discrete time to the next.  We can then generate a local estimate of the solution error by 
taking two time steps on the base mesh, characterized by discretization scale $h$, and then comparing
the result to what we get by advancing the retarded data by one step on a coarsened mesh with 
discretization scale $2h$. That is we compute 
\beq
 e^h \eq \left( Q^h Q^h - Q^{2h} \right) u^n\,,
\eeq
where the coarse grid update operator $Q^{2h}$ employs the same FDA as $Q^{h}$.  (This 
technique is directly analogous to the procedure commonly used in adaptive ordinary 
differential equation (ODE) integrators.)
This error estimation can be performed dynamically and 
makes use of the same FDA that is used in the basic time stepping procedure previously described.
We also note that the technique can be applied 
independently of the particular PDE being solved or the specific difference technique
that is being used.

Once the error estimate is calculated, we can determine those grid points where the error
estimate exceeds some predefined tolerance.  These points are then organized into clusters
and the regridding procedure is carried out.  This includes creation of a new grid hierarchy 
(or some portion of it), transfer of values from the old hierarchy to the new, and interpolation 
of grid function values from parental to child grids for those spatial locations where a 
given level of refinement did not previously exist.

A final key aspect of the Berger and Oliger algorithm involves the fine-to-coarse transfer 
of values from level $l$ child grids to their level $l-1$ parents once the level $l$ 
integration has been advanced to the time of the level $l-1$ solution.  Were this not done, 
the level $l-1$ values coinciding with refined regions---which by definition are not of 
sufficient accuracy to satisfy the error control criterion---would eventually ``pollute'' 
level $l-1$ values in regions that should,  in principle, {\em not} require refinement. 

Again, the reader who is interested in further details of this AMR algorithm can
consult~\cite{BO,fransp:phd}.

\section{Excision Techniques}

Starting in the 1960s, R.~Penrose and S.~Hawking proved a series of theorems which 
are now collectively called the {\em singularity theorems}.  The essential implication of
these theorems is that {\em physical} singularities are generic features of spacetime, and are not
a consequence, for example, of symmetries that a specific spacetime might possess.  In particular,
these mathematically rigorous results, as well as the results from detailed numerical 
simulation of gravitational collapse make it clear that black holes invariably contain 
physical singularities.  This fact has proven to be of enormous significance for the 
numerical simulation of black hole spacetimes.
At singularities physical quantities become undefined and it is meaningless to evolve spacetime 
there.~\footnote{In fact the definition of spacetime should not contain singular events.}
 It thus seems to be an unavoidable conclusion that numerical simulations must
somehow avoid black hole singularities.  The traditional approach in numerical relativity (used in virtually 
all calculations prior to the early 1990's)  was to use coordinate freedom to accomplish singularity 
avoidance.  For instance the maximal slicing condition (described in more detail in
Chap.~\ref{bs1d}), whereby the 
trace of the extrinsic curvature tensor is required to vanish at all times, generates slices that 
automatically ``slow down time'' in the vicinity of a black hole singularity, 
essentially freezing the evolution before the 
singularity is encountered.  Unfortunately, singularity-avoiding coordinate systems 
generically become pathological (i.e.\ {\em coordinate} pathologies develop) on dynamical
time scales, and this 
typically leads to a breakdown of simulations (i.e.\ the code ``crashes'' due to 
an inability to resolve extremely steep gradients etc.). Thus the use of singularity avoiding 
coordinates alone does not appear to provide an effective 
resolution to the problem of long-time evolution of black hole spacetimes.

An alternate approach to singularity avoidance that has become quite popular over the past decade or so is 
based on an idea due to Unruh (see \cite{excision} for a discussion of the method).  Unruh argued in the early 1980's 
that since the interiors of black holes are causally disconnected from
the exterior universe, it is not necessary (and is in fact wasteful) to evolve regions
inside black holes when our main interest is in the exterior region of spacetime.  
He thus proposed that one excise the interiors of black holes from the computational domain.
However, since the location of a black hole surface---the event horizon---requires knowledge 
of the complete spacetime, one cannot in principle excise a black hole until the global solution is 
in hand.  Unruh's second suggestion was thus to use the location of an
apparent horizon (which one can very loosely view as an instantaneous approximation to the intersection 
of an event horizon with a given spacelike hypersurface) as the excision surface.
Black hole excision was first explored numerically by Thornburg \cite{excision},
while the first successful implementation of the method in a dynamical context was due to Seidel
and Suen \cite{seidel_suen}.  Some other recent references that discuss excision
include~\cite{alcubierre:2001,yo:2001,alcubierre:2001b,scheel:2002,yo:2002,calabrese:2003,baumgarte:2003,shoemaker:2003,calabrese:2004}.  We will describe this method in more detail in section~\ref{excisiontech}.

\section{Spatial Compactification} \lab{spacom}

In this section we discuss issues related to compactification of a spatial domain.  This technique
is used in the construction of the stationary boson star solutions  described in 
Sec.~\ref{compactMI} and Sec.~\ref{IVP2d}.  For simplicity of presentation we assume here that the domain to be compactified is 
1-dimensional. 

The principal motivating factor for using a compactified spatial domain is the desire to
use exact, rather than approximate, boundary conditions in the solution of our PDEs. 
Specifically, for the problems considered in Sec.~\ref{compactMI} and Sec.~\ref{IVP2d}, we know the exact boundary conditions at
spatial infinity, $r=\infty$, whereas we generally know only the asymptotic behaviour of given
functions at any specific finite distance, $r_{\rm max}$.
Moreover, even if we know the exact form (e.g.\ precise falloff conditions) for our functions, 
it may be computationally advantageous to work with Dirichlet conditions (as the conditions at 
infinity tend to be), rather than the mixed (Robin) boundary conditions that typically result from 
the use of known falloff behaviour. 
We thus would like
to explore one class of transformations that allows us to perform compactification.

Suppose we want to compactify $r\in [0,\infty)$ to $\zeta \in [a,b]$ via the
following transformation:

\beq
 \zeta(r) = \fr{f(r)}{1+f(r)}\,,
\eeq

\noi
where $f(r)$ is some arbitrary function of $r$.  To ensure $\zeta(0) = a$ and $\zeta(\infty)=b$, we have

\bea \lab{RE1}
 f(0) &=& \fr{a}{1-a} \,, \\ \lab{RE2}
 f(\infty) &=& \fr{b}{1-b} \,.
\eea

\noi
We also wish the coordinate transformation to be regular, so that

\beq
 \fr{d\zeta}{dr} = \fr{f'}{(1+f)^2} \ne 0\,,
\eeq

\noi
or simply 

\beq \lab{RE3}
 f' \ne 0\,.
\eeq

\noi
In other words, $f$ should be a strictly increasing (or decreasing) function of $r$.
(\ref{RE1}), (\ref{RE2}) and (\ref{RE3}) are the only restrictions on a general 
compactifying function $f$.  
However, as we will see in Sec.~\ref{IVP2d} when 
we consider the generation of initial data representing rotating boson stars, 
once we have fixed coordinates,
we find that as the angular momentum of the 
star increases, the location of the extremum of the solution moves towards a 
boundary of the computational domain.   In addition, for larger angular momentum 
values, the solution function becomes increasingly peaked, which means that for 
fixed solution accuracy and coordinates, finer resolution is required as the angular 
momentum increases.  Thus, in 
addition to compactifying the domain, we attempt to map the location of the 
function extremum to the central region of the computational domain. 

We therefore impose the additional conditions that the coordinate transformation map some arbitrary
points, $r_i$, to some given values $\zeta_i$.  We then have

\beq \lab{RE4}
 f(r_i) = \fr{\zeta_i}{1-\zeta_i} \,,
\eeq

\noi
where $i = 0, \cdots, N$. If we assume $b=1$, then one such possibility is to write
\bea \nn
f(r) &=& \al_0 + \al_1 r + \al_2 r^2 + \cdots  \\
&=& \sum_i \al_i r^i\,, 
\eea

\noi
whereby $\al_i$ can be chosen to satisfy conditions (\ref{RE1}) and
(\ref{RE4}).
More specifically, if we want to compactify the radial coordinate to
$[0,1]$, and we want to bring one particular point $r=r_0$ (say the location 
where the grid function is a maximum) to $\zeta = 1/2$, then $\al_0 = 0$,
$\al_1 = 1/r_0$  and $\al_2 = \al_3 = \cdots = 0$; i.e.\ we have

\bea \nn
 \zeta(r) &=& \fr{\al_1 r}{1+\al_1 r} \\ \lab{transcentral}
 &=& \fr{r}{r_0+r} \,.
\eea

\noi
In other words, transformation~(\ref{transcentral}) maps $[0,\infty)$ to
$[0,1]$ with $\zeta(r_0) = \ha$.

  \resetcounters

\chapter{Boson Stars in Spherical Symmetry} \lab{bs1d}

In this chapter we study boson stars in spherical symmetry.
In spherically symmetric spacetime the equations of motions are greatly simplified,
and with a proper choice of coordinates the number of variables that must be evolved 
is significantly reduced. 
Therefore it is relatively easy to study the system numerically compared to
higher dimensional (2-dimensional and 3-dimensional) simulations.  At the same time, of all of the symmetries 
that could be imposed to reduce the field equations to a set of PDEs in one space 
dimension and time, spherical symmetry is clearly the most appropriate for the study of isolated,
gravitationally compact objects.

This chapter consists of two parts.  The first part concerns the construction of
static solutions  which represent 
spherically symmetric boson stars in the ground state.  These solutions provide the initial
data for the simulations of critical phenomena of boson stars in
spherical symmetry studied in the second part of this chapter, 
as well as for the simulation of critical phenomena of 
boson stars in axisymmetry in Chap.~\ref{bs2d}. 
Together with 
rotating solutions computed in the Newtonian limit, 
the solutions also provide initial guesses in the 
construction of the general relativistic, stationary, rotating boson star solutions
which will also be described in Chap.~\ref{bs2d}.  

The second part of the chapter concerns the dynamics of
spherical boson stars.  More specifically, 
in an extension of the work performed by Hawley and Choptuik~\cite{scott_matt:2000},
we study the critical phenomena of boson stars which are driven to the threshold of 
black hole formation via an external perturbing agent.
As in the previous work, our boson stars are perturbed by a real massless scalar field
and an overall amplitude of this real scalar field is used to tune to criticality.
However, we use a different coordinate system---maximal-isotropic coordinates---than that used 
in~\cite{scott_matt:2000}.  Although the accuracy of finite difference calculations in any
given coordinate system can in principle be estimated using intrinsic means (e.g.\
convergence tests), we feel that it is nonetheless useful to reproduce the calculations of
\cite{shawley:phd,scott_matt:2000} in a different coordinate system.  In addition, since maximal-isotropic coordinates
{\em can} penetrate black hole horizons (the polar-areal coordinates used in~\cite{scott_matt:2000} 
generally cannot), we can also implement black hole excision in our simulations.  
We emphasize however, that we feel that our excision calculations---which are restricted to
spherical symmetry and use an {\em  ad hoc} boundary condition at the excision surface 
for the lapse---are only a minor result of the thesis.

The principal new results presented in this chapter concern the long-time evolutions of
subcritical simulations in critical collapse of boson stars.  
We provide numerical evidence that---contrary to
the previous conjecture~\cite{scott_matt:2000} that subcritical solutions disperse most of 
the original mass of the boson star to large distances---the late time behaviour of subcritical 
evolution is characterized by oscillation about a stable boson star solution.
We also apply a linear perturbation analysis similar to
that in~\cite{scott_matt:2000} and confirm that the observed oscillation modes agree with
the fundamental modes given by perturbation theory.  (We use a code provided by S.~Hawley \cite{hawley:private} to
generate the frequencies from the perturbation analysis.)  We also observe that there 
is generically an overall lower-frequency modulation of the post-critical-phase oscillations.
At this time we have no explanation
for the origin of this additional oscillatory behaviour, although we can safely rule out the 
possibility that it originates from the beating
between the fundamental and first harmonic mode.

The outline of the remainder of the chapter is as follows.  In Sec.~\ref{sss} we derive the equations of motion
for boson stars in a general spherically symmetric spacetime. In Sec.~\ref{MIcoord} we
specialize the equations of motion to maximal-isotropic coordinates.
In Sec.~\ref{IVP1d} we derive systems of equations for the initial value problem in
maximal-isotropic coordinates, polar-areal coordinates and compactified maximal-isotropic
coordinates, and briefly describe the family of stationary solutions that we can generate 
using any of the three approaches.
In Sec.~\ref{critphenBS} we present the results of near-critical evolutions and, for the 
case of subcritical evolutions, compare the oscillatory behaviour observed at late times to 
that predicted from perturbation theory.  
Finally in Sec.~\ref{excisiontech} we study the implementation of
black hole excision technique applied to our system.
Throughout this chapter we denote $' \eq \pa_r$ and $\dot{} \eq
\pa_t$,  and $m = 1$ is used for the particle mass of the boson field.

\section{Spherically Symmetric Spacetime} \lab{sss}
Loosely speaking, we may regard a spherically symmetric spacetime as one that admits 
a preferred timelike observer such that the spacetime is spherically symmetric 
about every point on this special observer's
world-line.~\footnote{Spherically symmetric spacetime is defined as one which admits
the group $SO(3)$ as a group of isometries, with the group orbits spacelike
two-surfaces.}
The most general metric for a time-dependent spherically-symmetric spacetime can be 
written as \cite{scarroll,matt:phd,MTW}
\beq \lab{sssmetric}
 ds^2 = \left(  -\al^2 + a^2 \bt^2 \right) dt^2 + 2 a^2 \bt\, dt dr + a^2 dr^2 + r^2
b^2 d \Om^2 \,,
\eeq
\noi
where $d \Om^2 \eq d \te^2 + \sin^2\te\, d \vp^2$ is the metric on the unit 2-sphere,
$\bt$ is the $r$-component of the shift
vector, $\bt^i = \left( \bt, 0, 0\right)$, and $\al, \bt, a, b$ are functions of $t$
and $r$ only.
Corresponding to the above metric is an extrinsic curvature tensor with only two
independent components~\footnote{The form of the extrinsic curvature can 
be inferred by considering
the expression $\pa_t{\ga}_{ij} = -2 \al \ga_{ik}
K^k{}_j + \bt^k \ga_{ij,k} + \ga_{ik} \bt^k{}_{,j} + \ga_{kj} \bt^k{}_{,i}$.}
\beq
 K^i{}_j = {\mbox {diag}} \left( K^r\,_r, K^\theta\,_\theta,
K^\theta\,_\theta \right)\,,
\eeq
\noi
where $K^r{}_r$, $K^\te{}_\te$ are also functions of $t$ and $r$.  Together with $\al,\bt,a$
and $b$ we thus must deal with a maximum of 6 out of 16 possible geometrical variables in 
spherically symmetric calculations.
The 4-metric can be written in matrix form as
\beq \lab{spherical_symmetric_metric}
g_{ \mu \nu} = 
 \left[ 
\ba{cccc}
-\al^2 + a^2 \bt^2 &  a^2 \bt & 0  &  0  \\
a^2 \bt & a^2 & 0 & 0  \\
0        &  0     &  (r b)^2  & 0 \\
0 & 0 & 0 & ( r b  \sin \theta)^2
\ea
 \right] \,,
\eeq
\noi
from which it immediately follows that the inverse metric is given by

\beq  \lab{inverse_spherical_symmetric_metric} 
g^{ \mu \nu} =
 \left[ 
\ba{cccc}
- 1/\al^2 &  \bt/\al^2 & 0  &  0  \\
\bt/\al^2 &  1/a^2 - \bt^2/\al^2 & 0 & 0  \\
0        &  0     &  1/(rb )^2 & 0 \\
0 & 0 & 0 &  1/(rb \sin \theta)^2
\ea
 \right]\,.
\eeq
\noi
We also have $\sr{-g} = \al ab^2 r^2 \sin \te$, where $g$ is the determinant of the 
4-metric. Now considering the 3-metric, $\ga_{ij}$, we have $\ga_{ij} = g_{ij}$, so 
\beq 
\ga_{ij} = 
 \left[ 
\ba{ccc}
a^2 & 0 & 0  \\
0     &  (r b)^2  & 0 \\
0 & 0 & ( r b  \sin \theta)^2
\ea
\right] \,,
\eeq
\noi
and 
\beq 
\ga^{ij} = 
 \left[ 
\ba{ccc}
1/a^2& 0 & 0  \\
0     &  1/(r b)^2  & 0 \\
0 & 0 & 1/( r b  \sin \theta)^2
\ea
\right] \,.
\eeq
\noi
The non-vanishing Christoffel symbols constructed 
from $\ga_{ij}$ are
\bea \nn
 \Ga^r\,_{r\,r} &=& \fr{a'}{a} \,, \\ \nn
 \Ga^r\,_{\theta \theta} &=& -\fr{ \left( r^2 b^2\right)'}{2 a^2}\,, \\ \nn
 \Ga^r\,_{\ph \ph } &=& \sin^2 \theta \, \Ga^r\,_{\theta \theta}\,,\\ 
 \Ga^\theta\,_{r \theta} &=& \Ga^\theta\,_{\theta r} = \fr{\left( r b\right)'}{ rb}\,,\\ \nn
 \Ga^{\theta}\,_{\ph \ph} &=& - \sin \, \theta \cos \theta \,, \\ \nn
 \Ga^\ph\,_{r \ph} &=& \Ga^\ph\,_{\ph r} = \Ga^{ \theta}\,_{r \theta}\,, \\  \nn
  \Ga^\ph\,_{\theta \ph} &=& \Ga^\ph\,_{\ph \theta} = \cot \theta\, .
\eea

\noi
The non-vanishing components of the Ricci tensor $R^i\,_j$ are
\bea \lab{bs_Rrr}
 R^r\,_r &=& - \fr{ 2}{a r b} \left[ \fr{ \left( r 
b\right)'}{a}\right]' \,, \\
 R^\theta\,_\theta &=& \fr{ 1}{ a ( r
b )^2 } \left[ a - \left( \fr{ r b}{a } ( r b)' \right)' \right] \,, \\
 R^\ph\,_\ph &=& R^\theta\,_\theta  \,.
\eea
\noi
and the scalar curvature $R$ is
\bea \nn \lab{bs_R}
  R &=& R^r\,_r + R^\theta\,_\theta + R^\ph\,_\ph  \\ \nn
  & = & R^r\,_r + 2 R^\theta\,_\theta \\ \nn
  & = & - \fr{ 2}{a r b} \left[ \fr{ \left( r b\right)'}{a}\right]' + \fr{
2}{ a ( r b )^2 } \left[ a - \left( \fr{ r b}{a } ( r b)' \right)' \right]
\\  \lab{scalarR}
  & = & - \fr{2}{a r b} \left\{ \left[ \fr{ (r b)'}{a}\right]' + \fr{ 1}{rb
} \left[ \left( \fr{ rb}{a} (rb)'\right)'- a\right]\right\} \,.
\eea

\noi
On the other hand, from (\ref{Ctmunu}), (\ref{energydensity})
(\ref{momentumdensity}), (\ref{Sij_def}) and (\ref{S_def})
the non-vanishing components of the matter source
terms are 
\bea \lab{bs_rho}
 \rh &=& \fr{ |\Ph|^2+|\Pi|^2 }{ 2 a^2} + \fr{ m^2 |\ph|^2}{ 2} \,, \\
\lab{bs_jr} j_r &=&  - \fr{ \Pi^\ast \Ph + \Pi \Ph^\ast}{2 a} = a^2 j^r\,, \\
\lab{bs_Srr} S^r\,_r &=&  \rh - m^2 |\ph |^2 \,, \\
\lab{bs_Sthth}
 S^\theta\,_\theta &=&  \fr{ |\Pi|^2 - |\Ph|^2 }{2 a^2} - \fr{ m^2
|\ph|^2}{2}\,, \\
\lab{bs_Sphph} S^\ph\,_\ph &=&  S^\theta\,_\theta \,, \\ \lab{bs_S}
S &=& \fr{3|\Pi|^2-|\Ph|^2}{2a^2}-\fr{3}{2}m^2 |\ph|^2 \,.
\eea
\noi
Here, we have defined the auxiliary scalar-field variables
\bea \lab{Ph_def}
 \Ph &\eq& \ph' \,,\\ \lab{Pi_def}
 \Pi &\eq& \fr{a}{\al} \left( \dot{\ph} - \bt \ph'\right)\, ,
\eea
which are also useful in recasting the Klein-Gordon equation in first-order-in-time
form.

\noi
The Hamiltonian constraint~(\ref{hamiltonian_constraint}) now becomes
\bea \lab{HAMC1d}
- \fr{2}{a r b}\left\{ \left[ \fr{(rb)'}{a}\right]' + \fr{1}{rb} \left[
  \left( \fr{rb}{a} (rb)'\right)' - a \right] \right\}  + 4 K^r{}_r
K^{\theta}{}_{\theta} + 2 {K^{\theta}{}_{\theta}}^2 =
8 \pi\left[ \fr{|\Ph|^2 + |\Pi|^2}{a^2} + m^2 |\ph|^2 \right] \,\,
\eea
\noi
while the momentum constraint~(\ref{momentum_constraint}) is
\beq
 K^{\theta}{}_{\theta}\left.'\right. + \fr{(rb)'}{rb}\left(
K^{\theta}{}_{\theta}- K^r{}_r \right) = \fr{2 \pi}{a} \left( \Pi^{\ast} \Ph + \Pi
\Ph^{\ast} \right)\,.
\eeq
\noi
A straightforward but tedious calculation of the evolution equations gives
\bea \lab{adot}
 \dot{a} &=& - \al a K^r{}_r + \left( a \bt \right)'\,,\\ \lab{bdot}
\dot{b} &=& - \al b \ktt + \fr{\bt}{r} \left( rb\right)'\,. \\
\dot{\krr} &=& \bt \krr' - \fr{1}{a} \left( \fr{\al'}{a}\right)' + \al
\left\{ -\fr{2}{arb} \left[ \fr{(rb)'}{a}\right]' + K \krr - 4 \pi \left[ \fr{2
|\Ph|^2}{a^2} + m^2 |\ph|^2 \right]\right\}\,,\\
\dot{\ktt} &=& \bt \ktt' + \fr{\al}{(rb)^2} - \fr{1}{a(rb)^2} \left[ \fr{\al rb}{a}
\left( rb\right)'\right]' + \al \left( K\ktt-4 \pi m^2 |\ph|^2 \right)\,.
\eea
\noi
The definitions (\ref{Ph_def}), (\ref{Pi_def}) and the Klein-Gordon equation
(\ref{klein_gordon_eq}) leads to the evolution equations
for the scalar field and the 
auxiliary field
\bea \lab{kgeq1}
 \dot{\ph} &=& \fr{\al}{a} \Pi + \bt \Ph\,,\\ \lab{kgeq2}
\dot{\Ph} &=& \left( \bt \Ph + \fr{\al}{a} \Pi\right)' \,,\\ \lab{Pidot1d}
\dot{\Pi} &=& \fr{1}{(rb)^2} \left[ (rb)^2 \left( \bt \Pi + \fr{\al}{a} \Ph\right)
\right]' - \al a m^2 \ph + 2 \left[ \al \ktt - \bt \fr{(rb)'}{rb}\right] \Pi\,.
\eea
\noi
Equations~(\ref{HAMC1d})-(\ref{Pidot1d}) comprise the general system of equations 
for a general-relativistic complex scalar field in spherical symmetry.  
To study the system numerically, one
needs to choose specific coordinate conditions, impose boundary conditions and  
set up appropriate initial conditions.  In the following section we will 
specialize the above equations to the case of maximal-isotropic coordinates.

\section{Maximal-Isotropic Coordinates} \lab{MIcoord}
\subsection{The equations of motion}
The equations presented in the previous section are valid in any spherically symmetric 
coordinate system compatible with~(\ref{sssmetric}).  However, in order to perform numerical
simulations we need to choose a {\em specific} coordinate system, which, in the 3+1 
approach is equivalent to prescribing the lapse, $\alpha$, and the shift vector component,
$\beta$.  For a variety of reasons, we have chosen to adopt so-called {\em maximal-isotropic}
coordinates. First, these coordinates have been used in several previous calculations in 
numerical relativity (for example, see \cite{petrich:1985,choptuik:1991}) and have generally worked well.   Second, 
maximal-isotropic coordinates are the specialization to spherical symmetry of the coordinate system 
used in the axisymmetric code discussed in the next chapter.  This fact allows us, in
principle, to directly 
compare results from the spherical code to those from the axisymmetric code, when the latter is supplied with
spherically symmetric initial data.  Third, in contrast to the polar-areal coordinates used 
in~\cite{shawley:phd,scott_matt:2000}, maximal-isotropic coordinates 
{\em can} penetrate apparent horizons, 
and thus using them, we can more readily study the process of black hole formation.   In particular 
we can incorporate black hole excising techniques as discussed in Sec.~\ref{excisiontech}.

The ``maximal'' part of maximal-isotropic refers to the {\em maximal slicing} condition, which fixes
the lapse function.  A hypersurface is maximal if its mean extrinsic curvature vanishes; that is if
\beq
 K \equiv K^i{}_i = 0\, .
\eeq
\noi
Geometrically, it can be shown that $K=0$ implies that the volume of the hypersurface is maximized 
with respect to localized, infinitesimal deformations of the slice (analogous to the {\em minimization}
of the surface area by a soap film bounded by a wire frame).
Computationally, we implement maximal slicing by choosing initial data for the extrinsic curvature 
that  satisfies $K=0$, and then demanding that 
\beq
 \dot{K}(t,r) = 0\,.
\eeq
\noi
for all $t$ and $r$.
We note that the maximal condition allows us to eliminate one of the two non-trivial components of the 
extrinsic curvature from the computational scheme.  In particular, since 
\beq \lab{maxslice}
 K=0 \longrightarrow \ktt = -\ha \krr\,. 
\eeq
\noi
we choose to eliminate $\ktt$.  As we will see shortly, the maximal condition leads to a second order 
ODE that constrains the lapse function, $\alpha$ at all times.

Given maximal slicing, the remaining (spatial) coordinate freedom is fixed by the {\em isotropic 
condition} which requires 
\beq \lab{istro_cond}
a=b \eq \ps(t,r)^2\,,
\eeq
for some positive function $\ps(t,r)$, such that the spatial metric takes the ``isotropic'' (or 3-conformally
flat form)
\beq
	{}^{(3)}ds^2 = \ps^4 \left( dr^2 + r^2 d\Omega^2 \right) \,.
\eeq

\noi
This is implemented by choosing initial data such that $a(0,r)  \equiv b(0,r)$ and then demanding that 
\bea
 \dot{a}(t,r) = \dot{b}(t,r)\,,
\eea
for all $t$ and $r$.  As will be shown below, this condition leads to a first-order ODE that fixes the 
shift vector component, $\beta$, on each slice. 
\noi

Turning now to the equation for the lapse, $\alpha$, that is implied by $K=0$, we note that by taking 
the 3-trace of~(\ref{extrinsic_curvature_evolution_eq}) we have:
\beq
 \dot{K} = \bt K' - \fr{1}{a (rb)^2} \left[ \fr{(rb)^2}{a }\al'\right]' + \al
\left[ R + K^2 + 4 \pi \left( S-3 \rh\right)\right]\,,
\eeq
\noi
Using $K\equiv0$, as well as (\ref{hamiltonian_constraint}), (\ref{bs_S}) and (\ref{bs_rho}), for $R$, $S$ and $\rho$ respectively, we have 
\beq
 \al'' + \fr{a}{(rb)^2} \left( \fr{(rb)^2}{a}\right)' \al' + \left[4 \pi a^2 m^2
|\ph|^2  -8 \pi |\Pi|^2 - \fr{3}{2} a^2 {\krr}^2 \right] \al = 0\,.
\eeq
\noi
Using the isotropic condition, $a\equiv b\equiv \psi^2$, this can be further simplified to
\beq
 \al'' + \fr{\left[ (\ps r)^2\right]'}{(\ps r)^2}  \al' + \left[ 4 \pi \ps^4m^2 
|\ph|^2 - 8 \pi |\Pi|^2 - \fr{3}{2} \left( \ps^2 \krr\right)^2 \right] \al = 0\,,
\eeq
\noi or 
\beq \lab{maxcond}
 \al'' + \fr{2}{r \ps^2} \fr{d}{dr^2}\left( r^2 \ps^2\right)  \al' + \left[ 4 \pi
\ps^4m^2 
|\ph|^2 - 8 \pi |\Pi|^2 - \fr{3}{2} \left( \ps^2 \krr\right)^2 \right] \al = 0\,,
\eeq
\noi
where $d/dr^2$ is the derivative with respect to $r^2$.  We note that here and elsewhere,
and following Evans~\cite{evans:phd},
we often rewrite terms of the form $\partial_r f(t,r)$ as $pr^{p-1}\partial_{r^p} f(r,t)$ for 
functions $f$ satisfying $\lim_{r\to0} f(r,t) = r^p f_p(t) + O(r^{p+2})$.  This technique ensures 
that when standard second-order centered difference formulae are applied to such terms, the leading 
order regularity behaviour is preserved in the discrete domain as $r\to0$. 

The ODE that fixes the shift vector component $\beta$ (i.e.\ the ``isotropic condition
for the shift''), can be easily derived from~(\ref{metric_evolution_eq}), by equating the 
right hand sides of the
respective evolution equations for $a(t,r)$ and $b(t,r)$.  Doing this we find
\beq 
 r \left( \fr{\bt}{r}\right)' = \al \left( \krr - \ktt\right)\,,
\eeq
\noi
which can be further simplified using the maximal condition to yield
\beq \lab{isocond}
 r \left( \fr{\bt}{r}\right)' = \fr{3}{2} \al \krr\,.
\eeq
\noi
Both the slicing condition~(\ref{maxcond}) and the shift-component equation~(\ref{isocond}) must be 
supplemented by appropriate boundary conditions, but we will relegate these details to
App.~\ref{ap4}.

Having fixed the {\em kinematical} geometric variables (i.e.\ $\alpha$ and $\beta$),
we now consider the specific form of the equations of motion for the dynamical geometric
variables, as well as those for the scalar fields that are coupled to the gravitational 
field.

Due to our restriction to spherical symmetry (in which case the gravitational field 
has {\em no} dynamics that is not tied to the dynamics of a matter field), as well
as our choice of maximal-isotropic coordinates we can implement a so-called
{\em fully constrained}  scheme 
that uses only the constraint equations to update the dynamical geometric variables 
in time.  Given our coordinate choice, the only non-trivial geometric variables that remain are the 
(3-)conformal factor, $\psi$, and the extrinsic curvature component, $\krr$.
These can be determined from the Hamiltonian and momentum constraints, respectively.

Specifically, using expressions~(\ref{hamiltonian_constraint}), (\ref{scalarR}), (\ref{maxslice}),
(\ref{bs_rho}), (\ref{istro_cond}), (\ref{momentum_constraint}) and (\ref{bs_jr}), we have 
\bea \lab{hamc}
\fr{3}{\ps^5} \fr{d}{dr^3} \left( r^2 \fr{d \ps}{dr} \right) + \fr{3}{16}
{\krr}^2 &=& -\pi\left( \fr{|\Ph|^2 + |\Pi|^2}{\ps^4} + m^2 |\ph|^2 \right)  \,, \\ 
\lab{momc}
 {\krr}\left.'\right. + 3 \fr{( r \ps^2)'}{r \ps^2} K^r\,_r &=& -\fr{4 \pi}{\ps^2}
\left( \Pi^{\ast} \Ph + \Pi \Ph^{\ast} \right)\,.
\eea

\noi
(\ref{maxcond}), (\ref{isocond}), (\ref{hamc}) and (\ref{momc}) completely
determine the geometric variables $\al, \bt, \ps$ and $\krr$ in maximal-isotropic
coordinates.  

A caveat is in order here.  A fully-constrained evolution works well as long as 
we are not implementing black hole excision.  When excision {\em is} being performed, 
we must supply boundary conditions for $\psi$ and $\krr$ at the excision surface,
and these are provided by the {\em evolution} equations for those variables, which 
are
\bea \lab{psdot}
 \dot{\ps} &=& - \ha \al \ps \krr + \fr{(\ps^2 \bt)'}{2 \ps}\,, \\
\dot{\krr} &=& \bt \krr' - \fr{2 \al}{(r \ps^2)^2} + \fr{2}{r^2 \ps^6} \left[ \al r
\left( r \ps^2\right)'\right]' + 8 \pi m^2 \al |\ph|^2\,.
\eea

\noi
Now considering the complex-scalar matter field, and using~(\ref{kgeq1}), (\ref{kgeq2}), (\ref{Pidot1d})
and (\ref{istro_cond}), we have the following 
equations of motion:
\bea \lab{phdot}
 \dot{\ph} & = & \fr{ \al}{\ps^2} \Pi + \bt \Ph \,,\\
 \dot{\Ph} & = & \left( \bt \Ph + \fr{\al}{\ps^2} \Pi\right)' \,,\\
\lab{pidot}
 \dot{\Pi} & = & \fr{3}{\ps^4}  \fr{d}{d r^3}\left[ r^2 \ps^4 \left( \bt
\Pi + \fr{ \al }{ \ps^2} \Ph\right)\right]- \al \ps^2 m^2 \ph
 -  \left[ \al \krr+ 2 \bt \fr{(r \ps^2)'}{r \ps^2} \right] \Pi\,.
\eea

\noi
(\ref{maxcond}), (\ref{isocond}), (\ref{hamc}), (\ref{momc}),
(\ref{phdot})-(\ref{pidot}) constitute the basic set of equations that 
we use in our study of boson star dynamics in 
maximal-isotropic coordinates.  

As discussed previously, in order to gravitationally {\em perturb} boson 
stars, we will incorporate an additional, minimally coupled, massless real scalar 
field in our model.
It is straightforward to generalize the above equations of motion to account 
for such an addition, and here we will merely state the results.
To simplify notation we denote 
the real and imaginary parts of the complex scalar field as 
$\ph_1(t,r)$ and $\ph_2(t,r)$, respectively, while
$\ph_3(t,r)$ is the massless real scalar field: 

\beq
  \al'' + \fr{2}{r \ps^2} \fr{ d}{dr^2}\left( r^2 \ps^2 \right) \al' + \left[4 \pi
\ps^4 m^2 \sum_{i=1}^2 \ph_i^2 - 8 \pi \sum_{i=1}^3 \Pi_i^2 - \fr{3}{2}
(\ps^2 \krr)^2 \right] \al  =  0 \,,
\eeq
\beq \lab{shift_eqn}
 r \left( \fr{\bt}{r}\right)' = \fr{3}{2} \al \krr\,,
\eeq
\bea
\fr{3}{\ps^5} \fr{d}{dr^3} \left( r^2 \fr{d \ps}{dr} \right) + \fr{3}{16}
{\krr}^2
&=& - \pi\left( \fr{ \sum_{i=1}^3 \left( \Ph^2_i + \Pi^2_i\right)  }{\ps^4} +
m^2 \sum_{i=1}^2 {\ph_i}^2 \right) \,,\\
 {\krr}\left.'\right. + 3 \fr{( r \ps^2)'}{r \ps^2} {\krr} &=& -\fr{8
\pi}{\ps^2} \left( \sum_{i=1}^3 \Pi_i  \Ph_i \right)\,,
\eea
\bea 
 \dot{\ph}_i & = & \fr{ \al}{\ps^2} \Pi_i + \bt \Ph_i \,,\\
 \dot{\Ph}_i & = & \left( \bt \Ph_i + \fr{\al}{\ps^2} \Pi_i\right)' \,,\\
\nn
 \dot{\Pi}_i & = & \fr{3}{\ps^4}  \fr{d}{d r^3}\left[ r^2 \ps^4 \left( \bt
\Pi_i + \fr{ \al }{ \ps^2} \Ph_i\right)\right]- \al \ps^2 m^2 \ph_i
\left( 1 - \de_{i3} \right) \\
&&-  \left[ \al\krr + 2 \bt \fr{(r \ps^2)'}{r \ps^2} \right] \Pi_i\,.
 \eea

\subsection{The mass aspect function}
Before discussing the construction of initial data, we want to note that
for diagnostic purposes it is often useful to compute the {\em mass aspect
function $M(t,r)$}, which approaches the ADM mass as $r\to\infty$. 
In Schwarzschild-like coordinates $(T, R, \theta, \phi)$, 
($b\equiv1$, $\beta\equiv0$ in~(\ref{sssmetric})) the
mass aspect can be easily expressed in terms of
the metric function $a$, since for large $R$ (i.e.\ outside the support of any 
matter fields in the spacetime) we expect the metric
to be of the form
\beq \lab{slikemetric}
 ds^2 = - \left( 1-\fr{2M}{R}\right) dT^2 + \left( 1-\fr{2M}{R}\right)^{-1}
dR^2 + R^2 d \Om^2\,,
\eeq
\noi
where $M$ is a constant to be interpreted as the (ADM) mass of the system.  On
the other hand, the metric~(\ref{sssmetric}) specialized to the case 
$b\equiv1$, $\beta\equiv0$ is
\beq
 ds^2 = - \al^2 dT^2 + a^2 dR^2 + R^2 d \Om^2\, .
\eeq
\noi
We can thus \emph{define}
the mass aspect function $M(T,R)$ such that
\beq
\left( 1-\fr{2M(T,R)}{R}\right)^{-1} \eq   a(T,R)^2\,,
\eeq
\noi
or, solving for $M$,
\beq
 M(T,R) \eq \fr{R}{2} \left( 1-\fr{1}{a(T,R)^2}\right)\,.
\eeq
\noi
We can generalize this expression to the case of an arbitrary time-dependent
spherically-symmetric coordinate system as follows.  We first note that, in spherical 
symmetry, and at least in vacuum regions, $M$ is a geometric (i.e.\ coordinate-independent)
quantity, as is the areal radius $R$ (note that we generally have $R(t,r)\equiv rb(t,r)$ 
from~(\ref{sssmetric})).

We now consider the square of the gradient of the areal radius, 
$\nabla_\mu R \,\nabla^\mu R$, which is also a geometrical invariant.  In Schwarzschild coordinates 
we have
\bea \nn
 \na_\mu R\,\na^\mu  R &=& g^{\mu \nu} \na_{\mu} R\,\na_{\nu} R \\ \nn
&=& g^{RR} \na_R R \,\na_R R\\ \nn
&=& g^{RR} \\ 
&=& 1-\fr{2M}{R}\,.
\eea

\noi
Thus we have the coordinate-independent formula
\beq
 M = \fr{R}{2} \left( 1 - \na_{\mu} R \, \na^{\mu} R\right)\,,
\eeq
\noi
where, again, we emphasize that $R$ is the areal radius. 

In maximal-isotropic coordinates we have $R = \psi^2 r$, so 

\beq
 M(t,r) \eq \fr{\ps^2 r}{2} \left[ 1 - \na_{\mu} \left( \ps^2 r\right)
\na^{\mu} \left( \ps^2 r\right)\right]\,.
\eeq
\noi
With a 
little algebra the above expression can be written as
\beq
M(t,r) = \left( \fr{\ps^2 r}{2}\right)^3 {K^{r}{}_{r}}^2 - 2 \ps' r^2 \left( \ps + r
\ps'\right)\,.
\eeq

Finally, we note that 
we summarize all essential equations, boundary conditions, as well as the finite
differencing of the equations discussed in this section in App.~\ref{ap4}.

\section{The Initial Value Problem in Spherical Symmetry} \lab{IVP1d}
As we have discussed previously, we are particularly interested in the study of 
the dynamics of boson stars which are {\em stationary} (or equivalently {\em static}
in the current case of spherical symmetry) solutions of the coupled Einstein-Klein-Gordon 
equations.  Determination of initial data for the scalar and gravitational fields that 
represents such stars is a special case of the general initial value problem for 
our model. In fact, we have used three different methods to obtain 
spherically-symmetric boson star initial data.  The three approaches are summarized in the 
following sub-sections: each is conveniently labelled by the spatial coordinates used in the 
initial value computation, and each has its own specific advantages.

\subsection{Maximal-Isotropic Coordinates}

We first discuss the construction of boson star data within the context of the maximal-isotropic coordinate system described in the previous section.
To this point we have been considering the full dynamical equations describing 
the Klein-Gordon field coupled to gravity in spherical symmetry.  
To construct a star-like configuration we need to impose additional specific 
constraints on the scalar field.  Ideally one would like a ``star'' to be  
described by a localized, time-independent matter source that generates an
everywhere regular (i.e.\ non-singular) gravitational field.
However, for the case of a complex scalar field, it can be shown that such
regular, time-independent configurations do not exist~\cite{Friedberg:1987}.  Despite 
this fact, since the stress-energy tensor 
(\ref{Ctmunu}) depends only on the modulus of the scalar field (and the 
gradients of the modulus), one {\em can} construct scalar field configurations with harmonic 
time-dependence that produce time-independent metrics.  
Specifically, we adopt the following ansatz for boson
stars in spherical symmetry:
\beq \lab{bs1dansatz}
\ph(t,r) = \ph_0(r)\,e^{-i \om t}\,, 
\eeq
\noi
and then demand that the spacetime be static, i.e.\ we demand that the metric admits a timelike
Killing vector field $\ch$ which is orthogonal to the $t={\rm const.}$ surfaces.
Adapting coordinate time to the 
timelike Killing vector field, we have
\beq
 \bt = 0\,, 
\eeq
\noi
for all time $t$.  Additionally, we have that the time derivatives of any of the geometrical
variables identically vanish.  It then follows immediately from~(\ref{psdot}) that
\beq
 \krr = 0\,.
\eeq
\noi
As is necessary for the consistency of the ansatz~(\ref{bs1dansatz}), the 
isotropic condition for $\beta$~(\ref{isocond}) 
is automatically satisfied, and we are left with geometrical variables 
$\al(0,r)$, $\ps(0,r)$ and $\ph_0(r)$
that need to be determined from the maximal slicing condition (\ref{maxcond}),
the Hamiltonian constraint (\ref{hamc}) and the Klein-Gordon equation (\ref{pidot}),
respectively.
Before considering the solution of those three equations, 
we note that from the ansatz~(\ref{bs1dansatz}) we have
\bea
 \dot{\ph} &=& - i \om \ph_0 \,e^{-i \om t}\,, \\
  \ph' &=& {\ph_0}' e^{-i \om t}\,,
\eea
\noi
and hence
\bea
 \Pi &=& -i \fr{\ps^2}{\al}  \om \ph_0 e^{-i \om t}\,, \\
  \Ph &=& {\ph_0}' e^{-i \om t}\,. 
\eea
\noi
Therefore,
\beq
  \Pi^{\ast} \Ph + \Pi \Ph^{\ast} = 0\,,
\eeq
\noi and the momentum constraint (\ref{momc}) is satisfied for all times.

Using the ansatz~(\ref{bs1dansatz}) for $t=0$, as well as $\beta=K^r{}_r =0$, straightforward 
manipulation of the Hamiltonian constraint, the Klein-Gordon equation,
and the maximal slicing equation yields the following set of ODEs:
\bea \lab{bs1divp1}
 \ps' & = & \Ps \,,\\
 \Ps' & = & -  \fr{ 2\Ps}{r} - \pi \left[ \ps \Ph^2 + \ps^5 \left( \fr{
\om^2}{\al^2} + m^2 \right) \ph^2 \right] \,,\\
  \ph' & = & \Ph \,,\\
  \Ph' & = & - \left( \fr{ 2}{r} + \fr{ A}{\al} +  \fr{ 2\Ps}{\ps} \right)
\Ph + \ps^4  \left( m^2 - \fr{ \om^2}{\al^2}\right) \ph \,,\\
  \al' & = & A \,,\\\lab{bs1divp6}
 A' & = & - 2 \left( \fr{ 1}{r} + \fr{ \Ps}{\ps}\right) A + 4 \pi \ps^4
\al \left( \fr{ 2 \om^2}{ \al^2} - m^2 \right) \ph^2\,.
\eea
\noi
Here, in order to simplify notation, we have dropped the subscript ``0'', making the 
identifications $\ph(r) \equiv \ph_0(r)$ and $\Ph(r) \equiv \ph'(r) \equiv \ph_0'(r)$.
We have also introduced auxiliary variables $\Psi(r) \equiv \psi'(r)$, $\Phi(r) \equiv \phi'(r)$
and $A(r) \equiv \al'(r)$ in order to cast the above system of nonlinear ODEs in a 
canonical first-order form.
We assert that for any given value of $\ph(0) \equiv \ph_0(0)$,
the system~(\ref{bs1divp1})-(\ref{bs1divp6}) 
constitutes an eigenvalue problem with eigenvalue $\om = \om(\ph(0))$.  That is, for any specific 
value of $\ph(0)$ (which one can loosely view as being related to the central density of 
the star), a solution of~(\ref{bs1dansatz}) that satisfies the appropriate regularity and 
boundary conditions will only exist for some specific 
value of $\om$.~\footnote{
In fact, in general one expects an infinite, discrete set of eigenvalues, $\omega_i, \; i
=0,1,,\cdots$ (for any 
value of $\ph(0)$), corresponding to ``wave functions'' $\ph_i(r)$ with increasing number 
of nodes.  We restrict attention here to ``ground state'' boson stars, where $\ph_0(r)$ has 
no nodes.
}
The system~(\ref{bs1divp1})-(\ref{bs1divp6}) must be supplemented by boundary conditions, some of which 
are naturally applied at $r=0$, with the rest naturally set at $r=\infty$.  In particular,
regularity at $r=0$ implies 
\bea
\Ps(0) &=& 0\,, \\
\Ph(0) &=& 0 \,, \\
A(0) &=& 0 \,,
\eea
\noi
while at the outer boundary, we have
\bea \lab{psouterBC}
 \lim_{\rtoi}\ps(r) &=& 1 - \fr{C}{r} \,, \\ 
 \lim_{\rtoi}\ph(r) &\approx& 0\,,\\ \lab{alouterBC}
  \lim_{\rtoi}\al(r) &=& \fr{2}{\ps} - 1\,.
\eea
\noi
Here the second condition follows from the expectation that $\ph$ should decay exponentially
as $r\to0$.  (The outer boundary conditions for 
$\ps$ and $\al$ are derived in detail in 
App.~\ref{ap3}.)

We further note that due to the homogeneity and linearity of the slicing equation, we 
can always arbitrarily (and conveniently) choose the central value of the lapse via
\beq
 \al(0) = 1\,,
\eeq
\noi
and then, after integration of~(\ref{bs1divp1})-(\ref{bs1divp6}), can rescale $\al$ and $\om$ 
simultaneously to satisfy the outer boundary condition for $\al$:
\bea
 \al(r) &\longrightarrow& c\, \al(r) \,,\\
 \om &\longrightarrow& c\, \om \,.
\eea
\noi
where $c$ is given by 
\beq
 c = \fr{2/\ps(r_{\rm max}) - 1}{\al(r_{\rm max})}\,,
\eeq
\noi
and $r_{\rm max}$ is the radial coordinate of the outer boundary of the 
computational domain.  

As mentioned above, any solution of~(\ref{bs1divp1})-(\ref{bs1divp6}) can
be conveniently labelled by the central value of the modulus of the scalar 
field, $\ph_0(0)=\ph(0)$.  For any given value of $\ph_0(0)$, we must then determine the 
eigenvalue, $\omega$, and in the current case of maximal-isotropic coordinate, the 
central value of the conformal factor $\psi(0)$, so that all of the boundary conditions 
are satisfied.  In principle, we can compute pairs $[\omega,\psi(0)]$ as a function
of $\ph_0(0)$ using a two-parameter ``shooting'' technique~\cite{numrec,jason:phd}.

The process outlined above is the most direct approach for
finding boson star solutions in maximal-isotropic coordinates.  The principal
disadvantage lies in the fact
that we found it difficult to implement a convergent 2-parameter shooting method 
that allows one to compute values $[\omega,\psi(0)]$  given $\ph_0(0)$.

\subsection{Polar-Areal Coordinates} \lab{PAcoord}
In this subsection we describe a technique for generating boson star initial data 
in maximal-isotropic coordinates by first constructing the stars in so-called 
polar-areal coordinates, and then performing a coordinate transformation. 

Polar-areal coordinates, which have seen widespread use in spherically symmetric 
computations in numerical relativity, can be viewed as the generalization of the 
usual Schwarzschild coordinates to {\em time-dependent}, spherically symmetric 
spacetimes. As with maximal slicing, the slicing condition in this case---known
as polar slicing---is expressed as a condition on the mean extrinsic curvature:
\beq
 K =\krr\,.
\eeq
\noi
Since in general we have $K = K^i{}_i = K^r{}_r + 2K^\theta{}_\theta$, this condition 
is implemented by requiring 
\beq
	K^\theta{}_\theta(t,r) = {\dot K^\theta{}_\theta}(t,r) = 0  \,,
\eeq
for all $t$ and $r$.

The spatial coordinates are fixed by demanding that the coordinate $r$ measure 
proper surface area (i.e.\ that it be an {\em areal} coordinate), which, in terms 
of the general form~(\ref{sssmetric}) implies that 
\beq
 b = 1\,.
\eeq
\noi
It can be shown that this combination of polar slicing and areal spatial coordinate
induces further simplification in the 3+1 form of the metric---namely the shift 
vector component $\beta$ identically vanishes, so we are left with a metric 
of the form
\beq
 ds^2 = -\al^2 dt^2 + a^2 dr^2 + r^2 d \Om^2\,.
\eeq
\noi

As before, to construct star-like solutions, we adopt the time-harmonic ansatz~(\ref{bs1dansatz})
for the complex scalar field,
adapt the the time coordinate to the timelike Killing vector field, and require the
spacetime to be static.  We again find that the extrinsic curvature tensor vanishes 
identically (so that, for static data, the slicing is maximal as well as polar),
and that the momentum constraint~(\ref{momc}) is automatically satisfied.

Again, considering the Hamiltonian constraint, the Klein-Gordon equation, and the slicing condition
\beq
\dot{\ktt} = 0\,,
\eeq
\noi
at $t=0$, we have (dropping the subscript 0's  as before):
\bea \lab{mcsfivpeq1}
  a' & = & \ha \left\{ \fr{a}{r} \left( 1-a^2\right) + 4 \pi r a \left[ \ph^2 a^2
\left( m^2 + \fr{\om^2}{\al^2}\right) + \Ph^2 \right] \right\} \,, \\
\al' & = & \fr{\al}{2}\left\{ \fr{a^2-1}{r} + 4 \pi r \left[  a^2 \ph^2
\left(  \fr{\om^2}{\al^2} - m^2 \right) + \Ph^2 \right]\right\} \,, \\
  \ph' & = & \Ph \,, \\ \lab{mcsfivpeq4}
  \Ph' & = & - \left( 1+a^2 - 4 \pi  r^2 a^2 m^2\ph^2 \right) \fr{\Ph}{r}
- \left(  \fr{\om^2}{\al^2} - m^2\right) \ph a^2\,.
\eea
\noi
In this case, the regularity conditions are
\bea
 a(0) &=& 1 \, , \\
 \Ph(0) &=& 0\,,
\eea
\noi
while the outer boundary conditions are
\bea
 \lim_{\rtoi}\ph(r) &\approx & 0\,, \lab{phouterBC} \\
 \lim_{\rtoi}\alpha(r) &=& \frac{1}{a(r)} \,.
\eea
As before, we can convert the last condition to an {\em inner} condition on $\alpha$ by 
taking advantage of the linearity and homogeneity of the slicing equation.  Specifically, 
we can again choose $\alpha(0)=1$, and then after integration of (\ref{mcsfivpeq1})-(\ref{mcsfivpeq4})
simultaneously rescale $\al(r)$ as well as the eigenvalue, $\om$, so that the outer boundary 
condition for $\alpha$ is satisfied.

We again consider the family of boson star solutions parametrized by the central
value of the modulus of the scalar field, $\ph_0(0)$.  In this case, given a value of 
$\ph_0(0)$, and using the conditions $a(0)=1$, $\alpha(0)=1$, $\Phi(0)=0$, we need only 
adjust the eigenvalue $\omega$ itself in order to generate a solution with the appropriate 
asymptotic behaviour (i.e.\ so that $\lim_{r\to\infty} \phi(r) = 0$). This is a classic
1-parameter shooting problem, which is comparatively easier than the 2-parameter shooting 
method described in the previous section.

Once we have computed a solution in areal coordinates, we can perform a coordinate
transformation from areal coordinates to isotropic coordinates \cite{dinverno,LPPT} (recall
that the maximal and polar slices coincide for the static case).
Essentially this amounts to solving an ODE of the form

\bea  \nn
 \left. r\,\right|_{R=R_\mathrm{max}} &=& \left[\left(
\fr{1+\sr{a}}{2}\right)^2 \fr{R}{a}\right]_{R=R_\mathrm{max}}\,, \\
 \lab{ODEb} \fr{dr}{dR} &=& \, a \fr{r}{R}\,.
\eea

\noi
and the details of the
transformation are given in App.~\ref{ap3}.

\subsection{Compactified Maximal-Isotropic Coordinates} \lab{compactMI}

The two methods discussed thus far for determining boson star initial data both use 
a ``shooting'' method to fix one or more parameters (an eigenvalue, and possibly an
initial condition) so that the solution of a system of ODEs satisfies all appropriate 
regularity and asymptotic conditions corresponding to a star-like solution.  The 
shooting technique is inherently iterative and, particularly for the two-parameter 
case, one needs a fairly good initial estimate of the unknown parameters for the 
iteration to converge quickly.

An alternate method, which seems to be the best for computing boson star
initial data in maximal-isotropic coordinates (without resort to coordinate 
transformation) involves the use of a compactified spatial coordinate and the 
solution of the eigenvalue problem using standard linear algebra software. 
This approach has the further advantage of generalizing quite easily to problems
with more than one spatial dimensions; i.e.\ to boundary value PDEs with eigenvalues, although
it must be stressed that naive application of standard
linear algebra software for the eigenvalue problem is in general inefficient for higher dimensional
problems (see the discussions in \ref{challenges}).  (Such inefficiency can be 
remedied, however, by the use of a more efficient algorithm such as the multigrid
eigenvalue method discussed in Sec.~\ref{numstr}.)
Shooting methods, on the other hand, are generally restricted to systems of ODEs.

The key idea here is simply to introduce a compactified coordinate, $\zeta$,  as previously discussed 
in Chap.~\ref{num_method}:

\beq
 \zeta = \fr{r}{1+r} \,,
\eeq
\noi
where $\zeta \in [0,1]$.  Under the above transformation we have:
\bea
  r &=& \fr{\zeta}{1-\zeta} \,, \\
 \fr{d}{dr} &=& (1-{\zeta})^2 \fr{d}{d {\zeta}} \,,\\
  \na^2_r &=& (1- \ze)^4 \na_\zeta^2 \,,
\eea

\noi
where $\na^2_r \eq 3 \fr{d}{d r^3} \left( r^2 \fr{d}{d r}\right)$, and
$\na^2_\zeta \eq 3 \fr{d}{d \ze^3} \left( \ze^2 \fr{d}{d \ze}\right)$.

With this definition the system of ODEs~(\ref{bs1divp1})-(\ref{bs1divp6}) can be rewritten as 
\bea  \lab{bs1divpc1}
\nabla_\zeta^2 \ph + \left( \fr{A}{\al} + \fr{2 \Ps}{\ps}\right) \Ph -
\fr{\ps^4}{(1-{\zeta})^4} \left( m^2 - \fr{\om^2}{\al^2}\right) \ph &=& 0
\,, \\
\nabla_\zeta^2 \ps + \pi \left[ \ps \Ph^2 + \fr{\ps^5}{(1-{\zeta})^4} \left(
\fr{\om^2}{\al^2} + m^2 \right) \ph^2 \right] &=& 0 \,, \\ \lab{bs1divpc6}
\nabla_\zeta^2 \al + \fr{2 \Ps}{\ps} A - 4 \pi \fr{\ps^4 \al}{(1-{\zeta})^4}
 \left( \fr{2 \om^2}{\al^2 }
- m^2 \right) \ph^2  &=& 0 \,.
\eea
\noi
As explained in Chap.~\ref{num_method}, the advantage of using a compactified spatial coordinate is that 
we can precisely impose the boundary conditions at infinity in our numerical computations.
Specifically, in the current case we have $\left.\ph_0(\ze)\right|_{\ze=1} = 0$,
$\left.\ps(\ze)\right|_{\ze=1} = 1$ and $\left.\al(\ze)\right|_{\ze=1} = 1$.

After finite differencing~(\ref{bs1divpc1})-(\ref{bs1divpc6}) we are left with a standard nonlinear eigenvalue problem 
that can be treated numerically using standard linear algebra packages.  
Details of our implementation are described in App.~\ref{ap2}.

\subsection{Family of Stationary Solutions} 
\begin{figure}
\begin{center}
\epsfxsize=16.0cm
\includegraphics[width=16.0cm,clip=true]{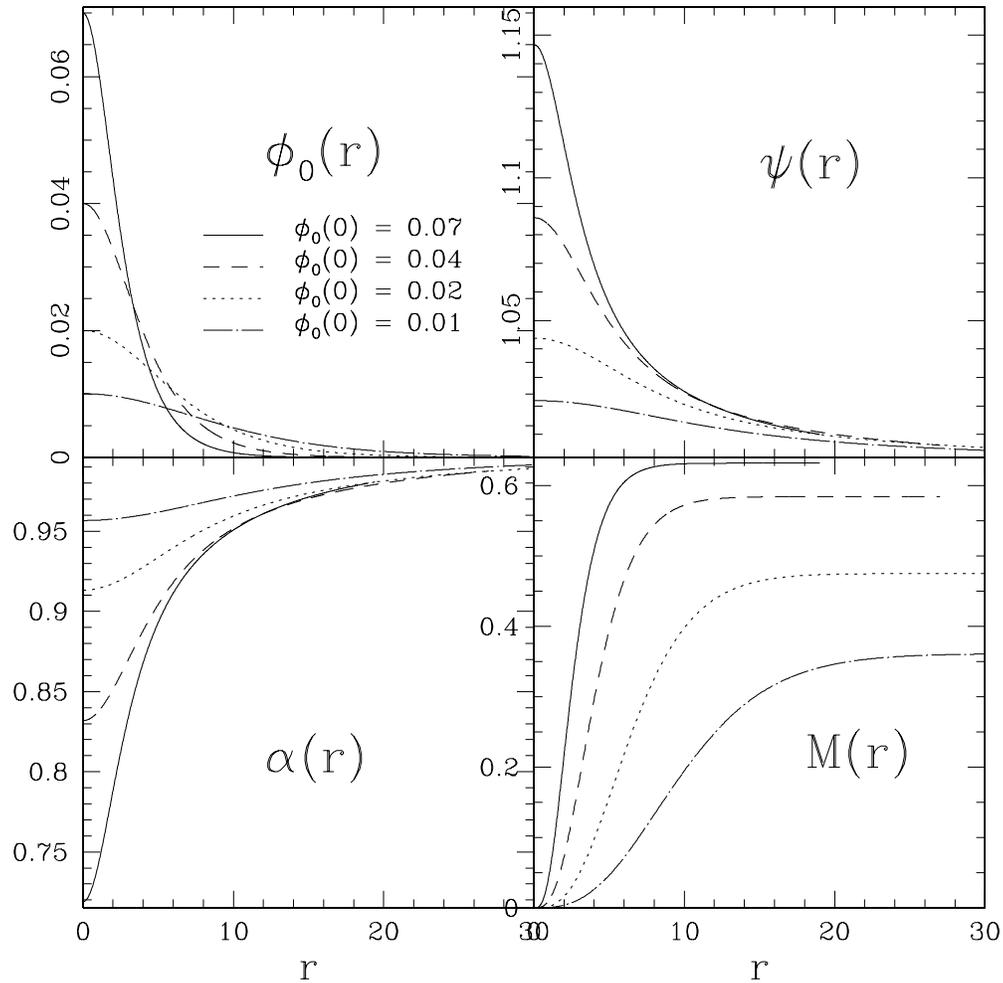}
\end{center}
\caption
[Static spherically symmetric boson stars]
{Static spherically symmetric boson stars. 
Plots of $\ph_0(r), \ps(r), \al(r)$ and $M(r)$ for boson stars with 
$\ph_0(0) = 0.01,0.02, 0.04$ and $0.07$ in maximal-isotropic coordinates.  In
general, for boson stars such as these which are on the stable branch, 
the configurations become more compact and of higher mean density as the 
central scalar field value increases.
}
\label{bs1divp}
\end{figure}

\begin{figure}
\begin{center}
\epsfxsize=8.0cm
\includegraphics[width=8.0cm,clip=true]{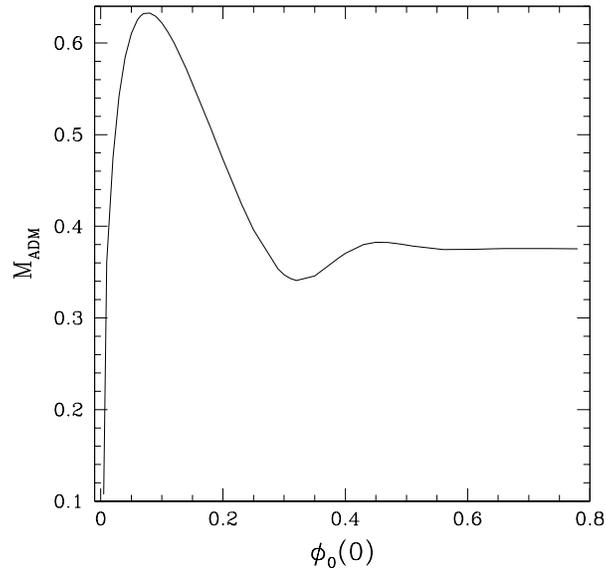}
\end{center}
\caption
[ADM mass $M_{\mathrm{ADM}}$ of spherically symmetric boson stars {\em vs} the central scalar field value $\ph_0(0)$]
{ADM mass $M_{\mathrm{ADM}}$ of spherically symmetric boson stars {\em vs} the central scalar field value $\ph_0(0)$.  
Note that there exists a maximum value of the ADM mass---$M_{\mathrm{max}} \approx 0.633 M_{\rm pl}^2/m$---above 
which there are no static solutions.  Note also that each extremum point satisfy $d M_{\rm ADM}(\ph_0(0))/d
\ph_0(0) = 0$, which corresponds to change in sign of the eigenvalue $\si^2$ of the pulsation equation
(See Sec.~\ref{stability_BS}).}
\label{bs1dfs}
\end{figure}

\begin{figure}
\centering
\includegraphics[width=7.3cm]{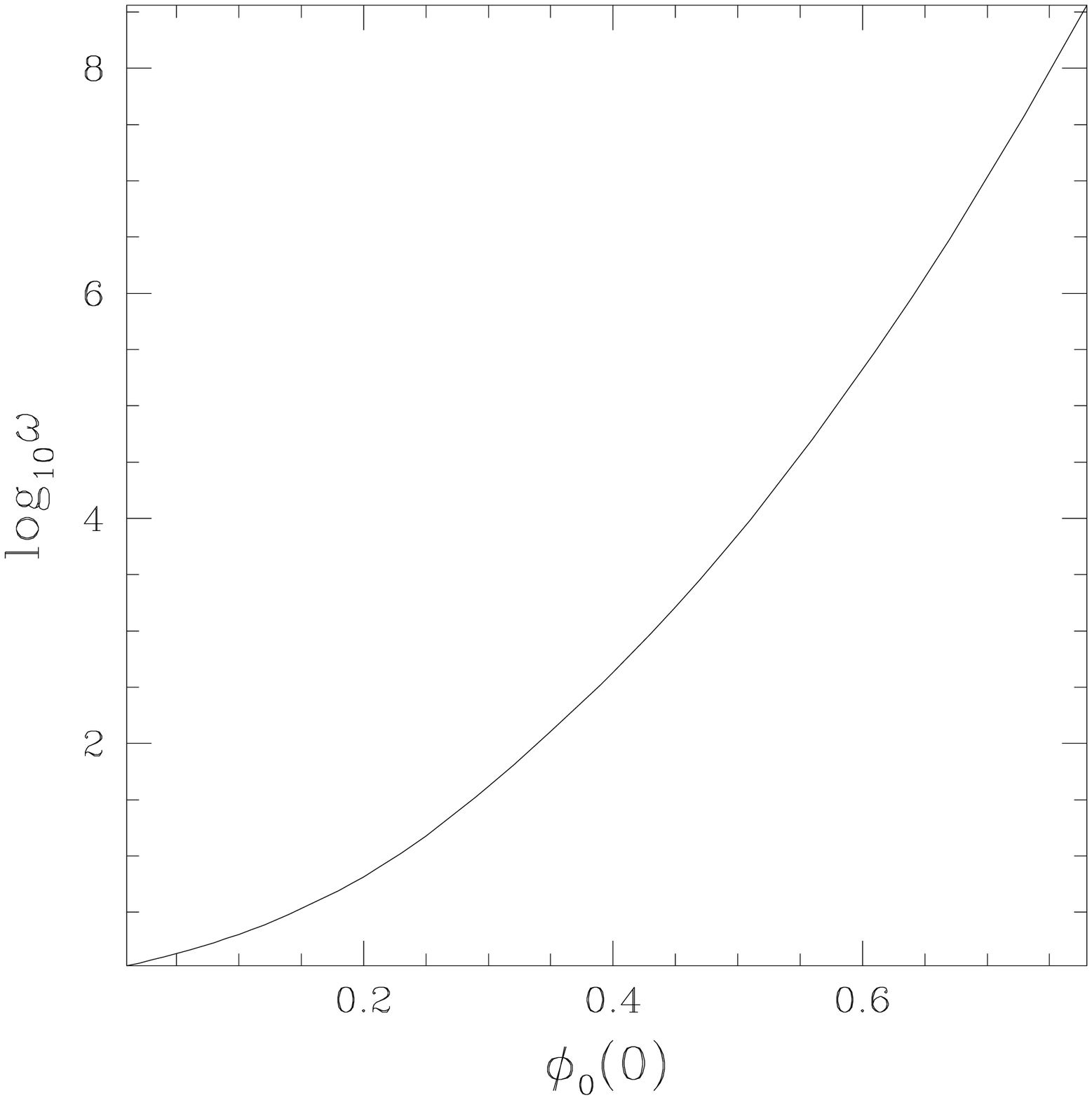}
\includegraphics[width=7.3cm]{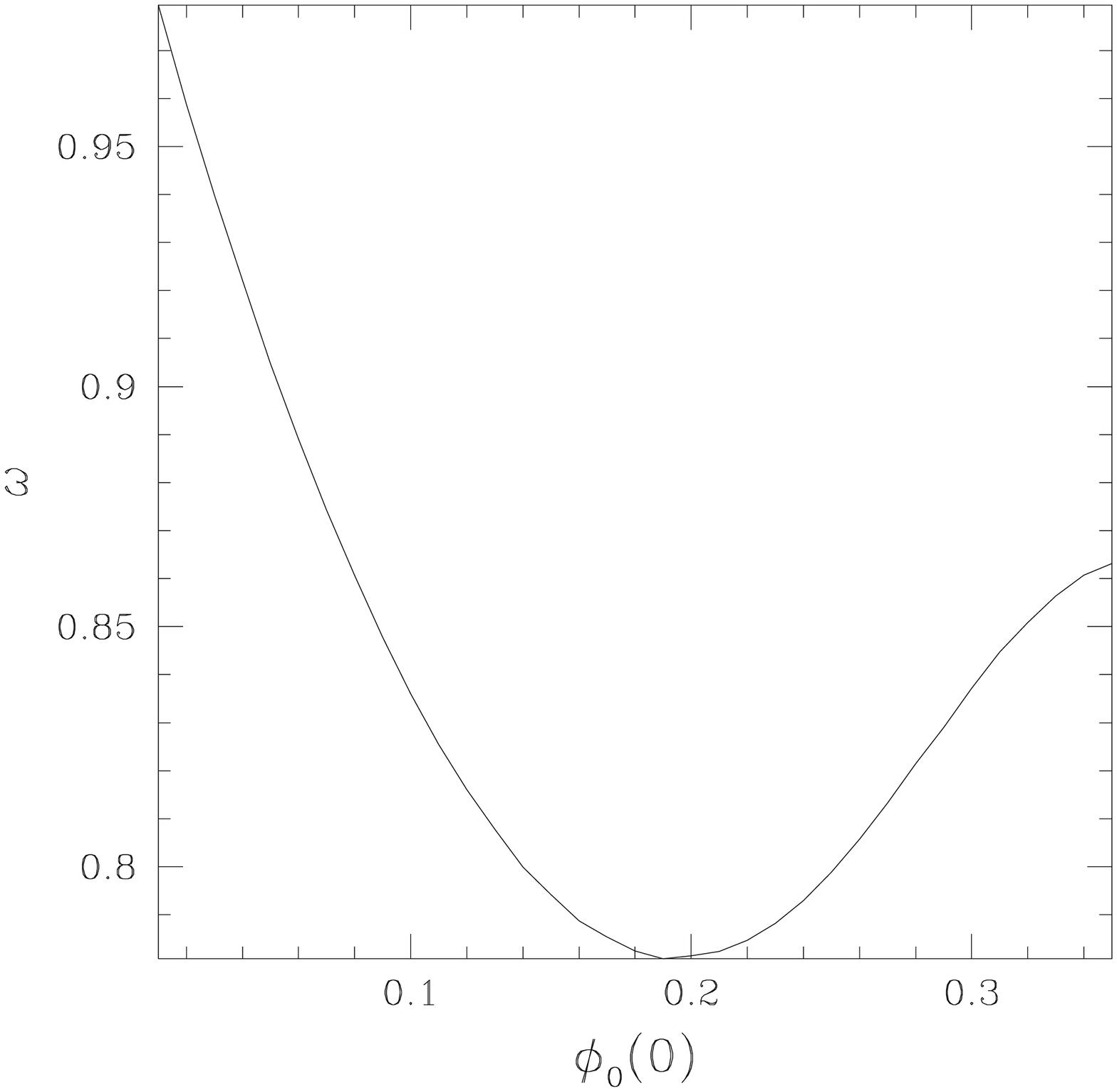}
\caption
[Eigenvalue, $\om$, of spherically symmetric boson stars {\em vs} the central scalar field value $\ph_0(0)$]
{Eigenvalue, $\om$, of spherically symmetric boson stars {\em vs} the central scalar field value $\ph_0(0)$, for the system
(\ref{mcsfivpeq1})-(\ref{mcsfivpeq4}). The figure on the left shows the value of
$\log_{10}(\om)$ before rescaling of $\al$.  The figure on the right shows the value of $\om$ after rescaling.}
\label{bs1dfs2}
\end{figure}

\lab{family_statsol}
For completeness, we display a few typical stationary solutions representing 
spherically symmetric boson stars in Fig.~\ref{bs1divp}.  Specific properties of the entire 
one-parameter family of stars are 
shown in Fig.~\ref{bs1dfs} (ADM mass {\em vs} central scalar field value $\phi_0(0)$) and 
Fig.~\ref{bs1dfs2} (eigenvalue $\omega$ {\em vs} $\phi_0(0)$). 
These diagrams will subsequently be compared with similar plots generated for
families of axisymmetric, rotating, stationary solutions in Chap.~\ref{bs2d}.

We remark that the boson stars become more compact (in other words, have
smaller and smaller effective radii) as $\phi_0(0)$ increases.  We also note that,
as is apparent from Fig.~\ref{bs1dfs}, there exists a maximum mass, $ M_{\mathrm{max}} \approx 0.633
M_{\rm pl}^2/m$, for the family of stars, corresponding to a central scalar
field value $\ph_0(0) \approx 0.08$.  Thus this model exhibits the analog of a 
Chandrasekhar mass limit, above which no static configurations exist, although
we should emphasize, that in contrast to the white-dwarf case, the existence of 
an upper bound on the mass of boson stars is purely a relativistic effect.  In particular,
no such upper bound exists for {\em non-relativistic Newtonian} boson stars.
Furthermore, perturbation analysis, as well as full simulations,  show that
configurations depicted in Fig.~\ref{bs1dfs} that lie to the left of the mass maximum
(i.e.\ for $\ph_0(0) < 0.08$) are {\em stable} to radial perturbations, 
while those to the right ($\ph_0(0) > 0.08$) are {\em unstable} to such 
perturbations.  Indeed, this instability will play a crucial role in our 
analysis of critical behaviour in the model which we consider in the next section.

\section{Critical Phenomena of Boson Stars} \lab{critphenBS}
In the remainder of this chapter we focus on black-hole threshold behaviour 
(or critical behaviour) in the Einstein-Klein-Gordon system.  In particular, 
following Hawley~\cite{shawley:phd} and Hawley and Choptuik~\cite{scott_matt:2000} we study black-hole 
threshold solutions that are generated by gravitationally ``perturbing'' stable 
boson stars using an additional massless scalar field as the perturbing agent.  
The massless field is {\em not} explicitly coupled to the complex, boson 
star field, but the two fields {\em can} and {\em do} interact through the 
gravitational field.  The previous calculations~\cite{shawley:phd,scott_matt:2000} have provided 
strong evidence that the black hole transition in this case is Type I (so 
that the black hole mass just above threshold is generically {\em finite}), with 
the critical solution itself being a perturbed boson star on the {\em unstable} branch
(i.e.\ lying to the right of the mass maximum in Fig.~\ref{bs1dfs}).   In addition, scaling laws 
of the form 
\beq
\label{scaling}
	\tau(p) \sim -\gamma \ln | p - p^\star | \,,
\eeq
were observed for near-critical evolutions, where $p$ is the family parameter, $p^\star$ is 
the critical parameter value that demarks the black hole threshold, $\tau$ is the length 
of time that the evolution remains ``close'' to the (static) critical solution (i.e.\ the 
``lifetime'' of the near-critical solution), and $\gamma$ is a scaling exponent that depends 
only on which of the infinitely many critical solutions is generated by the $p=p^\star$ evolution.
Furthermore, in accord with the now-familiar picture of one-mode instability of critical solutions,
the values of $\gamma$ determined from the scaling relation~(\ref{scaling}) were shown to 
be in agreement with the reciprocal Lyapunov exponents associated with the single unstable mode of the 
(unstable) boson star which was a best match to the mean motion of the near critical evolutions.
In this regard we should point out that with reference to Fig.~\ref{bs1dfs}, we expect 
a change of stability at each extremum of the plot of $M_{\rm ADM}$ {\em vs} ${\phi_0(0)}$.
In particular, those configurations to the right of the absolute maximum of the plot, but to 
the left of the subsequent local minimum (near $\phi_0(0) = 0.3$) each have a single unstable mode 
in perturbation theory, and thus each can act as an ``intermediate attractor'' in critical 
collapse.

As we will see,
our current calculations confirm these basic results (and do so in a different coordinate 
system than that used in the previous investigation), but also provide clear indications that the end-state of 
subcritical evolutions is quite different than that which was proposed in~\cite{shawley:phd,scott_matt:2000}.

\subsection{Setup of Numerical Experiments}
The PDEs solved in the simulations discussed here are those 
listed in Sec.~\ref{MIcoord}, including the equation of motion for the 
massless scalar field.  We also provide a summary of the equations of motion of the system,
the boundary conditions, and details
of the finite difference approximation used in Apps.~\ref{ap5} and \ref{ap4}.
Additionally, results of convergence tests of the code are discussed in App.~\ref{ap6}.

In order to study critical behaviour in the model we start with initial data for 
the complex field that represents a boson star on the stable branch (i.e.\ a star 
with a central scalar field value $\ph_0(0) < 0.08$).  We generally choose a configuration
that is reasonably relativistic, i.e.\ with $\ph_0(0)$ bounded away from 0, but not too 
close to the instability point, $\ph(0) \approx 0.08$. 

To drive the boson star to criticality, we implode a (spherical) shell of massless scalar
field on to it.  Specifically, we choose initial data for the massless field of the 
following ``gaussian'' form
\beq
  \label{ph3-gaussian}
  \ph_3(0,r) = A_3 \, \exp \left[ - \left( \fr{ r - r_0}{\si} \right)^2 \right] \, ,
\eeq
where $A_3, r_0$ and $\si$ are adjustable parameters, controlling the overall 
amplitude, position and width, respectively, of the imploding gaussian wave packet.  To ensure that 
the massless field is almost purely in-going at the initial time, we specify the 
``conjugate'' variable $\Pi_3 \equiv \ps^2 / \alpha \left( \dot{\phi_3}  - \bt {\ph_3}'\right)$ as follows:
\beq
 \Pi_3(0,r) = -\left(\Ph_3(0,r) + \fr{\ph_3(0,r)}{r}\right) \, .
\eeq
In all of our studies described here, we have fixed $r_0$ and $\si$ in~(\ref{ph3-gaussian})
to $r_0 = 40$ and $\si=5$.  This ensures that the support of the massless field is well 
separated from that of the complex field (i.e.\ from the boson star {\em per se}) at the initial 
time.  

A typical evolution of initial data of the form described above proceeds as follows.  Once 
we have fixed the boson star configuration, we complete the specification of the massless 
scalar field initial data by fixing the overall amplitude factor, $A_3$, and then 
start the simulation. 
Initially, the shell of massless scalar field implodes towards $r=0$ at the speed of light, while 
the boson star ``sits'' in its static state centered at the origin.  As the in-going massless shell
reaches the region of space occupied by the boson star, its contribution to the overall 
gravitational field tends to compress the boson star to a higher mean density and smaller radius.  The 
massless field passes through the origin and then ``explodes'' outward, eventually propagating 
off the computational domain.  Depending on the strength of the perturbation from the massless 
field, we find that the compressed boson star either relaxes to something resembling a stable boson
star with large-amplitude oscillations, or collapses to form a black hole.  Thus by adjusting the 
massless scalar amplitude factor, $A_3$---which we generically use as the adjustable parameter, $p$, in
our study of critical behaviour in the model---we can tune the evolution to the threshold of black 
hole formation.   In practice we use a bisection search to refine our estimate of the critical value,
$A_3^\star$, and can carry the search to machine precision (8-byte real arithmetic), so that 
$\Delta A_3 / A_3 \sim 10^{-15}$.  Typically we have chosen the number of mesh points $N_r =
1025$ on $0\le r\le 50$, a Courant factor $\De t/\De r = 0.3$, and the coefficient of
Kreiss-Oliger dissipation $\ep_d = 0.5$ (see App.~\ref{ap4} for the definition of $\ep_d$).
Note that there is a whole family of critical solutions (see Fig.~\ref{transit}) for different
initial data.  Therefore if we, for instance, changed $r_0$ from $r_0 = 40$ to $r_0 = 35$ in
(\ref{ph3-gaussian}), we would find that $A_3^\star$ changes, i.e.\ $A_3^\star = A_3^\star(r_0,\si)$,
and therefore, in general, the critical solution would also change.

In the following section we discuss results from detailed studies of black hole threshold solutions
generated from several distinct initial boson star states.  Table~\ref{table} summarizes the 
values of $\phi_0(0)$  that were used, the corresponding values of $A_3$ required to 
generate a critical solution, and the figures that display results associated with the 
respective calculations. 
Since we will not dwell on this point below, we note that all of our calculations confirm the basic picture 
previously reported that the black holes that form just above threshold in this type of collapse 
generically have {\em finite} mass (i.e.\ that the critical transition is Type I).

\vspace{0.5cm}
\begin{table}[htbp]
\begin{center}
\begin{tabular}[l]{| c | c | c |}
\hline
{\bf Fig.}             & ${\mathbf{ \ph_0(0)}}$ & $\mathbf{A_3}$ \\ \hline
\hline
\ref{dmdr}       & 0.05       & 0.0032     \\
\ref{modph}, \ref{transit}& 0.035, 0.04, 0.05     & 0.00471, 0.00342, 0.00316 \\
\ref{t_vs_lndp} & 0.02, 0.035, 0.04, 0.05 &  0.00915, 0.00471, 0.00342, 0.00316 \\
\ref{tmrphps} & 0.04 & 0.00342\\ 
\ref{bhexcise}, \ref{MdotM}, \ref{m_012outer_rmax150}, \ref{m_012outer_rmax300} \ref{dm_012outer} & 0.01       & 0.041      \\
\hline
\end{tabular}
\caption[Summary of parameters used to generate results displayed in Figs.~\ref{dmdr}-\ref{dm_012outer}]{
Summary of parameters used to generate the results displayed in Figs.~\ref{dmdr}-\ref{dm_012outer}.
Listed are the figure number, central amplitude of the complex field, $\ph_0(0)$, and the overall
massless scalar amplitude factor, $A_3^\star$ (see~\ref{ph3-gaussian}), that generates a marginally-critical 
solution in each case.  Other parameters defining the massless scalar initial profile~(\ref{ph3-gaussian})
are held fixed at $r_0 = 40$, $\si = 5$ for all simulations.
}
\end{center}
\label{table}
\end{table}

\subsection{Critical Phenomena: Results} \lab{CPresults}
We start by examining results from a critically perturbed boson star having an unperturbed central 
field value $\phi_0(0) = 0.05$.  As just described, the critical massless amplitude factor, 
$A_3^\star \sim 0.0032$ was determined by performing a bisection search on $A_3$, to roughly 
machine precision. (Recall that each iteration in this search involves the solution of the 
time-dependent PDEs for the model for a specific value of $A_3$, with all other parameters held 
fixed, and the criterion by which we adjust the bisection bracket is whether or not the simulation 
results in black hole formation. )  

\begin{figure}
\begin{center}
\epsfxsize=13.0cm
\includegraphics[width=13.0cm,clip=true]{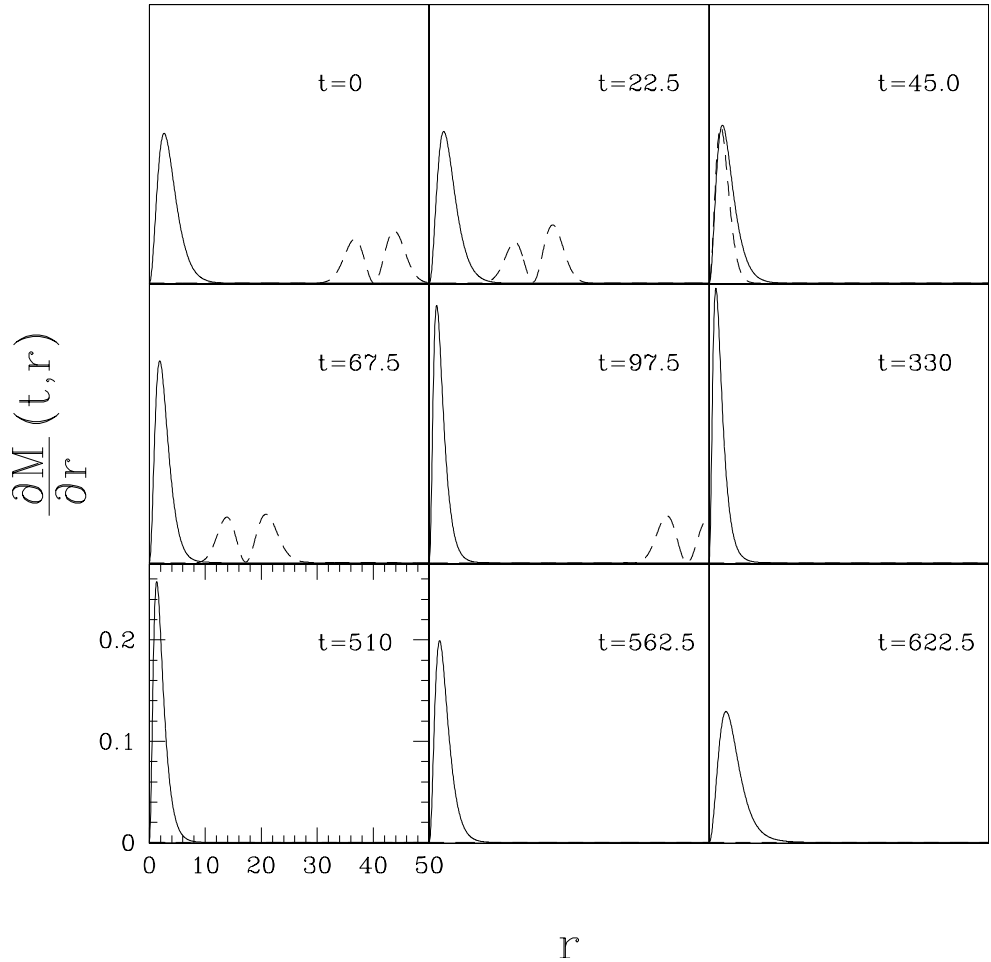}
\caption
[Critical evolution of a perturbed boson star with $\ph_0(0) = 0.05$
and mass $M_C = 0.62 M_{Pl}^2/m$]
{Critical evolution of a perturbed boson star with $\ph_0(0) = 0.05$
and mass $M_C = 0.62 M_{Pl}^2/m$.  This figure shows the time development of 
contributions to $\pa M/\pa r$ from the complex (solid line) and real (dashed
line) scalar fields. 
Note that the temporal spacing between successive snapshots is {\em not} constant---the time instants
displayed have been chosen to illustrate the key features of the near-critical evolution.
Also note that we have multiplied the value of $\pa M/\pa r$
for the {\em real} scalar field by a factor of 8 to aid in the visualization of that 
field's dynamics.
The evolution begins with a stable boson star centered
at the origin, and an in-going gaussian pulse (shell) of massless, real scalar field 
that is used to perturb the star.  The overall amplitude factor, $A_3$,  of the initial real scalar 
field profile (see~(\ref{ph3-gaussian}), is the control 
parameter for generating the one-parameter family of solutions that interpolates through the 
black hole threshold. For the calculation shown
here, $A_3$ has been tuned to a critical value $A_3^\star \approx 0.0032$ via a bisection search (and 
with a fractional precision of $\approx 10^{-15}$).  
The other parameters defining the gaussian initial
profile of the massless fields are $r_0 =40$ and $\si=5$.
The snapshots show that the real scalar field enters the region 
containing the bulk of boson star at $t\approx 22$, implodes through the origin
at $t \approx 45$, leaves the boson star region at $t\approx70$, and, finally, completely 
disperses from the computational domain at $t \approx 100$.  The boson star enters the critical
state at roughly the same time that the real field leaves the domain, and remains in that 
state for a period of time which is long compared to the crossing time of the massless field.
At $t\approx 510$, the boson star departs from the critical state. 
}
\label{dmdr}
\end{center}
\end{figure}

A series of snapshots of $\pa M(t,r)/\pa r$ (where $M(t,r)$ is the mass aspect function) for a marginally 
subcritical evolution is shown in Fig.~\ref{dmdr}.
Full analysis of the results of this simulation indicate that the boson star enters what
we identify as the critical
state at $t \approx 130$, and remains in that state until $t \approx
510$.  It is worth noting that the boson star actually completes its collapse into a more 
compact configuration well after the real scalar field has dispersed from the boson star 
region.   We also note that the amount of time, $\tau$, spent in the critical 
state---$\tau \approx 380$ in this case---is a function of how closely the control parameter
has been tuned to criticality.  Specifically, we expect $\tau$ to be linear in $\ln|A_3 -
A_3^\star|$
(see (\ref{scaling})), and we will display evidence for this type of scaling below.

\begin{figure}
  \begin{center}
\epsfxsize=12.0cm
\includegraphics[width=12.0cm,clip=true]{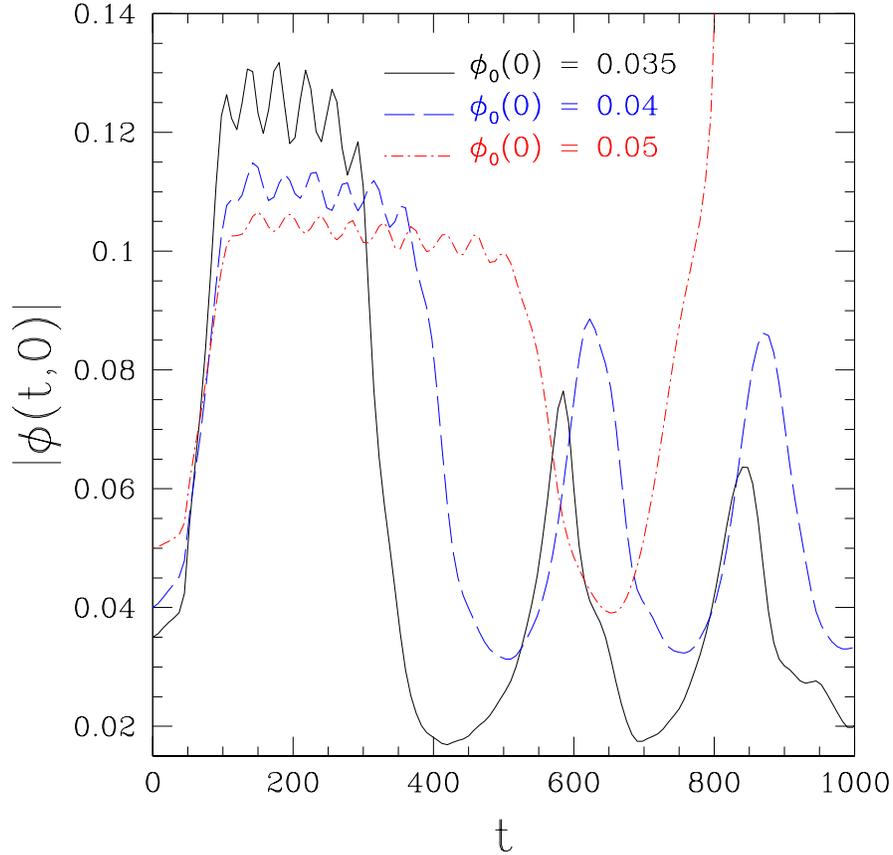}
    \caption
[Time evolution of central value of the modulus of scalar field for subcritical evolution of 
perturbed boson stars]
{Time evolution of central value of the modulus of scalar field for subcritical evolution of perturbed 
boson stars.
The figure shows the time evolution of $|\ph(t,0)|$ for marginally subcritical evolutions 
generated from boson star
initial states with $\ph_0(0) = 0.035, 0.04$ and
$0.05$.  See the text for a description of key features of this plot.
} 
    \label{modph}
  \end{center}
\end{figure}

Fig.~\ref{modph} shows the time evolution of the central modulus of the complex
scalar field for marginally subcritical evolutions generated from boson star 
initial states with $\ph_0(0) = 0.035, 0.04$ and
$0.05$.  From the figure we can see that in all three cases the perturbed stars enter an excited, critical 
state at $t\approx 100$ and remain in that state for a finite time which is a function of $\phi_0(0)$
(i.e.\ of the initial state).  Additionally, at least for the cases $\phi_0(0) = 0.035$, $\phi_0(0) = 0.04$,
the figure provides evidence that following the critical evolution phase, the excited stars relax
to states characterized by large amplitude oscillations of the complex field.  This behaviour will be 
examined in more detail below.  Finally, also apparent in the plot are the smaller-amplitude oscillations 
during the periods of  critical evolution. Previous work~\cite{shawley:phd,scott_matt:2000} indicated
that these oscillations can be interpreted as excitations of the (stable) first {\em harmonic} mode of 
the unstable boson star that is acting as the critical solution---the 
unstable {\em fundamental} mode is the one that determines whether or not the configuration will
evolve to a black hole.
Although we have not studied this matter in any detail, we assume that the same picture holds for 
our current calculations.

\begin{figure}
  \begin{center}
\epsfxsize=12.0cm
\includegraphics[width=12.0cm,clip=true]{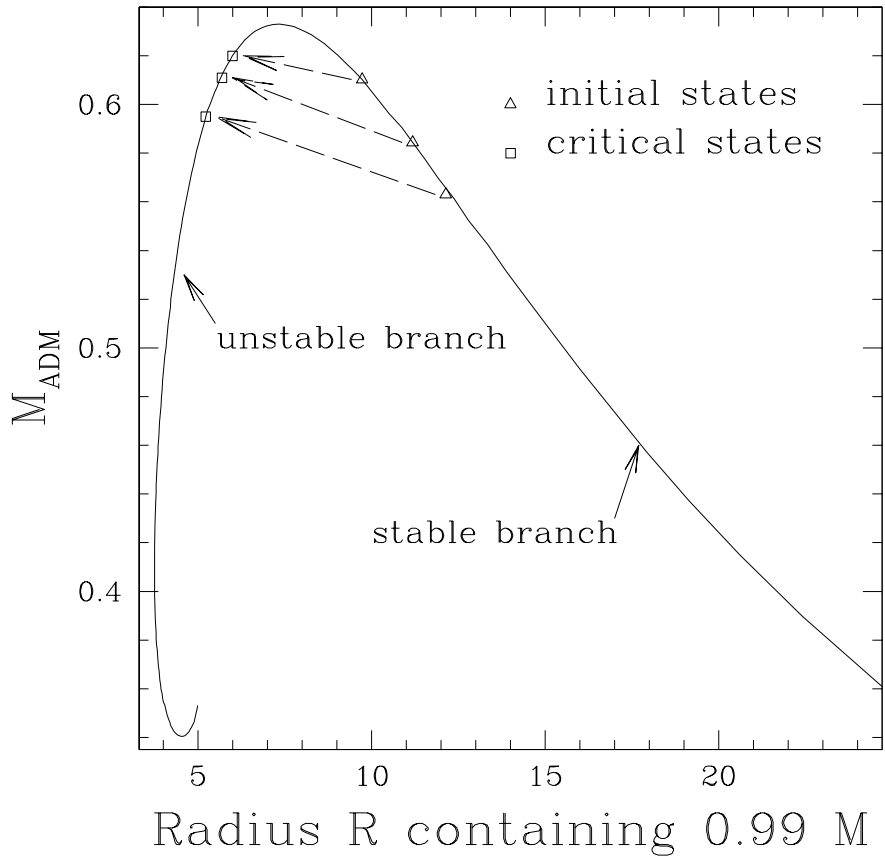}
    \caption
[Transition of perturbed boson stars in critical evolutions]
{Transition of perturbed boson stars in critical evolutions.  The solid curve shows the 
parametric mass {\em vs} radius plot of static boson stars (curve parameter, $\phi_0(0)$), where we 
have defined the stellar radius, $R$, so that $M(R) = 0.99 \, M(\infty) = 0.99 \, M_{\rm ADM}$.
Triangles label the initial configurations, squares show the corresponding critical 
solutions (identified as one-mode-unstable boson stars with oscillations---largely in
fundamental mode), and the dashed arrows represent schematically the transition between the initial and critical 
states. See the text for more details.}
    \label{transit}
  \end{center}
\end{figure}

The results from our simulations of critically perturbed boson stars are thus in agreement with 
previous studies~\cite{shawley:phd,scott_matt:2000} that identified the critical states as 
excited (primarily in the first harmonic mode), unstable boson stars.  Following that work
we can display an approximate correspondence between
the initial boson stars and the critical solutions as show in 
Fig.~\ref{transit}.  
The solid line shows the one-parameter family of static boson stars (parameterized as usual by
$\phi_0(0)$), where we have defined the radius, $R$, of a boson star so that $M(R) = 0.99 M_{\rm ADM} = 
0.99 M(\infty)$.  
The triangles indicate the initial stable boson star configurations, the squares indicate our 
best estimate of the corresponding unstable critical boson star states, and each arrow schematically 
depicts the transition between the two states that is induced by the perturbing scalar field. 
We note that to identify which unstable boson star is acting as the critical solution---which
is equivalent to identifying an effective value of $\phi_0(0)$---we time average 
the central modulus of the complex field, $\vert \phi(t,0) \vert$ during the period of critical evolution.
In addition, in accord with previous results, we observe that in all cases, the mass of the unstable 
critical state is {\em larger} than that of the progenitor boson star, indicating that a 
significant amount of mass-energy is extracted from the massless scalar field through its purely 
gravitational interaction with the complex field.

As discussed previously, for both subcritical and supercritical simulations,
the closer one tunes $A_3$ to the critical value $A_3^\star$, the longer the 
perturbed star will persist in the critical state.  Specifically, we observe scaling 
of the lifetime, $\tau$, of the critical evolution of the form
\beq
\label{scaling-2}
   \tau(A_3) \sim -\gamma \ln | A_3 - A_3^\star | \,,
\eeq
where we define the lifetime to be the lapse of coordinate time from the start 
of the evolution, $t=0$, to the time of first detection of an apparent horizon, and where $\gamma$ is 
a scaling exponent that depends on which of the infinitely many one-mode unstable boson stars acts as 
the critical solution in the particular scenario being simulated.  We note that 
the details of the definition of $\tau$ are not important to the determination of 
$\gamma$ in~(\ref{scaling-2}) since $\gamma$ actually measures the {\em differential} in lifetime 
with respect to changes in $A_3 - A_3^\star$, and this differential is insensitive 
to precisely how we define $\tau$,
at least as $A_3 \to A_3^\star$.  In addition, we note that in using coordinate time in our definition
of the scaling relationship~(\ref{scaling-2}), we are defining the scaling with respect to proper time 
at spatial infinity.  Another choice---arguably more natural---would be to define $\tau$ in terms of 
the proper time measured by an observer at rest at $r=0$ (central proper time).  Since the critical 
solutions are nearly static, the relation between these two different definitions of time would be 
a specific factor for each distinct value of $\phi_0(0)$, and would thus lead to a $\phi_0(0)$-dependent 
``renormalization'' of the scaling exponents, $\gamma$.

\begin{figure}
  \begin{center}
\epsfxsize=12.0cm
\includegraphics[width=12.0cm,clip=true]{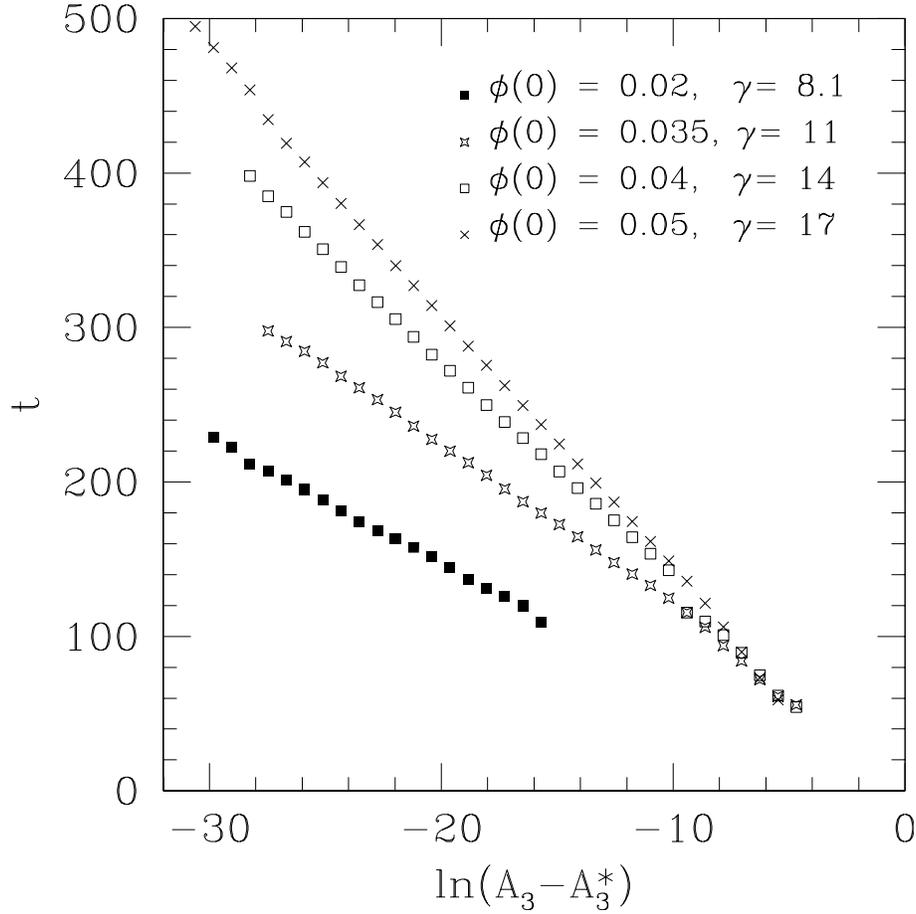}
    \caption[Measured lifetime scaling laws for critically perturbed boson stars]
{Measured lifetime scaling laws for critically perturbed boson stars.
This figure shows the measured lifetimes of various near-critical evolutions of 
perturbed boson stars as a function of $\ln|A_3 - A_3^\star|$, for cases with 
$\ph(0) = 0.02, 0.035, 0.04$ and $0.05$.  Quoted scaling exponents, $\ga$ 
(see~(\ref{scaling-2}) are computed from linear least-squares fits to the data.
The apparent convergence of the data for different $\ph_0(0)$ as $\ln\vert A_3-A_3^\star\vert \to 0$
is not significant, as it reflects calculations {\em  far} from criticality i.e.\ far from
the $\ln\vert A_3 - A_3^\star\vert \to -\infty$ limit. See the text for additional details.
}
    \label{t_vs_lndp}
  \end{center}
\end{figure}

Fig.~\ref{t_vs_lndp} shows measured scaling laws from supercritical evolutions of 
perturbed boson stars  defined by $\ph_0(0) = 0.02, 0.035, 0.04$ and 
$0.05$.  It is clear from
these plots that, at least as $A_3 \to A_3^\star$, we have lifetime scaling of the 
form~(\ref{scaling-2}).  Estimated values of $\gamma$---computed from linear least-squares 
fits to the plotted data---are $\ga = 8.1, 11, 14, 17$ for $\ph_0(0) = 0.02, 0.035, 0.04, 
0.05$, respectively.  We note that according to the now standard picture of critical 
collapse (see for example~\cite{gundlach:2003}), each value of $\ga$ can
be identified with the reciprocal Lyapunov exponent (i.e.\ growth factors) of the 
single unstable mode associated with the corresponding critical solution.  Again, the reason
that we observe different values of $\ga$ for different choices of initial boson star (different 
values of $\phi_0(0)$) is that distinct critical solutions are being generated in the 
various cases.  That is, we cannot expect universality (with respect to initial data) in this case 
because the model admits an entire family of one-mode unstable solutions that sit at the threshold 
of black hole formation.

\subsection{Final Fate of Subcritical Evolutions}
In previous work on the problem of critically perturbed spherically symmetric 
boson stars~\cite{shawley:phd,scott_matt:2000}, it was conjectured that the 
end state of subcritical evolution was characterized by {\em dispersal} of the boson 
star to large distances (relative to the size of the initial, stable star).
This conjecture was at least partially influenced by the behaviour observed, for 
example, in the collapse of a {\em massless} scalar field~\cite{choptuik}, where 
subcritical evolutions {\em do} involve complete dispersal of the field.  However, 
another key reason for what we claim is a misidentification of the true subcritical
end-state was that the simulations described in~\cite{shawley:phd,scott_matt:2000}
simply were not carried out for sufficient coordinate time to see the long-time
behaviour.  Our current simulations strongly suggest that subcritical evolutions
lead to a ``relaxation'' of the critically perturbed state to something that 
approximates a boson star (not necessarily the original star) undergoing large amplitude 
oscillations. As argued in the next sub-section, these oscillations
can largely be identified with the fundamental perturbative mode 
associated with the final boson star state. The numerical evidence also suggests that these
oscillating configurations eventually {\em  re-collapse} in general; a ``prompt" re-collapse
can be seen in the $\ph_0(0) = 0.05$ data in Fig.~\ref{modph}.

\begin{figure}
\begin{center}
\epsfxsize=12.0cm
\includegraphics[width=12.0cm,clip=true]{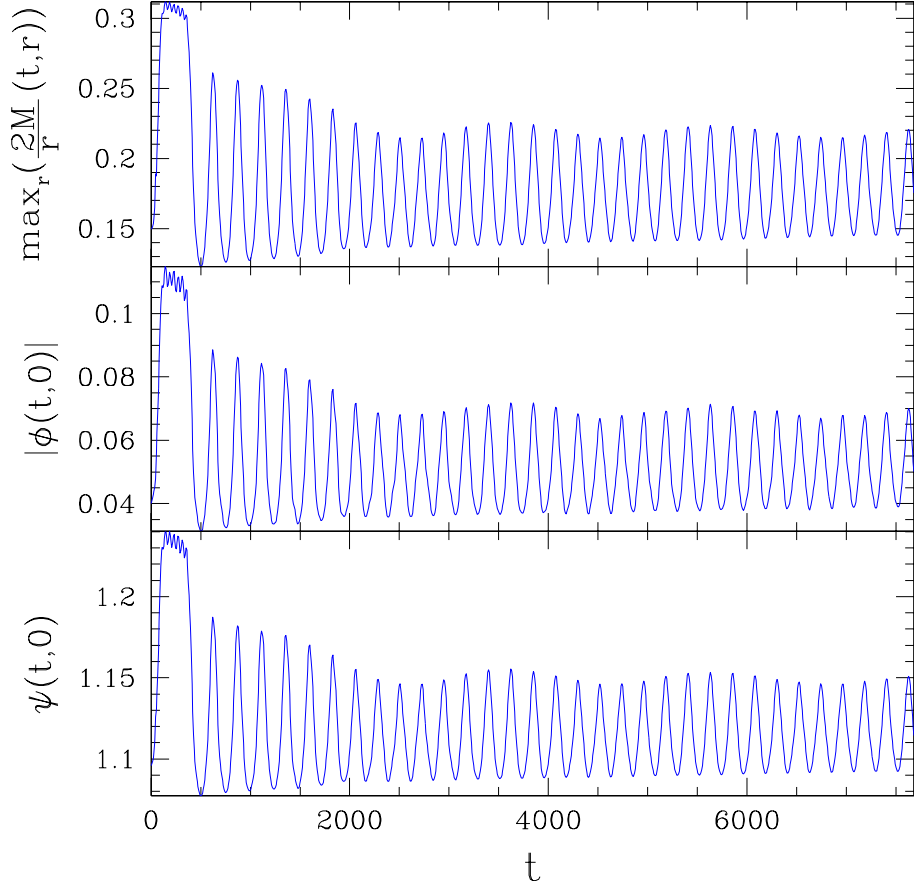}
\caption
[Long time behaviour of subcritical evolution for $\ph(0,0) = 0.04$]
    {Long time behaviour of subcritical evolution for $\ph(0,0) = 0.04$. 
This figure shows the long-time behaviour of $\max_r(2M(t,r)/r_S)$, $\vert \phi(t,0)\vert$ and $\psi(t,0)$ for 
a near-critically perturbed boson star ($\ph_0(0) = 0.04, A_3^\star \approx 0.00342$),
where  $r_S = \psi^2 r$ is the areal radius, and where, loosely speaking, 
$2M(t,r)/r_S = 1$ signals the surface of a black hole.
Note that this is a {\em subcritical} evolution, so that a black hole does {\em not} form.
The figure provides evidence that the final state of subcritical evolution is 
characterized by large amplitude oscillations about something approximating a boson
star on the stable branch, rather than dispersal of the complex field as suggested
in~\cite{shawley:phd,scott_matt:2000}.
Detailed calculation (see Sec.~\ref{perttheory}) shows that the 
pulsation frequency is approximately the fundamental mode frequency 
computed from perturbation theory about a background stable boson star solution
with $\ph_0(0)=0.051$.
Also note the overall lower-frequency modulation of the post-critical-phase oscillations. 
This effect is not yet understood, although one possible explanation---namely that the 
envelope modulation represents ``beating'' of the fundamental and first harmonic modes---appears to 
be ruled out. 
}
    \label{tmrphps}
  \end{center}
\end{figure}

Fig.~\ref{tmrphps} shows the long-time behaviour of $\mbox{max}_r(2M(t,r)/r_S)$, $\vert
\phi(t,0)\vert$ and 
$\psi(t,0)$ for a near-critically perturbed boson star ($\ph_0(0) =
0.04, A_3^\star \approx 0.00342$),
where  $r_S = \psi^2 r$ is the areal radius, and where, loosely speaking, 
$2M(t,r)/r_S = 1$ signals the surface of a black hole.
Note that this is a {\em subcritical} evolution, so that a black hole does {\em not} form.
As shown in more detail in previous figures, the boson star enters a critical
state (well approximated by an unstable boson star) shortly after the real scalar field leaves 
the computational domain ($t\approx100)$,
and while in the critical state, it oscillates with the frequency of the fundamental mode as
computed from perturbation theory using the unstable boson star state as the 
background (see \cite{shawley:phd,scott_matt:2000}).  At $t\approx 370$ the star leaves 
the more compact critical configuration, decreases in central density, expands
in size, and starts to pulsate with a different frequency.  Although at late
time the oscillation amplitudes are much larger than those seen in the critical phase of 
evolution, we will show in the following section that the oscillations can nonetheless be largely 
attributed to excitations of the fundamental perturbative mode
associated with the final boson star state.

\subsection{Perturbation Analysis of Subcritical Oscillations} \lab{perttheory}
We now proceed to an application of perturbation theory to the oscillations seen in 
long-time evolutions of marginally subcritical configurations, such as those shown in 
Fig.~\ref{tmrphps}.
Here we follow~\cite{glesier_watkins:1989} and \cite{scott_matt:2000},
and refer the interested readers to those sources for details of the approach 
that are not included here.  In particular, we emphasize that we have {\em not} carried out 
the complete perturbation analysis ourselves, but are simply using a computer code provided by
Hawley~\cite{hawley:private} to
analyze our current simulations.  Nonetheless, to make contact between the perturbative and 
simulation results, it is useful to briefly review the setup of the perturbative problem.

To formulate the equations for the perturbation analysis, we first rewrite the 
complex scalar field as:
\beq
 \ph(t,r) = \left( \ps_1(t,r)+i \ps_2(t,r)\right) e^{-i \om t}\,,
\eeq
\noi
(Note that this representation is distinct from $\ph = \ph_1 + i \ph_2$, and the 
reader should be careful not to confuse the $\ps$'s used here with the conformal metric variable, $\ps$.)
Additionally, the spacetime metric is written in Schwarzschild-like (polar-areal) 
coordinates:
\beq
 ds^2 = -e^{\nu(t,r)}dt^2 + e^{\la(t,r)} dr^2 + r^2\left(d \te^2+\sin^2 \te d
\vp^2\right)\,.
\eeq
\noi
We further introduce four perturbation fields $\de \la(t,r), \de \nu(t,r), \de \ps_1(t,r)$ and
$\de \ps_2(t,r)$ which represent the perturbations about the
equilibrium values $\la_0(r), \nu_0(r), \ph_0(r)$:
\bea
\la(t,r) &=& \la_0(r)+\de \la(t,r)\,,\\
\nu(t,r) &=& \nu_0(r)+\de \nu(t,r)\,,\\
\ps_1(t,r) &=& \ph_0(r) \left( 1+\de \ps_1(t,r)\right)\,,\\
\ps_2(t,r) &=& \ph_0(r)\, \de \ps_2(t,r) \,.
\eea
\noi
With the above definitions we can write the coupled  Einstein-Klein-Gordon field
equations as a set of PDEs for the functions $\de \la, \de \nu, \de \psi_1$ and $\de \psi_2$.
With some manipulation we can then eliminate $\de \nu$ and $\de \ps_2$ to produce 
a system of two coupled second-order PDEs for $\de \ps_1$ and $\de \la$:
\bea \nn
  \de {\ps_1}'' &=& - \left( \fr{2}{r} + \fr{{\nu_0}'-{\la_0}'}{2}\right) \de
{\ps_1}' -
\fr{\de \la'}{r {\ph_0}^2} + e^{\la_0-\nu_0} \ddot{\de \ps_1} \\ \nn
&& - \left[
\fr{{\ph_0}'}{\ph_0} \left( \fr{{\nu_0}'-{\la_0}'}{2} + \fr{1}{r}\right) +
\left( \fr{{\ph_0}'}{\ph_0}\right)^2 + \fr{1-r {\la_0}'}{r^2 {\ph_0}^2 } +
e^{\la_0-\nu_0} \om^2 - e^{\la_0} \right] \de \la \\ 
&& + 2 e^{\la_0} \left[ 1 + e^{-\nu_0} \om^2 + e^{-\la_0} \left(
\fr{{\ph_0}'}{\ph_0}\right)^2 + r \ph_0 {\ph_0}'\right] \de \ps_1\,, \\ \nn
  \de \la'' &=& -\fr{3}{2} \left( {\nu_0}' - {\la_0}'\right) \de \la' + 
\left[ 4 {{\ph_0}'}^2 + {\la_0}'' + \fr{2}{r^2} - \fr{ ({\nu_0}' - {\la_0}')^2}{
2} - \fr{ 2 {\nu_0}' + {\la_0}'}{r} \right] \de \la \\ \nn
&& + e^{\nu_0-\nu_0} \ddot{\de \la} - 4 \left( 2 \ph_0 {\ph_0}' - r e^{\la_0}
{\ph_0}^2\right) \de {\ps_1}' \\
&& - 4 \left[ 2 {{\ph_0}'}^2 - r e^{\la_0} {\ph_0}^2 \left( 2
\fr{{\ph_0}'}{\ph_0} + \fr{2 {\nu_0}' + {\la_0}'}{2}\right) \right] \de \ps_1\,.
\eea
\noi
Note that these equations involve only second time derivatives (i.e.\ there are 
no terms involving $\dot{\delta \psi_1}$ or $\dot{\delta \lambda}$), and that they 
are linear in the second time derivatives.   If we thus assume 
a harmonic time-dependence for the perturbed fields:
\bea \lab{harmonicdep1}
 \de \ps_1(t,r) &=& \de \ps_1(r) e^{i \si t}\,,\\ \lab{harmonicdep2}
 \de \la_1(t,r) &=& \de \la_1(r) e^{i \si t}\,,
\eea
\noi
then the equations for the perturbations contain $\si$ only in the form 
$\si^2$, and the sign of 
$\si^2$, as computed by solving a particular mode equation, determines the stability of that
mode.  (Note that the system can be shown to be self-adjoint so that the
values of $\si^2$ must be real.)  If any of the values of $\si^2$ 
are found to be negative, then the
associated perturbations will grow and the boson star will be unstable.  Moreover, as
the eigenvalues form an infinite discrete ordered sequence, examining the 
fundamental radial mode ${\si_0}^2$ determines the overall stability of any particular 
star with respect to radial perturbations.

In order to compare the simulation results with those given by perturbation
theory, we first observe
that there is a difference in the choice of the time coordinates used in the two calculations.
Specifically, 
in the perturbative analysis~\cite{glesier_watkins:1989,scott_matt:2000}, the lapse 
was chosen to be unity at the origin, so we have
\[
\sigma^2 \Bigl.\Bigr\vert_{\rm perturbative} \rightarrow
{\sigma^2 \over \alpha^2} \Bigl.\Bigr\vert_{\rm simulation}\,.
\]

\noi
We also note that there is a factor of 2 difference in the definitions of $T_{\mu
\nu}$ used in the two calculations, and that the 
definition of the complex field, $\ph(t,r)$, in the perturbative calculation includes a factor of $\sr{8 \pi}$.  We thus 
have
\[
\ph\Bigl.\Bigr\vert_{\rm perturbative} \rightarrow
\sr{4 \pi}\ph\Bigl.\Bigr\vert_{\rm simulation}\,.
\]

\noi
The numerical technique for obtaining the fundamental mode and first harmonic
mode frequencies of boson stars has already been described in \cite{scott_matt:2000}
and will not be repeated here; again we will simply quote and use results from that 
study.
From Fig.~\ref{tmrphps} we note that
there are 20 oscillations between $t = 620$ and $t=5200$, giving a period $T
\approx 229$.  Hence we have an oscillation frequency $\si = 2 \pi/T \approx 0.0274$.  
The time average of the lapse function $\langle \al(t,0) \rangle$ in the interval 
is 0.78, and so $\si^2/\al^2 \approx 0.0013$.  We also compute the time average of
$\ph(t,0)$ in the interval, and use the resulting value to identify the 
stable boson star solution about which we perform the 
perturbation analysis.  We find $\langle \ph_0(t,0) \rangle \approx
0.051 \times \sr{4 \pi} = 0.18$.  For a boson star with $\ph_0(0) = 0.18$, the perturbative 
calculations (see Fig.~7 of \cite{scott_matt:2000}) 
predict $\si_0^2 = 0.0014$, which is in reasonable agreement with the 
simulation results.  Hence the oscillations that occur in the post-critical
regime appear to be largely fundamental mode oscillations of a final-state, stable, boson star.  
We also remark that since the oscillations are of such large amplitude, it does not appear
possible to precisely identify an effective background state (i.e.\ an effective value 
of $\ph_0(0)$), so the level of agreement in the oscillation frequencies is possibly 
as good as one could expect.

We have also attempted to understand the nature of the slower frequency 
oscillation which modulates the envelope of the fundamental-mode oscillations, as
is visible in Fig.~\ref{tmrphps}.  
One hypothesis is that this modulation represents a ``beating'' effect of 
two oscillatory modes with frequencies $\sigma_1$ and $\sigma_2$ such 
that $|\sigma_1 - \sigma_2| \ll \sigma_{1,2}$.
We observe that period of the envelope of oscillation is roughly $T\approx 2000$, 
corresponding to a squared-frequency $\si=2 \pi/T \approx 0.00314$.  The average of the lapse function 
during one cycle is roughly 0.78, so $\si^2/\al^2 \approx 1.62\times
10^{-5}$.  However, the first harmonic frequency from perturbation analysis 
for  $\langle \ph_0(t,0)
\rangle \approx 0.18$ is $\si_1^2 = 0.0105$, much different from the
fundamental mode $\si_0^2 = 0.0014$.  Therefore the beating cannot be
produced from the superposition of the fundamental and first harmonic modes.  

\section{Black Hole Excision for Boson Stars} \lab{excisiontech}
In the last section of this chapter we describe our application of 
black hole excision techniques in the context of gravitational
collapse of boson stars.  As described in Chap.~\ref{num_method}, black hole excision methods, whereby the interiors
of black holes are excluded from the computational domain, have been the 
subject of intense study in numerical relativity over the past decade.  
We view our current implementation as a further proof-of-principle for the 
strategy, as well as providing some indications of the subtle interactions 
between inner (excision) and outer boundaries that can arise when using 
the excision method.  We also emphasize that we have not yet used calculations with 
excision in order to generate new results of physical significance.

To study excision techniques, we naturally need to evolve spacetimes which
eventually develop black holes.  Conveniently, we can consider the same model
and initial data configurations used in our previously-described study of critical phenomena, 
provided that a sufficiently large amplitude factor, $A_3$, for the scalar field
is used.  
In particular, we need to choose $A_3$ large enough so that the black hole 
that forms is large compared to the spatial discretization scale, $\Delta r$, used in the simulation.
For the purposes of illustrating our implementation of excision, we consider a boson star with 
$\ph_0(0) = 0.01$, and perturb it with an imploding massless scalar field having an 
initial profile of the form~(\ref{ph3-gaussian}) with $A_3 = 0.041$,
$r_0 = 40$ and $\si = 5$.  In fact, $A_3$ is large enough in this instance that almost all of 
the mass associated with both the real and complex fields ends up in the final black hole.

\begin{figure}
\begin{center}
\epsfxsize=16.0cm
\includegraphics[width=16.0cm,clip=true]{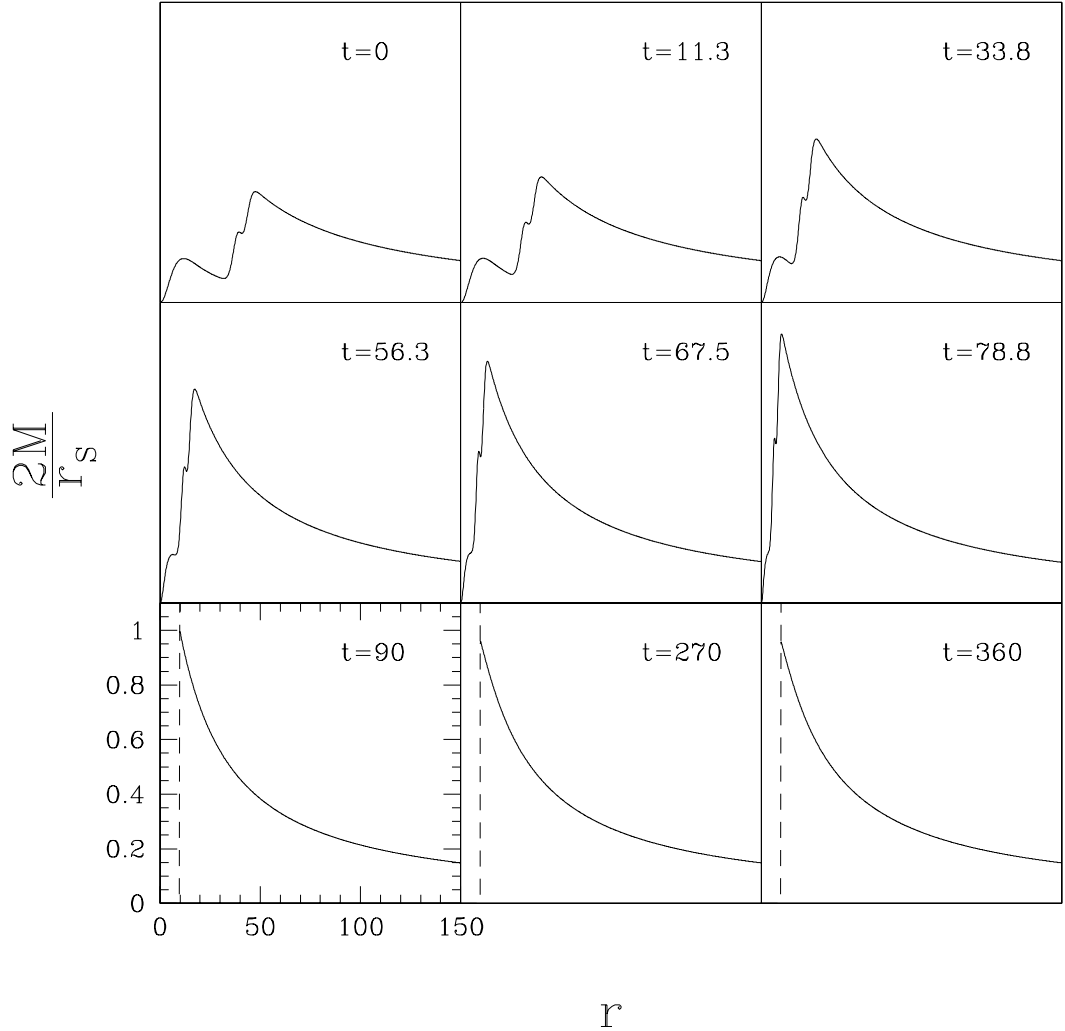}
\end{center}
\caption[
Evolution of a highly perturbed boson star ($\ph_0(0) = 0.01$)
using black hole excision]
{Evolution of a highly perturbed boson star ($\ph_0(0) = 0.01$) 
using black hole excision.  This figure shows snapshots
of $2M(t,r)/r_S$, where  $r_S = \psi^2 r$ is the areal radius, and where,
loosely speaking, 
$2M(t,r)/r_S = 1$ signals the surface of a black hole.  In snapshots with
a vertical dashed line, the region from $r=0$ to the radial position of the 
dashed line is excised from the computational domain.
The evolution starts with a stable 
boson star centered at the 
origin, and a high-amplitude, in-going pulse of real scalar field 
of the form~(\ref{ph3-gaussian}) ($A_3 = 0.041$, $\si=5$ and $r=40$).
At $t \approx 80$ the real field
reaches the origin, and an apparent horizon containing most of 
the combined mass of the real and complex scalar fields forms at 
$t = 80.51$.  
At late times, the exterior solution settles 
down to a configuration very close to a static Schwarzschild spacetime.
Note that the temporal spacing between successive snapshots
is not constant--the time instants displayed have been chosen to illustrate the key
features of the simulation.
}
\label{bhexcise}
\end{figure}

\begin{figure}
\begin{center}
\epsfxsize=10.0cm
\includegraphics[width=10.0cm,clip=true]{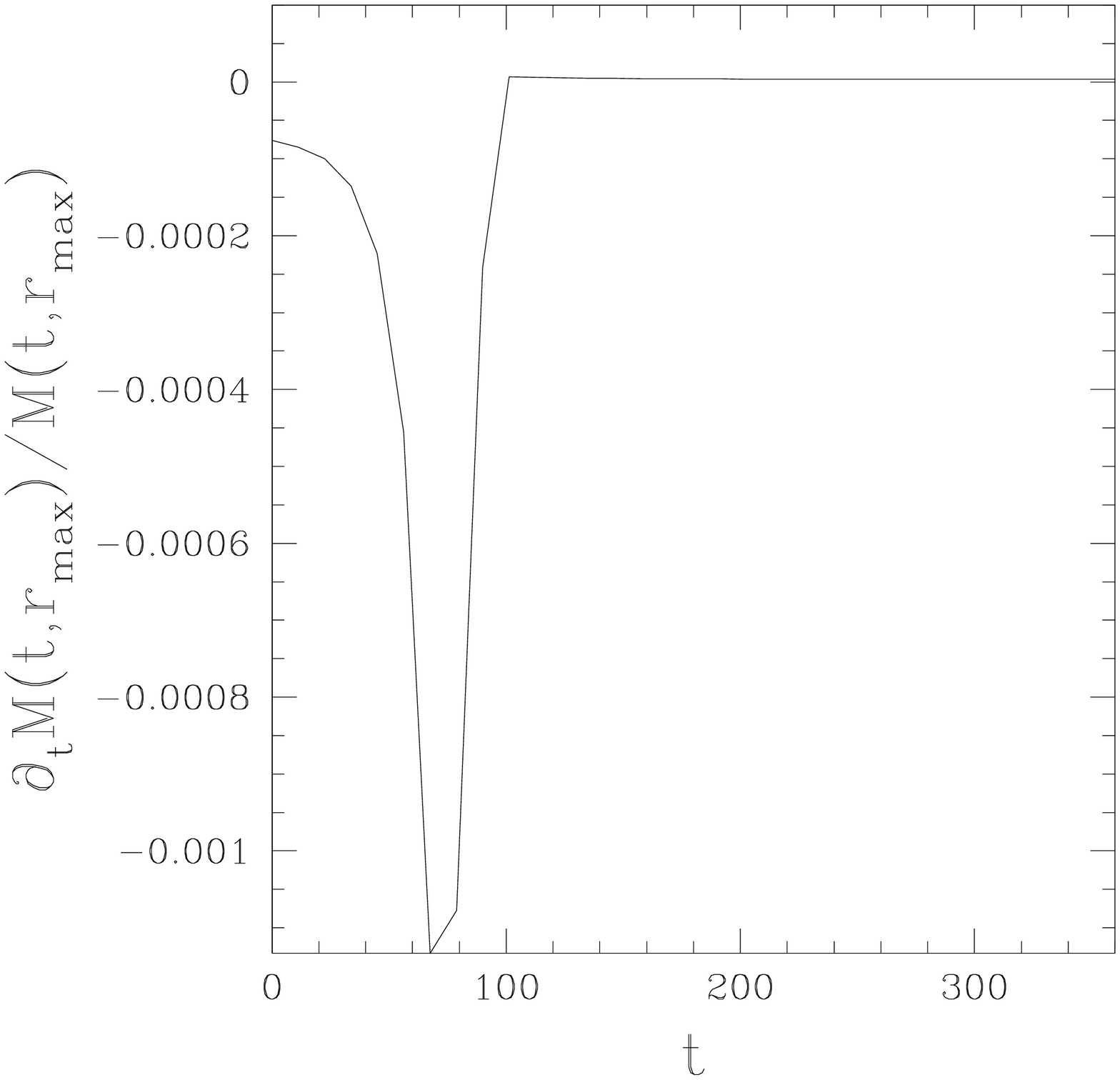}
\caption[
Relative time rate of change of the mass aspect function at outer boundary
for the evolution of a highly perturbed boson star
($\ph_0(0) = 0.01$) using black hole excision
]
{
Relative time rate of change of the mass aspect function at outer boundary, 
$\pa_t M(t,r_{\rm max}) /
M(t,r_{\rm max})$, for the evolution of a highly perturbed boson star
($\ph_0(0) = 0.01$) using black hole excision.
The figure shows the result for the same evolution in Fig.~\ref{bhexcise}, with $r_{\rm
max} = 150$ and $\De r \approx 0.146$.  At late time $t > 100$ the
relative change in the  mass aspect function at outer boundary is of the order of $10^{-6}$, indicating
almost no mass is passing through the outer boundary.
}
\label{MdotM}
\end{center}
\end{figure}

Fig.~\ref{bhexcise} shows a series of snapshots of $2M(t,r)/r_S$ during the resulting
evolution, where $M(t,r)$ is the mass aspect function and $r_S = \psi^2 r$
is the areal coordinate ($r$ is the isotropic radial coordinate).
During the evolution we monitor~(\ref{AH_eq}) to detect the appearance of an
apparent horizon---in this case an apparent horizon is first detected at 
$t = 80.51$ and at $r = 9.67$. Assuming cosmic censorship, the region $r \le 9.67$ 
lies within a black hole for all future times so can safely be excised from the simulation 
domain for $t \ge 80.51$.  (In Fig.~\ref{bhexcise}, the region
being excised lies to the left of the dashed vertical line in each sub-plot.)
Fig.~\ref{MdotM} shows the 
relative time rate of change of the mass aspect function at outer boundary, 
$\pa_t M(t,r_{\rm max}) /
M(t,r_{\rm max})$, for the same evolution.
At late time ($t > 100$) the
relative change in the mass aspect function at outer boundary is of the order 
of $10^{-6}$, indicating
almost no mass is passing through the outer boundary.

Once excision is enabled, the origin is no longer part of the computational domain and thus 
the inner boundary (regularity) conditions previously imposed at $r=0$ for the variables 
$\al, \ps$ and $\krr$, must be replaced by conditions imposed at the excision surface.
In the case of the dynamical geometrical variables, $\ps$ and $\krr$, we can use discrete 
versions of the evolution equations~(\ref{ps_evolution_eq}) and
(\ref{krr_evolution_eq}) to update values for $\ps$ and $\krr$ on the excision boundary.
For the case of the lapse function, $\al$, we have found that a simple strategy, whereby we 
`freeze'' the inner value of the lapse to whatever its value is when excision starts,
leads to stable, convergent results.  
(Also note that since our simulations {\em do} appear to converge using our simple strategy,
we have {\em not} experimented with the 
condition on $\al(r_e,t)$, where $r_e$ is the $r$-coordinate of the
excision surface. 
So, for example, we have not investigated whether the condition on $\al$ is perturbatively
stable.)  Finally, although the shift component, $\beta$, also satisfies a regularity 
condition, the boundary condition for~(\ref{shift_eqn}), which governs the shift, is naturally 
set at large $r$ and thus we do not need to specify a condition on $\beta$ at the excision
surface.

With regards to our simulations with excision, a
key observation is that after the coupled boson star/real scalar field 
configuration promptly collapses to form a black hole, the region exterior
to the black hole 
quickly settles down to an
essentially Schwarzschild spacetime. In particular, the mass aspect function outside 
the black hole approaches an almost-constant function, and a stable simulation of the 
black hole spacetime can be carried out for very long integration times.

\begin{figure}[h]
\begin{center}
\epsfxsize=12cm
\includegraphics[width=12.0cm,clip=true]{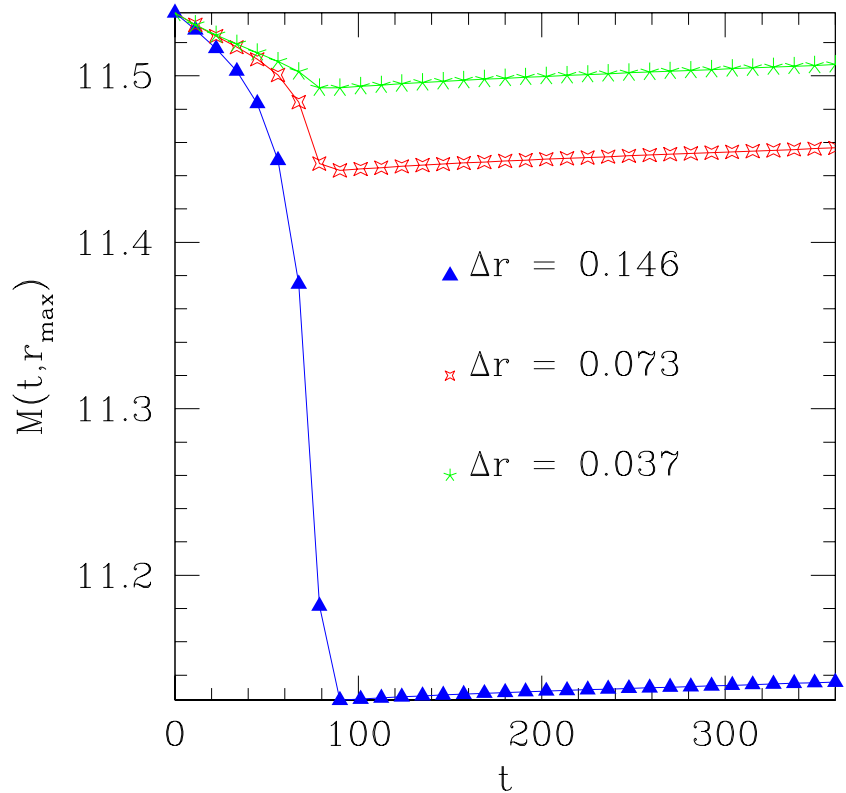}
\caption
[The mass aspect function $M(t,r_{\rm max})$ {\em vs} $t$ for $r_{\rm max} = 150$]
{The mass aspect function $M(t,r_{\rm max})$ {\em vs} $t$ for $r_{\rm max} = 150$ from
simulations with excision, and 
computed using three different finite difference resolutions, $\Delta r$,  in a 
4:2:1 ratio.
For resolution $\De r = 0.146$, (the coarse calculation), the mass decreases 
quite suddenly (and by a few percent) at $t \approx 80$, at which
time a black hole is detected and a region of spacetime
is excised.  Following this early transient associated with the enabling of excision,
the mass remains roughly constant (relative to the transient fluctuation), although a slight
constant rate of mass increase is observed.  As resolution increases ($\De r = 0.073$,
$\De r=0.037$), 
the early transient mass decrease
apparently converges away (possibly only as $\Delta r$).  However, the 
late-time mass increase, does not seem to vanish as $\Delta r \to 0$ 
as can be seen from the
fact that the slope of the plots for $t > 100$ remains roughly constant as the resolution
is varied.
}
\label{m_012outer_rmax150}
\end{center}
\end{figure}

However promising the excision technique appears, careful examination 
of the mass aspect function over sufficiently long integration
times {\em does} reveal certain numerical artifacts.
For example, 
Fig.~\ref{m_012outer_rmax150} plots the mass aspect function at the outer
boundary of the computational domain, $r = r_{\rm max} = 150$, against time, 
for $0\le t\le 360$.  During any period where there is no flux of real or complex 
mass/energy crossing the outer boundary, we should have 
$M(t,r_{\rm max}) = {\rm const.}$
From the figure we see that $M(t,r_{\rm max})$
decreases initially, mainly due to the escape of real scalar
field from the computational domain.  At $t\approx 80$, following the first 
detection of an apparent horizon, and the enabling of excision, 
there is a relatively sharp, and transient decrease in $M(t,r_{\rm max})$.  However, as 
can also be seen in the figure, all 
indications are that this particular decrease ``converges away'' in the continuum limit 
$\Delta r \to 0$.  Following the transient phase, although the mass generically remains
roughly constant, there is always a small rate of increase in $M(t,r_{\rm max})$ which
is largely {\em independent} of the grid resolution, $\Delta r$.
This is surely an artifact---since no mass is physically added to the system, we claim that 
the observed mass increase must be a purely numerical/computational effect, and we have 
thus investigated possible remedies.

\begin{figure}[h]
\begin{center}
\epsfxsize=12cm
\includegraphics[width=12.0cm,clip=true]{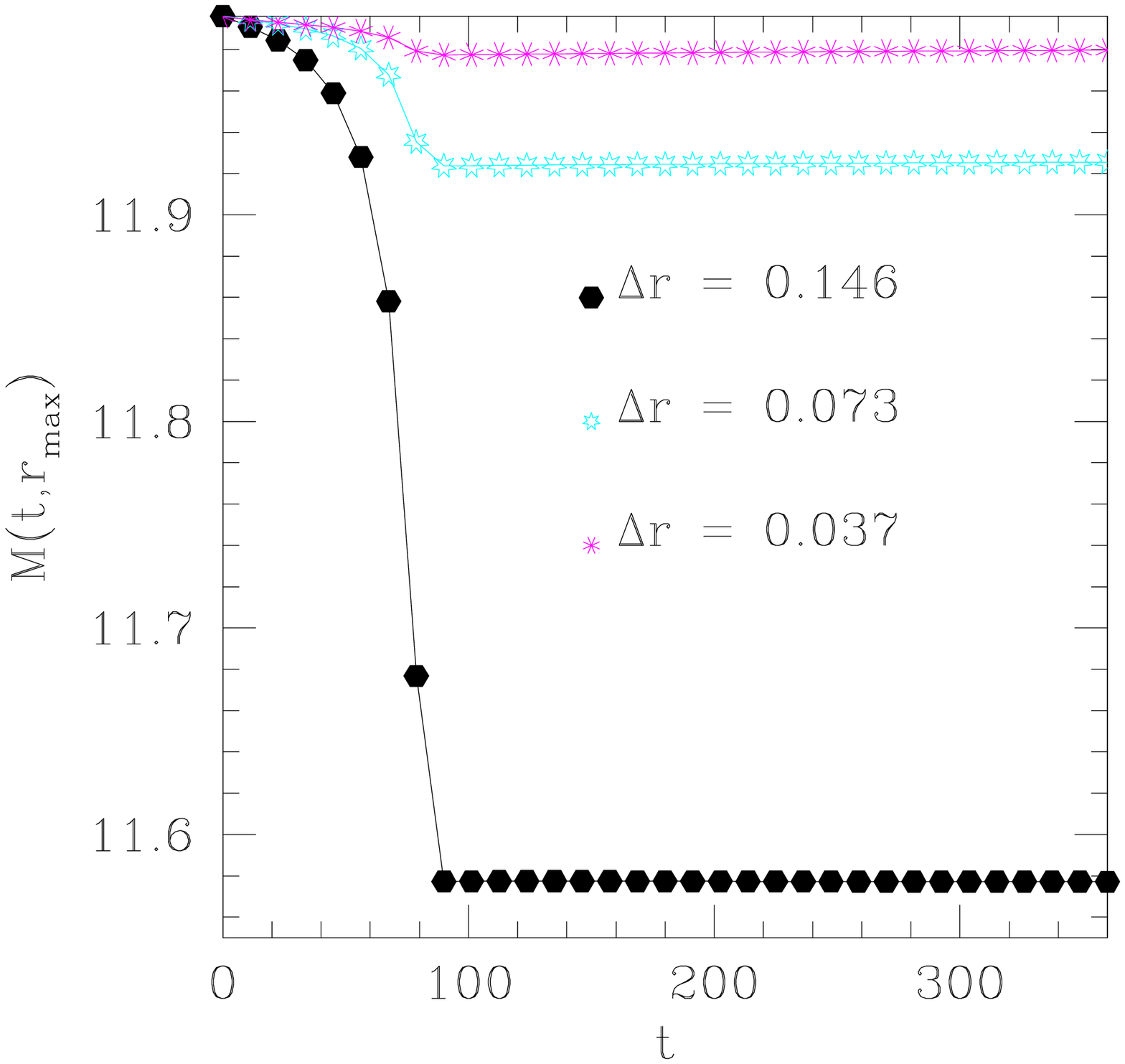}
\caption
[The mass aspect function $M(t,r_{\rm max})$ {\em vs} $t$ for $r_{\rm max} = 300$]
{The mass aspect function $M(t,r_{\rm max})$ {\em vs} $t$ for $r_{\rm max} = 300$ from
simulations with excision, and 
computed using three different finite difference resolutions, $\Delta r$,  in a 
4:2:1 ratio.  
As in Fig.~\ref{m_012outer_rmax150}, with increasing resolution, 
the magnitude of the early ($t \approx 80$) transient mass drop apparently converges 
away.
Moreover, the mass is more nearly conserved at late times in the calculation, relative 
to the simulation shown in Fig.~\ref{m_012outer_rmax150}. 
In other words, the spurious mass increase seems to be due to our treatment 
of the outer boundary conditions, and we can apparently suppress the 
resulting numerical artifacts by increasing the size of the computational domain.
}
    \label{m_012outer_rmax300}
\end{center}
\end{figure}

\begin{figure}[h]
\begin{center}
\epsfxsize=12cm
\includegraphics[width=12.0cm,clip=true]{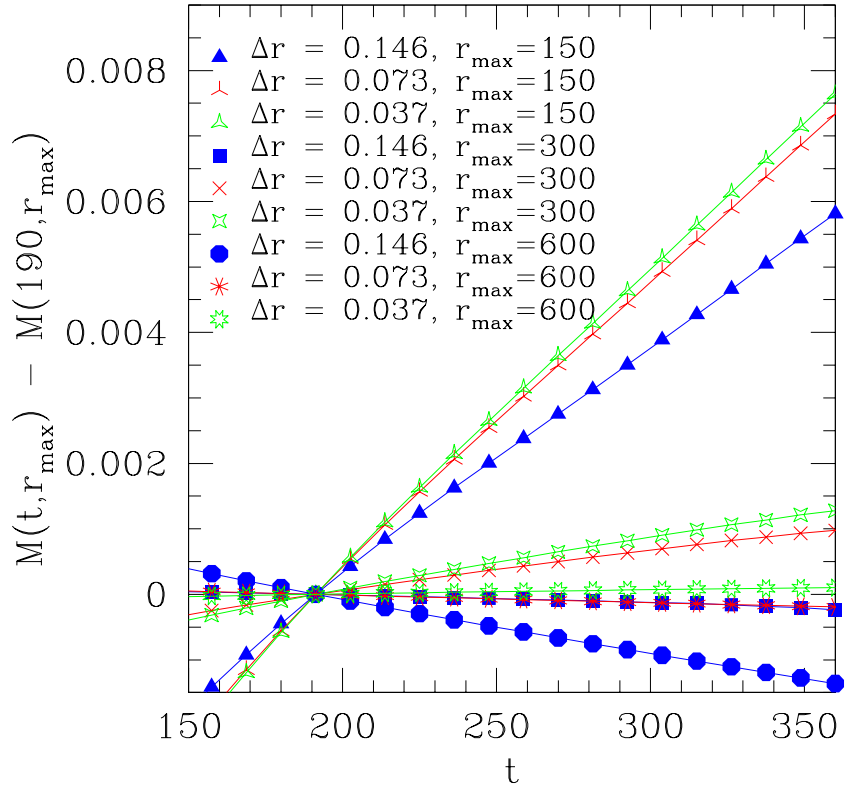}
\caption
[The deviation of the mass aspect function $M(t,r_{\rm max})$ from $M(190,r_{\rm max})$ {\em vs}
$t$ for $r_{\rm max} = 150, 300$ and $600$]
{The deviation of the mass aspect function $M(t,r_{\rm max})$ from $M(190,r_{\rm max})$ {\em vs}
$t$ for $r_{\rm max} = 150, 300$ and $600$
from simulations with excision, and 
computed using three different finite difference resolutions, $\Delta r$,  in a 
4:2:1 ratio.  
The deviation is measured relative to the mass aspect function at $t=190$, by which time the 
transients associated with the enabling of excision have settled down (see
Fig.~\ref{m_012outer_rmax150} and
Fig.~\ref{m_012outer_rmax300}).
The deviation is smaller when we increase the location of the outer boundary from $r_{\max}=150$ to
$r_{\max}=300$ and $r_{\max}=600$.  The rate of convergence is roughly
$r_{\rm max}^{-2}$.  See the text for further details.
}
 \label{dm_012outer}
\end{center}
\end{figure}

Because our simulations are performed on a finite radial domain, $0 \le r \le r_{\rm max}$,
it is natural to consider the effect of the placement of the outer boundary 
of the domain, $r_{\rm max}$, on the results. 
Fig.~\ref{m_012outer_rmax300} plots $M(t,r_{\rm max})$ {\em vs} $t$ from simulations 
using the same initial data described in Fig.~\ref{m_012outer_rmax150}, 
but where $r_{\rm max}$ has been doubled to $r_{\rm max} = 300$.
From this figure it can be seen that the transient mass decrease immediately 
following the enabling of excision still converges away as $\Delta r \to 0$. 
In addition, it is apparent that the level of mass increase {\em following} the 
transient phase is substantially reduced from the $r_{\rm max} = 150$ 
calculations.  
To illustrate this point more clearly, Fig.~\ref{dm_012outer} shows the deviation of the mass aspect function $M(t,r_{\max})$
from $M(190,r_{\max})$ {\em vs} time $t$ for $r_{\max} = 150, 300$ and $600$.
The deviation is measured relative to the mass aspect function at $t=190$, by which time
the transients associated with the enabling of excision have settled down.
The figure shows that the deviation is smaller when we increase the location
of the outer
boundary from $r_{\max} = 150$ to $r_{\max} = 300$ and $r_{\max}=600$.
The calculated value for $ \left( \De M(t,150) - \De M(t,300)\right)/
\left( \De M(t,300)-\De M(t,600)\right) \approx 5$, where $\De
M(t,r_{\max}) \eq M(t,r_{\max})- M(190,r_{\max})$, which indicates the rate
of convergence is roughly $r_{\max}^{-2}$.

We thus conjecture that the most dominant contribution to 
the small amount of post-excision ``mass inflation'' we observe is due to 
our (approximate) asymptotic treatment of the boundary conditions.  
We have done some preliminary investigations of improvements to the outer boundary 
conditions but have no significant findings to report at this time.

  \resetcounters

\def\graxi{{\tt graxi}}
\def\phik{{\phi_{(k)}}}
\def\phikm{{\phi^M_{(k)}}}
\chapter{Boson Stars in Axisymmetry} \lab{bs2d}

We now consider the evolution of the coupled Einstein-Klein-Gordon field 
equations in axisymmetry.  As time-dependent, spherically symmetric
spacetime is the most natural  1 space + 1 time-dimensional arena in which to approximate 
star-like objects, so is time-dependent axisymmetric spacetime the most natural 2+1 
context.  Crucially, the generalization to axisymmetry allows us to consider 
time-independent configurations with net angular momentum, which, for the case of boson stars turn 
out to be surprisingly non-trivial to construct.   In addition, with an 
axisymmetric code we can simulate scenarios such as the head-on collision of two or 
more boson stars, as well as axisymmetric (i.e.\ non-spherically symmetric) critical collapse. 
This chapter concerns the construction of {\em stationary} (time-independent~\footnote{
We emphasize that when we speak of {\em time-independent} solutions in this chapter, we 
mean that the {\em geometry} is time-independent.  In general, the complex scalar field,
$\phi(t,{\bf x})$,  will {\em always} be time-dependent, even if that dependence is 
``trivially'' harmonic.  Similarly, when we consider rotating boson stars, and speak of axially
symmetric solutions, we will also mean that the {\em geometry} is axially-symmetric; again, the 
particular ansatz that we will adopt for a rotating boson star will result in explicit $\varphi$ dependence
in $\phi(t,r,\theta,\varphi)$. 
}), {\em rotating} boson star 
solutions (which requires the solution of an elliptic quadratic eigenvalue problem in two spatial dimensions), as well 
as the construction of time-dependent axisymmetric solutions representing the evolution of 
{\em generic} (in principle) axisymmetric initial data for a complex scalar field.  Unfortunately, the stationary solutions 
that we construct in the first case are not time-independent in the coordinate system used 
in the latter~\footnote{
Specifically, the coordinate system $({\tilde t},{\tilde \rho},{\tilde
z},\tilde{\varphi})$
currently used in \graxi\ for evolving complex scalar fields with rotation~\cite{graxi:2004} is not adapted 
to the timelike Killing vector of our stationary solutions, and so a non-trivial coordinate transformation---itself
involving solution of a complex set of non-linear PDEs---must 
be performed to determine the initial data 
$[\phi({\tilde t},{\tilde \rho},{\tilde z},\tilde{\varphi}), \Pi({\tilde
t},{\tilde \rho},{\tilde z},\tilde{\varphi})]$ required by
\graxi.
}
and we
are thus unable to directly study the evolution of rotating boson stars by combining the two calculations. 
We are, however, able to reproduce and extend previous calculations of rotating  boson star configurations, as 
well as to study critical collapse of driven boson stars in axisymmetry through a completely original 
set of calculations.  Some of the work presented in this chapter was done in collaboration with Choi~\cite{choi:www},
who performed numerous tests on the solutions for stationary rotating boson stars in the 
Newtonian limit, as well as
tests of the modified {\graxi} code.  Results of these tests are available on-line~\cite{choi:www}
and will not be further described here. 

Before proceeding to a summary of the remainder of the chapter, we present an overview of some key features
of general relativistic, rotating boson stars (and previous work on the subject), and then an overview of 
related work on axisymmetric critical collapse as well as our general approach to that problem.

\subsection{Rotating Boson Stars}
The question of existence of rotating fluid stars (solutions of the stationary,
axisymmetric Einstein/hydrodynamics equations) has a clear answer on 
physical grounds.  The observation of rotating neutron stars---for example, in the
form of pulsars---tells us that relativistic, rotating fluid stars naturally occur
in our universe, and the theoretical study of such objects has been an active
area of research for decades~(for example, see
\cite{hartle:1967,friedman:1984,Ruderman:1974,owen:1998,Yoshida:2000,strohmayer:2002}).
Interestingly, it was not until fairly recently that the theoretical
existence of relativistic, rotating boson stars was
demonstrated~\cite{schunck_mielke:1996,yoshida_eriguchi:1997,Schunck:1998}.
This is perhaps not too surprising when one considers the lack of astrophysical
importance (to date at least!) of boson stars.  In addition, at a more
technical level, the regularity condition for a rotating bosonic star is
different from that for a rotating fermionic star, a fact which leads to
completely different topological shapes for the rotating stationary states
in the two cases.  For example, as has already been mentioned,
for boson stars with non-zero angular momentum, the stationary solutions
have mass-energy concentrated in a roughly toroidal (rather than spheroidal)
configuration.

Schunck \& Mielke~\cite{schunck_mielke:1996}, as well as 
Yoshida \& Eriguchi~\cite{yoshida_eriguchi:1997} have previously constructed general relativistic, stationary, rotating boson star solutions using numerical means.
One of the more interesting features of such solutions is that angular
momentum is quantized (or perhaps more properly, discretized).   Specifically, 
to generate time-independent, rotating configurations, one is naturally lead to an ansatz 
(in spherical-polar coordinates)~\footnote
{
	As noted in the introduction, this ansatz was apparently first written down, for the case 
   of {\em Newtonian} boson stars, by Silveira \& de~Sousa~\cite{sousa_silveira:1995}, and is {\em not} the 
   most general ansatz that could be considered.
}
\beq
\label{stat}
\phi(t,r,\theta,\varphi) \longrightarrow \phi_0(r,\theta) e^{i\omega t + i k \vp} \, .
\eeq
Now, $\omega$ can in principle take on arbitrary real values; $k$, however, must be {\em integral} for $\phi$ to be single valued.  
Since the specific choice of $k$ affects boundary conditions at $r=0$, the ansatz~(\ref{stat}) itself naturally leads us to 
consider a {\em series} of (elliptic eigenvalue) problems for $k = 1, 2, \cdots$ (the case $k=0$ describes the spherical boson stars 
of Chap.~\ref{bs1d}).

As in the spherical case, $k=0$, for any value of $k=1,2, \cdots$,
$\omega$ will be the eigenvalue $\omega(\phik)$, where $\phik$, is a generalization of the 
central modulus value, $\phi_0$, conventionally used to parametrize the family of 
stars in spherical symmetry.  In other words, as was the case in spherical symmetry, for each value 
of the angular momentum parameter (or azimuthal quantum number), $k$, we expect to 
find an entire {\em family} of rotating stars, which will be parametrized by
$\phik$.  

Given the clear analogy between~(\ref{stat}) and axisymmetric bound states in  quantum mechanics, 
it is perhaps not surprising that rotating boson stars have distinctive topological characteristics.
In fact, given the ansatz~(\ref{stat}) one can show that regularity as $r\to0$ implies
\beq
	\lim_{r\to0} \phi_0(r,\theta;k) = r^k f_k(\theta) + O(r^{k+2})\,.
\eeq
Thus, for $k > 0$ (i.e.\ the non-spherically symmetric cases) we have that $\phi$ and its first $k-1$ radial 
derivatives vanish on axis, and are relatively small close to the axis.   This means that for $k \ge 1$, the 
boson stars might be more accurately termed boson tori.

Solutions for rotating boson stars using the full general relativistic equations were first obtained by
Schunck \& Mielke~\cite{schunck_mielke:1996,Schunck:1998}.  
Specifically, configurations for  $2 \le k \le 10$ as well as for $k=500$ were found.
However, as pointed out by 
Yoshida \& Eriguchi~\cite{yoshida_eriguchi:1997} the solutions obtained were 
all close to Newtonian;
for instance, in \cite{Schunck:1998} a $k=1$ solution is presented which has
$M = 0.07331 M_{{\rm pl}}^2/m$, a mass that is only about 6\% of the maximum possible 
mass for the $k=1$ family.  Overall we estimate that none of the solutions 
constructed by Schunck \& Mielke (for any of the values of $k$ considered)
have masses in excess of 10\% of the maximum mass of the corresponding family.
As another example of the weak-field nature of the Schunck \& Mielke solutions, 
the quasi-Newtonian potentials that are displayed have almost no angular dependence, 
whereas the potentials corresponding to  highly relativistic stars do. 
Moreover, Schunck \& Mielke determined only a few configurations for any 
value of $k$, and thus did not 
study any specific family of solutions in sufficient detail to determine, for example, whether 
a ``Chandrasekhar limit" existed, as is the case for 
fluid stars, as well as for spherically symmetric boson stars.  Finally,
some of the solution plots in~\cite{schunck_mielke:1996} (such as Figs.~6.4 and 6.6) are not very 
smooth, indicating that the calculations may have been underresolved.

The work by Yoshida \& Eriguchi~\cite{yoshida_eriguchi:1997} represented a 
significant advance over Schunck \& Mielke's in several respects.
The solutions they constructed were smooth, and they were able to 
compute the entire family of solutions for the case $k=1$. 
Further, they showed that the plot of $M^{k=1}_{\rm ADM}(\phik)$  (where $\phik$ is again the 
family parameter), exhibited a maximum, and concluded that the maximum
mass for a $k=1$ rotating boson star is $M_{\rm max} = 1.31 M_{{\rm pl}}^2/m$.

Despite the considerable success of the calculations described in~\cite{yoshida_eriguchi:1997},
we have identified several unsatisfactory points about that work as well, and have at least
partially ameliorated the identified weaknesses in our current research.  First, the code used in~\cite{yoshida_eriguchi:1997}
breaks down before the maximum-mass star for $k=2$ can be computed, whereas the method we 
describe below can be used to compute the complete family for the $k=2$ case.  Second,
Yoshida \& Eriguchi's numerical approach uses a so-called ``self-consistent field" method in which the 
solutions are represented as two-dimensional surface integrals at every mesh point (the 
mesh itself is two-dimensional); this means the solution algorithm is quite expensive.
In contrast, we adopt a finite-difference approach that uses homotopic and multigrid 
techniques to solve the resulting non-linear, elliptic eigenvalue problem, and although 
we have no way of performing direct timing estimates, we feel that our approach is 
likely to be significantly more efficient computationally.  

Finally, although our 
calculations are currently limited by regularity problems (so that, for example, 
we cannot compute solutions for $k=3,4,\cdots$), we have reformulated our basic 
equations in a manner that we feel will largely solve those problems (see Sec.~\ref{system-b}).  
Although work on the modification of our algorithm to solve this new system is 
not complete, we feel that with it we will be able to compute families of 
solutions for $k>2$, and, importantly, will be able to do proper convergence 
tests to assess the accuracy of the results.  The lack of convergence tests or 
other means of estimating the accuracy of solutions is another shortcoming of 
both the Schunck \& Mielke and Yoshida \& Eriguchi work, and, indeed, our
regularity problems inhibit the complete convergence testing of our current
results as well.  We note that regularity problems near coordinate singularities 
in curvilinear coordinates have often been encountered in calculation in numerical
relativity.  In this case, we have no rigorous understanding of the origin of these
problems; however, we have strong evidence that the {\em  solutions} of the discrete
equations are {\em  not} regular as $r \to 0$ (i.e.\ our solution of the non-linear
eigenvalue problem is currently {\em  unstable}).

\subsection{Axisymmetric Critical Phenomena}
An investigation of axisymmetric critical phenomena very similar to the one described below was 
carried out by Rousseau~\cite{rousseau:master}, who studied critical phenomena associated with driven 
collapse of spherically symmetric boson stars in the context of the so-called conformally 
flat (or Isenberg-Mathews-Wilson) approximation to general relativity.  Using a massless 
scalar field as the perturbing agent, Rousseau was able to verify the existence of 
families of solutions that interpolate across the black hole threshold (the conformally 
flat approximation admits solutions with apparent horizons), and found some evidence 
of Type I behaviour---with the critical solutions being unstable boson stars---and indications 
of lifetime scaling of the form~(\ref{scaling3}).  However, Rousseau's calculations were {\em not}
fully relativistic, and were severely resolution-limited, so the results are neither very complete nor 
accurate.

All of the dynamical studies described below have been performed using a code, \graxi,
previously developed by Choptuik, Hirschmann Liebling \&
Pretorius~\cite{fransp:phd,graxi:2003},
that solves the coupled Einstein-Klein-Gordon equations in axisymmetry.
I was 
responsible for the implementation of the initial data interface for boson stars, as well as 
for the execution and analysis of the numerical experiments reported in this Chapter. 
We note that the dynamical calculations that we describe in Sec.~\ref{dynamics2d} are quite well resolved and 
accurate (estimated local accuracy of a few percent) 
by virtue of the fact that \graxi\ incorporates Berger-and-Oliger-style adaptive mesh 
refinement.  

We have used \graxi\ to investigate Type I critical behaviour of axisymmetric boson stars 
using two novel strategies.   The first involves the head-on collision of two identical 
spherically symmetric boson stars, where the family parameter is the initial momentum, $p_z$, imparted to each 
star.  The second involves perturbation of a single spherically symmetric boson star via 
the implosion of a non-spherically-symmetric pulse of massless scalar radiation.  In this 
case the family parameter is the overall amplitude of the initial massless scalar pulse.
In both instances we find strong evidence for Type I behaviour, as well as strong evidence
for lifetime scaling of the form~(\ref{scaling3}).   In addition, in some of the collision simulations
we observe interesting ``solitonic'' behaviour that has previously been seen in the collision
of Newtonian boson stars~\cite{dale}.

\subsection{Outline}
The remainder of this chapter is organized as follows.  We start in Sec.~\ref{saspacetime} with a review 
of the most general geometrical notion of ``time-independence'' for a generic spacetime
(stationarity) and a discussion of a suitably general, symmetry-adapted coordinate system 
in which to study {\em axisymmetric}, stationary solutions of the coupled Einstein-Klein-Gordon
equations.  We then proceed in Sec.~\ref{IVP2d} to a discussion of an approach to the construction
of solutions representing general relativistic stationary boson stars with angular momentum. As with 
the determination of spherical boson star data {\em per se}, this involves the solution of the 
initial value (constraint) equations for the gravitational field, but is {\em significantly}
complicated by the fact that we must simultaneously solve a {\em non-linear elliptic eigenvalue problem}
(an elliptic PDE) for the matter field.  In contrast to the previous work
of Yoshida \& Eriguchi~\cite{yoshida_eriguchi:1997}, which employed 
a so-called ``self-consistent field" technique, we proceed via finite-difference solution of the stationary PDEs for 
rotating boson stars and outline some of the 
computational challenges that result in Sec.~\ref{challenges}.  In Sec.~\ref{numstr} we then discuss details of the 
multigrid-based solver that we have designed and implemented to solve the eigenvalue problem. 
Results for rotating boson stars generated using the solver are presented in Sec.~\ref{MG_results} and we 
follow with remarks concerning the performance of our method and possible enhancements in 
Sec.~\ref{remarks}.  

As already mentioned, our solver for rotating boson stars is currently limited by regularity problems, and we are, in fact, unable 
to compute solutions for $k=3,4,\cdots$.  One promising approach to resolving this difficulty is to redefine 
variables (both scalar field and geometric) so that regularity in the discrete domain is more automatically 
enforced.  To that end we present a candidate reformulation of the equations in Sec.~\ref{system-b}.  Although
we have not yet implemented a solver for a discretization of this new system, we are confident that this task
will be straightforward, given the computational techniques that we have already developed.  We thus 
hope that the incomplete work described in Sec.~\ref{system-b} will soon give rise to a solver that will work
for $k>2$.

In Sec.~\ref{dynamics2d} we
turn attention to the {\em dynamics} of boson stars.
We discuss the results from a study of head-on collisions of two boson stars in Sec.~\ref{bincol} and 
finally, in Sec.~\ref{perturb2d}, we summarize results from axisymmetric generalizations of the spherical
critical collapse studies previously described in Sec.~\ref{critphenBS}.

\section{Stationary, Axisymmetric Spacetime} \lab{saspacetime}
As mentioned above, the concept of {\em stationarity} in general relativity
is meant to capture the most general geometric notion of ``time-independence''. 
A spacetime is stationary if it admits a timelike Killing vector field,
and operationally, we will equivalently define a stationary spacetime as 
one for which, at least locally, there exists a coordinate system in 
which all metric components are independent of the time coordinate, 
$t$.~\footnote{This will
be true if the time coordinate is adapted to the timelike Killing vector
field $\chi^\al = \left( {\pa}/{\pa t}\right)^\al$.  In this case the
Killing equation becomes $\mathcal{L_\ch}g_{\mu \nu} = {\pa g_{\mu
\nu}}/{\pa t} = 0$, i.e.\ the metric components are time-independent.}
In addition, we now wish to restrict 
attention to {\em axisymmetric} spacetimes, so that the spacetime also has a 
spacelike, azimuthal Killing vector (with closed orbits).  Again, we 
will assume that this means that we can choose a coordinate system with 
one of the coordinates, $\vp$, adapted to the symmetry, so that none 
of the metric coefficients in that coordinate system depend on $\vp$.
In other words, for stationary, axisymmetric spacetimes, if we adopt 
``natural'' (i.e.\ symmetry-adapted) spherical-polar coordinates, $(t,r,\theta,\vp)$, 
the metric will depend only on $r$ and $\theta$:

\[
 g_{\mu \nu}(t,r,\theta,\phi) \longrightarrow g_{\mu \nu}(r,\theta)\,\,.
\]

\noi
Since we are considering rotating configurations, our spacetimes will in 
general {\em not} exhibit time reversal
symmetry as $t\to-t$ will result in a configuration representing stars rotating in
a sense opposite to the original.  (In other words, stationary spacetimes 
are not necessarily {\em static}.)
However, our spacetimes {\em should} be invariant 
if we simultaneously reverse both time and the azimuthal
angle, i.e.\ the metric should have the discrete symmetry $(t,\vp) \rightarrow (-t,-\vp)$.  In the general expression for the line-element
in symmetry-adapted coordinates this fact immediately excludes the $dtdr$, $dtd\theta$,
$drd\vp$ and $d\theta d \vp$ terms as they do not preserve this discrete symmetry.  
Finally, we can perform a coordinate
transformation \cite{chandrasekhar} such that the metric of the 
2-dimensional $r$-$\theta$ subspace has the diagonal
form

\[
\ps^4 \left( dr^2 + r^2 d \theta^2 \right) \,,
\]

\noi
as long as the subspace has a positive-definite (or negative-definite)
signature.  With these considerations we can now write 
a suitably general form for the metric of a stationary, axisymmetric
spacetime:
\beq \lab{axigenmetric}
ds^2 = -\al^2 dt^2 + u\left( \bt dt + d \vp \right)^2 + \ps^4
\left( dr^2 + r^2 d \theta^2 \right) \,.
\eeq

\noi
To summarize, by an appropriate choice of coordinates, we can 
reduce the number of non-trivial metric components that are needed 
to describe stationary, axisymmetric spacetimes to four; here we identify those 
four quantities with the 
functions $\al, \bt, \ps$ and
$u$.  In what follows, we will also restrict attention to those stationary, 
axisymmetric configurations with equatorial symmetry (i.e.\ those that are symmetric under 
$\theta \to \pi - \theta$).  We thus can compute using the reduced angular range $0\le\theta\le\pi/2$.

\section{The Initial Value Problem for Rotating Boson Stars} \lab{IVP2d}

\subsection{The System of Equations in Quasi-Isotropic Coordinates (System A)} \lab{systemA}
Paralleling what was done in the case of spherical symmetry in the previous chapter,
we have adopted what we will call the quasi-isotropic (spatial) coordinate condition 
in choosing the form~(\ref{axigenmetric}) for the metric of a general axisymmetric,
stationary spacetime.  In fact, it is convenient to take $u$ in~(\ref{axigenmetric}) 
to be of the form
\beq
 u \eq \ps^4 r^2 \sin^2\te e^{2 \si}\,,
\eeq
\noi
so that the deviation of $\si(r,\theta)$ from zero reflects the deviation of the 3-metric from
conformal flatness.  Thus, the spacetime metric now becomes
\beq
ds^2 = -\al^2 dt^2 + \ps^4 r^2 \sin^2\te e^{2 \si} \left( \bt dt + d \vp \right)^2 + \ps^4
\left( dr^2 + r^2 d \theta^2  \right)\,,
\eeq
\noi
or
\beq
 ds^2 = \left( -\al^2+\ps^4 r^2 \sin^2 \te e^{2 \si} \bt^2\right)dt^2
+\ps^4 \left( dr^2 + r^2 d \te^2 + r^2 \sin^2 \te e^{2 \si} d \vp^2\right)
+2 \ps^4  r^2 \sin^2 \te e^{2 \si} \bt dt d \vp
\eeq
where, again, $\al, \bt, \ps$ and $\si$ are functions of $r, \te$ only.
To generate stationary solutions (having angular momentum in general) we adopt the following 
ansatz for $\ph(t,r,\te,\vp)$:
\beq
\label{rot-ansatz}
 \ph(t,r,\te,\vp)=\ph_0(r,\te) e^{-i(\om t + k \vp)} \, .
\eeq
\noi
As discussed above, and is the case for a quantum mechanical wave function, $k$ must take
on integer values, since $\phi$ must be single valued with respect to $\vp$:
\beq
 \ph(t,r,\te,\vp) = \ph(t,r,\te,\vp + 2 \pi)\,.
\eeq
As we have also previously remarked, this ``quantization'' of 
angular momentum
is one of the more novel features of rotating boson star solutions relative to 
rotating fluid configurations.

We can now proceed to derive the PDEs for stationary boson stars in quasi-isotropic 
coordinates from the Hamiltonian and momentum constraints, the Klein-Gordon equation,
the ansatz~(\ref{rot-ansatz}), and the assumption of stationarity.
Since the derivation is lengthy but straightforward,  we will 
simply state the resulting equations. 
For numerical convenience we choose $\ka = 1$ instead of the usual
convention $\ka = 8 \pi$ (see (\ref{eineq})) as in the previous chapter. 

First, from the Klein-Gordon equation we have the following
\beq \lab{2dkg}
\begin{split}
\ph_{,rr}+\fr{2}{r} \ph_{,r} + \fr{1}{r^2}\ph_{,\te\te}+\fr{\cot\te}{r^2} \ph_{,\te}
- \fr{k^2}{r^2 \sin^2 \te e^{2\si}} \ph 
 + \left( \si_{,r}+\fr{\al_{,r}}{\al}+ \fr{2 \ps_{,r}}{\ps}\right) \ph_{,r} \\ +
\left( \si_{,\te}+\fr{\al_{,\te}}{\al}+ \fr{2 \ps_{,\te}}{\ps}\right) \fr{\ph_{,\te}}{r^2}
+ \left[ \left( \fr{\om-\bt k}{\al}\right)^2 -m^2 \right] \ps^4 \ph = 0\,,
\end{split}
\eeq

\noi

\noi
and where for simplicity of notation we have dropped the subscript ``0'' so that $\ph_0 \to \ph$.
Note that from regularity considerations, the leading order behaviour of $\ph_0(r,\te)$ as
$r \to 0$ is $\ph_0(r,\te) = r^k \ph_{0k}(\te) + O(r^{k+2})$.  
Therefore, in principle the equilibrium solutions can be parametrized by the central value of the
$k$th radial derivative of the scalar field $\pa^k \ph_0(0,\te)/\pa r^k$.  In fact, numerically it may
be more reasonable to define a new function $\bar{\ph_0}(r,\te)$ such that
$\ph_0(r,\te) \eq r^k \bar{\ph_0}(r,\te)$ (see Sec.~\ref{system-b}).  However, our current approach
simply use $\ph_0(r,\te)$ as one of our fundamental variables.

The Hamiltonian constraint gives
\bea \nn 
\ps_{,rr}+\fr{\ps_{,\te \te}}{r^2}+ \left(
\fr{1}{r}-\fr{\al_{,r}}{\al}-\fr{\ps_{,r}}{\ps}\right) \ps_{,r}
 - \left( \fr{\ps_{,\te}}{\ps}+\fr{\al_{,\te}}{\al}\right) \fr{\ps_{,\te}}{r^2}
-\fr{\ps^5 r^2\sin^2 \te e^{2 \si}}{8 \al^2}\left(\bt_{,r}\,^2+\fr{\bt_{,\te}\,^2}{r^2}
\right) \\ \nn
 -\fr{\ps}{2r} \left[ \left( 1+r \si_{,r}\right)\fr{\al_{,r}}{\al}+\fr{1}{r}\left( \cot \te +
\si_{,\te}\right) \fr{\al_{\te}}{\al}\right]
= \\ 
- 2 \pi \ps \left[ \left( \ph_{,r}\,^2+\fr{\ph_{,\te}\,^2}{r^2}\right)+ \ph^2 \left(
\fr{\ps^4}{\al^2}(w-\bt k)^2 - \fr{k^2}{r^2 \sin^2 \te e^{2 \si}}\right)\right]
\,.
\eea

\noi
The $r$ and $\te$ components of the momentum constraint vanish identically, while
the $\vp$-component of the momentum constraint gives

\beq
 \bt_{,rr}+\fr{\bt_{,\te\te}}{r^2} + \left(\fr{4}{r}+3 \si_{,r}-\fr{\al_{,r}}{\al}+6
\fr{\ps_{,r}}{\ps}\right) \bt_{,r} +
\left(3 \cot \te+3 \si_{,\te}-\fr{\al_{,\te}}{\al}+6
\fr{\ps_{,\te}}{\ps}\right) \fr{\bt_{,\te}}{r^2} = -16 \pi \ph^2\fr{k(\om-\bt k)}{r^2\sin^2
\te e^{2 \si}}\,.
\eeq

\noi
From the assumption that the solutions are stationary, and because we work in a coordinate 
system adapted to the timelike Killing vector, the metric components
must all be $t$-independent.  As a result, the left and right hand sides of the 
evolution equations for the metric~(\ref{metric_evolution_eq}) and 
extrinsic curvature~(\ref{extrinsic_curvature_evolution_eq})
must vanish.  We can thus construct equations for the remaining variables, $\al$ and
$\si$, by taking linear combinations of the right hand sides of those equations.
The combination 
$\dot{\krr}+\dot{\ktt} + \dot{\kpp} \equiv {\dot K} = 0$ gives
\beq
\begin{split}
\al_{,rr}+\fr{2}{r} \al_{,r} +
\fr{1}{r^2}\al_{,\te\te}+\fr{\cot\te}{r^2} \al_{,\te} 
 + \left( \fr{2 \ps_{,r}}{\ps}+\si_{,r}\right) \al_{,r} + \left( 2 \fr{\ps_{,\te}}{\ps}
+ \si_{,\te}\right) \fr{\al_{,\te}}{r^2} \\
- \fr{\ps^4 r^2 \sin^2 \te e^{2 \si}}{2 \al}
\left( \bt_{,r}\,^2+\fr{\bt_{,\te}\,^2}{r^2}\right) + 
4 \pi \al \ps^4 \ph^2 \left( m^2 -
\fr{2 (\om - \bt k)^2}{\al^2}\right) = 0\,.
\end{split}
\eeq
\noi
Finally, the
combination
$\dot{\krr}+\dot{\ktt} - \dot{\kpp} = 0$ gives
\beq 
\begin{split}
\si_{,rr}+\fr{\si_{,\te\te}}{r^2} +
\left(\si_{,r} +4 \fr{\ps_{,r}}{\ps} + \fr{2
\al_{,r}}{\al} + \fr{3}{r}\right) \si_{,r} + 
\left(\si_{,\te} +4 \fr{\ps_{,\te}}{\ps} + \fr{2
\al_{,\te}}{\al} + 2\cot \te\right) \fr{\si_{,\te}}{r^2} \\ 
 +\fr{4}{\ps} \left[ \left( \fr{\ps_{,r}}{\ps}+\fr{\al_{,r}}{\al}+\fr{1}{r}\right) \ps_{,r} +
\left( \fr{\ps_{,\te}}{\ps}+\fr{\al_{,\te}}{\al}+\cot \te\right) \fr{\ps_{,\te}}{r^2}\right]
+\fr{2}{\al r} \left( \al_{,r}+\cot \te \fr{\al_{,\te}}{r}\right) \\ 
+\fr{3 \ps^4 r^2\sin^2 \te e^{2 \si}}{4 \al^2} \left(
\bt_{,r}\,^2+\fr{\bt_{,\te}\,^2}{r^2}\right) \\ \lab{2dsieq}
+4 \pi \left[ \left( m^2+\fr{3k^2}{\ps^4 r^2\sin^2 \te e^{2 \si}}-\left(\fr{\om-\bt
k}{\al}\right)^2 \right) \ph^2 \ps^4 - \left( \ph_{,r}^2+\fr{\ph_{,\te}^2}{r^2}\right)\right]
= 0 \,.
\end{split}
\eeq

\noi
Again, the leading order behaviour of $\si(r,\te)$ as $r \to 0$ is $\si(r,\te) =  r^2 \si_2(\te) + O(r^4)$. 
(To properly impose the regularity condition it would be more appropriate to define a new function
$\sigmabar(r,\te)$ such that $\sigma(r,\te) = r \sigmabar(r,\te)$.)

(\ref{2dkg})-(\ref{2dsieq}) comprise our system of equations for stationary,
axisymmetric boson stars.  For any specific choice of $k = 0, \pm 1, \pm2, \cdots$ these equations constitute a 
coupled, nonlinear eigenvalue problem for the five unknown functions $\al, \bt, \ps, \si$ and 
$\ph$ (with eigenvalue $\omega$).  For convenience we will refer to (\ref{2dkg})-(\ref{2dsieq}) as 
{\em System A}.

\subsection{Challenges and Comments} \lab{challenges}
Before discussing our strategies for solving the above system, we would like to
address some key challenges we face in solving System A. 
The discussion here motivates the particular computational strategies that we adopt in the following
subsection. 

\begin{enumerate}
\item System A is nonlinear: numerical solution of nonlinear systems typically
proceeds via iteration.  Success of an iterative method is often contingent 
on the existence of a good initial guess with which to start the iteration. 
The definition of a ``good initial guess'' depends on both the method of solution
being used, as well as the problem being solved.  Our experience with the numerical
solution of System A suggests that the basin of attraction for a given solution is often
very small.  Thus, we need to have the means of providing initial estimates which 
are already quite close to the solution.

\item System A is a 2D eigenvalue problem:  We could proceed to solve this problem 
using a direct method that generalizes the strategy outlined in Sec.\ref{compactMI} for the case 
of spherically symmetric boson stars.  This would ultimately involve using linear algebra 
software to solve a general full matrix eigenvalue problem, which would require $O(N^3)$ 
operations, where $N$ is the dimension of the matrix.  For a 2D problem with modest resolution,
say $N_r = N_\theta = 100$, the
dimension of the matrix is $N=10^4$, so the eigenvalue solution would require $O(10^{12})$ 
operations. This is obviously an expensive approach, even with current computers, not to mention 
that we would still have to iterate the eigenvalue solution.  The direct technique 
also requires a large amount of memory to store the full
matrix.  We therefore need to focus on devising a more efficient algorithm for solving
the eigenvalue problem.

\item System A is a quadratic eigenvalue problem: 
the eigenvalue $\om$  appears both linearly and quadratically in the equations of System A.
Our solution algorithm must therefore be able to treat quadratic eigenvalue problems.  

\item System A has mixed (Robin) boundary conditions when the outer boundary of 
the computational domain is located at a finite radius.
Dirichlet boundary condition are the simplest conditions to implement, but 
we may need to impose conditions such as (\ref{psouterBC}).
\end{enumerate}

\subsection{Numerical Strategies} \lab{numstr}
Having reviewed the difficulties we encountered in solving System A, we will
now outline the solution strategies we have adopted. 
Since the specific form of the equations we eventually solve is lengthy, we 
focus here on a high-level description of the solution technique, and 
refer the interested reader to App.~\ref{ap7} for more details.

\begin{enumerate}
\item {\em  Compactification of coordinates}:

We introduce a compactified radial coordinate $\zeta$, as well
as a new angular coordinate,  $s$:
\bea
 \zeta &=& \fr{r}{1+r} \,, \\
 s &=& \cos \theta \,.
\eea
\noi
Thus we have $\zeta \in [0,1]$ and $s\in [0,1]$.  This transformation simplifies
some of the
boundary conditions.  For instance~(\ref{psouterBC}) becomes
\beq
 \left.\ps(\zeta,s)\right|_{\zeta=1} = 1\,.
\eeq
\noi
Importantly, compactification not only allows us to work with (algebraically) 
simpler boundary conditions, it permits us to use {\em exact} boundary 
conditions rather than the asymptotic conditions that are needed when an
arbitrarily truncated computational domain is used.
 
\item {\em Reduction to a standard eigenvalue problem and the use of a multigrid eigenvalue solver}:

We note that in principle we can solve the quadratic eigenvalue problem as
follows~\cite{komzsik:2001}.  Suppose we write equation~(\ref{2dkg}) schematically as
\beq
 \left( A \om^2 + B \om + C \right)  \ph = 0\,.
\eeq
\noi
where $A$, $B$ and $C$ are differential operators.
Then we can cast the above equation into a standard {\em linear} form via the 
introduction of an auxiliary function, $\Phi$ defined by
\beq
	\Phi \equiv \omega \phi \,.
\eeq
Specifically, we have
\beq
 \om 
   \left[ 
      \begin{array}{cc}
      A & 0\\
      0 & I
      \end{array}
   \right]
   \left[ 
      \begin{array}{c}
       \Ph \\ \ph
      \end{array}
   \right]
+
   \left[ 
      \begin{array}{cc}
      B & C\\
      -I & 0
      \end{array}
   \right]
   \left[ 
      \begin{array}{c}
       \Ph \\ \ph
      \end{array}
   \right]
= 0\,.
\eeq
\noi
Hence, our task is now to find an efficient algorithm for solving a standard 2D eigenvalue
problem~\cite{kohn:1995,Costiner:1995}.

As discussed in Chap.~\ref{num_method}, multigrid techniques generally provide the means for very 
efficient solution of multi-dimensional elliptic problems, and with a slight modification 
of a standard multigrid algorithm, we can produce a solver for the eigenvalue problem.
A pseudo-code of the top-level structure of our  multigrid eigenvalue solver for the
generalized eigenvalue problem
\beq \lab{eigen_eq}
 L(\om,\om^2) \ph \eq \left( A \om^2 + B \om + C\right)\ph = 0\,,
\eeq
\noi
is given in Fig.~\ref{table:MGEVS}.  
Again note that due to the regularity conditions
the scalar field must vanish on the axis for all $k>0$.  
As a result, for any $k$, the corresponding family of solutions 
could be naturally parametrized by the modulus of the $k$-th order radial derivative 
of the scalar field evaluated at the origin.
However, following ~\cite{yoshida_eriguchi:1997b} we have instead 
chosen to parametrize each family of 
solutions by specifying the value of $\ph_{(k)}$ at some 
arbitrary $k$-dependent point $(\zeta,s) = (\zeta_0(k),0)$ on the equatorial plane,
i.e.\ $\ph_{(k)} \eq \ph_0(\ze_0(k),0)$.
The other equations of System A are solved using a standard multigrid iteration
(see Fig.~\ref{table:BS2DIVP}).  We also note that we use a FAS (Full Approximation
Storage)
approach to deal with the nonlinearity of our elliptic system.

\begin{figure}
\fboxsep=0.4cm
\centerline{\fbox{\parbox{5.5in}{
\begin{tabbing}
\hspace*{0.5cm}\= \hspace{0.5cm} \= \kill	
\textbf{Let} $\ph \ne 0$ be an initial guess \\[0.25cm]
\textbf{repeat}\\[0.25cm]
\>Solve $\langle \ph,  L(\om, \om^2) \ph \rangle = 0$ for $\om$ \\[0.25cm]
\> Rescale $\ph$ so that $\ph(\ze_0,0) = \ph_{(k)}$ \\[0.25cm]
\> Perform $N_{V}$ multigrid $V$-cycles on $L(\om, \om^2) \ph = 0$  \\[0.25cm]
\textbf{until} $\|L(\om, \om^2) \ph \| < \ep_\ph$
\end{tabbing}
}}}
\caption[Pseudo-code of the multigrid eigenvalue solver]
{Pseudo-code of the multigrid eigenvalue solver for (\ref{2dkg}), represented here using the
general form (\ref{eigen_eq}).  $\langle \ph,  L \ph \rangle$ denotes the inner product
of $\ph$ and $L \ph$.  Specifically, if the finite difference representations on
the mesh $(\ze_i,s_j)$ of 
$\ph$ and $L \ph$ are $\ph_{i,j}$ and $(L \ph)_{i,j}$, respectively, 
then $\langle \ph,  L \ph \rangle \eq \sum_{i,j} \ph_{i,j} (L \ph)_{i,j}$.
$\ph_{(k)}$ is the family parameter for a particular value of $k$ (see text for additional
discussion).
$\ep_\ph$ is a controllable tolerance parameter for the overall 
algorithm.  $N_V$ is another adjustable parameter that controls how many
multigrid $V$-cycles are applied to the discretization of~(\ref{eigen_eq})
at each cycle of the iteration. 
This iteration is 
an essential part of the algorithm for solving system A, which is itself 
described in pseudo-code form in Fig.~\ref{table:BS2DIVP}.
}
\label{table:MGEVS}
\end{figure}

\item {\em A homotopic (continuation) method with initialization by Newtonian solutions}:

As mentioned above, in constructing any particular solution of System A,
we will generally need a good initial estimate to ensure convergence of the overall 
iteration (note that Fig.~\ref{table:MGEVS} is a pseudo-code for the solution of  the $\ph$ equation
only, with values of $\alpha, \beta, \psi$ and $\sigma$ considered fixed).  
We can generate good initial guesses using a simplified version of the 
homotopy (continuation) method \cite{burden}.  The idea is to embed the system to be solved (System
A) within an {\em expanded} system.  More specifically, if we want to solve
a $d$-dimensional non-linear system

\beq
\label{f0}
 {\mbf F}({\mbf x}) \eq {\mbf F}(x_1,\cdots,x_d) \eq 
\left( F_1(x_1,\cdots,x_d), F_2(x_1,\cdots,x_d), \cdots ,
F_d(x_1,\cdots,x_d) \right) = 
{\mbf 0}\,,
\eeq

\noi
where the $F_i$ are non-linear functions of $(x_1,\cdots,x_d)$ and ${\mbf 0}$ is the
$d$-dimensional zero vector,
we consider a family of problems with $p$ parameters, $\la_i$, 
$i = 1\cdots p$ that each take values in a range $[a_i,b_i]$ for 
specified constant values $a_i$ and $b_i$.  We then replace the 
above equation with

\beq
\label{g-cont}
 {\mbf G}({\mbf x}; \la_1, \la_2, \cdots, \la_p) = {\mbf 0}\,,
\eeq

where ${\mbf G}$ is defined in obvious analogy with~(\ref{f0}),
such that when $\la_i = a_i$ for all $i$, ${\mbf G}({\mbf x};\la_i) = {\mbf 0}$ 
has a known solution, while
when $\la_i = b_i$, ${\mbf G}({\mbf x};\la_i) = {\mbf 0}$ reduces to the system to be solved, 
namely,
${\mbf F}({\mbf x}) = {\mbf 0}$.  Therefore, by varying the $\la_i$ from $a_i$ to $b_i$ 
sufficiently slowly~\footnote{
In general, some of the parameters $\la_i$ may {\em not} be continuous variable;
for instance, one of the parameters might be the number of grid points $N_x$ in a given 
coordinate direction, which can only be changed by integral
amounts.  Nevertheless, we have found that variation of such parameters by the 
admissible small amounts has proven useful in obtaining the final solutions.} 
and by always using the solution from the previous calculation (previous values of $\la_i$) as 
our initial estimate for the new calculation, we can, in principle, obtain the solution 
to~(\ref{f0}) from the known solution to (\ref{g-cont}) (i.e.\ the solution to (\ref{g-cont})
for $\la_i = a_i$).

In practice, the continuation parameters, $\la_i$, that we use are $N_\zeta$,
$N_s$, $\ph_{(k)}$, $m$, $k$ and $\om_{\mathrm{UR}}$, where
$N_\zeta$ and $N_s$ are the number of grid points in the $\zeta$ and
$s$ coordinate directions, respectively, and $\om_{\mathrm{UR}}$ is the under 
relaxation parameter defined below.
Note that although physically
$k$ can only take integral values, in the continuation we often
find it useful to vary it as if it can take on an arbitrary real value.
In fact, in some cases we can even start
from a spherical solution, i.e.\ with $k=0$, and then vary the value of $k$ to get a rotating
solution.

With the continuation method and the other strategies mentioned above, the
overall pseudo-code for the solution of System A is 
given in Fig.~\ref{table:BS2DIVP}.  (See the caption for the parameters used.)
Fig.~\ref{HCrep} shows a schematic representation of the homotopic method
in Fig.~\ref{table:BS2DIVP} for two typical continuation parameters $\la_i$
and $\la_j$.

\begin{figure}[!h]
\fboxsep=0.4cm
\centerline{\fbox{\parbox{5.5in}{
\begin{tabbing}
\hspace*{0.5cm}\= \hspace*{0.5cm}\= \hspace{0.5cm} \= \kill	
\textbf{Partition} the set $\{\la_i\}$ into subintervals $\mathcal{P} \eq
\{\left.\la_{i,n_i}\right| n_i = 1\cdots N_i\}$ \\
where $\la_{i1}=a_i$ and $\la_{iN_i}=b_i$ \\[0.40cm]
$j = 0$\\[0.05cm]
\textbf{Foreach} ${\boldsymbol \la}=\{\la_{1 n_1}, \la_{2 n_2},\cdots,\la_{m
n_m}\} \in \mathcal{P}$ \\[0.05cm]
\> \textbf{Call} IVP\_Solver(${\boldsymbol \la},\mathcal{S}^j$) \\[0.05cm]
\> $j = j + 1$\\[0.05cm]
\textbf{End}\\[0.40cm]
\textbf{Subroutine} IVP\_Solver(${\boldsymbol \la}, \mathcal{S}^j$)\\[0.05cm]
\>\textbf{Initialize} solutions to $\mathcal{S}^j$ \\[0.05cm]
\>$l=0$\\[0.05cm]
\>\textbf{Repeat}\\[0.05cm]
\>\>$l=l+1$\\[0.05cm]
\>\>\textbf{Update} $\ph^l$ and $\om$ according to Fig.~\ref{table:MGEVS}\\[0.05cm]
\>\>\textbf{Update} $\ps^l$ using standard multigrid\\[0.05cm]
\>\>\textbf{Update} $\bt^l$ using standard multigrid\\[0.05cm]
\>\>\textbf{Update} $\al^l$ using standard multigrid\\[0.05cm]
\>\>\textbf{Update} $\si^l$ using standard multigrid\\[0.05cm]
\>\>\textbf{If} $l\ne0$ \textbf{then}\\[0.05cm]
\>\>\>\textbf{Call} UR\_Update($l$)\\[0.05cm]
\>\>\textbf{End If}\\[0.05cm]
\>\textbf{Until} $\|u^l-u^{l-1}\|/\|u^l\| < \ep$ \,,\;\;\;\; for
$u\in\{\ph,\ps,\bt,\al,\si \}$\\[0.05cm]
\>\textbf{Output} Solutions to $S^{j+1}$\\[0.05cm]
\textbf{End of Subroutine} \\[0.40cm]
\textbf{Subroutine} UR\_Update($l$)\\[0.05cm]
\>\textbf{Foreach} $u\in \{\ph,\ps,\bt,\al,\si\}$ \\[0.05cm]
\>\> $u^l = \om_{\rm UR} u^l + \left( 1-\om_{\rm UR}\right)
u^{l-1}$ \\[0.05cm]
\>\textbf{End} \\[0.05cm]
\textbf{End of Subroutine}
\end{tabbing}
}}}
\caption[Pseudo-code of the initial value solver for axisymmetric rotating boson stars]
{Pseudo-code of the initial value solver for axisymmetric rotating boson stars.  The continuation parameters
$\la_i$ that are used are $N_\ze, N_s, \ph_{(k)}, m, k$ and $\om_{\rm UR}$. $S^0$ is 
initialized either to a Newtonian solution or to a spherically symmetric solution obtained using 
one of the techniques described in Chap.~\ref{bs1d}.
$\ep$ is an adjustable convergence parameter for the top-level iteration of the 
algorithm, and is typically set to $10^{-8}$.  
The underrelaxation routine UR\_Update, which involves the adjustable underrelaxation
parameter, $\omega_{\rm UR}$, is used to ensure convergence---see the text for 
additional details.
The typical number of overall iterations (loop index $l$)
is between 10 to 50, while the number of $V$-cycles for each multigrid solver 
is less than 100.  
(The actual numbers depend on how the
partitions of $\la_i$ are chosen, i.e.\ on how good the initial guess $S^0$ is.)}
\label{table:BS2DIVP}
\end{figure}

\begin{figure}[h]
\begin{center}
\includegraphics[width=10cm,clip=true]{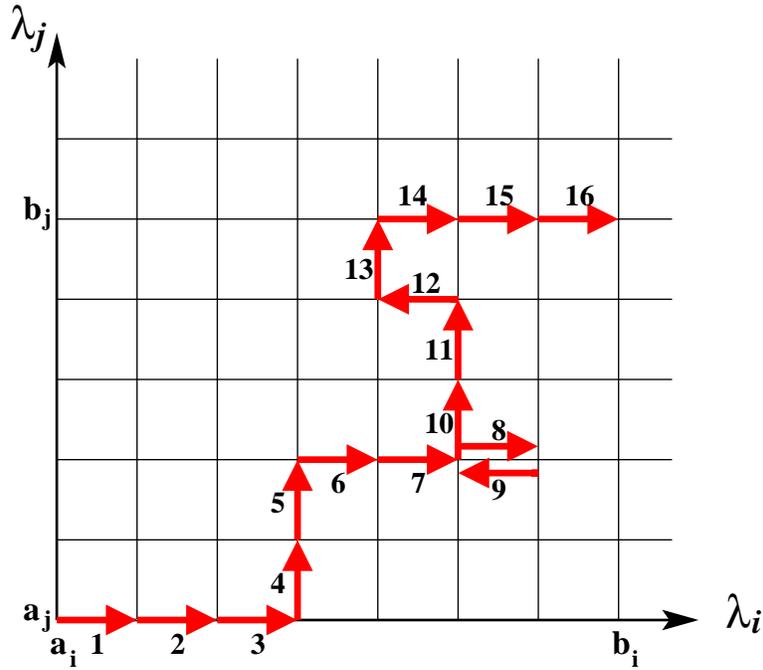}
\end{center}
\caption
[A schematic representation of the homotopic method in Fig.~\ref{table:BS2DIVP} for continuation
parameters $\la_i$ and $\la_j$.]
{A schematic representation of the homotopic method shown in pseudo-code form in
Fig.~\ref{table:BS2DIVP}, for the case of two continuation
parameters, $\la_i$ and $\la_j$.
The figure shows a {\em typical} path
in the parameter space of $\la_i$ and $\la_j$.
The arrows show the stepping-directions through parameter space,
while the numbers show the order of the stepping sequence.  In the first 3 steps
we increase the parameter $\la_i$ with stepping size
$\la_{i,n_{i+1}}-\la_{i,n_i}$
(see Fig.~\ref{table:BS2DIVP} for the meaning of $\la_{i,n_i}$, and note that
the stepping size is in general different for different $i$ and/or $n_i$).
At step 4, a further step in the $\la_i$ direction produces a problem that does not converge, so we 
then step
forward in the $\la_j$ direction.  After step 5 we find that we 
can again increase $\la_i$
until the end of step 8, when we realize that increasing either
$\la_i$ or $\la_j$ produces divergence.  We thus ``backtrack'' in 
step 9 and find that we can increase in the $\la_j$ direction again.  We continue with
this strategy until we end at step 16, with $\la_i=b_i$ and $\la_j=b_j$.
}
\label{HCrep}
\end{figure}

Finally, we  note that the homotopy method cannot be used 
without a known initial solution, i.e.\ a solution to
\beq \lab{initialcond}
 {\mbf G}({\mbf x}; a_1, a_2, \cdots, a_p) = {\mbf 0}\,.
\eeq
\noi
Although we do not have such a solution, we can tune the parameters $\la_i$
to the weak field limit 
$\ph_{(k)}\approx 0$, where Newtonian gravity should be a good approximation 
to general relativity, and can then use the solution of the Newtonian problem
as the initial guess for (\ref{initialcond}).   (Alternately, we can sometimes initialize 
using a spherically symmetric solution constructed using one of the algorithms 
described in Chap.~\ref{bs1d}.)

As derived in App.~\ref{ap8}, the (non-relativistic) Newtonian limit of the coupled Einstein-Klein-Gordon
system, yields the following coupled Poisson-Schr\"{o}dinger system:
\footnote{As far as we can ascertain, there is no complete derivation of this 
limit in the literature.  In Ref.~\cite{guenther} Sec.~2.2 an attempt was made to 
derive the Newtonian limit; however, some equations used there are incorrect, and 
hence the derivation is invalid.}
\bea
  i \dot{\ph}(t,{\bf x}) &=&  -\fr{1}{2 m} \na^2 \ph + m V \ph \,,\\
  \na^2 V(t,{\bf x}) &=& 4 \pi m^2 \ph \ph^* \,,
\eea
\noi
where $V$ is the Newtonian potential.  In the stationary case, and 
assuming an ansatz
\bea
	\phi(t,{\mathbf x}) \longrightarrow e^{-i(E+m)t} \phi({\mathbf x}) \,,
\eea
the above system reduces to 
\bea
\label{newt}
  E {\ph} &=&  -\fr{1}{2 m} \na^2 \ph + m V \ph \,,\\
  \na^2 V &=& 4 \pi m^2 \ph \ph^* \,.
\eea
Numerical experiments reveal that
it is easier to solve~(\ref{newt}) than System A.
In fact, using an algorithm similar to that described above for System A, 
we can usually find solutions of~(\ref{newt}) using suitably chosen 
gaussian profiles for $\phi$ and $V$.
Once the Newtonian solutions (possibly rotating) are in hand, they can be used to start the 
continuation process for the general relativistic solver.

As a final remark, we mention that the derivation of the Newtonian limit 
(see App.~\ref{ap8}) was very useful in identifying which approximations
had to be satisfied in the weak field limit.  In particular, this allowed
us to determine whether a particular Newtonian solution was likely to 
be a good initial guess for the full Einstein-Klein-Gordon system.

\item{{\em Underrelaxation:}}

Finally, for any value of the continuation parameters, 
our solution of the 5 coupled elliptic initial value equations 
proceeds via iteration, such that at each cycle (the {\bf Repeat} 
loop of Fig.~\ref{table:BS2DIVP}) each of the 5 unknown grid functions
is updated in turn.  In the notation of Fig.~\ref{table:BS2DIVP} we have
$\phi^{l-1}\to\phi^l$, $\psi^{l-1}\to\psi^{l}$, etc. in passing from the 
$l-1$-st iteration to the $l$-th.  Empirically, and 
following Yoshida \& Eriguchi~\cite{yoshida_eriguchi:1997}, we have found 
it generally useful---and in some cases essential---to adopt an
{\em underrelaxation} strategy, whereby each grid function, $u^l$ is 
updated using
\beq
u^l = \om_{\mathrm{UR}} {\hat u}^l + \left( 1-\om_{\mathrm{UR}}\right) u^{l-1} \,,
\eeq
where $\om_{\mathrm{UR}}$ is the adjustable underrelaxation parameter
and ${\hat u}^l$ is the ``bare'' solution computed during the course 
of the $l$-th pass through the main loop of ``IVP\_solver''.  In
many cases we find that we need $\om_{\mathrm{UR}} < 1$ for convergence
(hence the terminology ``underrelaxation''), although in some situations
$\om_{\mathrm{UR}} > 1$ can accelerate convergence (as in the usual
case of the {\em successive overrelaxation} technique for the solution of 
discretized elliptic equations).  In general, to achieve optimal
convergence, $\om_{\mathrm{UR}}$ must be adjusted on a problem-by-problem
basis.

\end{enumerate} 

\subsection{Results} \lab{MG_results}
We now present some results computed using the method described in the 
previous section for the specific cases $k=1$ and $k=2$.

\begin{figure}[htp]
\begin{center}
\ifthenelse{\equal{\highQ}{true}} {
\includegraphics[width=7.4cm,clip=true]{eps/bs2divp/phi.eps}\\
\includegraphics[width=6.9cm,clip=true]{eps/bs2divp/psi.eps}
\hspace{0.5cm}
\includegraphics[width=6.9cm,clip=true]{eps/bs2divp/alpha.eps}
\hspace{0.5cm}
\includegraphics[width=6.9cm,clip=true]{eps/bs2divp/beta.eps}
\hspace{0.5cm}
\includegraphics[width=6.9cm,clip=true]{eps/bs2divp/sigma.eps}
} {
\includegraphics[width=7.4cm,clip=true]{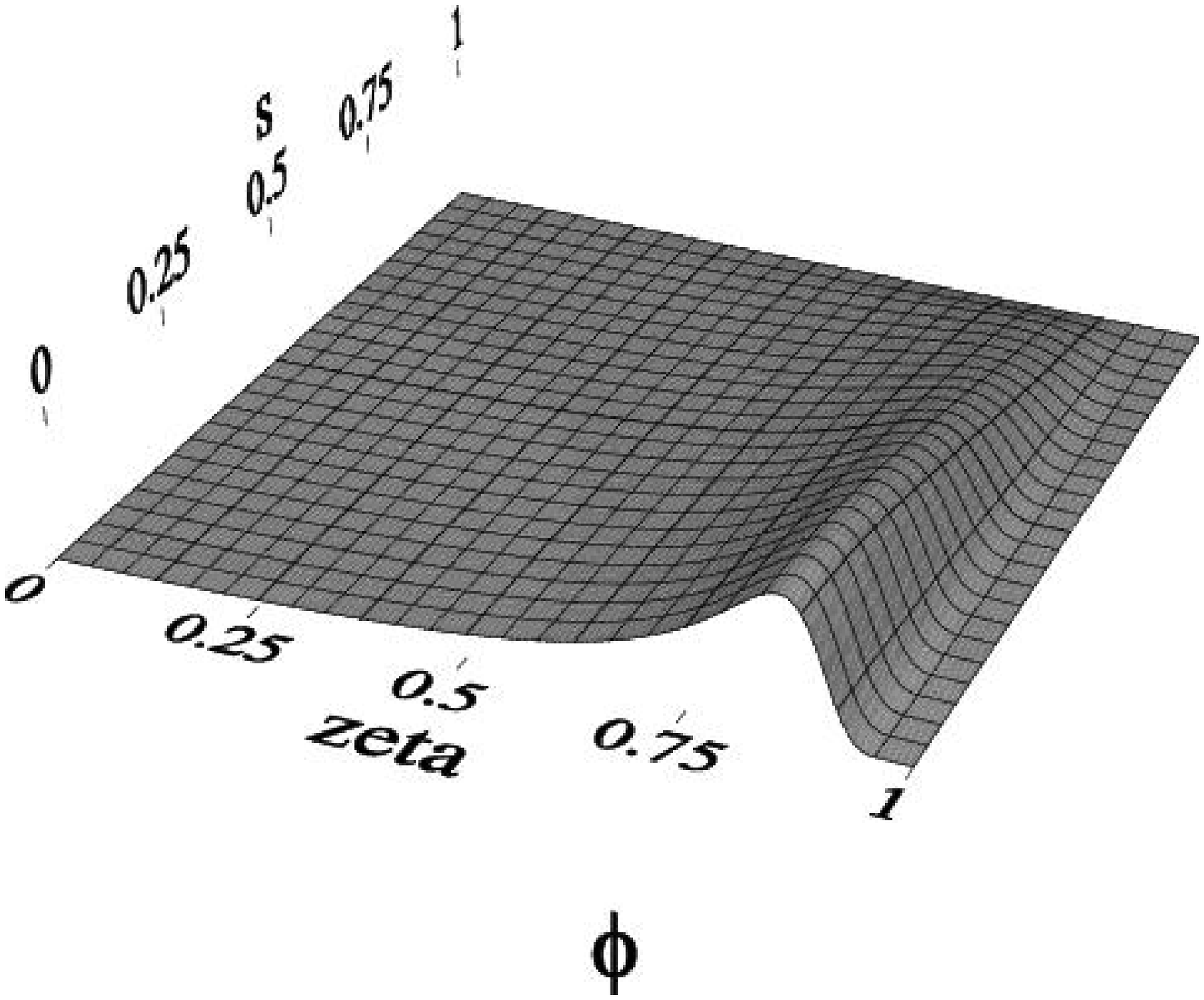}\\
\includegraphics[width=6.9cm,clip=true]{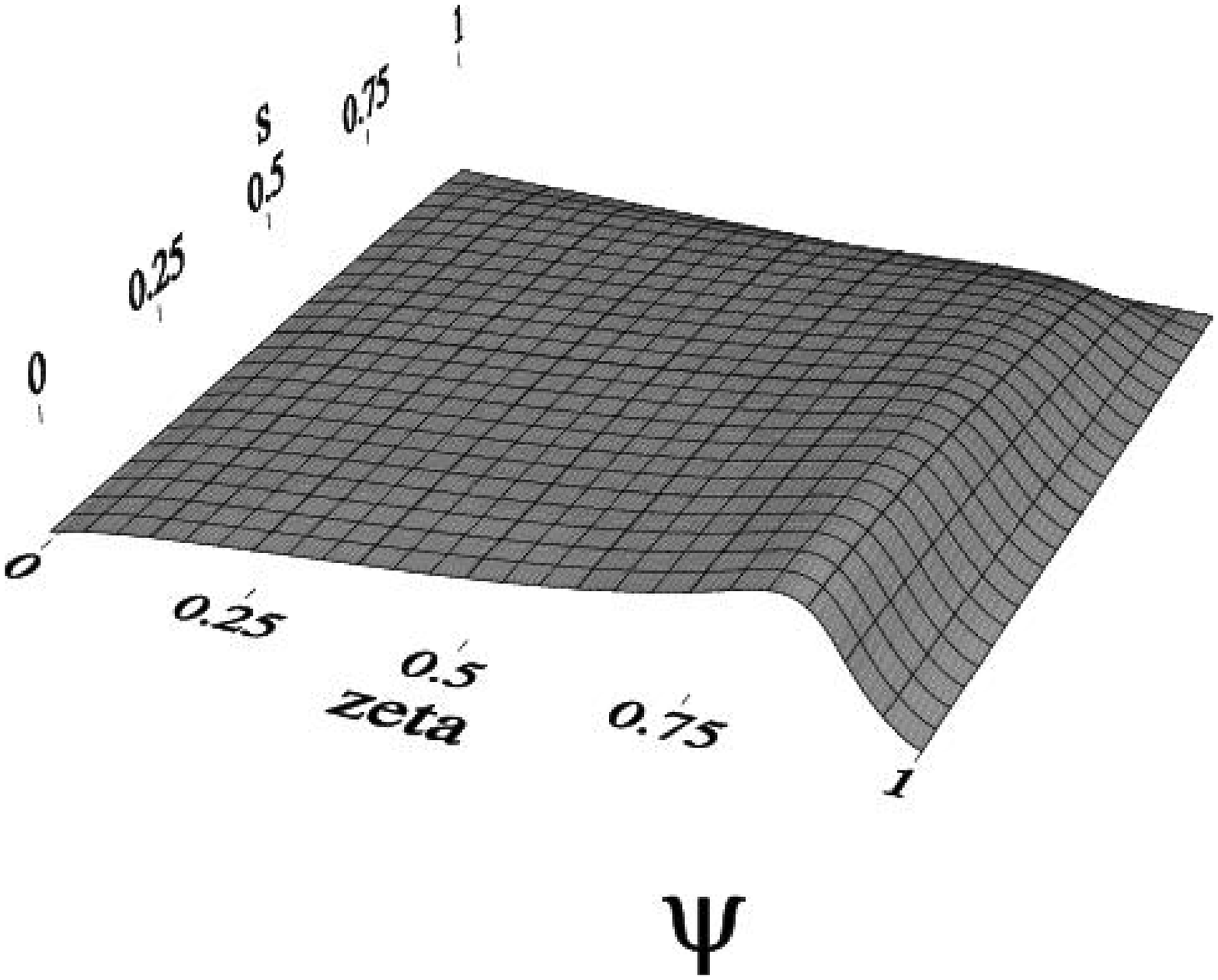}
\hspace{0.5cm}
\includegraphics[width=6.9cm,clip=true]{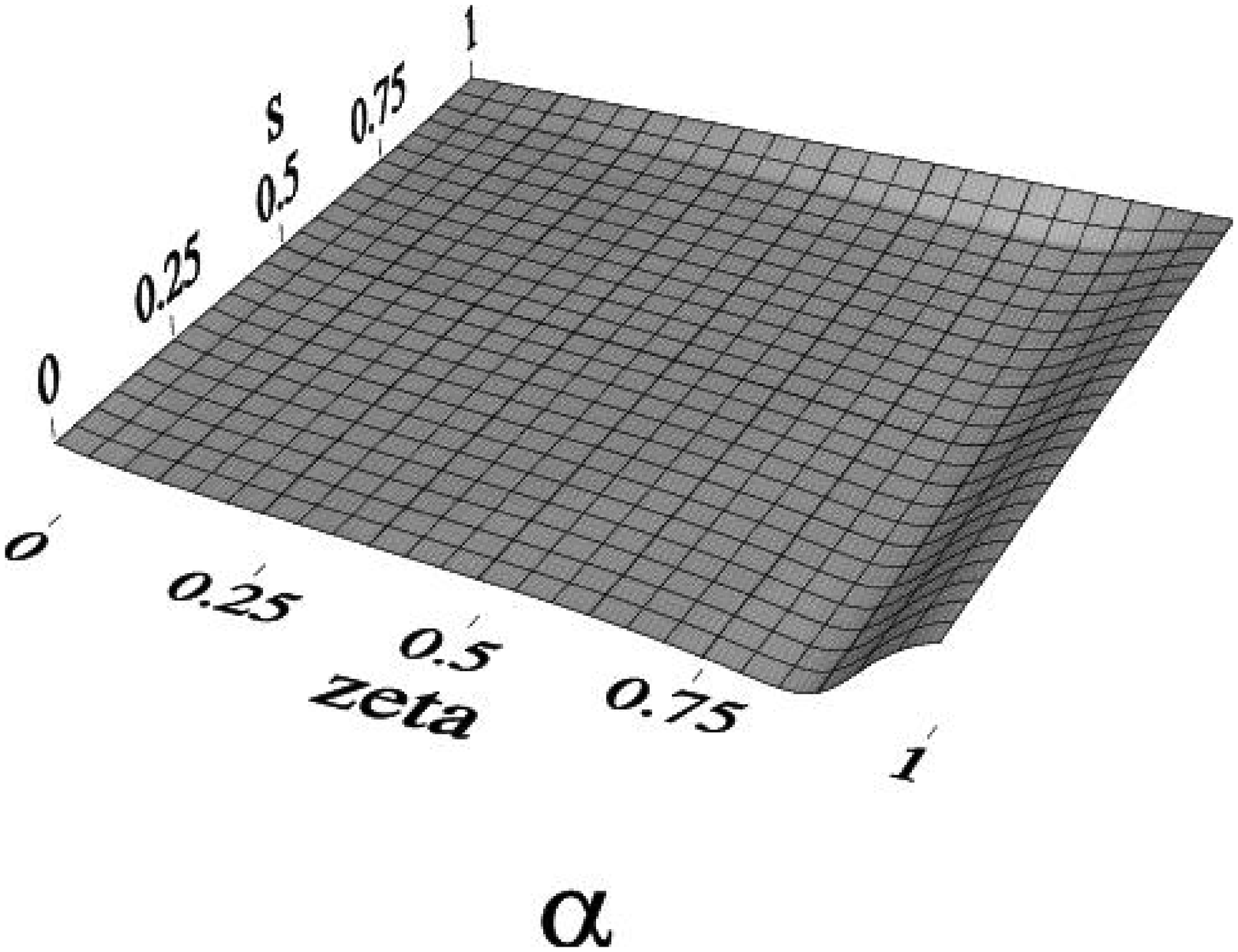}
\hspace{0.5cm}
\includegraphics[width=6.9cm,clip=true]{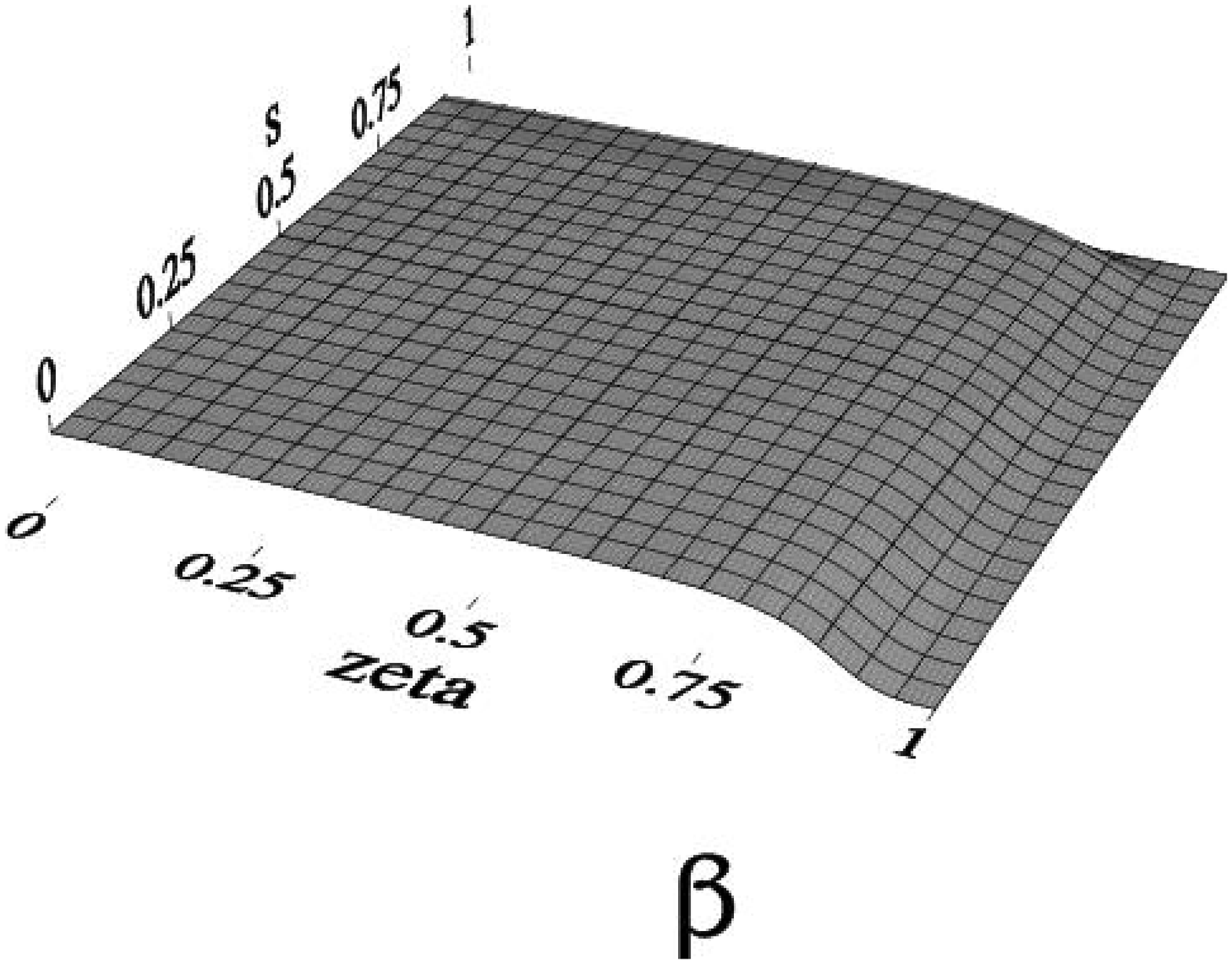}
\hspace{0.5cm}
\includegraphics[width=6.9cm,clip=true]{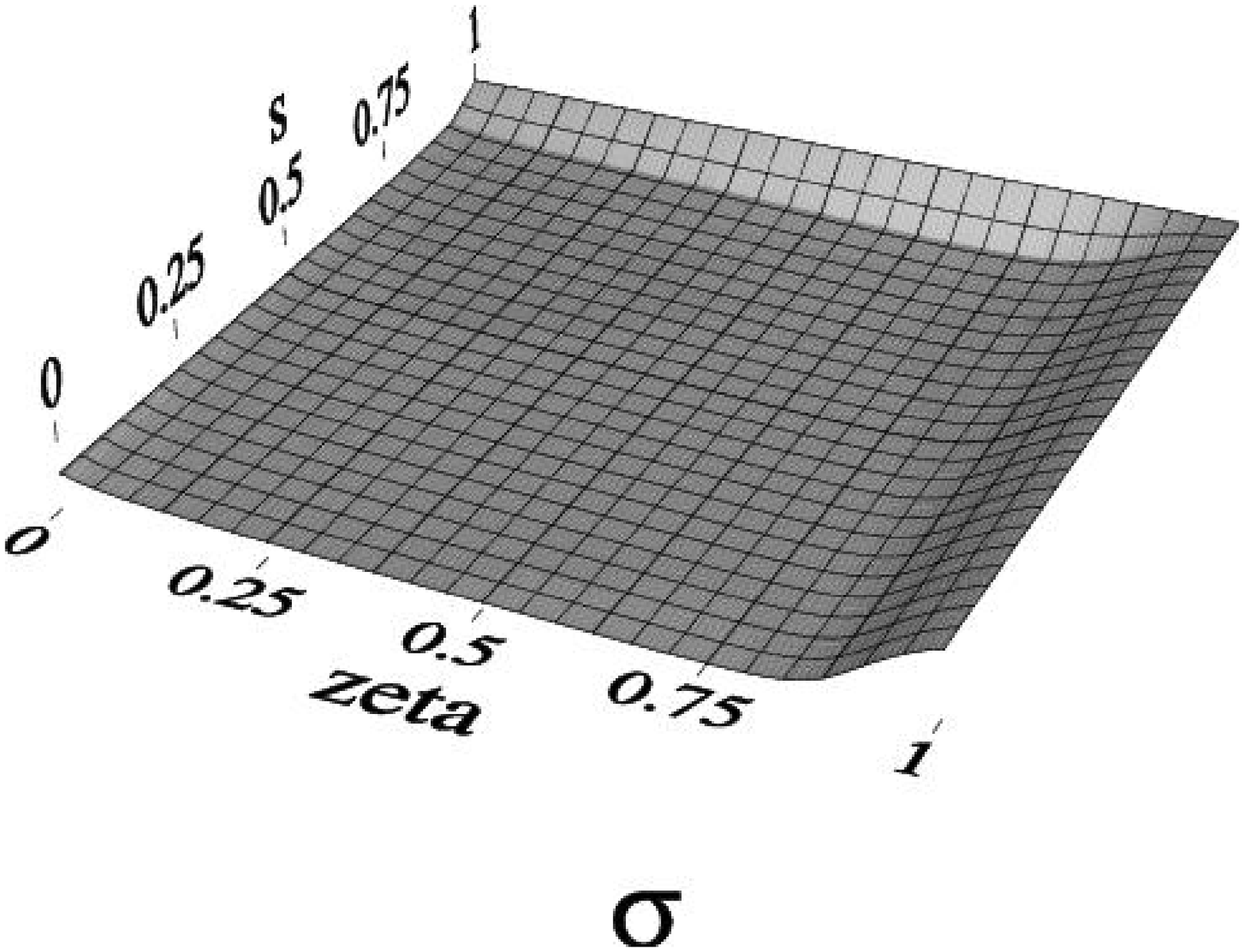}
}
\end{center}
\caption[Typical stationary rotating boson star solution with angular momentum parameter, $k=1$]
{Typical stationary rotating boson star solution with angular momentum parameter, $k=1$.
The top plot shows the configuration of the scalar field, which has
toroidal level surfaces.
The bottom plots show various metric functions. 
Note that the solutions are symmetric under reflection through the equatorial plane,  $s=0$.
See the text for additional details concerning the numerical solutions shown here.
In lieu of $z$-axes and labels on the above plots  (as well as on the other ``surface'' plots that appear in
subsequent figures), we quote the $z$-extrema for the various data sets visualised here:
$0.0\le \ph_0(\ze,s)\le 0.17$, 
$1.0\le \ps(\ze,s)\le 1.1$, 
$0.81\le \al(\ze,s)\le 1.0$, 
$0.0\le \bt(\ze,s)\le 0.014$, 
$-0.028\le \si(\ze,s)\le 0.0$.
}
\label{bs2divpk1}
\end{figure}

Fig.~\ref{bs2divpk1} shows a typical result for the stationary 
solution of a boson star with angular momentum parameter $k=1$.  
This solution was computed on a finite difference grid with $N_\zeta = 129$ and 
$N_\theta = 15$, and the convergence criterion for the iteration described in Fig.~\ref{table:BS2DIVP}
was ${\rm \ep} = 10^{-8}$.
We remind the reader that for $k=1, 2, \cdots$, the complex field vanishes at 
the origin, and we thus cannot use the central field value to parameterize the 
continuous family of solutions that exists for each value of $k$.  
Rather, we
parameterize the solutions by choosing (arbitrarily) some point $r=r_0$ in the 
equatorial plane (or, equivalently, $\ze=\ze_0$), and then specifying 
the value of the modulus of the scalar field there.  Specifically, for the 
solution shown in Fig.~\ref{bs2divpk1} we chose the point $\zeta = 0.5$, $s=0$, and 
set $\phi_{(1)} \equiv \phi_0(0.5,0) = 0.03$.
The plot at the top of~Fig.~\ref{bs2divpk1}
shows the configuration of the scalar field, $\phi_0(\zeta,s)$, while the plots 
below show the various metric components $\al, \bt, \ps$ and
$\si$.
In all of the plots, 
the axis labeled 
$\zeta$ corresponds to the equatorial plane, $s=0$, or $\te=\pi/2$, while 
the far edge of each plot, parallel to the $\zeta$-axis, is $s=1$ and is the 
azimuthal symmetry axis.  (As mentioned in the introductory section of this 
chapter, we are restricting attention to equatorially-symmetric solutions, 
and thus only need to solve our equations on a single hemisphere.)
The figure shows that the matter distribution for the rotating boson star 
has---in contrast to spherically symmetric boson stars or rotating fluid
stars---toroidal level surfaces
(the toroidal nature of a related configuration is clearer in the 
top plot of Fig.~\ref{bs2divpnonc} which 
uses the usual cylindrical coordinates $(\rho,z)$).
We also note that the 
scalar field in this case has a maximum at $\zeta = 0.85$, or $r = \zeta/(1-\zeta) = 5.7$.
We will see that for $k=2$, the maximum value of the scalar field is attained at a 
larger value of $\ze$, which implies greater difficulty in constructing solutions at 
fixed resolution, relative to $k=1$.
See Sec.~\ref{remarks} for further details.

\begin{figure}[htp]
\begin{center}
\ifthenelse{\equal{\highQ}{true}} {
\includegraphics[width=6.9cm,clip=true]{eps/bs2divp/phik2.eps}\\
\includegraphics[width=6.5cm,clip=true]{eps/bs2divp/psik2.eps}
\hspace{0.5cm}
\includegraphics[width=6.1cm,clip=true]{eps/bs2divp/alphak2.eps}
\hspace{0.5cm}
\includegraphics[width=6.4cm,clip=true]{eps/bs2divp/betak2.eps}
\hspace{0.5cm}
\includegraphics[width=6.5cm,clip=true]{eps/bs2divp/sigmak2.eps}
}{
\includegraphics[width=6.9cm,clip=true]{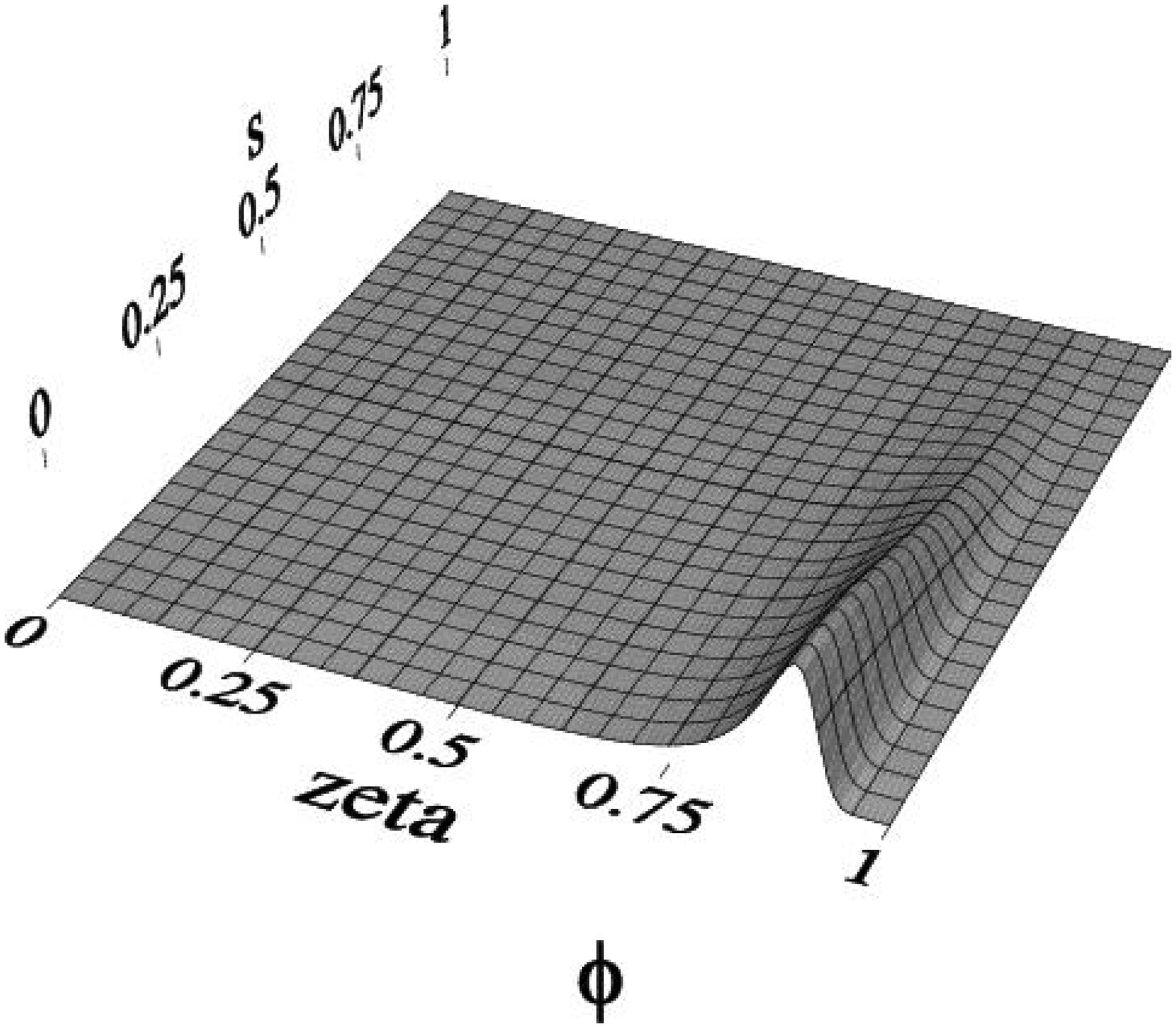}\\
\includegraphics[width=6.5cm,clip=true]{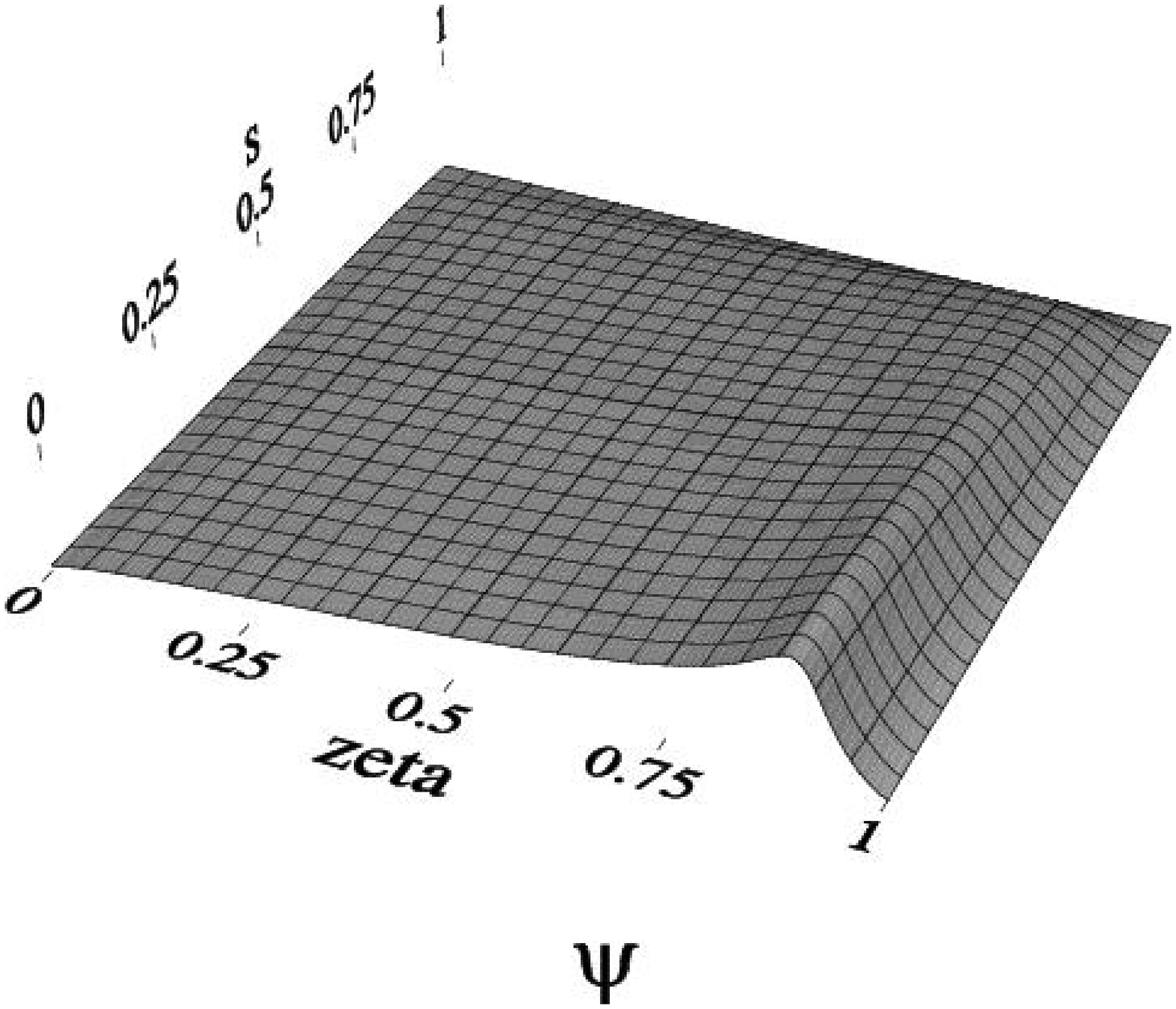}
\hspace{0.5cm}
\includegraphics[width=6.1cm,clip=true]{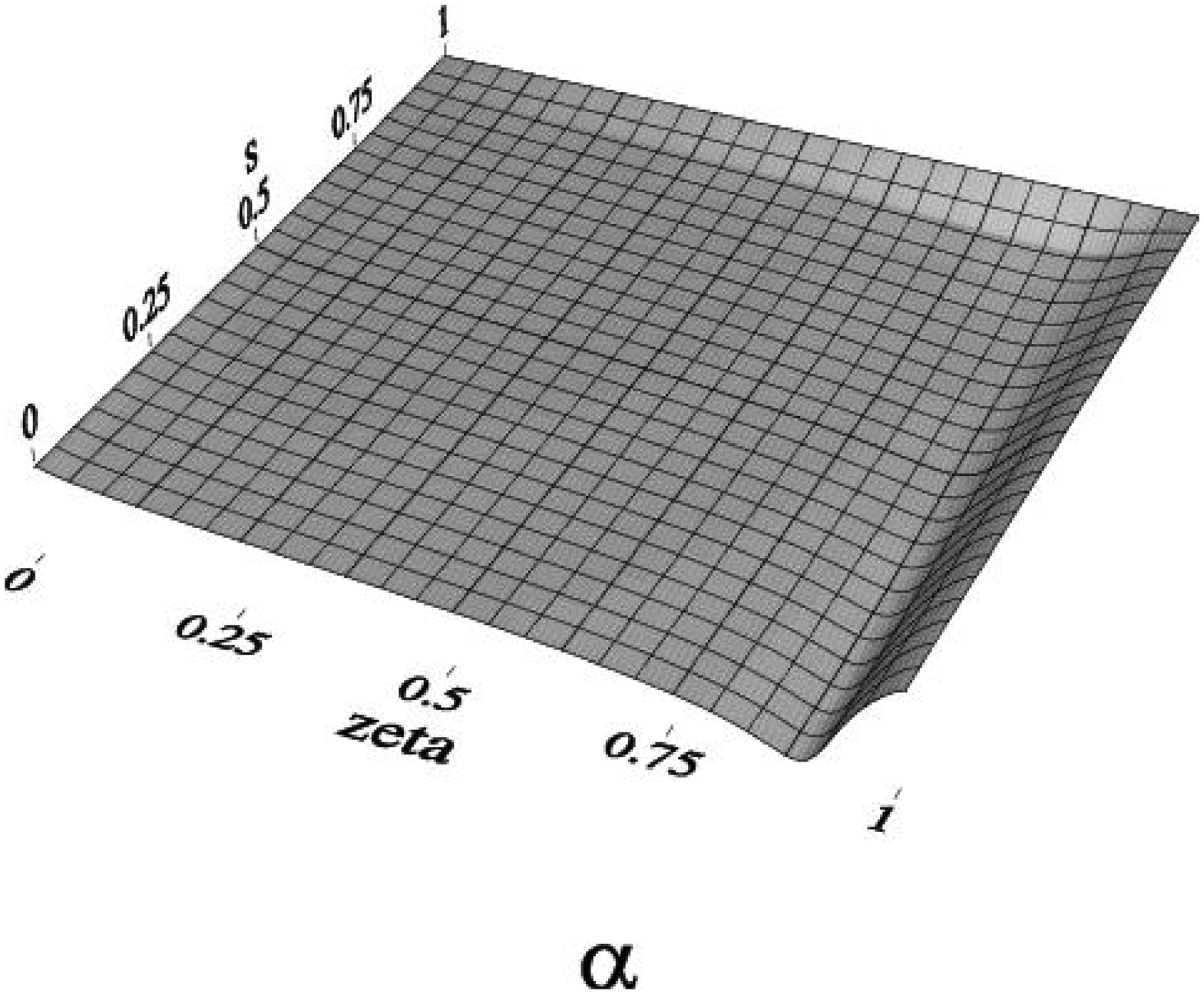}
\hspace{0.5cm}
\includegraphics[width=6.4cm,clip=true]{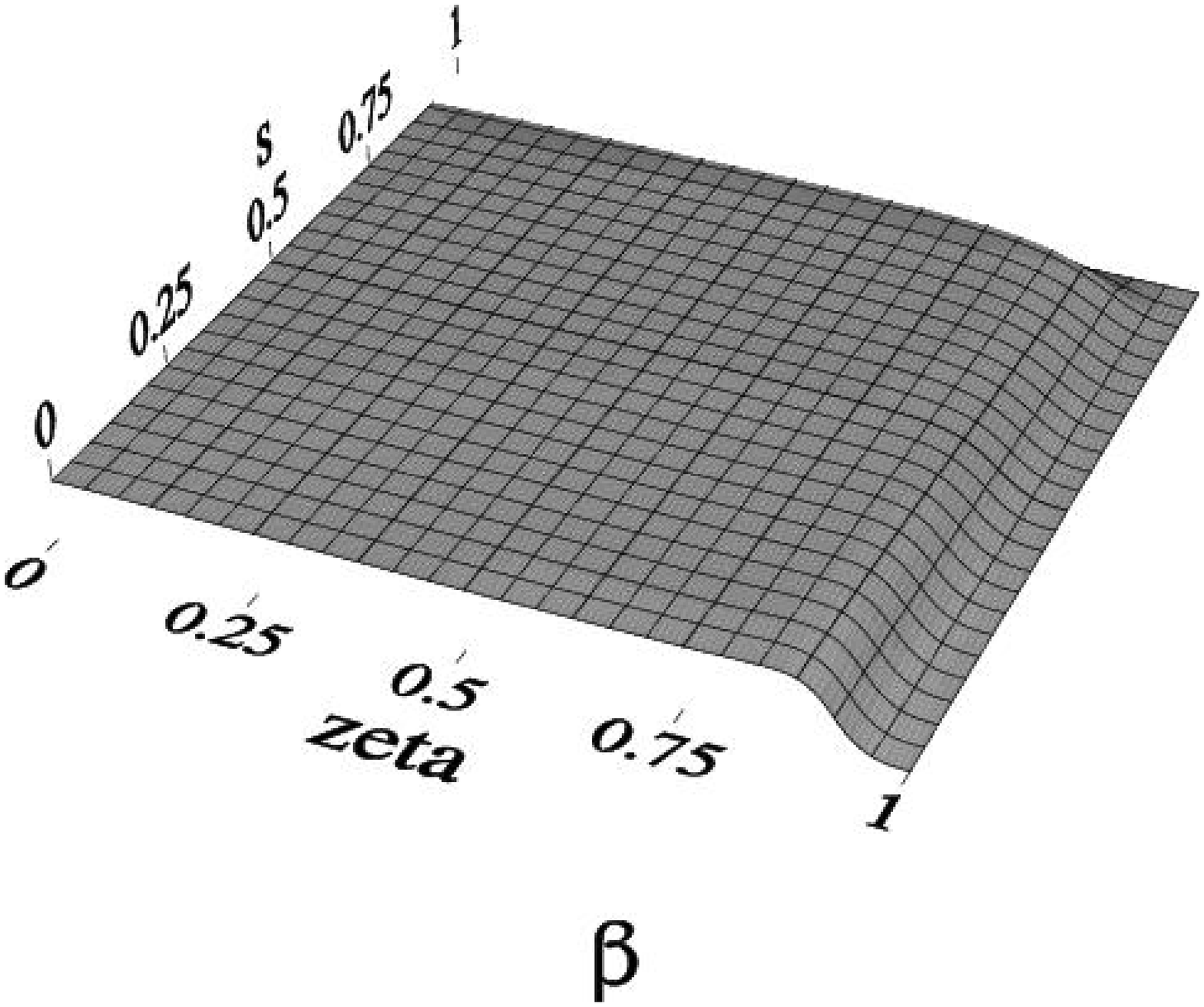}
\hspace{0.5cm}
\includegraphics[width=6.5cm,clip=true]{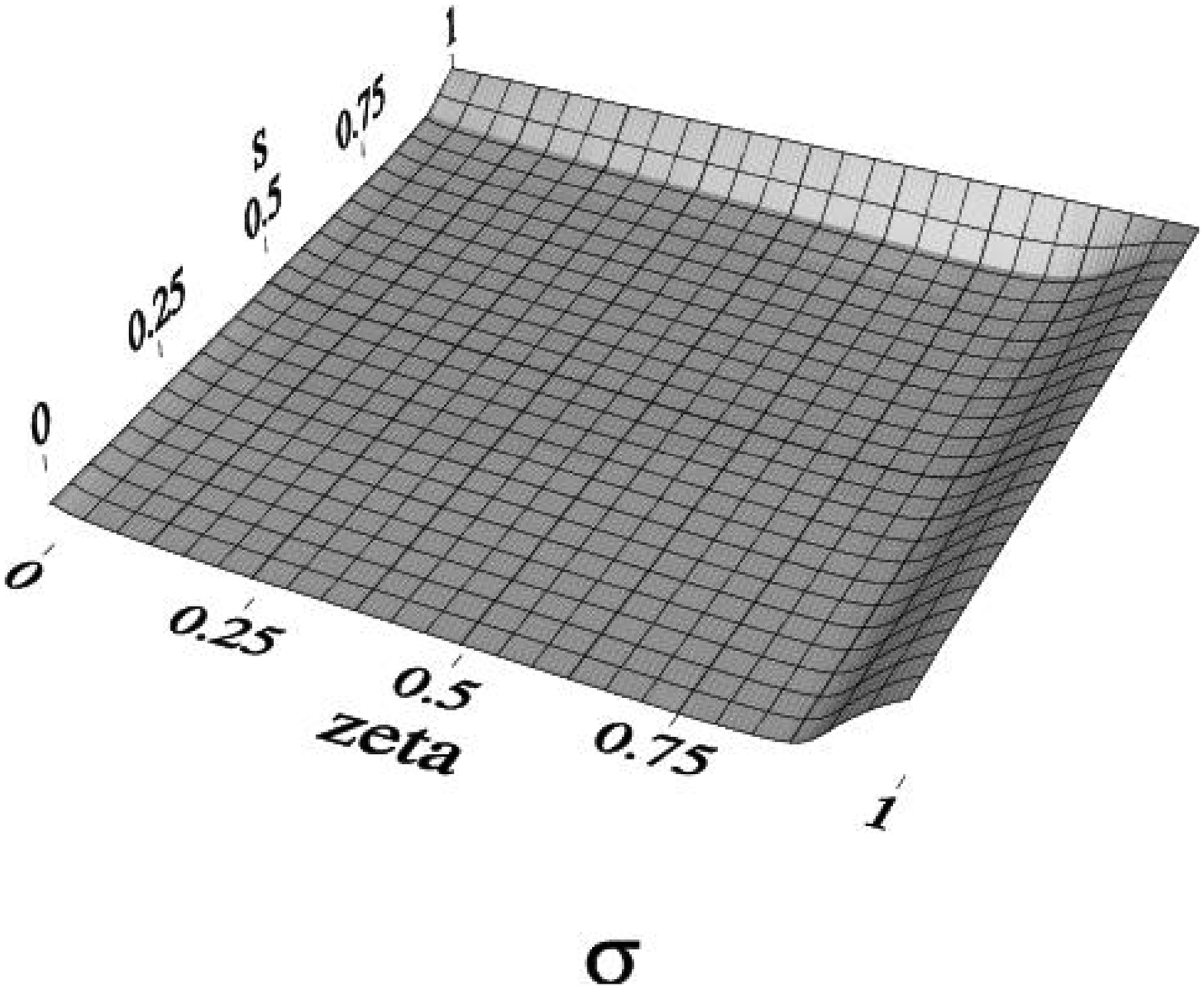}
}
\end{center}
\caption[Typical stationary rotating boson star solution with angular momentum parameter, $k=2$]
{Typical stationary rotating boson star solution with angular momentum parameter, $k=2$.
The solutions are symmetric about the equatorial plane.
Again, the top plot shows the configuration of the scalar field, which has
toroidal level surfaces,
while the bottom plots show various metric functions.  
As with the $k=1$ case, the solutions are symmetric under reflection through the equatorial plane,  $s=0$.
$0.0\le \ph_0(\ze,s)\le 0.18$, 
$1.0\le \ps(\ze,s)\le 1.2$, 
$0.75\le \al(\ze,s)\le 1.0$, 
$0.0\le \bt(\ze,s)\le 0.014$, 
$-0.033\le \si(\ze,s)\le 0.0$.
}
\label{bs2divpk2}
\end{figure}

\begin{figure}[htp]
\begin{center}
\ifthenelse{\equal{\highQ}{true}} {
\includegraphics[width=9.0cm,clip=true]{eps/bs2divp/phi_cy.eps}
\includegraphics[width=9.0cm,clip=true]{eps/bs2divp/phik2_cy.eps}
}{
\includegraphics[width=9.0cm,clip=true]{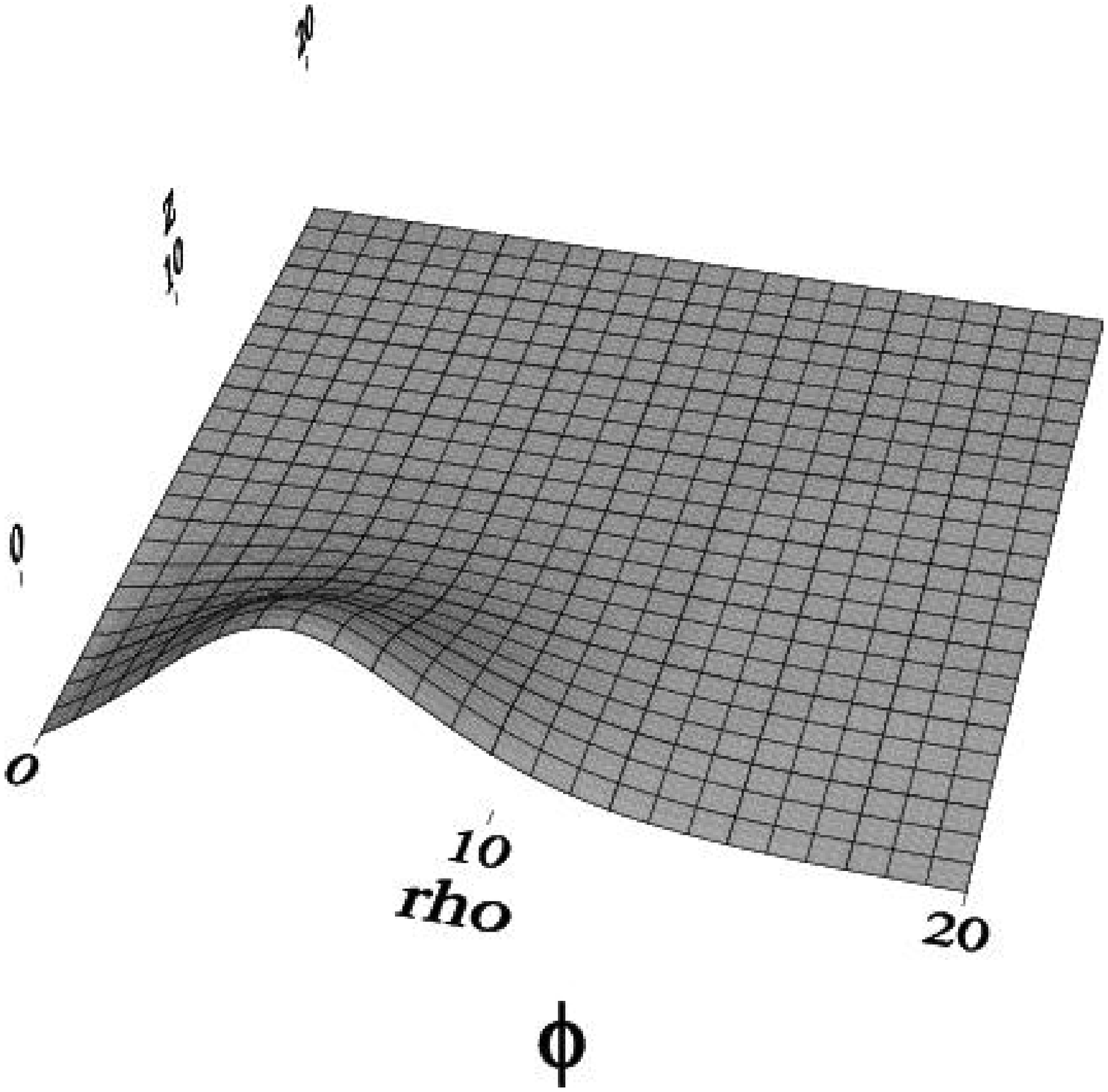}
\includegraphics[width=9.0cm,clip=true]{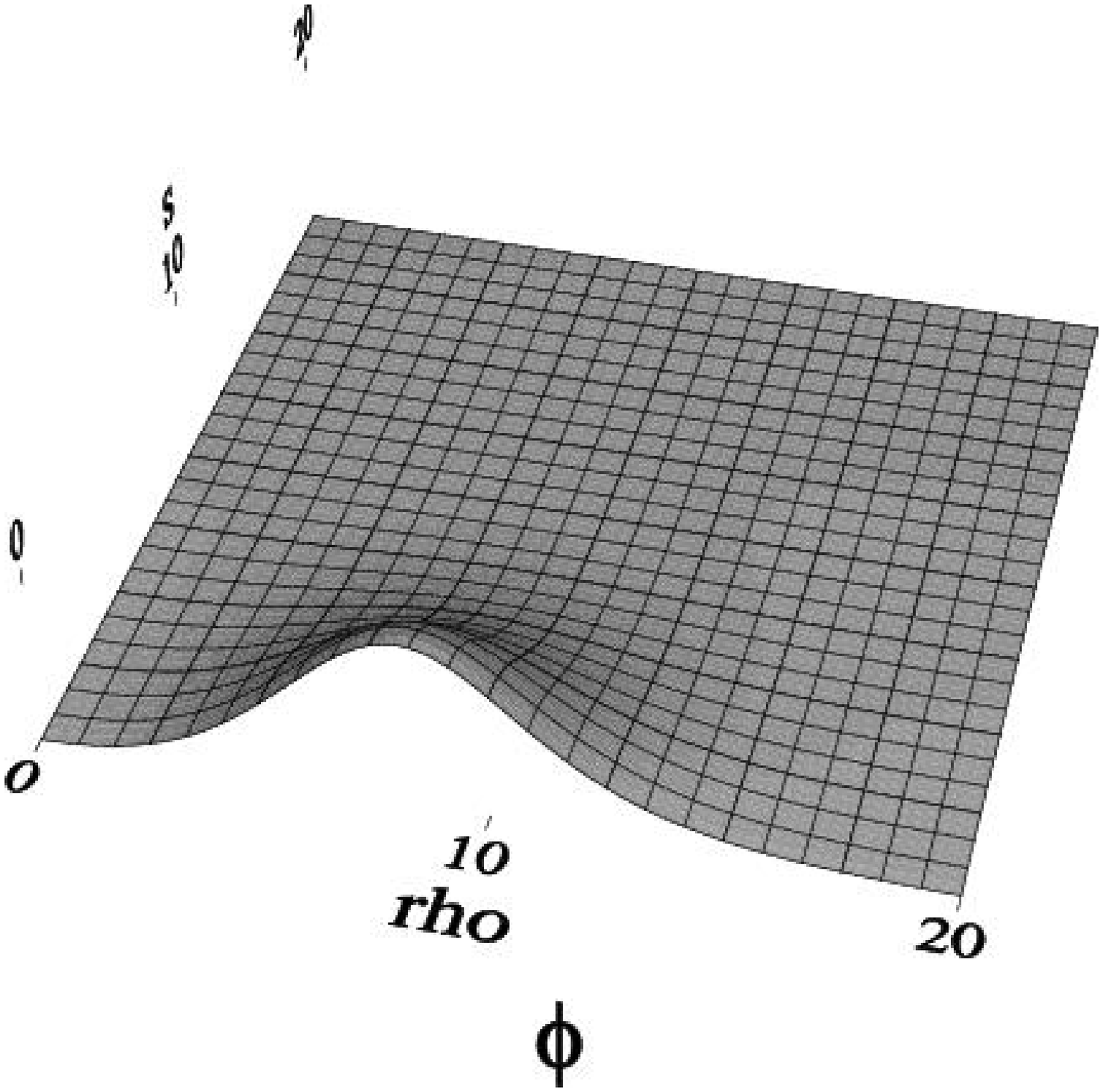}
}
\end{center}
\caption[Comparison of scalar field configurations for $k=1$ and $k=2$]
{Comparison of scalar field configurations for $k=1$ and $k=2$.  These plots 
show the scalar field solutions previously displayed in Fig.~\ref{bs2divpk1} and Fig.~\ref{bs2divpk2}, but 
now plotted in cylindrical coordinates $(\rh,z)$.  The top and bottom plots show the 
$k=1$ and $k=2$ results, respectively.
We note that the maximum amplitudes of the two scalar
fields are comparable for the two cases ($\ph_{\mathrm{max}} \approx 0.17$).
$0.0\le \ph_0(\rh,z)\le 0.17$ for $k=1$, 
$0.0\le \ph_0(\rh,z)\le 0.18$ for $k=2$. 
}
\label{bs2divpnonc}
\end{figure}

Fig.~\ref{bs2divpk2} shows a typical stationary boson solution computed for the case
$k=2$.  In this case the scalar field {\em and} its first radial derivative vanish as
$r\to0$, and the stellar configuration is more compact, in the sense that the support
of the modulus of the scalar field on the equatorial plane, $\left\{ \ze:
|\ph_0(\ze,0)| > \ep \right\}$ for some small number $\ep$, is of a smaller interval, 
with a maximum value located 
further from $r=0$ (but still along the equator) than for $k=1$.
Thus the scalar field at $(\zeta,s)=(0.5,0)$ (the point at which we fix $\phi$ for 
the $k=1$ calculations) will be small and, in fact, 
may be comparable to the level of numerical error.  In other words, the relative error in $\ph_{(2)}$,
if fixed at $(0.5,0)$, would be 
large compared to the $k=1$ case.  
Clearly, it is more sensible to
choose the reference point for $\ph_{(2)}$ closer to the point where the maximum of a typical
solution in the $k=2$ family is attained.
In this case, we have fixed the solution by specifying 
$\phi_{(2)}\equiv \phi_0(0.875,0) = 0.16$. 
(We note that the maximum value, $\max_{\ze,s}\vert \ph_0(\ze,s)\vert$, of the configuration shown in Fig.~\ref{bs2divpk1}
is attained at $\zeta = 0.89$, or $r = \zeta/(1-\zeta) = 8.1$.)

From the top plot in the figure, we see that the boson field again takes on a 
toroidal shape, with a more completely evacuated ``hole'' relative to the $k=1$ case.
As before, the figures at the bottom show the
metric functions, which appear qualitatively the same as the $k=1$
case except that their extrema are located further away from the axis, as we might expect.

It is also instructive to view the scalar field configurations in ordinary
cylindrical coordinates, $(\rh,z)$.  Fig.~\ref{bs2divpnonc}
shows the two scalar field configurations previously displayed, but now
plotted in $(\rh,z)$ coordinates.  
Since the vertical scales of the two plots are very nearly the same, we can see from the 
figure that
the scalar field has roughly the same maximum amplitude ($\ph_{\mathrm{max}} \approx 0.17$) in the 
two calculations.  It is also clear from the plots that 
the configuration which ostensibly has the larger angular momentum (i.e.\ the $k=2$ calculation, 
bottom plot) has a larger extent than the $k=1$ configuration.

As for the case of spherical symmetric boson stars, we can study some of the 
properties of our rotating configurations as a function of the family parameter 
(the quantity $\phi_{(k)}$ in this case).   In particular, we can produce plots 
of total mass {\em vs} $\phi_{(k)}$ that are completely analogous to~Fig.~\ref{bs1dfs}.
To compute the total mass we use Tolman's 
expression \cite{tolman:1930} (modulo our different convention for the metric signature):

\bea \nn
 M_{\rm Tolman} &=& \int \left( -2 T^0{}_0 + T^\al{}_\al \right) \sr{|g|} \,d^3 x \\ \nn
   &=& \int \fr{\ps^6 r^2 \sin \te e^{\si}}{\al} \,\ph^2  \left[ 2 \om \left( \om -
\bt k \right) - m^2 \al^2 \right] dr\, d \te\,d \vp  \\ \nn
  &=& 4 \pi \int^\infty_0 dr \int^{\fr{\pi}{2}}_0 d \te \,
    \fr{\ps^6 r^2 \sin \te e^{\si}}{\al} \,\ph^2  \left[ 2 \om \left( \om -
\bt k \right) - m^2 \al^2 \right] \\ \nn
 &=& 4 \pi \int^1_0 d\zeta \int^0_1 d s \,\fr{dr}{d\zeta}\, \fr{d \te}{ds}
    \fr{\ps^6 e^{\si}}{\al} \, \left( \fr{\zeta}{1-\zeta}\right)^2  \sr{ 1-s^2}\,\ph^2  
\left[ 2 \om \left( \om -
\bt k \right) - m^2 \al^2 \right] \\ \nn
&=& 4\pi \int^1_0 d\zeta \int^1_0 ds \,\fr{\ps^6 e^{\si}}{\al} \,\fr{\zeta^2}{(1-\zeta)^4}
\ph^2 
\left[ 2 \om \left( \om -
\bt k \right) - m^2 \al^2 \right] \,.
\eea

\noi
Note that for any stationary, asymptotically flat spacetime foliated with spacelike hypersurfaces
that are asymptotically orthogonal to the timelike Killing field, the Tolman mass expression
is equivalent to the ADM mass \cite{wald,Ashtekar:1979}.
Also note that
\cite{schunck_mielke:1996} shows that we have $J = k N$ where $J$ is the total angular
momentum defined by
\beq
 J = \int  T^0{}_3 \sr{-g}\, d^3 x \,,
\eeq
\noi
and $N$ is the total particle number.  Since $N$ has the same qualitative behaviour 
as a function of $\phik$ as $M_{\rm Tolman}$ (and, in fact the calculations of 
Yoshida \& Eriguchi~\cite{yoshida_eriguchi:1997}---see Fig.~3 of that
reference---show that the numerical values of $N$ and $M_{\rm Tolman}$ are 
quite similar),
with maxima of both quantities achieved at the same value of $\ph_{(k)}$, we can at least 
qualitatively compare the angular momenta of stars with varying $k$ using the $M_{\rm Tolman}$ {\em vs} $\ph_{(k)}$ curve
and the relation $J=kN$.

\begin{figure}[htp]
\begin{center}
\epsfxsize=9.0cm
\includegraphics[width=9.0cm,clip=true]{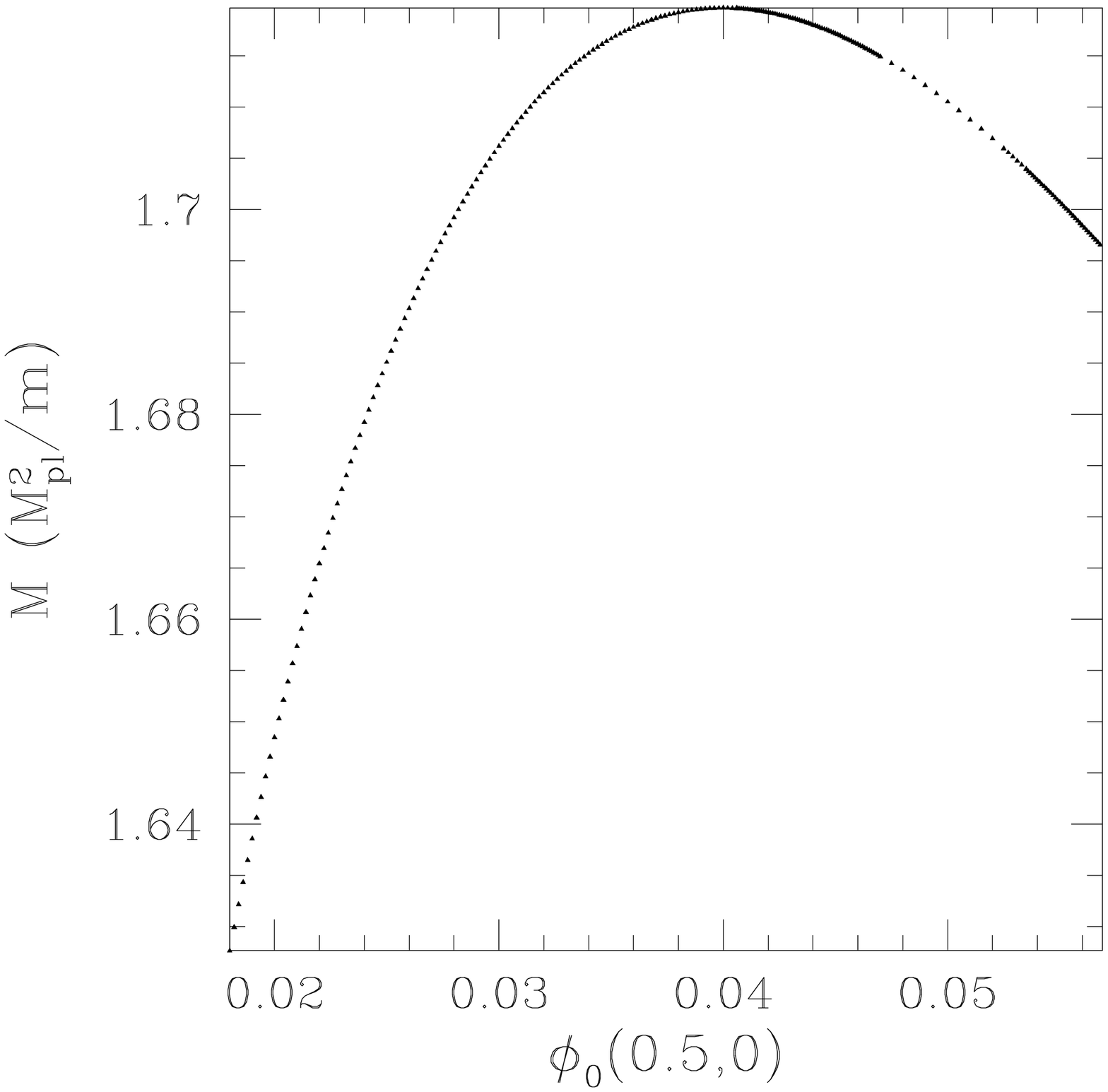}
\epsfxsize=9.0cm
\includegraphics[width=9.0cm,clip=true]{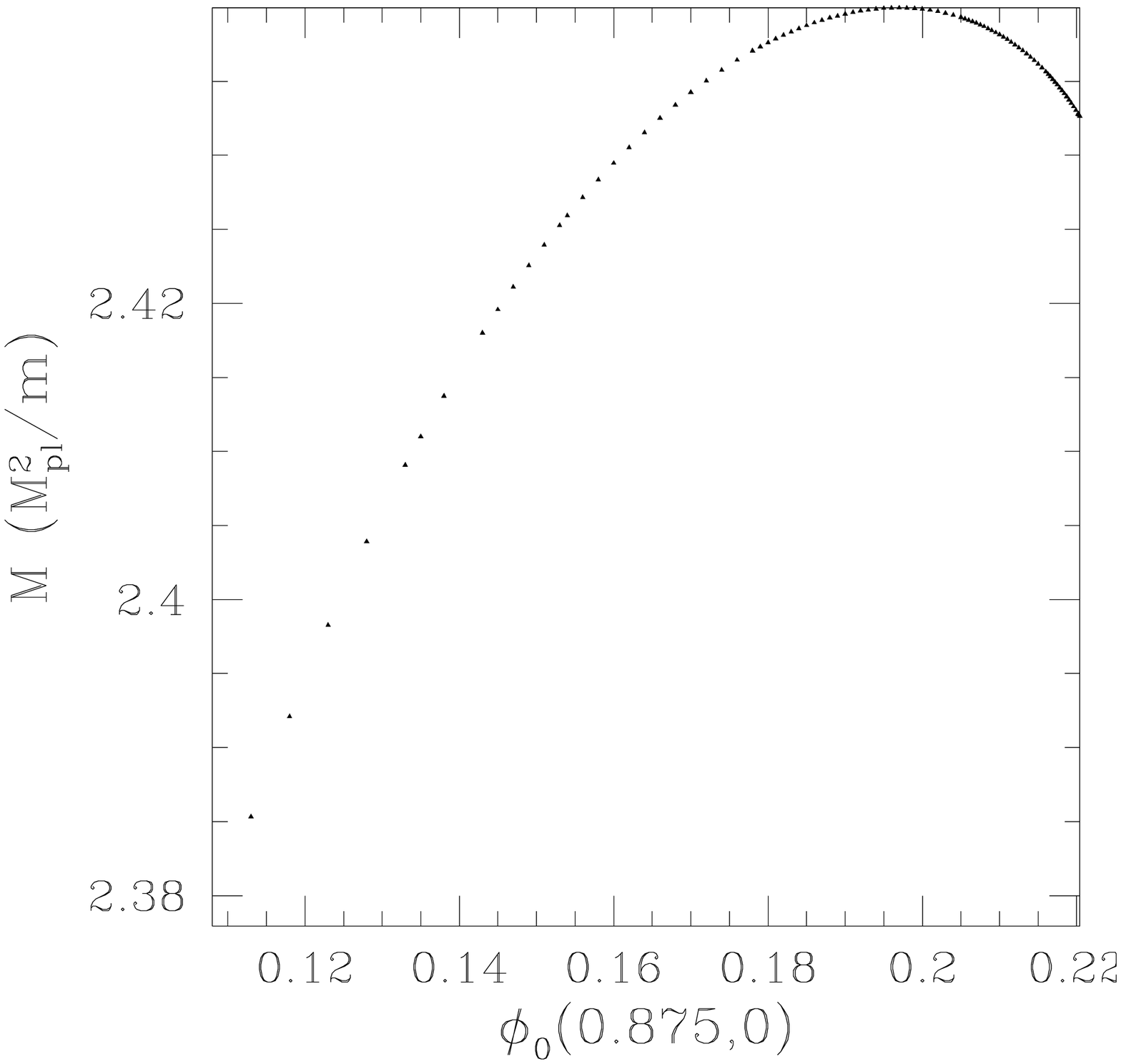}
\end{center}
\caption
[The Tolman mass $M_{\rm Tolman}$ of axisymmetric rotating boson stars with $k=1$ and $k=2$ as a function of family  
parameter $\ph_{(k)}$]
{The Tolman mass $M_{\rm Tolman}$ of axisymmetric rotating boson stars with $k=1$ and $k=2$ as a function of 
family parameter $\ph_{(k)}$.  
For $k=1$ (top) we fix the scalar field value, $\ph_{(k)}=\ph_{(1)}$ at $(\zeta,s)=(0.5,0)$, while
for $k=2$ (bottom), where the maximum of $\phi$ tends to be further from the axis, we
fix $\ph_{(k)}=\ph_{(2)}$ at $(\zeta,s)=(0.875,0)$.
As in the
spherically symmetric case, there exists a maximum mass limit above which no stationary solutions exist.}
\label{Mvsph}
\end{figure}

Fig.~\ref{Mvsph} shows plots of the total mass $M_{\rm Tolman}$ 
{\em vs} the family parameter $\ph_{(k)}$ for rotating boson stars with $k=1$ (top) and 
$k=2$ (bottom).
Again, for $k=1$ we fix the scalar field value, $\ph_{(k)} = \ph_{(1)}$ at $(\zeta,s)=(0.5,0)$, while
for $k=2$, where the maximum of $\phi$ tends to be further from the axis, we 
fix $\ph_{(k)} = \ph_{(2)}$ at $(\zeta,s)=(0.875,0)$.
From the plots we can see that in each case, and as for the 
spherically symmetric calculations, there exists a maximum mass limit 
above which no stationary solutions exist.  We also note that this mass limit 
appears to increase with the angular momentum parameter, $k$, which is in accord 
with naive expectations that centrifugal effects due to rotation should partially
offset gravitational attraction.

\subsection{Some Remarks} \lab{remarks}
Before ending this section we would like to make a few additional remarks:

\begin{itemize}
\item While Yoshida \& Eriguchi's code~\cite{yoshida_eriguchi:1997} broke down before the 
maximum-mass star for 
$k=2$ could be computed,  we encounter no difficulties in computing the maximum-mass solution (and beyond)
in this case~(see Fig.~\ref{Mvsph}).  
Moreover, we feel that the use of multigrid makes our solution strategy potentially more efficient than previous 
implementations.  
In particular, since the 
multigrid method is capable of solving elliptic PDEs in $O(N)$ time (where $N$ is the number of
grid points), in principle there should be no problems in performing high resolution calculations,
convergence tests etc.   Unfortunately, due to other
difficulties we encounter (see below), we are currently unable to perform meaningful
convergence tests at this time. 

\item As discussed in Sec.~\ref{systemA}, the leading order behaviour for $\ph_0(r,\te)$ as $r \to
0$ is $\ph_0(r,\te) = r^k \ph_{0k}(\te) + O(r^{k+2})$.  Therefore $\ph_0(\ze,s)\vert_{\ze = 0} = 0$ 
and $\pa^{i} \ph_0(\ze,s) / \pa \ze^{i}\vert_{\ze = 0} = 0$ for $i=1,\cdots k-1$.
In the multigrid routine we only impose the first condition.
Similarly for $\si(r,\te)$ the leading order behaviour as $r \to 0$ is $\sigma(r,\te) =  
r^2 \si_{2}(\te)+ O(r^4)$.  In other words $\si(\ze,s)\vert_{\ze =0} = 0$ and $\pa \si(\ze,s)/\pa \ze\vert_{\ze=0} =
0$.  Again we only impose the first condition in the multigrid iteration.  Moreover, a closer examination of
the $r\to0$ limit of System A, performed via Taylor series expansion of the variables about $r = 0$ (for example, we
write $\ps(r,\te) = \ps(0,\te) + r^2 \pa^2 \ps(0,\te)/\pa r^2 + O(r^4)$ and similarly for other
variables, and then substitute the expressions into the governing equations (\ref{2dkg})-(\ref{2dsieq})) shows that there are some 
remaining irregular terms (proportional to $1/r$) that do not explicitly cancel as $r\to 0$.
Thus it is perhaps not surprising that our solutions show irregularity near $r=0$ (or equivalently near $\ze=0$).
The fact that we {\em do} have some problems with regularity can be seen 
through close examination of the solutions, which reveals non-smoothness  near
the $r=0$ ``axis'' (see Figs.~\ref{bs2divpirreg} and \ref{irreg_sigma}).
In particular, Fig.~\ref{irreg_sigma} plots the function $\pa \si(\ze,s)/\pa \ze$ close to
the origin $\ze = 0$ for various constant-$s$ slices, $s= 0, 0.29, 0.64$ and 0.93. 
The function $\si(\ze,s)$ was computed in a calculation with $\phi_{(1)} \eq
\ph_0(0.5,0) = 0.03$, and using a mesh with $N_{\ze}=129$ and $N_s = 15$.
Ideally we expect $\pa \si(\ze,s)/ \pa \ze \to 0$ as $\ze \to 0$. However, the ``spikes"
in the region $0\le \ze \le 0.05$ clearly show that the regularity condition is
not satisfied.

\begin{figure}[htp]
\begin{center}
\ifthenelse{\equal{\highQ}{true}} {
\includegraphics[width=9.0cm,clip=true]{eps/bs2divp/sigma_cy.eps}
}{
\includegraphics[width=9.0cm,clip=true]{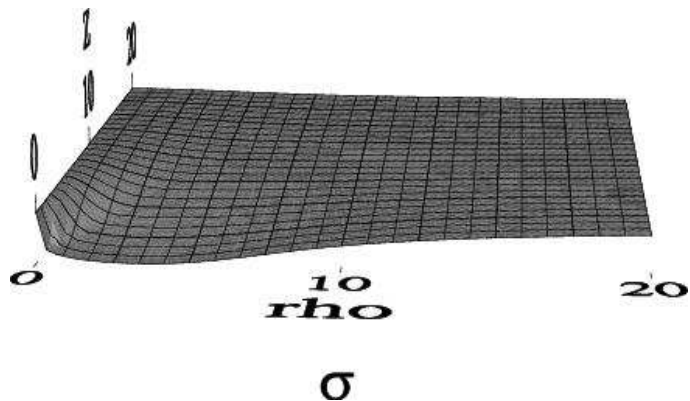}
}
\end{center}
\caption[Regularity problem close to the symmetry axis for $k=1$]
{Regularity problem close to the symmetry axis for $k=1$.  This figure shows $\si$ 
for the same solution shown in 
Figs.~\ref{bs2divpk1} and~\ref{bs2divpnonc}, and plotted in 
cylindrical coordinates $(\rh,z)$.  The figure shows that the function is not
completely smooth for small $\rh$ (i.e.\ close to the symmetry axis on the left).
$-0.028\le \si(\ze,s)\le 0.0$.
}
\label{bs2divpirreg}
\end{figure}

\begin{figure}[htp]
\begin{center}
\epsfxsize=10.0cm
\includegraphics[width=10.0cm,clip=true]{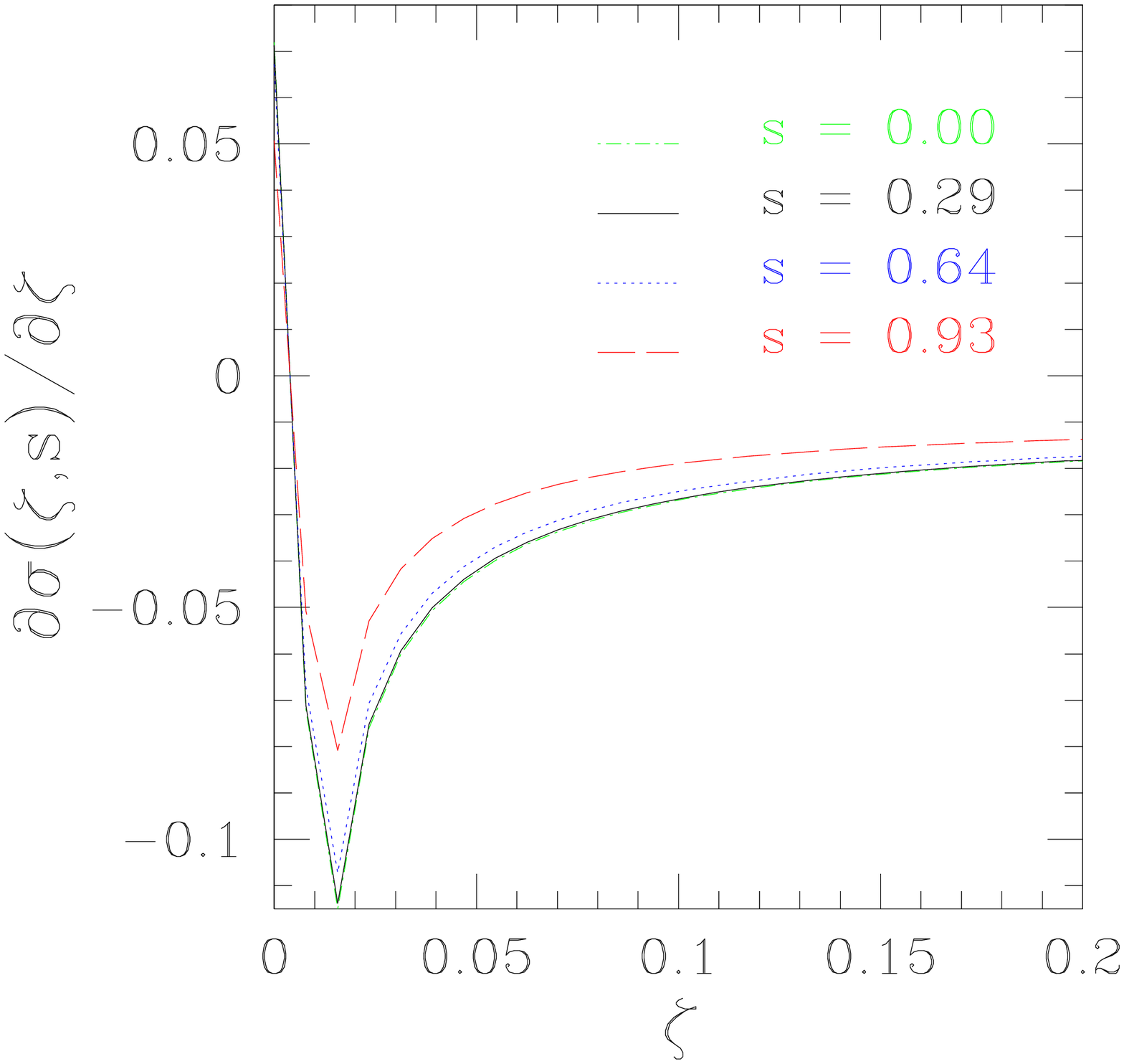}
\caption
[Irregularity of the metric component $\si(\ze,s)$ as $\ze\to 0$ for $k=1$]
{Irregularity of the metric component $\si(\ze,s)$ as $\ze\to 0$ for $k=1$.
Ideally, these curves should pass linearly through $(0,s)$ as $\De \ze\to0$.
The solution displayed here was computed with
$\phi_{(1)} \eq \ph_0(0.5,0) = 0.03$ on a computational domain with 
$N_{\ze} = 129$ and $N_{s} = 15$.
The figure shows the derivative of $\si(\ze,s)$ with respect to $\ze$
for constant $s$ slices $s= 0, 0.29, 0.64, 0.93$. The ``spikes" in the 
region $0\le \ze \le 0.05$ clearly show that the function $\si(\ze,s)$ does
not satisfy the regularity condition $\pa \si(0,s)/\pa \ze = 0$.
}
\label{irreg_sigma}
\end{center}
\end{figure}

\item As we have seen in Figs.~\ref{bs2divpk1} and~\ref{bs2divpk2}, the
larger the value of $k$, the larger the value of $r$ (or $\ze$)
at which $\ph_0(\ze,s)$ is a maximum, and the more concentrated in $(\ze,s)$ space the
stars become.  Notice that this increased concentration for increased $k$ is largely
a function of our use of the radially compactified coordinate $\ze$.  As $k$ increases, so
do the overall diameters of the toroidal regions, and as $r$ increases the corresponding
resolution $\De r(\ze)$ decreases.
Therefore, as $k$ increases, higher resolution is needed to sufficiently resolve
the stationary configurations.
A way to remedy this situation is to
use a more sophisticated compactification as discussed in Sec.~\ref{spacom}, as well
as in the next section.
This will increase the complexity of the system of equations that need 
to be solved, but will map the location of the scalar field maximum to the central 
region of the computational domain.  In turn, this should partly ameliorate the 
need for increased resolution for larger values of the angular momentum parameter $k$.

\item The maximum Tolman masses $M_{\mathrm{max}} \equiv \max_{\phi_{(k)}} M_{\rm Tolman}(\phi_{(k)})$ we find 
for rotating boson stars are $M_{\mathrm{max}} = 1.72$ for $k=1$ and
$M_{\mathrm{max}} = 2.44$ for $k=2$. 
In \cite{yoshida_eriguchi:1997b} Yoshida \& Eriguchi found that $M_{\mathrm max} =
1.31$ for $k=1$ and $M_{\mathrm max} \ge 2.40$ for $k=2$. 
Since their solutions show no irregularity at $r=0$.
the discrepancy (which is rather large)
may be due to the regularity problems described above.

\item 
As we have discussed several times, following \cite{yoshida_eriguchi:1997b} we parametrize each family of 
solutions for $k=1, 2, \cdots$ by specifying the value of $\phik$ at some 
arbitrary point $(\zeta,s) = (\zeta_0,0)$.
Within the context of our current particular finite difference solution of System A,
this approach has the advantage that it is very easy to implement.
However, from a theoretical point of view it seems more natural
to instead specify the $k$-th radial derivative of the scalar field.
This could be implemented by defining a new scalar function
$\bar{\ph}_0(\ze,s)$ such that~\cite{choptuik:private}
\beq
 \ph_0(\ze,s) = \zeta^k \bar{\ph_0}(\ze,s)\,,
\eeq
\noi
and rewriting System A for $\bar{\ph_0}(\ze,s)$.  In this new system we would 
then specify $\partial^{(k)}_\zeta \ph_0(\zeta,s)|_{\zeta = 0}$ as the 
family parameter, eliminating the {\em ad hoc} parameter $\zeta_0$ from
the algorithm.  We discuss the initial steps of the implementation of 
such an approach in the next section.

\item We can replace the individual updates of each variable
in the main iteration of our initial data solver 
(the {\bf Repeat} loop in Fig.~\ref{table:BS2DIVP}) by 
combining the independent multigrid updates shown in Figure~\ref{table:BS2DIVP} into a 
single multigrid solve for the five fields.
More specifically, in the new solver, a basic relaxation step would 
visit a given grid location $(\zeta_i,s_j)$ and simultaneously update
the 5 grid function values
$\phi_{i,j}, \ps_{i,j} \cdots \sigma_{i,j}$ using a 5-dimensional 
Newton method.
Thus, each pointwise update in this case would require us to set up and solve
a $5\times5$ linear system, rather than 5 scalar linear 
equations.  The $5\times5$ solve is more costly computationally, as well
as being more tedious to set up since the full Jacobian must be computed. 
A key advantage is that the underrelaxation (UR) part of the algorithm 
is no longer needed for convergence.

\item As in the study of dynamics of boson stars in spherical symmetry, we could study 
      the dynamics of boson stars with angular momentum in axisymmetry, using the
      solutions constructed in this section as initial data for the dynamical code.  However, as 
      mentioned in the introductory section of this chapter, the coordinate conditions adopted 
      in 
      {\tt graxi} (see~(\ref{2p1p1metric})) are not compatible with our ansatz
      (\ref{rot-ansatz}).  More precisely, if we assume the metric has the form 
      (\ref{2p1p1metric}) and the boson star has the form (\ref{rot-ansatz}),
      then the real and imaginary parts of the Klein-Gordon equation (\ref{klein_gordon_eq})
      will yield {\em different} equations for $\ph_0(r,\te)$.  The incompatibility
		can be traced to the coordinate choice made in adopting the form~(\ref{2p1p1metric})
      whereby the metric component
      $g_{t \vp}(\rh,z)$ is required to vanish~\cite{graxi:2004}.  To easily study the dynamics of the 
      rotating boson stars constructed here, 
      we would need to adopt a different form for the metric ($g_{t
\vp}(\rh,z) \ne 0$),
      which in turn would require substantial modifications to {\tt graxi}.  We have not 
      yet pursued this possibility.
\end{itemize}

\subsection{A Proposal for an Improved Rotating Boson Star Solver (System B)}
\label{system-b}

In the previous section  we discussed the difficulties we encountered in
constructing equilibrium solutions of rotating boson stars: 
we have regularity problems near $\ze=0$,
the maximal values of the scalar field
are achieved at larger $\ze$ on the equatorial plane for increasing $k$, and 
the configurations are increasingly concentrated in the $\zeta$-$s$ plane 
for increasing $k$. 
In this section we suggest a way of rewriting System A to treat these problems.
A new code based on this approach is under development, but at this point we do not have
any results to report.

Following \cite{yoshida_eriguchi:1997,ryan:1997} we write the metric in the form
\beq
 ds^2 = -e^{\et + \de} dt^2 + e^{2 \ps} \left( d\rh^2 + dz^2\right) + e^{\et - \de} \rh^2
\left( d \vp - \bt dt\right)^2\,,
\eeq

\noi
where $\et, \de, \ps$ and $\bt$ are functions of the cylindrical 
coordinates $\rh$ and $z$ only. 
(We note that~\cite{yoshida_eriguchi:1997,ryan:1997} use spherical-polar coordinates.)
The 
motivation for the above specific form is that a detailed analysis (again,
by Taylor series expansion of the variables about $\rh = 0$ with subsequent substitution of the 
expansions in the equations given below) shows that all equations
described below are manifestly regular as $\rh \to 0$, provided that the $\rho=0$ conditions 
demanded by local flatness on-axis are imposed.
Also motivated by regularity considerations, we define a new scalar
field variable $\bar{\ph_0}(\rh,z)$ by
\beq
 \ph_0(\rh,z) = \rh^k \bar{\ph_0}(\rh,z)\,,
\eeq
so that we explicitly factor out the leading order $\rho\to0$ behaviour of the 
field.
The Klein-Gordon equation, Hamiltonian constraint, momentum constraint and the
equations for $\et$ and $\de$ are derived in a manner similar to that used in 
defining the
corresponding equations for System A.  
For notational simplicity, we drop
the ``0" subscript and the bar so that $\bar{\ph_0}(\rh,z) \to \ph(\rh,z)$.
We also choose $\ka = 1$ instead of $8 \pi$ as previously, and continue to set $m=1$.
We then have {\em System B}:
\beq
\begin{split}
 \ph_{,\rh \rh} + \ph_{,zz} + \left( \et_{,\rh} + \fr{1+2k}{\rh}\right) \ph_{,\rh} + \et_{,z} \ph_{,z}
+ \left( e^{2 \ps - \et - \de} (\om + \bt k)^2 - e^{2 \ps}  + \fr{\et_{,\rh}k }{\rh } + \fr{
1-e^{2 \ps - \et + \de} }{\rh^2 }k^2 \right) \ph \,,
\end{split}
\eeq
\beq
\begin{split}
 \ps_{,\rh \rh} + \ps_{,zz} - \fr{1}{4} e^{-2 \de} \rh^2 \left(\bt_{,\rh}^2+\bt_{,z}^2  \right)
+\fr{1}{4} \left( \de_{,\rh}^2 + \de_{,z}^2 - \et_{,\rh}^2 - \et_{,z}^2 \right) - \fr{1}{2 \rh}
\left( \et_{,\rh} + \de_{,\rh}\right) \\
+ \fr{\rh^{2k}}{2} \left( \ph_{,\rh}^2 +
\ph_{,z}^2\right) + \rh^{2k-1} k \ph \ph_{,\rh} + \fr{\ph^2 \rh^{2k}}{2} \left[  
\left( \om + \bt k\right)^2 e^{2 \ps - \de - \et} +  \fr{k^2 }{\rh^2} \left( 1-e^{2 \ps + \de
- \et}\right)\right] = 0\,,
\end{split}
\eeq
\beq
 \bt_{,\rh \rh} + \bt_{,zz} + \left( -2 \de_{,\rh}+ \et_{,\rh} + \fr{3}{\rh}\right)
\bt_{,\rh} + \left( \et_{,z} - 2 \de_{,z} \right) \bt_{,z}
- 2 e^{2 \ps - \et + \de} \rh^{2 k - 2} \ph^2 k (w+\bt k) = 0\,,
\eeq
\beq
 \et_{,\rh \rh} + \et_{,zz} + {\et_{,\rh}}^2 + \fr{2 \et_{,\rh}}{\rh}
+ {\et_{,z}}^2 + \ph^2 \rh^{2k} \left[ e^{2 \ps} -e^{2 \ps -\et-\de} \left(\om +\bt k \right)^2
 + \fr{k^2 e^{2 \ps -\et + \de}}{\rh^2}  \right] = 0\,,
\eeq
\beq
\begin{split}
 \de_{,\rh \rh}+\de_{,zz} + \left( \et_{,\rh}+\fr{1}{\rh}\right) \de_{,\rh}
+ \et_{,z} \de_{,z} + \ph^2 \rh^{2k} \left[ -e^{2 \ps -\et-\de} \left( \om + \bt k\right)^2
- \fr{k^2 e^{2 \psi - \et  + \de }}{\rh^2} \right] \\
-e^{-2 \de} \left( {\bt_{,\rh}}^2 + {\bt_{,z}}^2 \right) \rh^2 - \fr{\et_{,\rh}}{\rh} = 0\,.
\end{split}
\eeq

Further, we compactify both coordinates via coordinate transformations $\rho \to \zeta(\rho)$ and
$z\to\mu(z)$ given by 
\bea \lab{rhtrans}
 \ze &=& \fr{\rh}{\rh_0 + \rh}\,, \\ \lab{ztrans}
  \mu &=& \fr{z}{z_0 + z}\,.
\eea

\begin{figure}[htp]
\begin{center}
\ifthenelse{\equal{\highQ}{true}} {
\includegraphics[width=8.0cm,clip=true]{eps/bs2divp/phi_rho0z01.eps}
\vspace{0.5cm}
\includegraphics[width=8.0cm,clip=true]{eps/bs2divp/phi_rho0z010.eps}
}{
\includegraphics[width=8.0cm,clip=true]{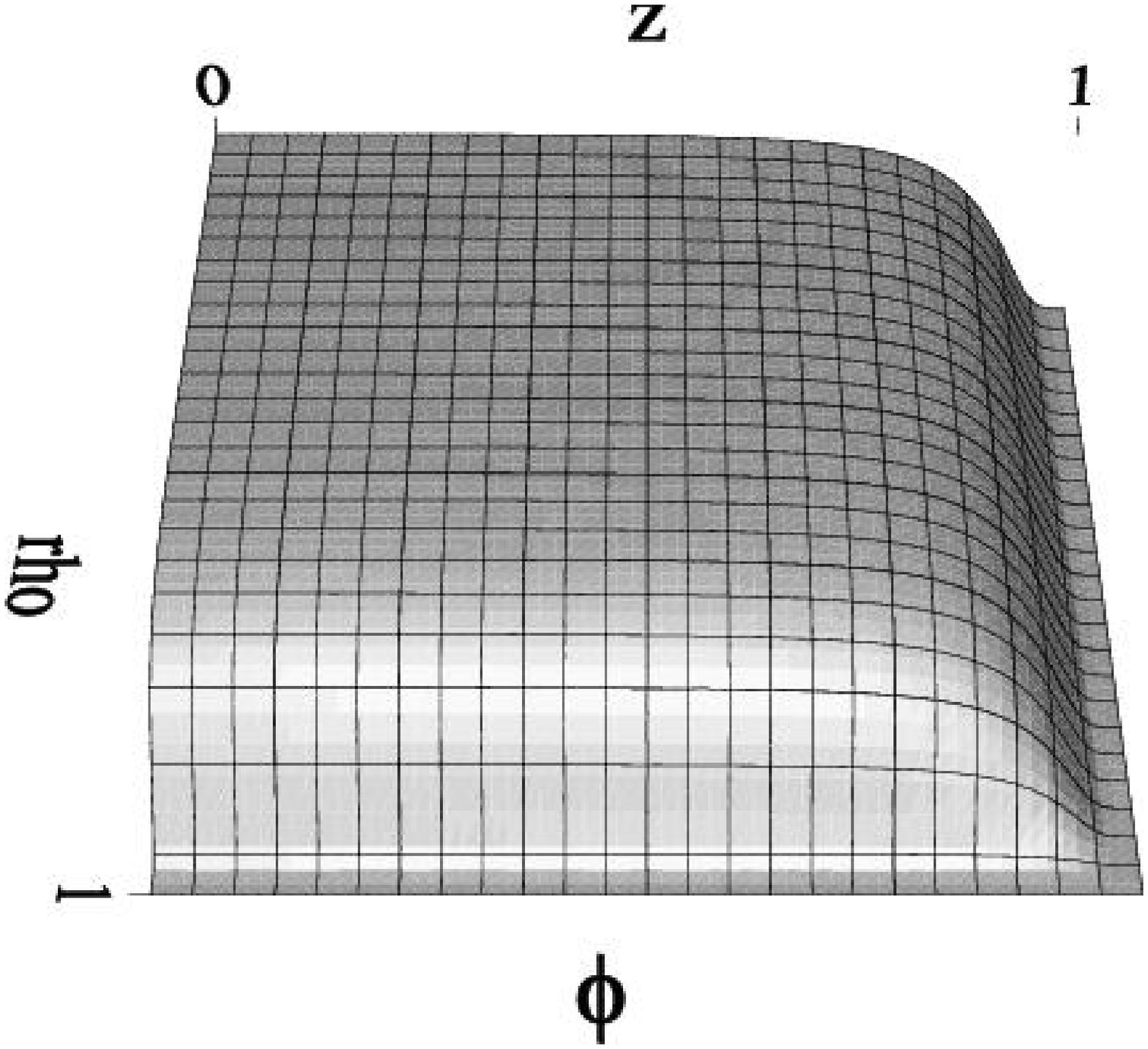}
\vspace{0.5cm}
\includegraphics[width=8.0cm,clip=true]{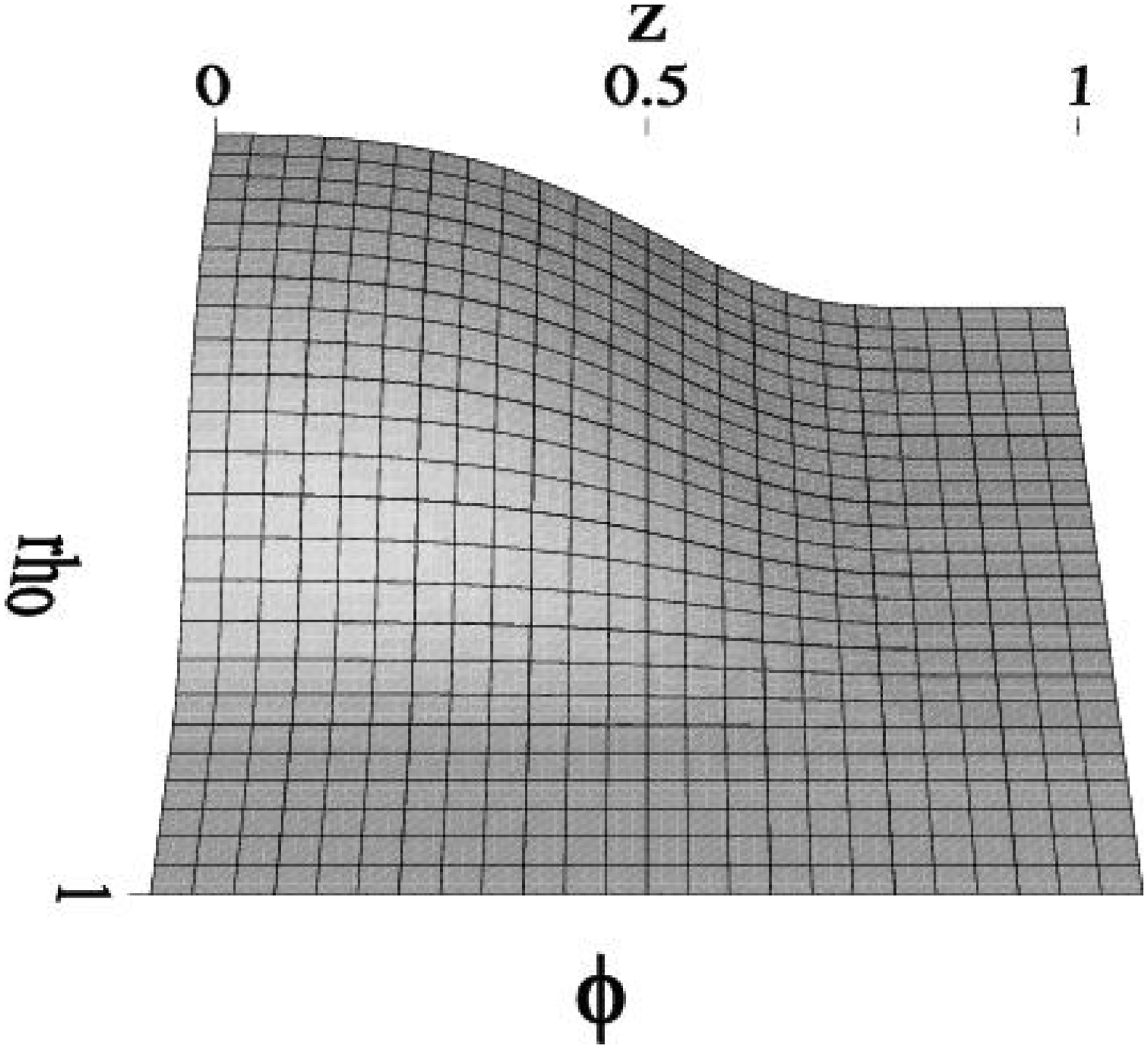}
}
\end{center}
\caption[Demonstration of the effect of adjusting the coordinate-transformation
parameters $\rh_0$ and $z_0$]
{Demonstration of the effect of adjusting the coordinate-transformation 
parameters $\rh_0$ and $z_0$.
The top figure shows a typical initial configuration of $\bar{\ph_0}(\ze,\mu)$
with $\rh_0=1$ and $z_0 =1$.  The lower figure shows the same configuration, but now
with $\rh_0=10$ and $z_0=10$.
$0.0\le \ph_0(\rh,z)\le 0.01$.
}
\label{bringin}
\end{figure}

The new parameters $\rh_0$ and $z_0$ can be chosen to map arbitrary coordinate lines $\rh =\rh_0$ and
$z = z_0$ to $\ze = 1/2$ and $\mu = 1/2$, respectively, in the new coordinates (see Sec.~\ref{spacom}).
Fig.~\ref{bringin} illustrates the effect of such a transformation.  Rewriting System B in $(\zeta,\mu)$ 
coordinates
is straightforward but leads to lengthy expressions that will not be given here.

To ensure System B is regular at $\rh =0$ (or equivalently, $\ze = 0$)
 we apply the locally-flat condition 
$2 \ps(0,\mu) - \et(0,\mu) + \de(0,\mu) = 0$.  The other regularity conditions are
\beq
 \ph_{,\ze}(0,\mu) = \ps_{,\ze}(0,\mu) = \bt_{,\ze}(0,\mu) = \et_{,\ze}(0,\mu)
= \de_{,\ze}(0,\mu) = 0 \,,
\eeq

\noi
and the other boundary conditions are:
\beq
\ph(1,\mu) = \ps(1,\mu) = \bt(1,\mu) = \et(1,\mu) = \de(1,\mu) = 0\,,  
\eeq
\beq
\ph(\ze,1) = \ps(\ze,1) = \bt(\ze,1) = \et(\ze,1) = \de(\ze,1) = 0\,,  
\eeq
\beq
 \ph_{,\mu}(\ze,0) = \ps_{,\mu}(\ze,0) = \bt_{,\mu}(\ze,0) = \et_{,\mu}(\ze,0) =
\de_{,\mu}(\ze,0)  = 0 \,.
\eeq

We can now parametrize the family by the value of $\ph(\ze,\mu) \eq \bar{\ph_0}(\ze,\mu)$ at $\ze =0$ and
$\mu =0$.  The multigrid solver will use 
point-wise simultaneous relaxation of all five unknowns defined at each grid point (using
a 5-dimensional Newton method)
as described in Sec.~\ref{remarks}.

  \section{Dynamics of Axisymmetric Non-rotating Boson Stars} \lab{dynamics2d}

\def\graxi{{\tt graxi}}

In this section we focus on the dynamics of axisymmetric {\em non-rotating} boson
stars.  In the first subsection we present the equations of motion that are 
to be solved, and that have been derived using 
the (2+1)+1 formalism outlined in Chap.~\ref{MathForm}.
We note that this derivation is not original to this work, nor is the basic 
code (\graxi) that implements the discrete version of the equations~\cite{graxi:2003,fransp:phd}.
However, as mentioned previously, we {\em have}  extended \graxi\ so that it can 
be initialized with data for the complex scalar field representing one or 
more spherical boson stars in an overall axisymmetric configuration.

In subsequent subsections we 
perform two classes of numerical experiments.  The first class involves the 
head-on collision of equal-mass boson stars, where each star can be 
boosted towards the other via an adjustable initial linear momentum parameter, $p_z$.
The second class involves perturbation of stable boson stars with a massless real scalar 
field, generalizing the calculations in spherically symmetry that were described
in Chap.~\ref{bs1d}.  In both instances, we study the critical behaviour that 
arises at the threshold of black hole formation.  We find evidence for Type I 
transitions in both cases, as well as evidence for scaling of the lifetime, $\tau(p)$, 
of near-critical configurations 
\beq
\label{scaling3}
   \tau(p) \sim -\gamma \ln | p - p^\star | \,,
\eeq
that one normally associates with Type I behaviour ($p$ is, as usual, the family 
parameter that is tuned to generate the critical solution).  As far as we are aware, 
this represents the first instance of the observation of Type I critical phenomena in 
an axisymmetric, general relativistic model.  In addition, we show evidence for 
interesting ``solitonic'' behaviour of interacting boosted boson stars, similar 
to that previously seen in Newtonian computations~\cite{dale}.

\subsection{The equations of motion} \lab{eom2d}

We choose cylindrical coordinates $(t,\rh,z,\vp)$, impose spatial 
coordinate conditions so that the 2-dimensional spatial metric (i.e.\ the metric in the 
$(\rho,z)$ plane) is conformally flat, and adopt 3-maximal slicing (i.e.\ the 
trace of the 3-dimensional curvature tensor $K_{ij}$ vanishes).  As mentioned 
above, we also restrict attention to the non-rotating case. The spacetime metric 
then has the form~\cite{graxi:2003}

\beq \lab{2p1p1metric}
 ds^2 = \left\{ -\al^2 + \ps^4 \left[ (\bt^\rh)^2 + (\bt^z)^2 \right]\right\} dt^2
+ 2 \ps^4 \left( \bt^\rh d \rh+ \bt^z dz \right) dt + \ps^4 \left( d \rh^2 + dz^2 + \rh^2 e^{ 2 \rh
\bar{\si}} d \vp^2\right)\,,
\eeq 

\noi
where $\al$, $\bt^\rh$, $\bt^z$, $\ps$, and $\bar{\si}$ are functions of
$\rh, z$ and $t$ only.  
The reason for using $\bar{\si}$ (in contrast to the seemingly more
natural variable $\si \eq \rh \sigmabar$) as a fundamental metric function is motivated by
regularity considerations. In particular the leading order behaviour of $\bar{\si}$
in the limit as $\rh \to 0$ is $\bar{\si}(\rh,z) = \rh \si_1(z) + O(\rh^3)$, which can be implemented
numerically as a Dirichlet boundary condition.  On the other hand the leading order behaviour of
$\si$ is $\si(\rh,z) = \rh^2 \si_2(z) + O(\rh^4)$, and is much more difficult to
maintain via finite-differencing of the type used in this thesis.

From (\ref{2p1p1metric}) we see that the squared-norm, $s^2$, of the rotational 
Killing vector (see~(\ref{s2_def})) is given by 

\beq
 s^2 = \ps^4 \rh^2 e^{2 \rh \sigmabar}\,.
\eeq

\noi
Without angular momentum, the twist vector $\om^\al$ (see \ref{om_def}) vanishes.

We write the evolution equation for $\bar{\si}$ in
first-order-in-time form by introducing an auxiliary variable $\bar{\Om}$
defined by

\bea \lab{omegabar_def}
 \rh \bar{\Om} &\eq& - 2\,{}^{(2)}K^\rh\,_\rh - \,{}^{(2)}K^z\,_z \\
		&=&  \fr{2}{s} n^a \pa_a s + \fr{\bt^z\,_{,z} -
\bt^{\rh}\,_{,\rh}}{2 \al} \,,
\eea

\noi
where $n^a$ is the future-directed, unit-norm, timelike vector orthogonal to the $t={\rm const.}$ 
hypersurfaces of the dimensionally reduced 2+1 spacetime, and
${}^{(2)}K^{\rh}\,_\rh, {}^{(2)}K^z\,_z$ are the components of the extrinsic curvature on those
hypersurfaces. 
Again, the choice of this particular form for $\bar{\Om}$ was motivated by regularity
concerns; certain terms that behave as $1/\rh$ as $\rh\to0$, and which would otherwise 
appear in the equations of motion for the variable naturally conjugate to $\sigmabar$,  manifestly cancel
with the definition~(\ref{omegabar_def}).
We also note that the leading order behaviour of $\bar{\Om}$ in the limit as $\rh\to0$ is $\bar{\Om}(\rh,z) =
\rh\bar{\Om}_1(z) + O(\rh^3)$.

We now adopt a slightly modified form for the 
stress-energy tensor of the complex scalar field 

\beq  \lab{phi4SET}
T_{\mu \nu} \equiv T^\phi_{\mu \nu} \eq 
\left[  \left( \na_{\mu} \ph \na_{\nu} \ph^\ast + \na_{\nu} \ph
\na_{\mu} \ph^\ast \right)
- g_{\mu \nu} \left( \ \na^\al \ph \na_\al \ph^\ast + m^2 |\ph|^2
+ 2 \la |\ph|^4 \right)\right] \,.
\eeq

\noi
(Note that the definition differs from that in~(\ref{Ctmunu}) by an overall factor
of 2).  In particular, we have added a quartic self-interaction term, 
$2\la |\ph|^4$ to the scalar field Lagrangian, where $\lambda$ is an adjustable 
coupling constant that will satisfy $\la=0$ for some of the simulations shown
below and $\la=1$ in other instances.  We note that in addition to the parameter (the particle
mass) $m$ which
sets the length scale of the system, the addition of 
the self-interaction term introduces a new dimensionless parameter, $\lambda$,
into the model, and that our results will generally now depend on the specific 
value of $\lambda$ that is used.  In what follows, we have used non-vanishing 
self-interaction primarily as a vehicle to allow us to probe the black hole 
threshold in certain cases.  Additional details will be provided in
Sec.~\ref{bincol}.

We also define the auxiliary complex scalar variable $\Pi$
via
\beq
 \Pi \eq n^a \pa_a \ph \,,
\eeq
so that the Klein-Gordon equation for $\ph$ can be recast as a set of first-order-in-time
equations.

In Sec.~\ref{perturb2d} we use a real scalar field to perturb the boson stars,
analogously to what was done in Sec.~\ref{CPresults} in the spherical case.  
Therefore we again introduce a real massless scalar field $\ph_3(t,\rho,z)$ with a 
stress energy tensor 
\beq
 T_{\mu \nu}^{\ph_3} \eq 2 \na_{\mu} \ph_3 \na^{\mu} \ph_3 - g_{\mu \nu} \na^{\al}\ph_3 \na_{\al}
\ph_3\, .
\eeq
The equation of motion for this field is 
\beq
\label{massless-eq}
 \na^\mu \na_{\mu} \ph_3 = 0\,,
\eeq
\noi
and, once more, we define an auxiliary field $\Pi_3$
\beq \lab{def_Pi3}
 \Pi_3 \eq n^a \pa_a \ph_3\,,
\eeq
so that~(\ref{massless-eq}) can be cast in first-order form.

We note that $\ph, \Pi, \ph_3$ and $\Pi_3$ are functions of $\rh, z$ and $t$ only.  Since the scalar field
$\ph = \ph_1 + i \ph_2$
and its conjugate momentum $\Pi = \Pi_1 + i \Pi_2$ are complex, we now have 12 fundamental real variables 
$\al, \bt^\rh, \bt^z, \ps, \bar{\si}, \bar{\Om}, \ph_1, \ph_2, \ph_3, \Pi_1, \Pi_2$ and $\Pi_3$ that must be 
evolved.

As mentioned above (and as for the case of the study of dynamical spherically symmetric boson
stars in Chap.~\ref{bs1d}), we choose
to implement maximal slicing, so that the trace of the 3-dimensional extrinsic curvature tensor of $t =
{\mbox {\rm const.}}$ slices within the 4-dimensional manifold (not to be confused with the trace of the
2-dimensional extrinsic curvature tensor ${}^{(2)}K$ in the 2+1 dimensionally reduced spacetime) is zero
\beq
 {}^{(3)} K = 0\,.
\eeq
\noi
This condition provides the
following elliptic equation for $\al$
\bea \lab{al_eqn}
2 \left( \rh \al_{,\rh} \right)_{,\rh^2} + \al_{,zz}
        + \al_{,\rh} \left(  2 \fr{ \ps_{,\rh} }{ \ps }
                               + \left( \rh \sigmabar \right)_{,\rh}
                               \right)
        + \al_{,z   } \left(  2 \fr{ \ps_{,z   } }{ \ps }
                               + \left( \rh \sigmabar \right)_{,z   }
                               \right)
\nn\\
   - \fr{\ps^4}{2\al}
       \left[  (  \bt^\rh{}_{,\rh} - \bt^z{}_{,z}      )^2
             + (  \bt^\rh{}_{,z   } + \bt^z{}_{,\rh} )^2 \right]
   - \fr{\ps^4}{6\al} \left[
                                    2 \al \rh \omegabar
                                  + \bt^\rh{}_{,\rh}
                                  - \bt^z{}_{,z}
                                 \right]^2 \nn \\ 
+ 8 \pi  \ps^4 \al \left( m^2 |\ph|^2  + \la |\ph|^4 - 2 |\Pi|^2 \right)
- 16 \pi \al \Pi_3^2 
= 0 \,.
\label{slicing_eqn}
\eea

The ADM decomposition in the dimensionally reduced 2+1 spacetime will
result in one Hamiltonian constraint and two momentum constraints, which,
when combined with the demand that the $(\rho,z)$ 2-spaces be conformally
flat, provides elliptic equations for $\ps$, and for $\bt^\rh$ and $\bt^z$, respectively.  They are

\begin{eqnarray}
          8 \fr{ \ps_{,\rh \rh} }{ \ps}
       +  8 \fr{ \ps_{,z    z   } }{ \ps}
       + 16 \fr{ \ps_{,\rh^2} }{ \ps }
       +  8 \left( \rh \sigmabar \right)_{,\rh}
              \fr{ \ps_{,\rh  } }{ \ps }
       +  8 \left( \rh \sigmabar \right)_{,z   }
              \fr{ \ps_{,z     } }{ \ps }
\nn\\
   +   {\ps^4 \over 2 \al^2}
       \left[  (  \bt^\rh{}_{,\rh} - \bt^z{}_{,z}      )^2
             + (  \bt^z{}_{,\rh}    + \bt^{\rh}{}_{,z} )^2 \right]
   + {\ps^4 \over 6\al^2}
        \left[ 2\al\rh\bar{\Om}
          + \bt^\rh{}_{,\rh}
          - \bt^z{}_{,z}   \right]^2
\nn\\
 =
             - 16 \pi \left( \ps^4 \left( m^2 |\ph|^2 + \la |\ph|^4 + |\Pi|^2 \right)
	       + |\ph_{,z}|^2 + |\ph_{,\rh}|^2 \right)
            -  6 \left( \rh^2 \left(\rh \sigmabar\right)_{,\rh} \right)_{,\rh^3}
\nn\\
            -  2 \left( \left( \rh \sigmabar \right)_{,\rh} \right)^2
            -  2 \left( \rh \sigmabar \right)_{,zz}
            -  2 \left( \left( \rh \sigmabar \right)_{,z   } \right)^2
-16 \pi \left( \Pi_3^2 + \ph_{3,\rh}^2 + \ph_{3,z}^2 \right) \,,
\label{hc_eqn}
\end{eqnarray}

\beq
\begin{split}
   \fr{2}{3}\bt^{\rh}{}_{,\rh\rh}
 + \bt^{\rh}{}_{,zz}
 + \fr{1}{3} \bt^z{}_{,z\rh}
 + 32\pi {\al \over \ps^2} \Pi_{3,\rh}
  + 16 \pi \al \left( \ph_{,\rh} \Pi^{\ast} + \ph_{,\rh}^{\ast} \Pi \right)
\\
 - \left(    \left(    \fr{ \al_{,z} }{ \al }
                   - 6 \fr{ \ps_{,z}   }{ \ps   } \right)
           - \left( \rh\sigmabar                   \right)_{,z}
         \right)
   \left(    \bt^z{}_{,\rh}
           + \bt^\rh{}_{,z}
         \right)
\\
 - \fr{2}{3} \left(     \fr{ \al_{,\rh} }{ \al }
                      - 6 \fr{ \ps_{,\rh}   }{ \ps   }
                     \right)
     \left[
               \bt^\rh{}_{,\rh}
             - \bt^z{}_{,z}
    \right]
 - \fr{8}{3} \al \bar{\Om}
 -\fr{2 \al \rh }{ 3 } \left[
                                     6 \omegabar \fr{\ps_{,\rh}}{ \ps }
                                   + \omegabar_{,\rh}
                                   + 3 \omegabar \left(\rh \sigmabar   \right)_{,\rh}
                                   \right]=0,
\label{brho_eqn}
\end{split}
\eeq

\noi
and
\begin{eqnarray}
   \bt^z{}_{,\rh\rh}
 + \fr{4}{3}\bt^z{}_{,zz}
 - \fr{1}{3} \bt^{\rh}{}_{,z\rh}
   + \left[
        \left( \rh \sigmabar \right)_{,\rh}
      + \fr{ 2 \al }{ \ps^6} \left(
                                           \fr{\rh \ps^6}{\al}
                                         \right)_{,\rh^2}
     \right]
       \left( \bt^\rh{}_{,z} + \bt^z{}_{,\rh} \right)
\nn\\
   + \left[   2 \left( \rh \sigmabar \right)_{,z}
            - \fr{4}{3} \left(    \fr{ \al_{,z} }{ \al }
                                - 6 \fr{ \ps_{,z}   }{ \ps   } \right)
           \right]
       \left( \bt^z{}_{,z} - \bt^\rh{}_{,\rh} \right)
 - \fr{2 \al\rh}{3} \left(
                                  6 \omegabar \fr{\ps_{,z}}{\ps}
                                + \omegabar_{,z}
                               \right)
\nn\\
+ 32\pi {\al \over \ps^2} \Pi_{3,z}
 + 16 \pi \al \left( \ph_{,z} \Pi^{\ast} + \ph_{,z}^{\ast} \Pi \right)
 - 2\al(\sigmabar_{,z})\rh^2\omegabar=0 \,.
\label{bz_eqn}
\end{eqnarray}
\noi
The evolution equation for $\sigmabar$ is, with the definition of
$\omegabar$
(\ref{omegabar_def}), given by 

\begin{equation}\label{sigmabar_eqn}
\dot{\sigmabar} =
        2 \bt^{\rh} \left( \rh\sigmabar \right)_{,\rh^2} + \bt^z
        \sigmabar_{,z}
     -  \al\bar{\Om}
     - \left[ {\bt^{\rh} \over \rh} \right]_{,\rh} \,,
\end{equation}
where, as usual, an over-dot
denotes a time derivative.
The evolution equation for $\omegabar$ is
\begin{eqnarray}
\dot{\bar{\Om}} =
     2 \bt^{\rh} \left(\rh\bar{\Om}\right)_{,\rh^2}
    +  \bt^z \bar{\Om}_{,z}
    - {1\over 2\al\rh} \left(
                                   \bt^{z}{}_{,\rh}{}^2
                                 - \bt^{\rh}{}_{,z}{}^2
                                 \right)
    + {1\over \ps^4} \left( {\al_{,\rh} \over \rh} \right)_{,\rh}
\nn\\
    + {\al \over \ps^6}\left(
               {(\ps^2)_{,\rh} \over \rh} \right)_{,\rh}
    - {2\al \over \ps^4}
        \left(   4 \fr{ \ps_{,\rh^2} }{ \ps }
               + \left( \rh \sigmabar \right)_{,\rh^2}
              \right)
        \left(   \fr{\al_{,\rh}    }{ \al }
               + \fr{ 2\ps_{,\rh} }{ \ps   }
              \right)
\nn\\
- {\al \over \ps^4} \left[
                                 \sigmabar_{,z} \left(
                                           \fr{   \al_{,z} }{ \al }
                                         + \fr{  2\ps_{,z}   }{ \ps   }
                                         \right)
                               + \rh \sigmabar_{,z}{}^2
                               + \sigmabar_{,zz}
                               \right]
    + 64\pi{\al\over \ps^4} \rh (\ph_{3,\rh^2})^2
+16 \pi \fr{\al}{\rh \ps^4} \vert\ph_{,\rh}\vert^2  \,.
\label{omegabar_eqn}
\end{eqnarray}
The definition of $\Pi$ and the Klein-Gordon equation give
\begin{equation}
\dot{\ph} =
       \bt^{\rh} \ph_{,\rh} + \bt^z \ph_{,z}
            + {\al } \Pi\,,
\end{equation}
and
\begin{eqnarray}
\dot{\Pi} =
       \bt^{\rh} \Pi_{,\rh} + \bt^z \Pi_{,z}
- \al \ph \left( 2 \la |\ph|^2 + m^2 \right)
+ \fr{2 \al}{\ps^5} \left( \ps_{,\rh} \ph_{,\rh} + \ps_{,z} \ph_{,z}\right)
\nn\\
     + \fr{1}{\ps^4} \left[ \al \left( 
                                     \left( \rh \sigmabar \right)_{,\rh}
\ph_{,\rh}
                                   + \left( \rh \sigmabar \right)_{,z   }
\ph_{,z}
\right)
+ \ph_{,\rh \rh} + \ph_{,zz} + \fr{\ph_{,\rh}}{\rh} + \al_{,\rh} \ph_{,\rh} +
\al_{,z} \ph_{,z}
                                  \right]
\,.
\end{eqnarray}

The definition of $\Pi_3$ (\ref{def_Pi3}) and the wave equation for $\ph_3$ (\ref{massless-eq}) gives
\begin{equation}\label{phi_eqn}
\dot{\ph_3} =
       \beta^{\rho} \ph_{3,\rho} + \beta^z \ph_{3,z}
            + {\alpha \over \psi^2 } \Pi_3,
\end{equation}
and
\begin{eqnarray}
\dot{\Pi_3} =
       \beta^{\rho} \Pi_{3,\rho} + \beta^z \Pi_{3,z}
     + {1\over3} \Pi_3 \left(   \alpha\rho\bar{\Omega}
                            + 2 \beta^{\rho}_{,\rho}
                            + \beta^z_{,z}
                     \right)
\nonumber\\
     + \frac{1}{\psi^4} \left[
                         2\left( \rho \alpha \psi^2 \ph_{3,\rho} \right)_{,\rho^2}
                        + \left(      \alpha \psi^2 \ph_{3,z}    \right)_{,z}
                              \right]
     + \frac{\alpha}{\psi^2} \left[
                                     \left( \rho \sigmabar \right)_{,\rho} \ph_{3,\rho}
                                   + \left( \rho \sigmabar \right)_{,z   } \ph_{3,z}
                                  \right]
\,.
\label{pi_phi_eqn}
\end{eqnarray}

The regularity conditions on the symmetry axis are
\beq
\al_{,\rh}(t,0,z) = \ps_{,\rh}(t,0,z) = \bz_{,\rh}(t,0,z) = \brho(t,0,z) = \sigmabar(t,0,z) = \omegabar(t,0,z) =0\,,
\eeq
\beq
 \ph_{1,\rh}(t,0,z) =\ph_{2,\rh}(t,0,z) =\ph_{3,\rh}(t,0,z) = \Pi_{1,\rh}(t,0,z) =
\Pi_{2,\rh}(t,0,z) =\Pi_{3,\rh}(t,0,z) = 0\,.
\eeq
The outer boundary conditions we use, applied at $\rh=\rh_{\rm max}$,
$z=z_{\rm max}$ and $z=z_{\rm min}$, are
\begin{eqnarray}
\al-1 + \rh \al_{,\rh} + z \al_{,z} &=& 0 , \nn\\
\ps-1 + \rh \ps_{,\rh} + z \ps_{,z} &=& 0 , \nn\\
\bz + \rh \bz_{,\rh} + z \bz_{,z} &=& 0 , \nn\\
\brho + \rh \brho_{,\rh} + z \brho_{,z} &=& 0 , \nn\\
r \sigmabar_{,t} + \rh \sigmabar_{,\rh} + z \sigmabar_{,z} + \sigmabar&=&0 , \nn\\
r \omegabar_{,t} + \rh \omegabar_{,\rh} + z \omegabar_{,z} + \omegabar&=&0 , \nn\\
r \ph_{1,t} + \rh \ph_{1,\rh} + z \ph_{1,z} + \ph_1&=&0 , \nn\\
r \ph_{2,t} + \rh \ph_{2,\rh} + z \ph_{2,z} + \ph_2&=&0 , \nn\\
r \ph_{3,t} + \rh \ph_{3,\rh} + z \ph_{3,z} + \ph_3&=&0 , \nn\\
r \Pi_{1,t} + \rh \Pi_{1,\rh} + z \Pi_{1,z} + \Pi_1&=&0 , \nn \\
r \Pi_{2,t} + \rh \Pi_{2,\rh} + z \Pi_{2,z} + \Pi_2&=&0 , \nn \\
r \Pi_{3,t} + \rh \Pi_{3,\rh} + z \Pi_{3,z} + \Pi_3&=&0 ,
\label{obc_sum}
\end{eqnarray}

\noi
where $r \eq \sr{\rh^2+z^2}$.

Thus, the
system of equations (\ref{al_eqn})-(\ref{obc_sum}) consists of 4 elliptic PDEs (the
equations for $\al, \ps, \bt^\rh$ and $\bt^z$), and 8 first-order-in time evolution
equations (also PDEs) that are hyperbolic in nature.  The finite differencing of the above equations 
is given in \cite{fransp:phd}, with trivial modifications for the extra 
terms involving the components $\ph_1, \ph_2$ of the complex scalar fields, 
and their conjugates $\Pi_1, \Pi_2$.  $\ph_3$ is identified with the original scalar field
described in \cite{graxi:2003}. The finite difference scheme also incorporates 
Kreiss-Oliger style dissipation, with a dissipation coefficient, $\ep_d$, similar to
that defined in (\ref{KOdiss}). Again the interested reader can find full
details in \cite{fransp:phd}.

Following modification to accommodate boson star initial data, 
the \graxi\ code was subjected to convergence tests to ensure that 
the discrete equations of motion had been implemented correctly. 
Results from one such test are described in App.~\ref{ap9}.

\begin{figure}[htp]
\vspace{0.5cm}
\begin{center}
\epsfxsize=8.5cm
\ifthenelse{\equal{\highQ}{true}} {
\includegraphics[width=8.5cm,clip=true]{eps/graxi_csf/bs.1s.N/phi.eps}
}{
\includegraphics[width=8.5cm,clip=true]{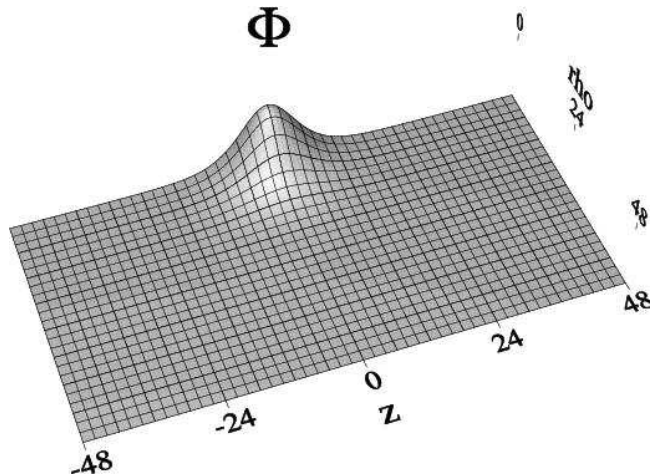}
}
\caption
[A typical scalar field configuration, $\ph_0(\rh,z)$ (denoted as $\Ph$), given by interpolation of a spherically 
symmetric (1D) boson star solution to a 2D finite difference mesh
in cylindrical coordinates, $(\rho,z)$]
{A typical scalar field configuration, $\ph_0(\rh,z)$ (denoted as $\Ph$), given by interpolation of a spherically 
symmetric (1D) boson star solution to a 2D finite difference mesh
in cylindrical coordinates, $(\rho,z)$.
$0.0\le \ph_0(\rh,z)\le 0.014$.}
\label{ph1dto2d}
\end{center}
\end{figure}

Since we study dynamics of non-rotating boson stars in this section, we can
use static spherically symmetric boson star solutions generated using 
one of the codes described in Chap.~\ref{bs1d}, as the basis for setting 
up axisymmetric initial data.  However, even for the seemingly trivial case 
where we evolve a {\em single} spherical star using \graxi, the initial
data setup is slightly complicated by the fact that we need to transform 
the spherically symmetric solution $\phi_0(r)$ to the $(\rho,z)$ coordinates 
used in \graxi.  We thus generally need to interpolate radially symmetric 
data onto the discrete domain $(\rho_i,z_j)$, possibly translating the data
so that the center of the star lies at some specified location along the $z$-axis.
Here we use standard polynomial (Lagrange) interpolation which is fourth 
order in the mesh spacings $(\Delta \rho, \Delta z)$.
Fig.~\ref{ph1dto2d} shows a typical
result of the interpolation process.  
Also note that all simulations in this section are
done with the adaptive mesh refinement (AMR) capability provided 
by \graxi, in order to effectively resolve
the various length and time scales encountered during the dynamical evolution of 
the scalar fields.  In particular, the complex field tends to form configurations 
that are considerable more compact at late times than at the initial time,
especially in cases where black holes form.

\subsection{Head-on Collisions of Boson Stars---Setup of Numerical Experiments} \lab{bincol}

The PDEs solved in the simulations discussed here are those listed in Sec.~\ref{eom2d}.
The initial data for the (first) boson star, always chosen from the stable
branch, is given by interpolation of a static spherically symmetric star as 
just described. 
The interpolated data, $\ph_0^{(1)}(\rh,z)$, are then scaled by
\beq
 \ph_0^{(1)}(\rh,z) \to \fr{1}{\sr{2}}\ph_0^{(1)}(\rh,z) \,,
\eeq
\noi
since the definitions (\ref{Ctmunu}) and (\ref{phi4SET}) differ
by a factor of 2.  The star is then
``boosted'' (roughly speaking, given an initial momentum) via (see \cite{guenther,dale})
\beq
\ph^{(1)}(\rh,z;p_z^{(1)}) \eq \ph^{(1)}_0(\rh,z) \,e^{i p^{(1)}_z z }\,,
\eeq
\noi
where $p_z^{(1)}$ is the parameter that controls the magnitude of the boost, and which
we will refer to as the {\em initial momentum parameter}, or, more loosely, as the 
{\em initial momentum}.

The boosted boson star is then
translated so that the center of the star
is located at some specified point, 
$(0,z_1)$, on the axis of symmetry 
\beq
 \ph^{(1)}(\rh,z; p_z^{(1)},z_1) \eq  T\left( \ph^{(1)}(\rh,z;p_z^{(1)})\,;z_1\right)\,, 
\eeq
\noi
\noi
\noi
where $T$ is the translation operator defined by
\beq
 T(f(\rh,z)\,;z_1) \eq 
\left\{
\ba{ll}
f(\rh,z-z_1)\;\;  &  z-z_1 \in [z_{\rm min},z_{\rm max}] \\
0 \;\; & \rm{otherwise}
\ea
\right.
\,,
\eeq
and $z_1$ is another adjustable parameter.

Similarly we construct another boosted boson star centered at $(0,z_2)$ via
\beq
 \ph^{(2)}(\rh,z; p_z^{(2)},z_2) \eq  T\left( \ph^{(2)}(\rh,z;p_z^{(2)})\,;z_2\right)\,, 
\eeq
\noi
and, finally, the complex scalar field {\em per se} is initialized to be 
the {\em sum} of the configurations representing the two boosted, translated stars
\beq
 \ph(0,\rh,z) =  \ph^{(1)}(\rh,z; p_z^{(1)},z_1) +  \ph^{(2)}(\rh,z; p_z^{(2)},z_2)\,.
\eeq

In the simulations described below we have always considered identical stars 
(i.e.\ $\ph_0^{(1)}(\rh,z) \eq
\ph_0^{(2)}(\rh,z) \eq \ph_0(\rh,z)$), each of which is boosted towards the other with the 
same initial ``speed", (i.e.\ $-p_z^{(2)} \eq p_z^{(1)} \eq p_z$).   We note that the
solutions generated from such initial data are symmetric with respect to the equatorial 
plane, but that we do not make explicit use of this fact in the calculations.  We also
note that there is nothing in principle that prevents us from carrying out simulations 
with $\ph_0^{(1)}(\rh,z) \ne \ph_0^{(2)}(\rh,z)$ and/or $-p_z^{(2)} \ne p_z^{(1)}$.

In addition to the complex field, $\phi$, we must also supply initial values for 
the conjugate variable $\Pi$.   To this end we define a contribution, $\Pi^{(1)}(\rh,z;p_z^{(1)})$,
associated with the first star
\beq
\Pi^{(1)}(\rh,z;p_z^{(1)}) =  \om_1 \ph^{(1)}(\rh,z;p_z^{(1)})  -\bt^{\rh}(\rh,z) \pa_{\rh} \ph^{(1)}(\rh,z;p_z^{(1)}) -
\bt^z(\rh,z)   \pa_z \ph^{(1)}(\rh,z;p_z^{(1)}) \,,
\eeq
\noi
where $\om_1$ is the eigenvalue of the first boson star.
We similarly define a contribution, $\Pi^{(2)}(\rh,z;p_z^{(2)})$, for the second star
\beq
\Pi^{(2)}(\rh,z;p_z^{(2)}) =  \om_2 \ph^{(2)}(\rh,z;p_z^{(2)})  -\bt^{\rh}(\rh,z) \pa_{\rh} \ph^{(2)}(\rh,z;p_z^{(2)}) -
\bt^z(\rh,z)   \pa_z \ph^{(2)}(\rh,z;p_z^{(2)}) \,,
\eeq
and then initialize $\Pi$ via
\beq
\Pi(0,\rh,z) =
 T\left( \Pi^{(1)}(\rh,z;p_z^{(1)}) \,;z_1\right) +
 T\left( \Pi^{(2)}(\rh,z;p_z^{(2)}) \,;z_2\right) \,.
\eeq

The remaining freely specifiable variables 
$\si(t,\rh,z)$ and $\bar{\Om}(t,\rh,z)$, 
(which can be associated with the
gravitational wave content of the spacetime), 
are initialized to zero:
\beq
 \si(0,\rh,z) = \bar{\Om}(0,\rh,z) = 0\,.
\eeq 
Finally, initial values for the remaining geometric variables, $\alpha, \beta^\rho, \beta^z$ and $\psi$,
are determined from the constraint and slicing equations. 

In the simulations we typically use a Courant factor $\De t/\De \rh = \De t/\De z =
0.2$ or $0.3$, and the dissipation coefficient is $\ep_d = 0.5$. 

\subsection{Head-on Collisions---Solitonic Behaviour} 

In the head-on collisions of Newtonian boson stars (i.e.\ in simulations involving the 
solution of a coupled Poisson-Schr\"{o}dinger system---see App.~\ref{ap8}) it has 
been observed that the stars sometimes exhibit ``solitonic'' behaviour~\cite{dale}.  This 
behaviour is a function of the Newtonian analog of the initial momentum parameter, 
$p_z$, defined above.
Specifically, for small initial momenta the stars
simply merge together to form a single star, while for 
sufficiently large initial momenta, the stars pass through each other
as if there was no interaction between them.  We have carried out numerical
experiments similar to those described in~\cite{dale}, but using the fully coupled
general-relativistic equations of motion. 
We find that solitonic behaviour also occurs in this case for large values 
of the initial momentum.

\begin{figure}
\begin{center}
\epsfxsize=15.0cm
\ifthenelse{\equal{\highQ}{true}} {
\includegraphics[width=15.0cm,clip=true]{eps/graxi_csf/soliton/modphpz0.4.eps}
}{
\includegraphics[width=15.0cm,clip=true]{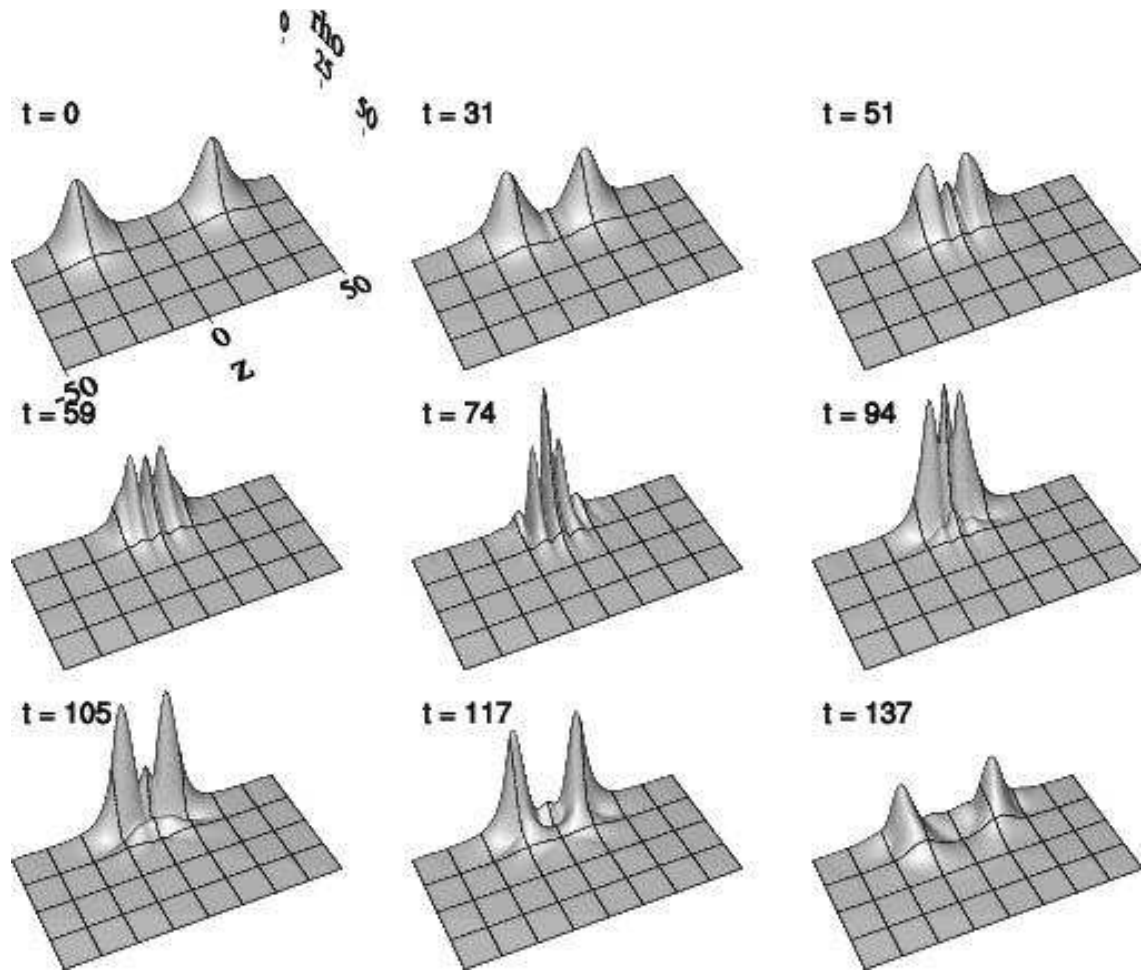}
}
\caption
[Typical solitonic behaviour in a head-on collision of boson stars]
{Typical solitonic behaviour in a head-on collision of boson stars.
The stars are initialized via interpolation, translation and boosting 
of a spherically symmetric solution with 
$\ph_0(0) = 0.02$. 
The stars are initially centered at $(0,-25)$ and $(0,25)$, with initial momenta
$-p_z^{(2)}=p_z^{(1)}=p_z=0.4$. The two stars start to overlap at
$t\approx 31$, interfere with each other, and then separate at $t\approx 117$.
Other simulation parameters are: $\De t/\De z=\De t/\De \rh = 0.2$,
$\De \rh = \De z = 0.78$.
Note that the temporal spacing between successive snapshots
is not constant---the time instants displayed have been chosen to illustrate the key
features of the simulation.
$0.0\le |\ph(t,\rh,z)| \le 0.03$.}
\label{soliton}
\end{center}
\end{figure}

In Fig.~\ref{soliton} we show results from a typical simulation of 
a head-on collision of two boosted boson stars with 
no self-interaction ($\la = 0$). The boson stars are initially centered at
$(0,z_1) = (0,-25)$ and $(0,z_2) = (0,25)$ and have initial momenta
$-p_z^{(2)} = p_z^{(1)} = 0.4$.  The computational domain is $0\le\rh\le 50$,
$-50 \le z \le 50$.  From the figure we see that the stars seem to pass 
through each other as if there were little net interaction between them.

\subsection{Head-on Collisions---Critical Behaviour} 

We now proceed to study critical behaviour in the context of head-on collisions 
of boson stars with equatorial-plane symmetry.  
For sufficiently large values of $\ph_0(0)$ (for instance, $\ph_0(0) = 0.02$), 
variation of the initial momentum parameter, $p_z$, generates a family of solutions that 
interpolates through the black hole threshold; however, in this case the sense of
$p_z$ is reversed from the normal situation---i.e.\ for small (large) $p_z$ black holes
do (do not) form.
Once we have determined an initial bracket for $p_z$ that is 
known to span the threshold, we can, in principle, tune the parameter to the 
critical value $p^\star_z$ that generates a critical solution. 

However, in our early tuning experiments---which involved {\em no} self-interaction
for the complex field---we found it difficult to determine whether or not a black 
hole would form in the evolution as $p_z$ approached the critical value. 
\footnote{More specifically, we found it difficult to distinguish between black hole
formation associated with the supercritical evolutions from black hole formation associated
with recollapse of subcritical evolutions.}
The addition of the quartic self-interaction term provides the stars with more 
binding, and reduces the distortion in their shapes during the collision.  This 
allows us to identify whether a particular calculation is subcritical or supercritical,
and we are then generally able to determine a value of $p^\star_z$ to 
machine precision.   We note that we only incorporate the interaction term
into the {\em dynamical} equations; i.e.\ we do {\em not} compute static boson 
stars that include the quartic term.~\footnote{This would certainly be possible
to do---see, for example, \cite{colpi:1986}.}
Thus, with self-interaction turned on, the initial data prescription described above
will no longer generate a static solution for a single boson star, and in 
fact, we find that in the evolution of data computed with $\lambda\ne0$ the
central amplitude of the complex field tends to grow
in time.  However, within the time scale for black hole formation in near-critical
evolutions ($t \approx 300$), and for the value $\la=1$ that we 
have adopted for the simulations described in this section, this growth in 
amplitude appears to be insignificant.

Also, we need to remark at this point that when we speak of the black hole threshold in
the head-on collision 
calculations described here, we are referring to what we might call ``prompt'' black hole 
formation, i.e.\ whether a black hole forms during the initial merger phase 
of the collision.  Calculations which are subcritical with respect to this 
definition are generated by relatively large values of $p_z$, and are characterized
by the re-appearance of two star-like configurations following the initial 
merger phase (see Fig.~\ref{subcrit2s}).  Subsequently, however, the stars 
will re-merge and then typically form a black hole (again, see Fig.~\ref{subcrit2s}).
However, this fact that subcritical simulations may {\em eventually} form black holes
does not interfere with our ability to study the threshold defined with respect
to prompt collapse.

We now discuss the results from the study of a specific family of 
head-on collisions in which Type I critical phenomena are 
observed.  As in the calculation discussed in the previous sub-section,
initial data for any member of the family of solutions describe two identical 
stars ($\ph_0(0) = 0.02$) centered at $(0,-25)$ and $(0,25)$, and with initial momenta
$-p_z^{(2)}=p_z^{(1)}=p_z$.  The stars are thus well separated from each
other at $t=0$.  
The computational domain is $-50\le z\le 50$, $0\le \rh \le 50$,
and the base resolution is $(N_\rho,N_z)=(129,257)$, yielding base mesh
spacings $\De \rh = \De z = 0.390625$.  
As many as 6 additional levels of 2:1 refinement
are employed in the calculations, so that the finest available resolution 
is  $\De \rh = \De z = 0.006104$.  The critical parameter value for this 
family of data is $p^\star_z \approx 0.21$.

\begin{figure}
\begin{center}
\epsfxsize=15.0cm
\ifthenelse{\equal{\highQ}{true}} {
\includegraphics[width=15.0cm,clip=true]{eps/graxi_csf/bs.ad.2sboost.cp/pz0modph.eps}
}{
\includegraphics[width=15.0cm,clip=true]{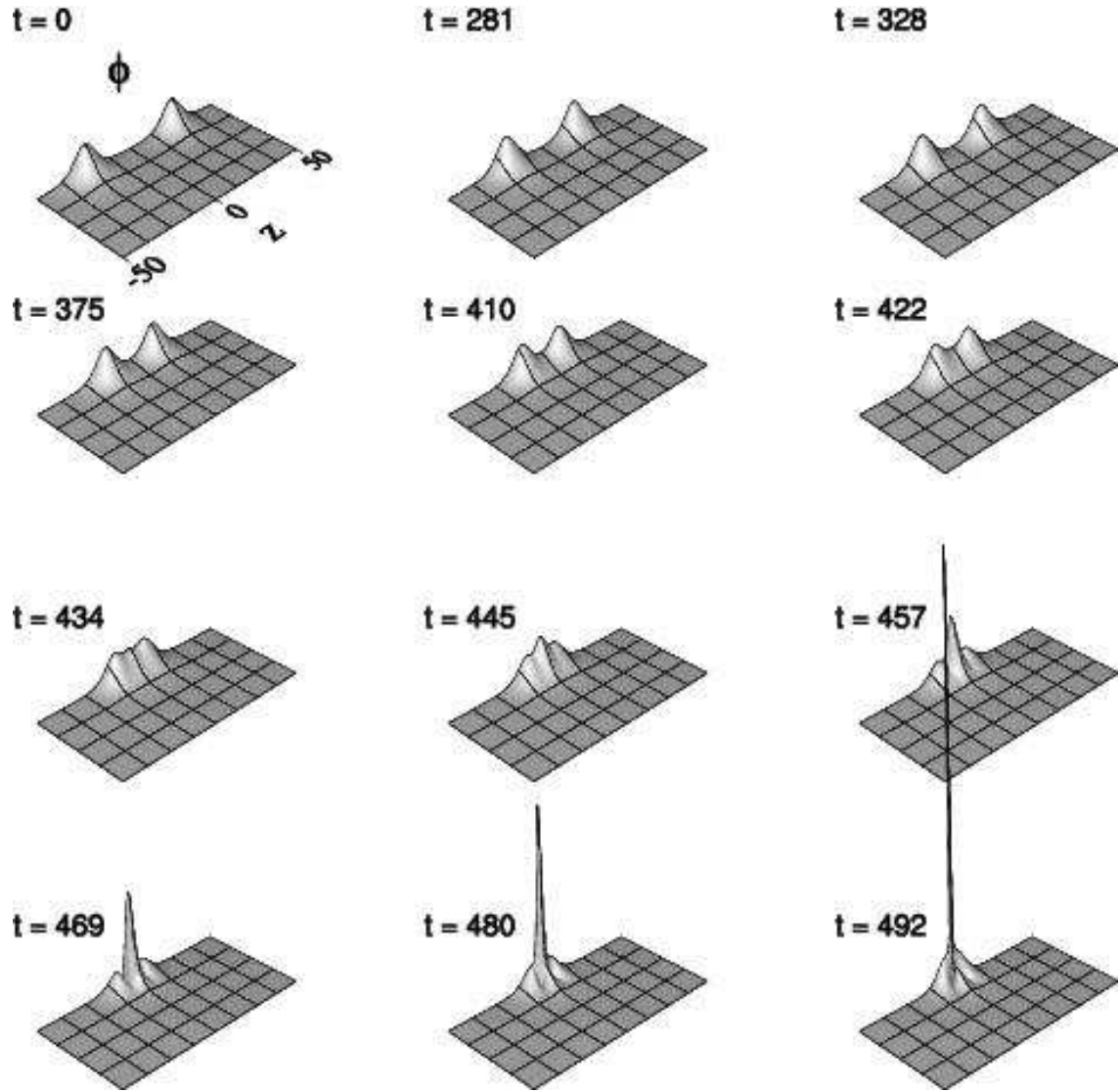}
}
\caption
[Head-on collision of boson stars with $p_z=0$]
{Head-on collision of boson stars with $p_z=0$.  Plotted here is 
the modulus of the complex field, $|\phi(t,\rho,z)|$.
The identical stars ($\ph_0=0.02$) are initially centered at $(0,-25)$ and $(0,25)$.  
They move towards one another due to their mutual gravitational 
attraction, then start to merge at $t\approx 400$.  The resulting 
configuration 
oscillates and radiates energy during the merger phase, and then collapses 
to form a black hole---an apparent horizon is first detected at $t=492$ (see Fig.~\ref{bs2sAH}).
See the text for additional details of the numerical parameters of the simulation.
$0.0\le |\ph(t,\rh,z)| \le 0.19$.}
\label{2sboost0}
\end{center}
\end{figure}

Fig.~\ref{2sboost0} shows snapshots of $|\phi(t,\rho,z)|$ 
from a simulation with $p_z = 0$, so that the stars are initially at rest.
The time development clearly shows that the two boson stars attract each 
other through their mutual gravitational interaction.  At
$t\approx 400$ the stars begin to merge, and subsequently they form
a configuration which approximates that of a single star.  
Following the merger, the 
configuration oscillates---emitting  some energy in the form of 
outgoing scalar radiation in the process---then collapses and forms a black hole.

\begin{figure}
\begin{center}
\epsfxsize=9.0cm
\includegraphics[width=9.0cm,clip=true]{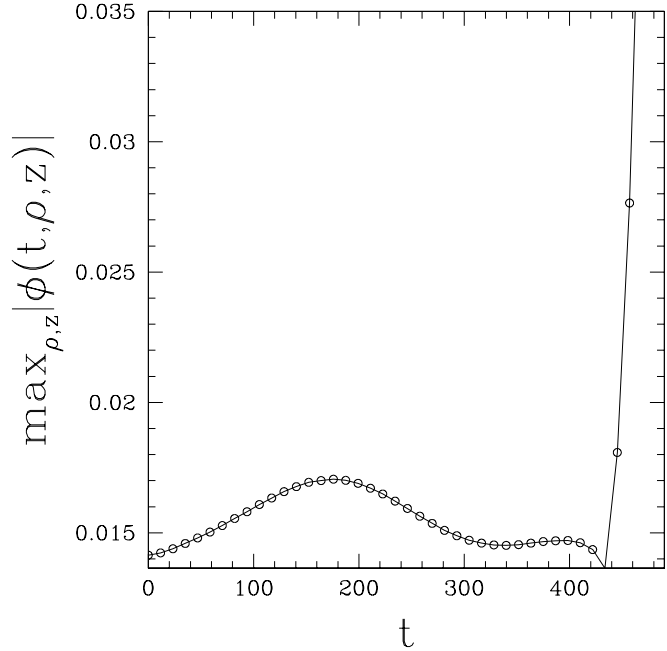}
\caption
[Maximum value of scalar field modulus $|\phi|_{\rm max}(t) \equiv \max_{\rh,z}|\ph(t,\rh,z)|$ {\em vs} 
time for a head-on boson star collision with $p_z=0$]
{
Maximum value of scalar field modulus $|\phi|_{\rm max}(t) \equiv \max_{\rh,z}|\ph(t,\rh,z)|$ {\em vs} 
time for a head-on boson star collision with $p_z=0$.
$|\phi|_{\rm max}(t)$ oscillates during the in-fall phase (presumably due to 
gravitationally induced distortions of the stars
and/or 
the quartic self-interaction),
then starts to grow
rapidly during the merger phase, $t > 430$, signalling the collapse of the configuration
(analogous to central density increasing in fluid collapse) and the imminent formation
of a black hole.
}
\label{maxmodphbs2spz0}
\end{center}
\end{figure}

\begin{figure}
\begin{center}
\epsfxsize=9.0cm
\includegraphics[width=9.0cm,clip=true]{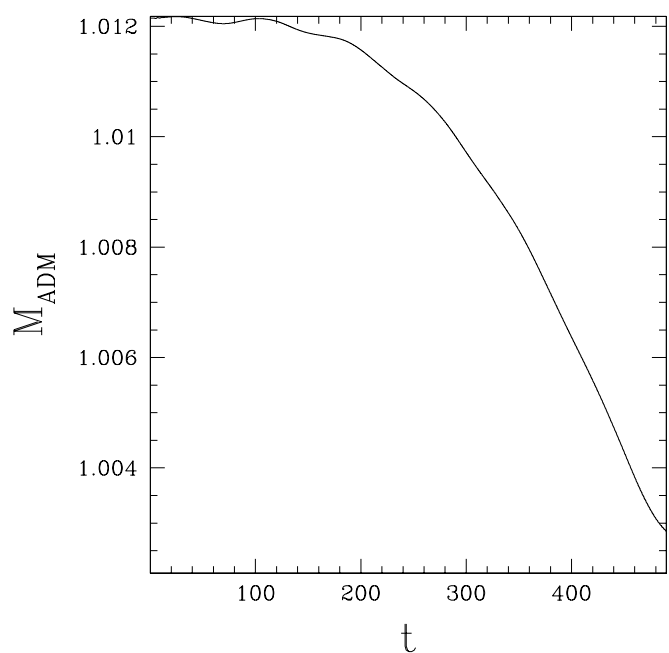}
\caption
[ADM mass {\em vs} time for a head-on boson star collision with $p_z=0$]
{ADM mass {\em vs} time for a head-on boson star collision with $p_z=0$.
Although there are slight oscillations in this quantity at early times 
(most
likely related to outer boundary effects), the mass shows an overall 
decrease in time, indicating that energy is being radiated from the system.}
\label{Mbs2spz0}
\end{center}
\end{figure}

\begin{figure}
\begin{center}
\epsfxsize=12.0cm
\includegraphics[width=12.0cm,clip=true]{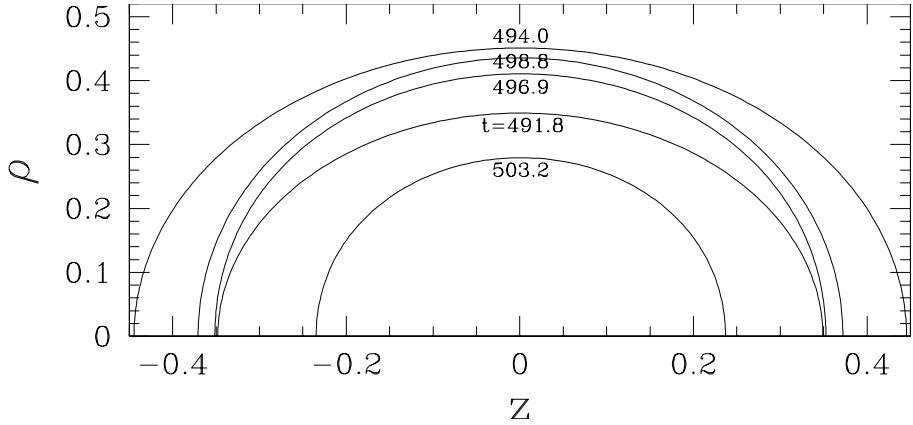}
\caption
[Development of the apparent horizon for a head-on collision of 
boson stars with $p_z=0$]
{Development of the apparent horizon for a head-on collision of
boson stars with $p_z=0$.
Five snapshots of the coordinate location of the apparent horizon are shown 
for $t=491.8, 494.0, 496.9, 498.8$ and $503.2$ (an apparent horizon 
is first detected at $t=492$). As discussed in the text, much 
of the ``dynamics'' of the horizon that is visible here is likely
attributable to coordinate effects.  In particular, contrary 
to what the figure suggests, we expect the area of the apparent horizon
to be a monotonically {\em increasing} quantity.
}
\label{bs2sAH}
\end{center}
\end{figure}

Fig.~\ref{maxmodphbs2spz0} shows the variation of the maximum value of the modulus of the 
scalar field, $|\phi|_{\rm max}(t) \equiv \mathrm{max}_{\rh,z}|\ph(t,\rh,z)|$, {\em vs} $t$, again for the 
case $p_z=0$.  This function initially has 
a value of $0.014$ (note that since (\ref{Ctmunu}) and (\ref{phi4SET}) differ
by a factor of 2, the value here is {\em not} 0.02).  As the evolution
proceeds it exhibits oscillations 
(presumably due to 
gravitationally induced perturbations of each of the stars
and/or 
the quartic self-interaction).
until $t\approx 430$, when the stars begin to merge.
At this time, $|\phi|_{\rm max}(t)$ begins to rapidly increase, and this 
behaviour signals the imminent formation of a black hole (see Fig.~\ref{bs2sAH}).  
Fig.~\ref{Mbs2spz0} shows the variation of the 
ADM mass, $M_{\mathrm{ADM}}$ {\em vs} $t$, for the same calculation ($M_{\rm ADM}$
is computed using (2.67) of \cite{fransp:phd}).  
The figure indicates that a small amount of mass is radiated from the system
during the simulation.
 
Finally, 
Fig.~\ref{bs2sAH} shows the time development of the apparent horizon detected 
in the simulation at late times---an apparent horizon is first found at $t=492$.
We note that this figure simply shows the location of the apparent horizon
in coordinate space for various times and that much of the ``dynamics''
that is visible is probably attributable to coordinate effects.  In
particular, the proper surface areas of the apparent horizons 
(which we expect ultimately to be closely tied to the black hole mass)
are {\em not} 
directly correlated to the coordinate areas covered by each of the contours.
Computation of the former requires evaluation of an integral involving 
$\psi(t,\rho,z)^4$, and $\psi$ is a highly dynamical quantity, that tends 
to rapidly increase in the central regions of the domain at late times.
Thus, although we have not explicitly computed the area of the apparent 
horizon, we anticipate that it is non-decreasing.

\begin{figure}[htp]
\begin{center}
\epsfxsize=15.0cm
\ifthenelse{\equal{\highQ}{true}} {
\includegraphics[width=15.0cm,clip=true]{eps/graxi_csf/bs.ad.2sboost.cp/pzsupercritmodph.eps}
}{
\includegraphics[width=15.0cm,clip=true]{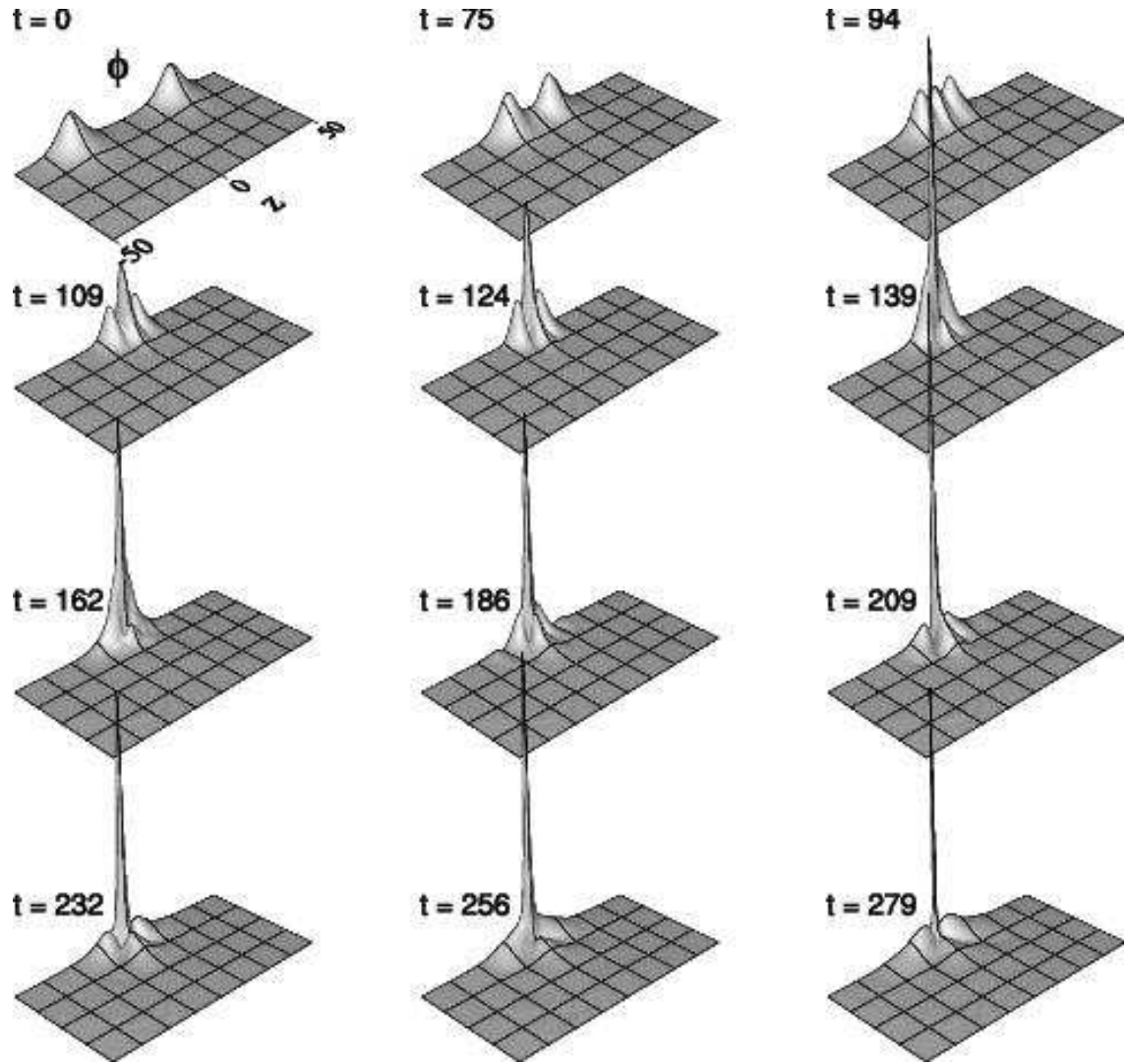}
}
\caption
[Marginally supercritical, head-on boson star collision]
{Marginally supercritical, head-on boson star collision.  The time 
development of $|\phi(t,\rho,z)|$ is plotted for a supercritical
calculation with $p_z$ tuned to roughly machine precision of the 
critical value $p^\star_z \approx 0.21$.  The stars merge at $t\approx 90$, 
significantly earlier than in the simulation shown in Fig.~\ref{maxmodphbs2spz0}.
The resulting merged configuration remains in the critical state for
$\De t\approx 130$, then collapses to form a black hole.  An apparent horizon is first
detected at $t=280$.
$0.0\le |\ph(t,\rh,z)| \le 0.13$.}
\label{supercrit2s}
\end{center}
\end{figure}

\begin{figure}[htp]
\begin{center}
\epsfxsize=15.0cm
\ifthenelse{\equal{\highQ}{true}} {
\includegraphics[width=15.0cm,clip=true]{eps/graxi_csf/bs.ad.2sboost.cp/pzsubcritmodph.eps}
}{
\includegraphics[width=15.0cm,clip=true]{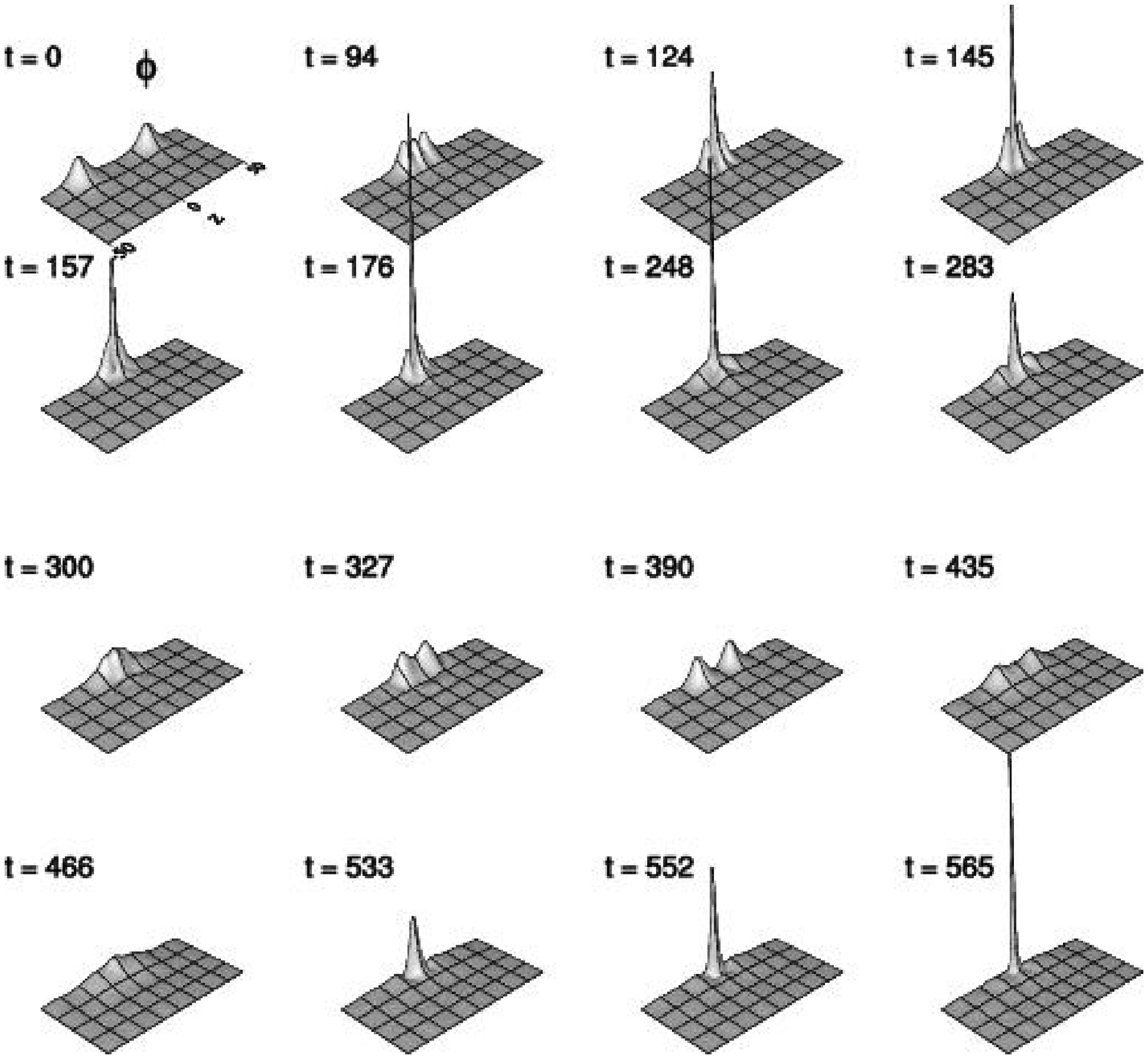}
}
\caption[Marginally subcritical, head-on boson star collision]
{Marginally subcritical, head-on boson star collision.  The time
development of $|\phi(t,\rho,z)|$ is plotted for a subcritical
calculation with $p_z$ tuned to roughly machine precision of the
critical value $p^\star_z \approx 0.21$. 
At early times the evolution here is indistinguishable from that shown in Fig.~\ref{supercrit2s}.
However, in contrast to the supercritical case, at
$t\approx 280$ two star-like configurations begin to reappear and are clearly
distinguishable as separate stars at $t \approx 390$.  Since the initial momenta of the stars
is moderate, the overall system is gravitationally bound and
the two stars re-merge and form a black hole.
An apparent horizon is detected at $t=570$.
$0.0\le |\ph(t,\rh,z)| \le 0.14$.
}
\label{subcrit2s}
\end{center}
\end{figure}

Fig.~\ref{supercrit2s} shows the time development of $|\phi(t,\rho,z)|$ for 
a marginally {\em supercritical} head-on collision.  Here, the momentum 
parameter, $p_z \approx 0.21$ has been tuned to criticality to within 
a relative precision $\De p_z/p_z = O(10^{-15})$.  In this calculation, 
(and due to the non-zero initial boost) the stars merge at a significantly
earlier time---$t\approx 90$---than for $p_z=0$ (compare with Fig.~\ref{maxmodphbs2spz0}).
Following the merger, we estimate that the configuration enters the critical
state at $t \approx 140$, and remains in that state (with some oscillations)
until a black hole forms.  An apparent horizon is detected at $t=280$, and 
we note that the final black hole will contain most of the mass-energy 
of the configuration at that time.  That is, the smallest-mass black 
hole that can be formed via variation of $p_z$ has {\em finite} mass, as 
is characteristic of a Type I transition.

Similarly, 
Fig.~\ref{subcrit2s} shows the time development of $|\phi(t,\rho,z)|$ for
a marginally {\em subcritical} head-on collision. Again, the momentum
parameter, $p_z \approx 0.21$ has been tuned to criticality to within
a relative precision $\De p_z/p_z = O(10^{-15})$. 
At early times the evolution here is indistinguishable from that of the supercritical data. 
However, in contrast to the supercritical case, at
$t\approx 280$ two star-like configurations begin to reappear and are clearly 
distinguishable as separate stars at $t \approx 390$.  Since the initial momenta of the stars
is moderate, the overall system is gravitationally bound and
the two stars re-merge and form a black hole.
An apparent horizon is detected at $t=570$.

\begin{figure}
\begin{center}
\epsfxsize=10.0cm
\includegraphics[width=10.0cm,clip=true]{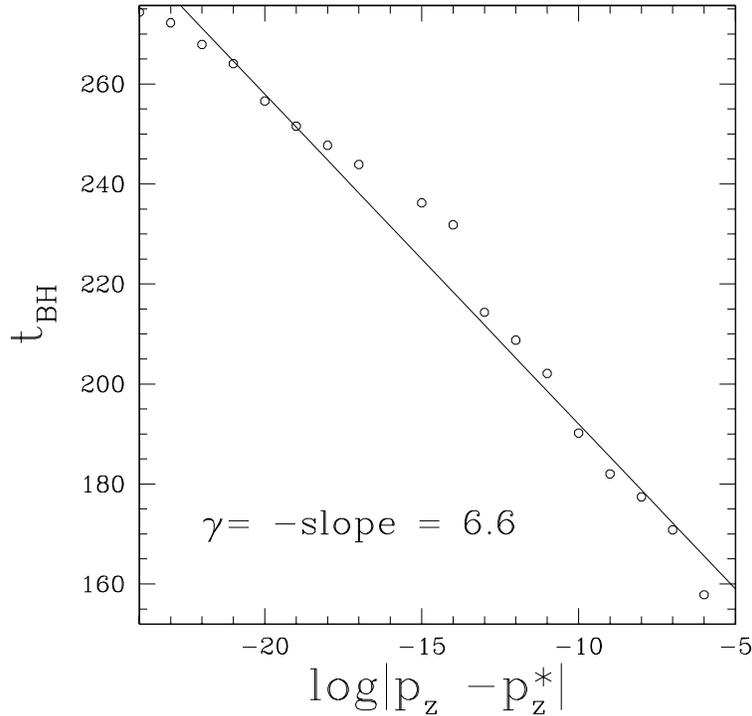}
\caption
[Scaling law for near critical evolutions of head-on boson star collisions]
{Scaling law for near critical evolutions of head-on boson star collisions
The time of black hole formation (time of first appearance of 
an apparent horizon), $t_{\mathrm{BH}}$, is plotted {\em vs}
$\log|p_z- p_z^{\ast}|$.  The linearity of the data provides evidence 
for lifetime scaling associated with a Type I transition, as 
previously seen in the spherically symmetric calculations 
described in Chap.~\ref{bs1d}.}
\label{scalinglaw2s}
\end{center}
\end{figure}

Finally, 
Fig.~\ref{scalinglaw2s} shows the time of black hole formation 
(time of first appearance of an apparent horizon)
$t_{\mathrm{BH}}$ plotted as a function of $\log|p_z-p_z^{\ast}|$.
The plot provides strong evidence for scaling of 
the lifetime, $\ta$, of the critical
configuration of the form
\beq
 \ta = -\ga \log\left|p_z -p_z^{\ast} \right|\,,
\eeq
\noi
with a value $\ga = 6.6$ computed from a least squares fit. Again,
observation of such scaling is consistent with a Type I transition 
in the model.   

There remains the question as how the critical 
solution observed in this study can itself be characterized.
Although we have not yet studied this matter in detail, we 
hypothesize that the solution can be described in terms of 
a static boson star on the unstable branch, superimposed with 
(largely) normal-mode oscillations that are not necessarily 
spherically symmetric.

\subsection{Perturbation of Boson Stars by an Aspherical Real Scalar Field} \lab{perturb2d}
In this section we present results from a second study of critical phenomena involving
boson stars in axisymmetry.  In this case the calculations generalize those 
described in Chap.~\ref{bs1d}, where a massless scalar field was used to 
induce collapse of the boson star.  In particular, the massless scalar field configuration
is {\em not} spherically symmetric in the simulations described below.

For this class of experiment, the initial data for the complex field, $\phi$, is 
simply a single, stable boson star centered at the origin, $(\rho,z)=(0,0)$.  As 
previously, we use a star with $\ph_0(0) = 0.02$.
The initial data for the real scalar field takes the form of a ``generalized 
gaussian''
\beq
\label{ph3-elliptical}
\ph_3(0,\rh,z) = A_3 \exp \left[
-\left(\fr{\sr{(\rh-\rh_0)^2 + \ep ( z-z_0)^2}-R_0}{\De}
\right)^2\right]\,,
\eeq
where $A_3$, $\rh_0$, $z_0$, $\De$, $R_0$ and $\ep$ are all adjustable parameters.
In the calculations described below we set $\rh_0=0$, $z_0=0$, $\Delta=4$, 
$\ep=0.4$ and $R_0=25$.  We note that for $\rh_0=z_0=0$, $\ep$ controls the 
``ellipticity'' of the initial pulse, while $R_0$ controls how far the pulse 
is from the origin at $t=0$.  The overall amplitude factor, $A_3$ is the 
parameter that is tuned to generate a critical solution, and we find 
$A^\star_3 \approx 0.013$.   We also note that we specify initial data for 
the conjugate variable $\Pi_3$ so that the massless field is almost purely 
incoming at the initial time.

The computational domain is $-75\le z\le 75$, $0\le \rh \le 75$,
and the base resolution is $(N_\rho,N_z)=(193,385)$, yielding base mesh
spacings $\De \rh = \De z = 0.390625$.  We again allow up to 6 additional
levels of 2:1 refinement, so that the finest available resolution
is  $\De \rh = \De z = 0.006104$.

\begin{figure}[htp]
\begin{center}
\ifthenelse{\equal{\highQ}{true}} {
\includegraphics[width=14.0cm,clip=true]{eps/graxi_csf/bs.ad.real/modph0.01289460949673069.eps}
\vspace{0.5cm}
\includegraphics[width=14.0cm,clip=true]{eps/graxi_csf/bs.ad.real/phiN0.01289460949673069.eps}
}{
\includegraphics[width=14.0cm,clip=true]{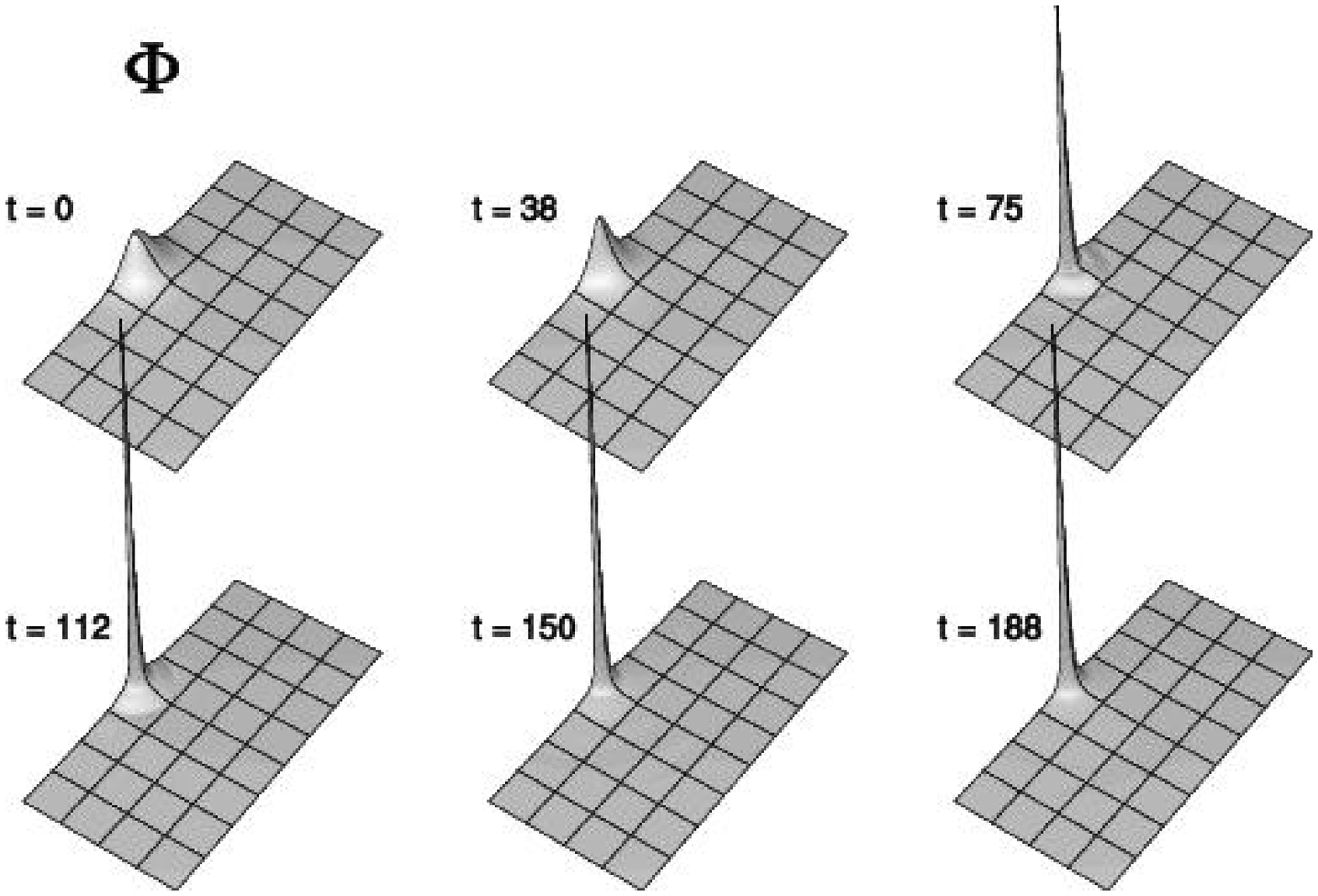}
\vspace{0.5cm}
\includegraphics[width=14.0cm,clip=true]{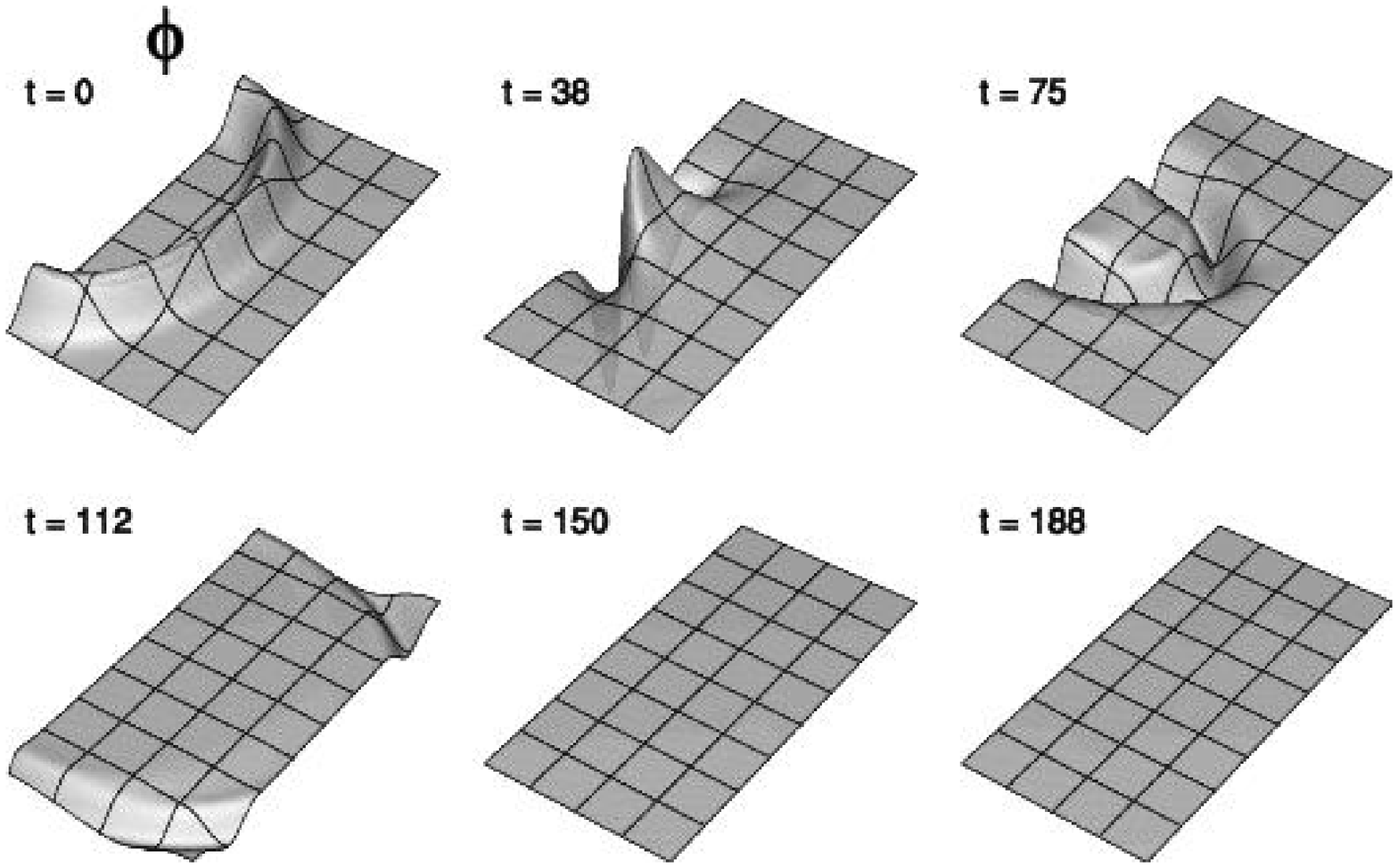}
}
\caption
[Perturbation of a boson star by an elliptical in-going real scalar field---supercritical evolution]
{Perturbation of a boson star (denoted $\Ph$) by an elliptical in-going real scalar field (denoted $\ph$).
This figure shows the time development of $|\phi(t,\rho,z)|$ and $\phi_3(t,\rho,z)$ for a 
marginally supercritical simulation, with tuning parameter $A_3 \approx
0.013$ (see~(\ref{ph3-elliptical})). 
While the simulation domain is $0\le \rh\le 75$, $-75\le z\le 75$,  for clarity only the region 
$0\le \rh\le 50$, $-50\le z\le 50$ is shown. 
After the massless scalar field implodes through the boson star, the boson star enters its critical state, 
oscillates about it for a while, and then collapses to form a black hole. 
An apparent horizon is first detected at $t\approx 220$.
$0.0\le |\ph(t,\rh,z)| \le 0.13$ (denoted as $\Ph$),
$-0.056\le \ph_3(t,\rh,z) \le 0.021$ (denoted as $\ph$).
}
\label{bsrealsuper}
\end{center}
\end{figure}

\begin{figure}
\begin{center}
\ifthenelse{\equal{\highQ}{true}} {
\includegraphics[width=14.0cm,clip=true]{eps/graxi_csf/bs.ad.real/modph0.01289460949673067.eps}
\vspace{0.5cm}
\includegraphics[width=14.0cm,clip=true]{eps/graxi_csf/bs.ad.real/phiN0.01289460949673067.eps}
}{
\includegraphics[width=14.0cm,clip=true]{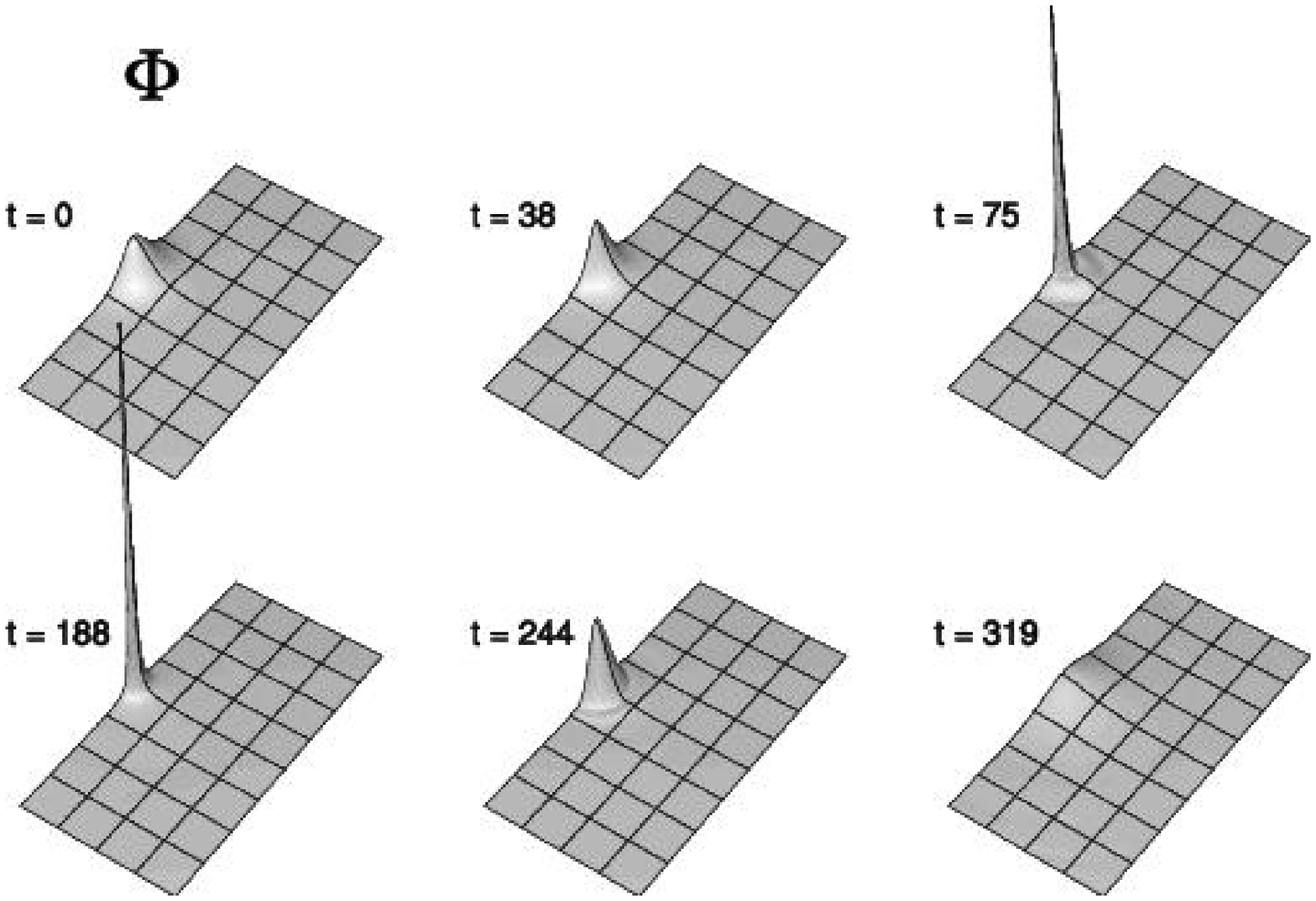}
\vspace{0.5cm}
\includegraphics[width=14.0cm,clip=true]{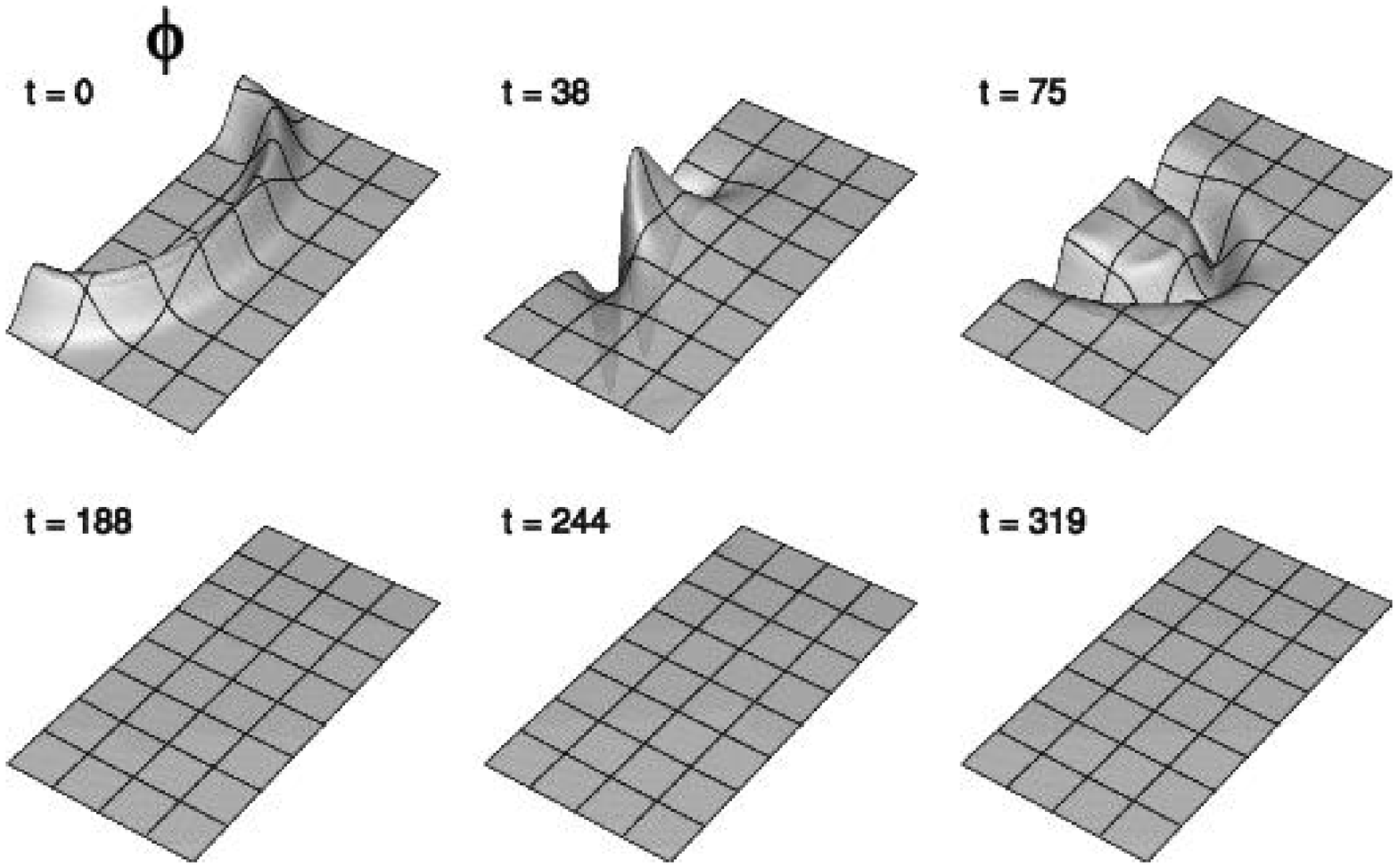}
}
\caption
[Perturbation of a boson star by an elliptical in-going real scalar field---subcritical evolution]
{This figure shows the time development of $|\phi(t,\rho,z)|$ and $\phi_3(t,\rho,z)$ for a
marginally subcritical simulation (i.e.\ for $A_3$ of the order of machine epsilon smaller than
the value used for the simulation shown in~Fig.~\ref{bsrealsuper}). In this case there are indications 
that the late time configuration will be a stable boson star with large amplitude oscillations, as 
seen in the spherical calculations of Chap.~\ref{bs1d}.
$0.0\le |\ph(t,\rh,z)| \le 0.13$ (denoted as $\Ph$),
$-0.056\le \ph_3(t,\rh,z) \le 0.021$ (denoted as $\ph$).
}
\label{bsrealsub}
\end{center}
\end{figure}

Fig.~\ref{bsrealsuper} shows the time development of $|\phi(t,\rho,z)|$ and 
$\phi_3(t,\rho,z)$ for
a marginally {\em supercritical} perturbed boson star. 
For clarity the subplots display only a portion of the computational
domain, namely $-50\le z\le 50$, $0\le \rh \le 50$.
The top 6 subplots in the figure depict the evolution of the boson star, which approximates 
the critical solution 
for a period of $\De t \approx 70$, before forming a black hole.  An apparent
horizon is first detected at $t\approx 220$.  The bottom 6 subplots show the corresponding
evolution of the real scalar field, which leaves the computational domain at $t\approx 110$,
at which time the boson star is still in a near-critical configuration.  
Fig.~\ref{bsrealsub} displays corresponding results from a marginally subcritical evolution,
in which there are indications that the end state of evolution is a stable boson star 
with large amplitude oscillations. 

Thus, we see that the type of non-spherical perturbation used here leads to the same type 
of behaviour seen in the spherically-symmetric calculations, at least at the 
qualitative level.  Among other things, this suggests that the unstable static boson stars 
do not develop {\em additional} unstable axisymmetric modes when the restriction to 
spherical symmetry is relaxed (or, if such modes {\em do} exist, that their growth rates 
are quite small in comparison to that of the unstable spherical mode).

\begin{figure}
\begin{center}
\epsfxsize=10.0cm
\includegraphics[width=10.0cm,clip=true]{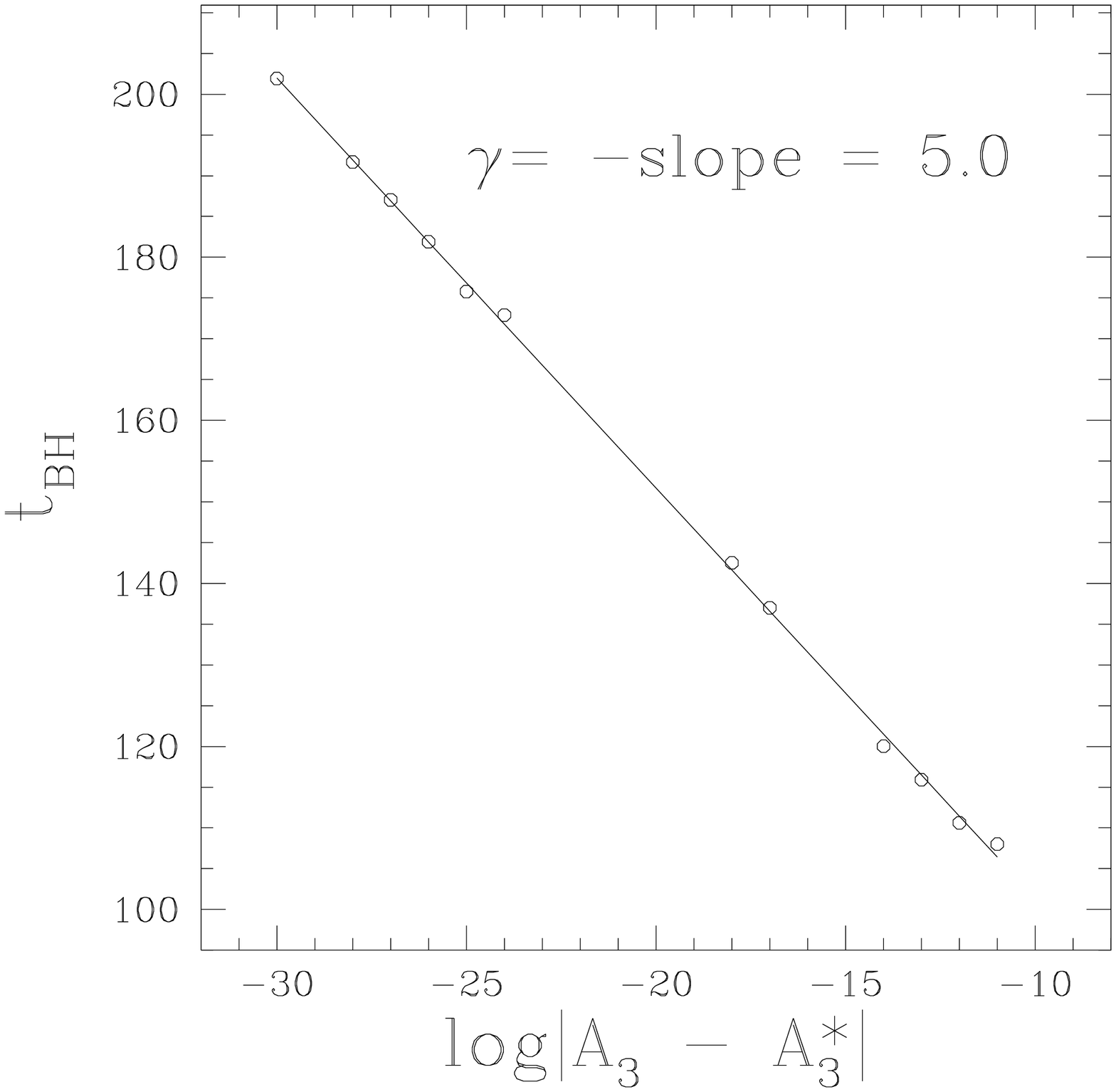}
\caption
[Lifetime scaling law for near critical evolutions for boson stars perturbed by
non-spherical real scalar field]
{Lifetime scaling law for near critical evolutions for boson stars perturbed by
non-spherical real scalar field.  The
time of black hole formation $t_{\mathrm{BH}}$ is plotted as a function
of
$\log|p - p^{\ast}|$.   The expected (Type I) scaling of the lifetime 
of the critical configuration is again observed.
}
\label{scalinglawreal}
\end{center}
\end{figure}

We can again plot the time of
black hole formation against $\log|p -p^{\ast}|$ to measure the expected lifetime 
scaling for near critical evolutions, as shown in Fig.~\ref{scalinglawreal}.
Once more we verify that the expected Type I scaling holds.

Before concluding this chapter, we want to make two additional observations
concerning the axisymmetric evolutions that we have performed.
The first one concerns non-radial oscillations of perturbed
boson stars, while the second one concerns the linear momentum of
such stars.  We note that we have not studied either matter in much detail, 
deferring such work to the future.

\begin{figure}
\begin{center}
\epsfxsize=10.0cm
\ifthenelse{\equal{\highQ}{true}} {
\includegraphics[width=10.0cm,clip=true]{eps/graxi_csf/bs.ad.real2/phiNamp0.01795688677531872.eps}
}{
\includegraphics[width=10.0cm,clip=true]{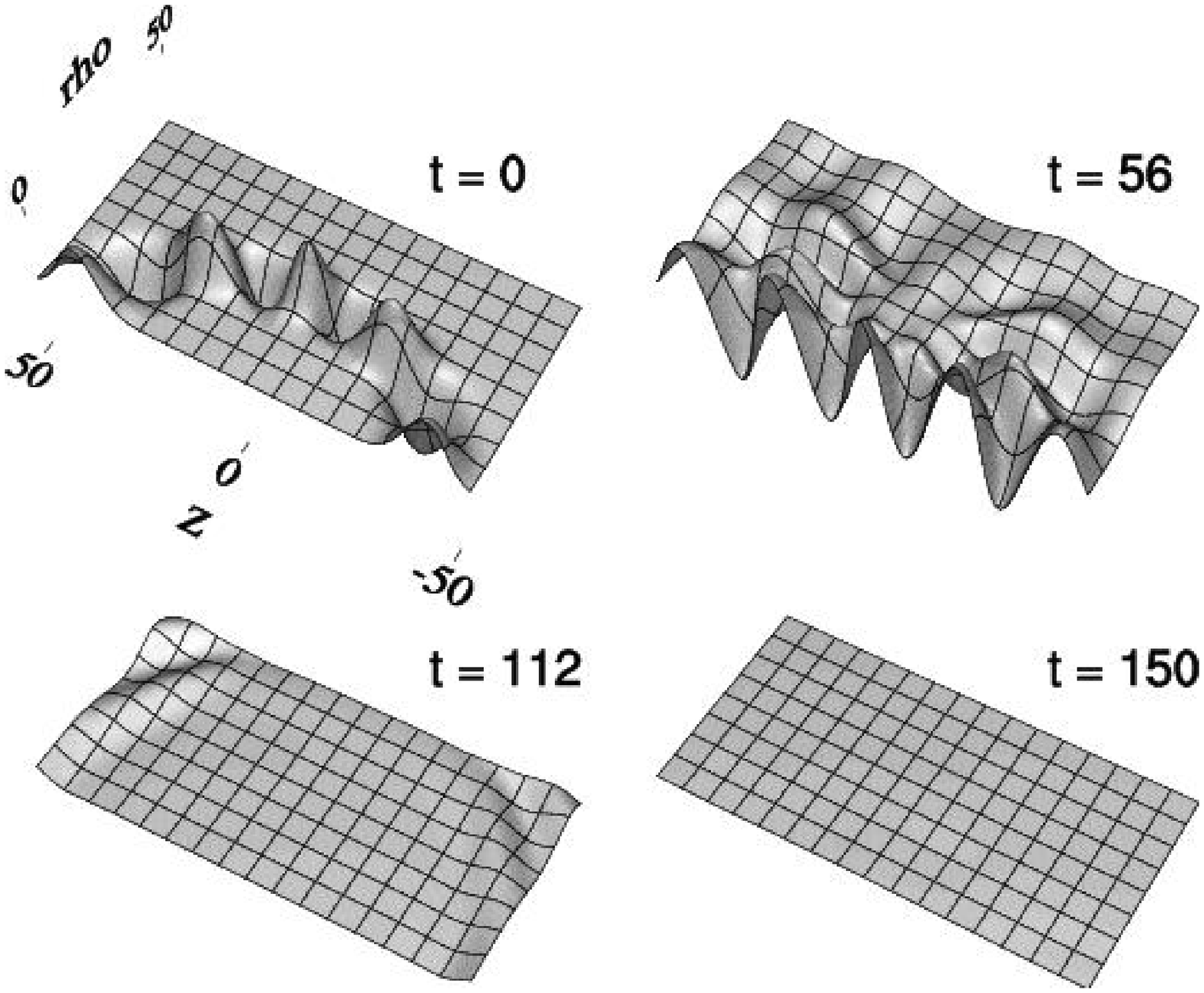}
}
\caption
[Dynamical evolution of non-radial distribution of perturbing real scalar field]
{Dynamical evolution of non-radial distribution of perturbing real scalar field.  The
initial configuration of the real scalar field is given by (\ref{nonradialID}).
$-0.023\le \ph_3(t,\rh,z) \le 0.022$.
}
\label{bsreal2phiN}

\vspace{0.5cm}
\ifthenelse{\equal{\highQ}{true}} {
\includegraphics[width=15.0cm,clip=true]{eps/graxi_csf/bs.ad.real2/modphamp0.01795688677531872.eps}
}{
\includegraphics[width=15.0cm,clip=true]{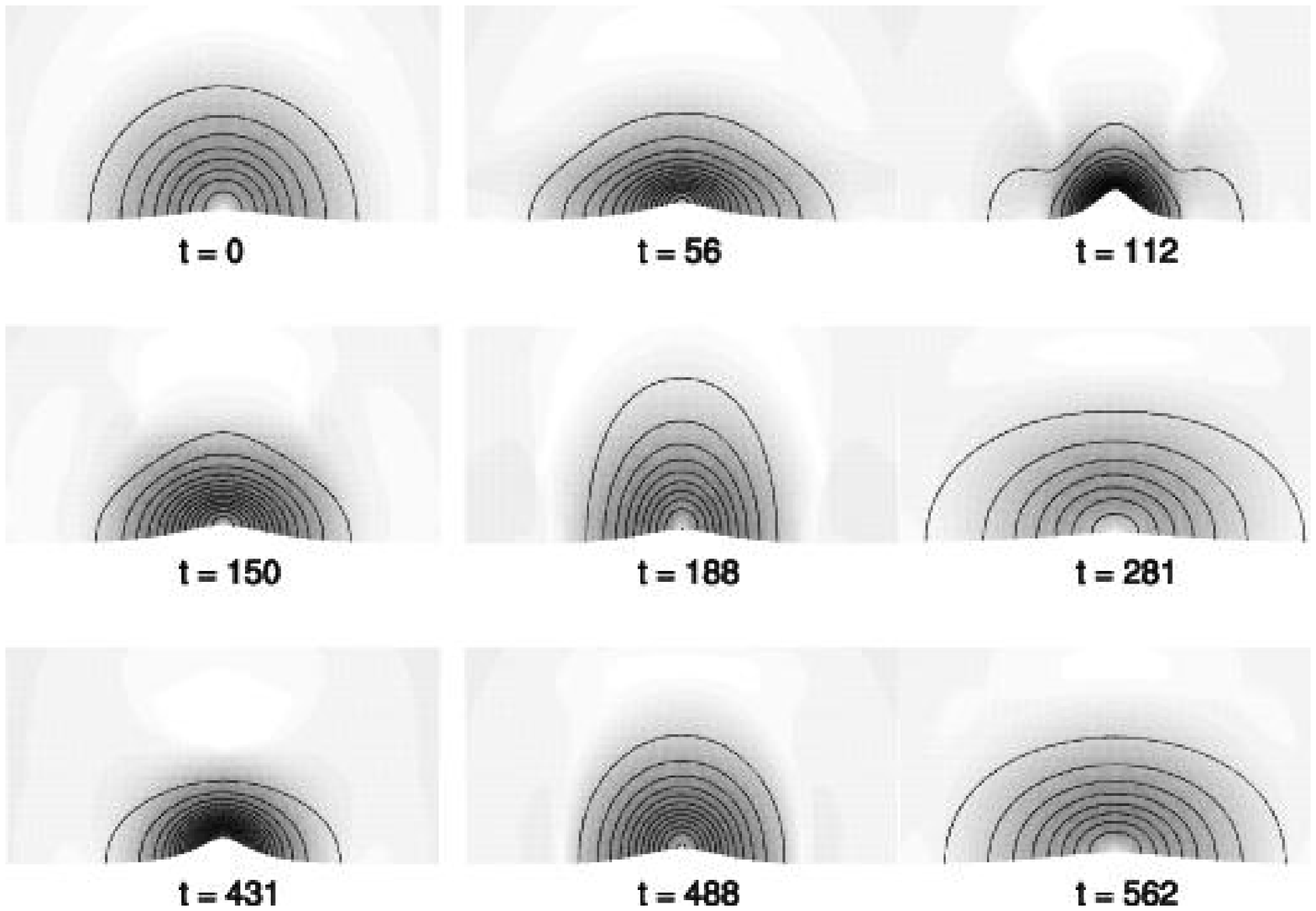}
}
\caption
[Dynamical evolution of boson star as a result of non-radial perturbation by a real
scalar field]
{Dynamical evolution of boson star as a result of non-radial perturbation by a real
scalar field.  The snapshots show both radial and non-radial oscillations of the star.
$0.0\le |\ph(t,\rh,z)| \le 0.049$.
}
\label{bsreal2modph}
\end{center}
\end{figure}

When boson stars are perturbed in a non-spherical manner, we 
observe that, as expected,  the stars exhibit oscillations which are not strictly radial.
Not surprisingly, non-radial oscillations are more pronounced when we
use a very aspherical distribution of real scalar field as the perturbing
agent.
In Fig.~\ref{bsreal2phiN} we show the evolution of such a highly non-radial
configuration of real scalar field.  We can see from the figure that the
initial data consists of several distinct pulses located at different positions
in the $\rho$-$z$ plane.
Specifically, the initial configuration of the massless field is given by
\beq \lab{nonradialID}
\ph_3(0,\rh,z) = A_3 \exp \left[
-\left(\fr{\sr{(\rh-\rh_0+\rh_1-\rh_2-\rh_3)^2 + \ep ( z-z_0)^2}-R_0}{\De}
\right)^2\right] \cos n \te\,,
\eeq
\noi
where
\bea 
 \rh_1 &=& \tilde{A} \exp \left[ - \left( \fr{z}{\tilde{\De}}\right)^2\right] \,, \\
 \rh_2 &=& \tilde{A} \exp \left[ - \left( \fr{z-z_c}{\tilde{\De}}\right)^2\right] \,, \\
 \rh_3 &=& \tilde{A} \exp \left[ - \left( \fr{z+z_c}{\tilde{\De}}\right)^2\right] \,,
\eea
\noi
and $\tan \te = \rh/z$.  
We have chosen $\tilde{A} = 5$, $\tilde{\De}=8$, $z_c=35$,
$n=8$, and have set all other parameters to the values used in the previously
described set of computations.  
For clarity,  we again display only a portion of the solution domain, namely
$-50\le z \le 50$ and $0\le \rh \le 50$.  Fig.~\ref{bsreal2modph}
displays the corresponding evolution of the boson star, where now the part of the
solution domain given by $-20\le z \le 20$ and $0\le \rh \le 20$ is shown.  
The figure plots various contour lines of $|\phi(t,\rho,z)|$ as a function
of time, and non-radial oscillations of the star are clearly visible.

\begin{figure}
\begin{center}
\epsfxsize=15.0cm
\ifthenelse{\equal{\highQ}{true}} {
\includegraphics[width=15.0cm,clip=true]{eps/graxi_csf/bs.ad.real3/modphamp0.025390625.eps}
}{
\includegraphics[width=15.0cm,clip=true]{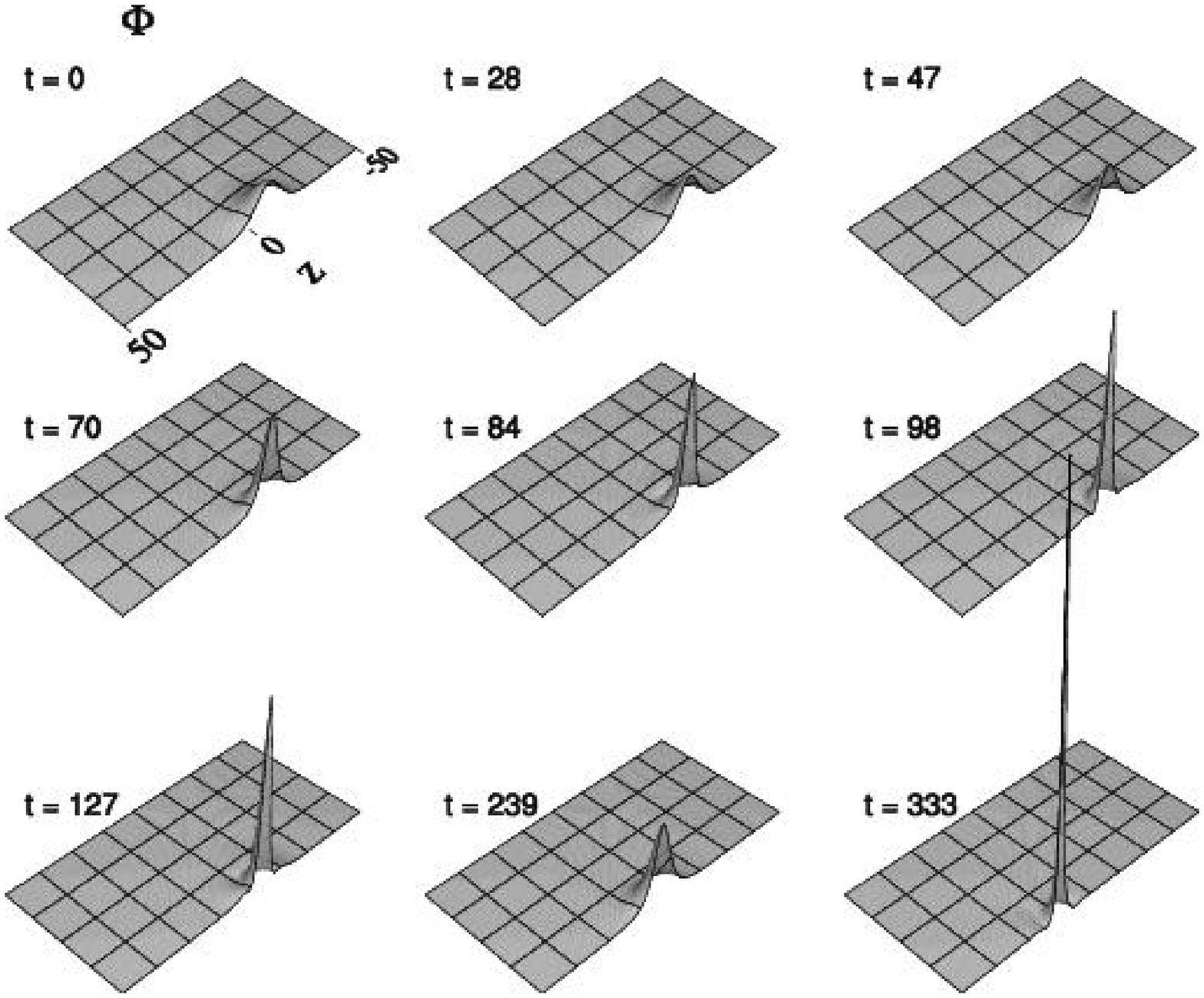}
}
\caption
[Evolution of a boson star as a result of perturbation by a non-radial perturbing real
scalar field]
{Evolution of a boson star as a result of perturbation by a non-radial perturbing real
scalar field, that is {\em not} symmetric about the equatorial plane.  The boson star appears to remain at
the same location $z\approx -8$ until $t\approx 80$, when most of the real scalar field
has passed through the star. The star then appears to  move in the positive $z$ direction 
at a speed of 0.09.
$0.0\le |\ph(t,\rh,z)| \le 0.14$ (denoted as $\Ph$).
}
\label{bsreal3modph}
\end{center}
\end{figure}

\begin{figure}
\begin{center}
\epsfxsize=15.0cm
\ifthenelse{\equal{\highQ}{true}}
{ \includegraphics[width=15.0cm,clip=true]{eps/graxi_csf/bs.ad.real3/phiNamp0.025390625.eps} }
{ \includegraphics[width=15.0cm,clip=true]{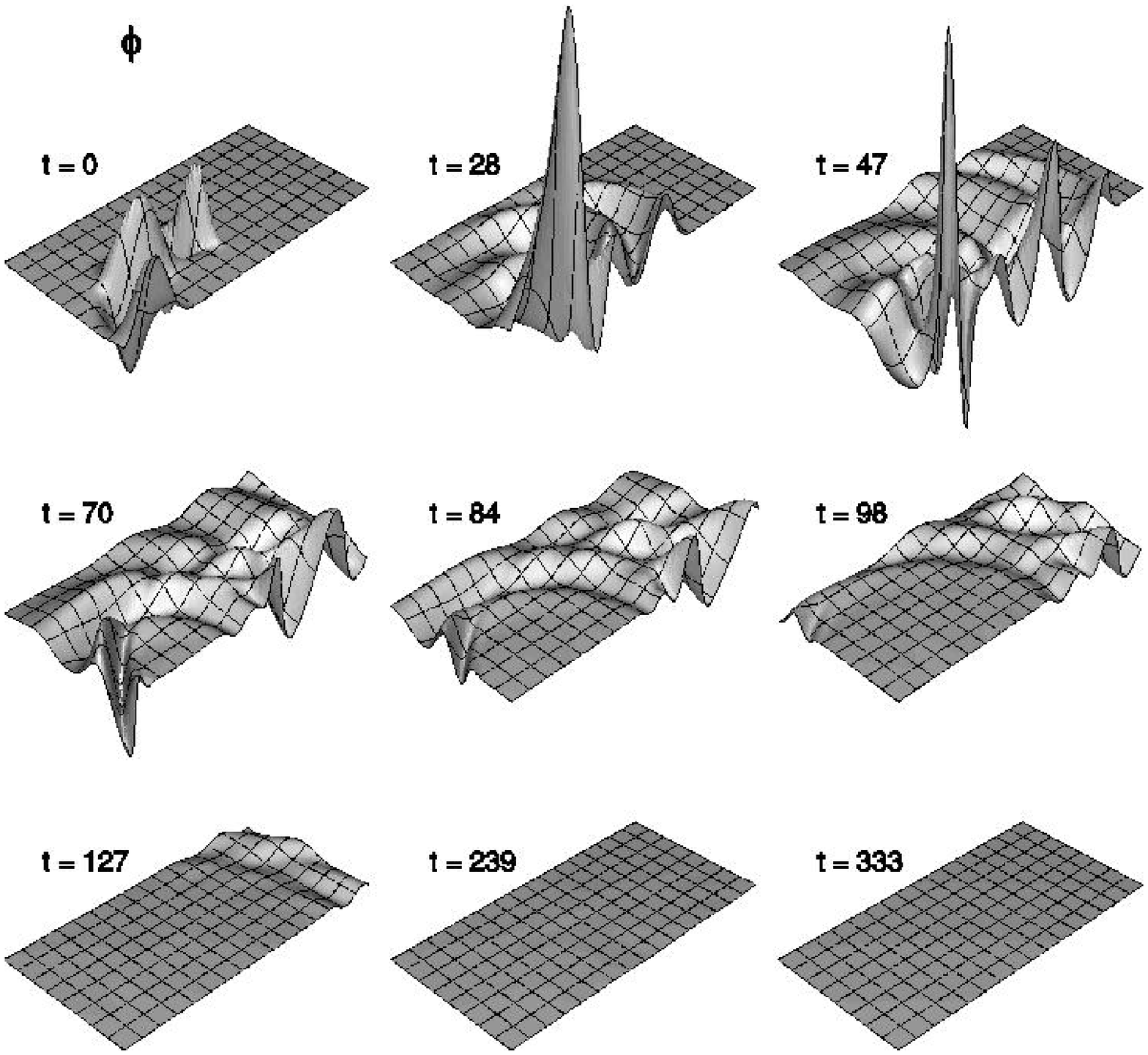} }
\caption
[Evolution of non-radial, non-equatorially-symmetric distribution of perturbing real scalar field]
{Evolution of non-radial, non-equatorially-symmetric distribution of perturbing real scalar field.
Here the real field,
whose support at the initial time is confined to $z>0$, 
passes through the boson star at $t\approx 80$, and leaves
the computational domain at $t\approx 130$.
$-0.047\le \ph_3(t,\rh,z) \le 0.090$ (denoted as $\ph$).
}
\label{bsreal3phiN}
\end{center}
\end{figure}

In Figs.~\ref{bsreal3modph} and~\ref{bsreal3phiN} we show another evolution of 
a boson star perturbed by an in-going non-radial distribution of real scalar field.
The boson star is initially located at $(0,-7.8125)$.  Again the evolution is solved on
a domain $-75\le z \le 75$,  $0 \le \rh \le 75$, while the figures show a portion
of the domain, $-50\le z \le 50$, $0 \le \rh \le 50$.  The initial configuration for 
the real scalar field is the same as for the previous calculation, except that we choose 
$n=9$ and restrict the support of the real scalar field to $z \ge 0$.   Thus the 
real scalar field is {\em not} symmetric about the equatorial plane in this case.
Fig.~\ref{bsreal3modph} shows that at early times, although the modulus of the complex 
field oscillates, the boson star remains at a fixed coordinate location.
At $t\approx 80$ most of the real scalar field has passed through the boson
star, and the boson star apparently starts to move in the positive $z$ direction,
with the amplitude of the complex field still oscillatory.  Fig.~\ref{bsreal3phiN} shows the 
corresponding evolution of the real scalar field.

\begin{figure}
\begin{center}
\epsfxsize=8.0cm
\includegraphics[width=8.0cm,clip=true]{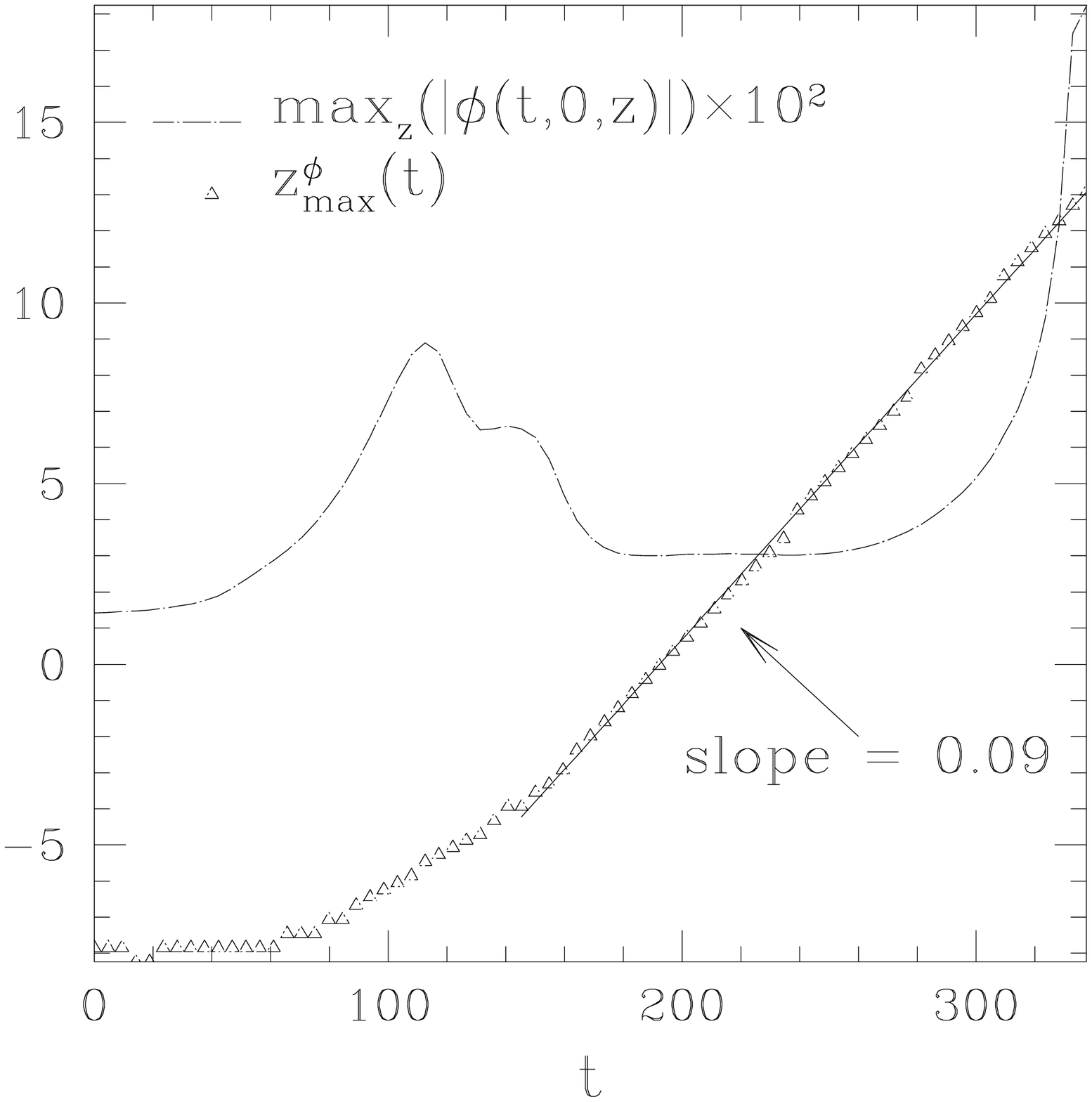}
\caption
[The maximum value of the boson star modulus on-axis $\max_z\left(|\ph(t,0,z)|\right)$
($\times 10^2$) and the $z$ location of that maximum as a function of time]
{The maximum value of the boson star modulus on-axis $\max_z\left(|\ph(t,0,z)|\right)$
($\times 10^2$) and the $z$ location of that maximum as a function of time.  The boson star starts
to move at $t\approx 80$, and attains a constant final speed of $0.09$. }
\label{bsreal3drift}
\end{center}
\end{figure}

Fig.~\ref{bsreal3drift} shows the maximum of the modulus of the complex scalar field,
$\max_z|\ph(t,0,z)|$, on-axis as a function of time $t$, 
as well as the $z$-coordinate of
the maximum $z^{\ph}_{\max}(t)$ defined by
\beq
 \ph(t,0,z^{\ph}_{\rm{max}}(t)) = \max_z|\ph(t,0,z)|\,. 
\eeq

\begin{figure}
\begin{center}
\epsfxsize=8.0cm
\includegraphics[width=8.0cm,clip=true]{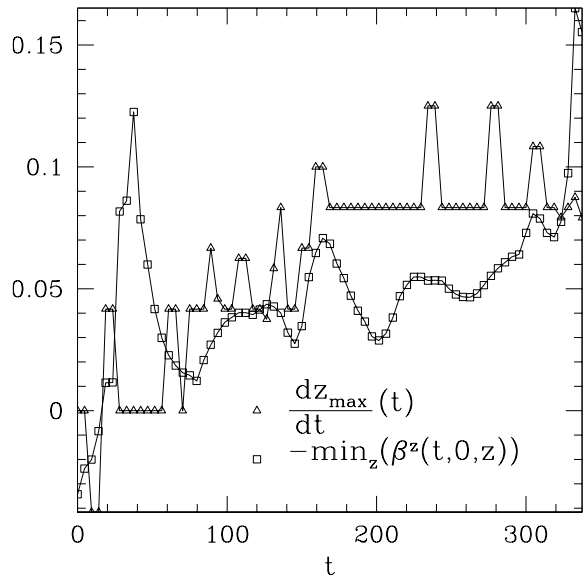}
\caption
[Velocity of boson star and the negative of the minimum of the shift
component $\bt_z(t,0,z)$
as  a function of time $t$]
{Velocity of boson star and the negative of the minimum of the shift
component $\bt_z(t,0,z)$
as  a function of time $t$.  The two functions are of the same order of magnitude,
indicating that the movement of the boson star depicted in Fig.~\ref{bsreal3drift} is 
primarily due to the particular coordinates we have chosen.}
\label{bsreal3drift2}
\end{center}
\end{figure}

From the graph we can see that the maximum remains at $z\approx -8$ until
$t\approx 80$, at which point it acquires a non-zero velocity which apparently
settles down to a speed of about 0.09.
However, this apparent motion of the boson star seems to be primarily a coordinate effect.  
Fig.~\ref{bsreal3drift2}
shows the value of $-\min_z \bt^z(t,0,z)$ as a function of time $t$,
as well as the apparent velocity of the boson star (which exhibits several jumps 
since the value $\max|\ph(t,0,z)|$ is not smooth).  The two
functions appear to be of the same order of magnitude, consistent with the conjecture
that the motion is mainly due to the particular coordinates that we have chosen.  

  \resetcounters

\chapter{Conclusions and Future Work}

In this thesis we have performed numerical simulations of general relativistic 
boson stars in both spherically and axially symmetric spacetime.
The principal new results are as follows.  

In the spherically symmetric case, we studied Type I critical phenomena associated 
with boson stars.  In particular, contrary to some previous claims, we found that 
the end state of subcritical evolution is a stable boson star executing 
large amplitude oscillations, that can largely be understood as excitations of 
the fundamental normal mode of the end-state star.
For the particular example that we examined in detail, the  oscillation frequency
of the ``post-critical'' state was estimated to be $\si^2/\al^2 \approx 0.0013$, in good agreement
with the frequency of the fundamental mode, $\si^2_0 = 0.0014$,
An overall modulation of the envelope of the normal mode oscillations was observed, and 
the origin of this phenomenon remains unclear.

In the axisymmetric case we developed an efficient algorithm for constructing the
equilibrium configurations of rotating boson stars.  The algorithm was based on an
eigenvalue multigrid method combined with homotopic/continuation techniques.  Complete
families of solutions for angular momentum parameter values $k=1$ and $k=2$ were found.  
The maximum mass computed for the two families were 
 $1.7 M_{{\rm pl}}^2/m$ and $2.4 M_{{\rm pl}}^2/m$, respectively.  The value for 
$k=1$ is only in moderate agreement (at the 25\% level) with previous work by Yoshida \& Eriguchi,
who quoted $1.314 M_{{\rm pl}}^2/m$. We are unsure of the origin of the discrepancy at this 
time, although it may be at least partly attributable to the problems with regularity that
we encountered with our current code.   The value for $k=2$ is a new result. 

We also demonstrated the existence of Type I critical phenomena in axisymmetry using two classes 
of computations involving non-rotating boson stars. 
The first involved the head-on collision of boson stars, while the second used 
perturbations of the boson star from a non-spherical distribution of massless 
scalar field. In both cases we measured lifetime scaling from near-critical evolutions
of the type expected for Type I collapse.

We also displayed some
typical non-radial oscillations of boson stars induced by the gravitational
interaction with non-spherical distribution of massless scalar waves.

As for the future, there are several directions in which we can extend our work.  For 
the case of rotating boson stars, more work first needs to be done to properly implement  
the regularity conditions. Combined with a modification of the system of equations as suggested   
in Sec.~\ref{spacom} that will map the maxima of the unknown functions to the interior of the
computational domain, this should allow us to find families of solutions with higher values 
of the angular momentum parameter.  More importantly, all of our numerical evolutions of boson stars are 
currently restricted to the non-rotating case.  It will be very interesting to see how angular momentum affects 
the behaviour of the system.  On the other hand, in this thesis we have focused on the dynamical evolution 
of axisymmetric boson stars within the context of critical phenomena.  More work can be
done to understand the non-radial oscillatory modes of boson stars, and to compare with 
the corresponding modes for neutron stars or white dwarfs.

\bibliography{phd}
\begin{appendix}
  \resetcounters

\chapter[Finite Difference Operators]
{}\label{ap5}
\vspace{-0.45in}
{\bf {\huge 
Finite Difference Operators
}}
\vspace{0.52in}

\noi
Here we list the definitions of all finite difference operators in 1D that are used in the thesis 
(the finite difference operators in 2D are defined in a similar way):

\vspace{1.0cm}
\centerline{\fbox{\parbox{4.5in}{
\begin{eqnarray*}
\Dtp u\nj &\eq& 
\fr{ u\npoj - u\nj }{\De t}  \\
\Drz u\nj &\eq& 
\fr{u\njpo - u\njmo}{2 \De r} \\
\Drzb  u\nj &\eq& \fr{3 u\nj - 4 u\njmo + u\njmt}{2 \De r}  \\
\De^{^{\scr r}}_{_{\scr \pm}} u\nj &\eq& 
\fr{ \pm u^{^{\scr n}}_{_{\scr j \pm 1}} \mp
 u\nj }{ \De r} \\
\Dhrz u\nj &\eq& \fr{ u\njph - u\njmh }{\De r}  \\
\Drsz u\nj &\eq& \fr{  u\njpo - u\njmo }{
r_{_{\scr j+1}}\,^2 - r_{_{\scr j-1}}\,^2 }  \\
\Drcz u\nj &\eq& \fr{  u\njpo - u\njmo }{
r_{_{\scr j+1}}\,^3 - r_{_{\scr j-1}}\,^3}  \\
\Dhrcz u\nj &\eq& 
\fr{ u\njph - u\njmh}{
r_{_{\scr j+\ha}}\,^3 - r_{_{\scr j-\ha}}\,^3} \\
\mu^{^{\scr t}}_{_{\scr \pm}} u\nj &\eq&  \ha \left( u^{^{\scr
 n \pm 1}}_{_{\scr j}} + u\nj
 \right) \\
\mu^{^{\scr r}}_{_{\scr \pm}} u\nj &\eq& \ha \left( u^{^{\scr
 n}}_{_{\scr j \pm 1}} + u\nj
 \right)
\end{eqnarray*}
}}}
\begin{table}[h]
\caption[Finite difference operators used in the thesis]
{Finite difference operators used in the thesis.}
\end{table}
\noi
We also define $\bar{\mu}^{^{\scr r}}_{_{\scr \pm}}$ which has the same
definition as $\mu^{^{\scr r}}_{_{\scr \pm}}$ but which has a higher precedence
than other algebraic operations, e.g.,
\[
\bmurp \left( \fr{ f g^2}{h}\right)^n_j = \fr{(\bmurp f\nj) (\bmurp
{g\nj})^2}{\bmurp h\nj}\,.
\]

  \resetcounters

\chapter[EOM for the Spherically Symmetric EKG System]
{}
\label{ap4}
\vspace{-0.45in}
{\bf {\huge 
Equations of Motion and Finite Difference Formulae for the Spherically Symmetric 
Einstein-Klein-Gordon System in Maximal-Isotropic Coordinates
}}
\vspace{0.52in}

\noi
In this appendix we summarize the system of equations governing the spherically 
symmetric Einstein-Klein-Gordon model in maximal-isotropic coordinates, as 
well as the finite difference approximations (FDAs) that are used to 
discretize the system.

In maximal-isotropic coordinates the spacetime metric is 
\beq
  d s^2 = \left( -\al^2 + \ps^4 \bt^2 \right) dt^2 + 2 \ps^4 \bt\, dt\,
dr + \ps^4 \left( dr^2 + r^2 d \Om^2 \right) \,.
\eeq
\noi
We also define auxiliary fields, $\Phi_i$ and $\Pi_i$
\bea
  \Ph_i &\eq& \ph_i'\,, \\
  \Pi_i &\eq& \fr{\ps^2}{\al} \left( \dot{\ph_i } - \bt \ph_i'\right)\,,
\eea
\noi
where $i=1,2$ correspond to the real and imaginary parts of the massive complex
scalar field (boson stars), while $i=3$ corresponds to the massless real scalar
field that is used to gravitationally perturb boson stars in our study of 
critical collapse of those stars.
The 
Hamiltonian constraint, momentum constraints,
Klein-Gordon equations, and the maximal-isotropic coordinate conditions give
\bea 
\fr{3}{\ps^5} \fr{d}{dr^3} \left( r^2 \fr{d \ps}{dr} \right) + \fr{3}{16}
\krr\,^2 
&=& - \pi\left( \fr{ \sum_{i=1}^3 \left( \Ph^2_i + \Pi^2_i\right)  }{\ps^4} +
m^2 \sum_{i=1}^2 {\ph_i}^2 \right) \,,\\
 \krr\left.'\right. + 3 \fr{( r \ps^2)'}{r \ps^2} \krr &=& -\fr{8
\pi}{\ps^2} \left( \sum_{i=1}^3 \Pi_i  \Ph_i \right) \,,
\eea
\bea 
 \dot{\ph}_i & = & \fr{ \al}{\ps^2} \Pi_i + \bt \Ph_i \,, \\
 \dot{\Ph}_i & = & \left( \bt \Ph_i + \fr{\al}{\ps^2} \Pi_i\right)' \,, \\ \nn
 \dot{\Pi}_i & = & \fr{3}{\ps^4}  \fr{d}{d r^3}\left[ r^2 \ps^4 \left( \bt
\Pi_i + \fr{ \al }{ \ps^2} \Ph_i\right)\right]- \al \ps^2 m^2 \ph_i
\left( 1 - \de_{i3} \right) \\ \lab{klein_gordon_eq3}
&& -  \left( \al K^{r}{}_{r} + 2 \bt \fr{(r \ps^2)'}{r \ps^2} \right) \Pi_i
\,,
 \eea
\beq 
  \al'' + \fr{2}{r \ps^2} \fr{ d}{dr^2}\left( r^2 \ps^2 \right) \al' + \left( 4 \pi
m^2 \ps^4 \sum_{i=1}^2 \ph_i^2 - 8 \pi \sum_{i=1}^3 \Pi_i^2 - \fr{3}{2} (\ps^2
K^r{}_r)^2 \right) \al  =  0\,,
\eeq
\beq 
 r \left( \fr{\bt}{r}\right)' = \fr{3}{2} \al K^r{}_r\,.
\eeq
\noi
The regularity conditions are
\bea
 \ps'(t,0) &=& 0\,, \\
 \krr(t,0) &=& 0 \,, \\
 \al'(t,0) &=& 0\,, \\
 \ph_i'(t,0) &=& 0\,, \\
 \Pi_i'(t,0) &=& 0\, .
\eea
\noi
The outer boundary conditions are
\bea \lab{psbc1}
\lim_{r\to\infty}\ps(t,r) &=& 1 + \fr{C(t)}{r} + O(r^{-2}) \,, \\
\lim_{r\to\infty}\al(t,r) &=& \lim_{r\to\infty}\fr{2}{\ps(t,r)}-1 = 
1 - \fr{2C(t)}{r} + O(r^{-2}) \,,\\
\lim_{r\to\infty}\bt(t,r) &=& \fr{D(t)}{r} + O(r^{-2}) \,,
\eea
\bea
 \dot{\Ph_i}+ \Ph_i'+\fr{\Ph_i}{r} &=& 0\,, \\ \lab{last_BC}
 \dot{\Pi_i}+ \Pi_i'+\fr{\Pi_i}{r} &=& 0\,,
\eea
\noi
for some functions $C(t)$ and $D(t)$.  The mass aspect function is defined by
\beq 
 M(t,r) =  \left( \fr{ \ps^2 r}{2} \right)^3 {K^r{}_r}^2 - 2 \ps' r^2 \left( \ps
+ r \ps'\right) \,,
\eeq
\noi
and the location of the apparent horizon is given by that value of $r$ (if any) such 
that 
\beq \lab{AH_eq}
 4 r \ps' + 2 \ps + \krr \ps^3 r = 0\;.
\eeq
\noi
In addition, 
the following evolution equations are used for setting inner boundary conditions
when black hole excision is used:
\bea \lab{ps_evolution_eq}
 \dot{\ps} &=& - \ha \al \ps \krr + \fr{(\ps^2 \bt)'}{2 \ps}\,, \\ \lab{krr_evolution_eq}
\dot{\krr} &=& \bt \krr' - \fr{2 \al}{(r \ps^2)^2} + \fr{2}{r^2 \ps^6} \left[ \al r
\left( r \ps^2\right)'\right]' + 8 \pi m^2 \al |\ph|^2\,.
\eea

The finite difference approximations of the 
Hamiltonian constraint, momentum constraints,
Klein-Gordon equations, and the maximal-isotropic coordinate choice are given by
\beq
\fr{ 3}{(\ps\nj)^5} \, \Dhrcz 
\left( r_{_{\scr j}}^2 \, \Dhrz \ps\nj \right) + \fr{ 3}{16}
(\krr)\nj \,^2
= - \pi\left( \fr{ \sum_{i=1}^3 \left( \Ph^2_i + \Pi^2_i\right)  }{\ps^4} +
m^2 \sum_{i=1}^2 \ph_i^2 \right)^n_j \,,
\eeq
\beq
\bmurm(\ps\nj)^2 \Drm (\krr)\nj + 3\, \Drm (r \ps^2)\nj 
\, \bmurm \left( \fr{\krr}{r}\right)^n_j
= \bmurm \left[ -8
\pi \sum_{i=1}^3 \Pi_i  \Ph_i  \right]^n_j\,,
\eeq
\bea
\Dtp (\ph_i)\nj &=& \mutp
\left( \fr{\al}{\ps^2} \Pi_i + \bt \Ph_i \right)^n_j \,, \\
\lab{fd_klein_gordon_eq_2}
 \Dtp (\Ph_i)\nj &=& \mutp \Drz
\left( \bt \Ph_i + \fr{\al}{\ps^2} \Pi_i\right)^n_j \,, \\ \nn
\Dtp (\Pi_i)\nj & = & \mutp \left\{ \fr{3}{(\ps^4)\nj}  \Drcz \left[ r^2 \ps^4 \left( \bt 
\Pi_i + \fr{ \al }{ \ps^2} \Ph_i\right)\right]^n_j - 
\left[ \al \ps^2 m^2 \ph_i
\left( 1 - \de_{i3} \right)\right]^n_j \right. \\
&& \left.  -  \left[ (\al \krr)\nj + 2 \bt\nj \, \fr{\Drz (r
\ps^2)\nj }{(r \ps^2)\nj } \right] (\Pi_i)\nj \right\}
\,,
\lab{fd_klein_gordon_eq_3}
\eea
\beq 
\Drp \Drm \al\nj + 
\fr{2}{(r \ps^2)\nj } \Drsz \left( r^2 \ps^2
\right)^n_j \Drz \al\nj + \left[ 4 \pi
m^2 \ps^4 \sum_{i=1}^2 \ph_i^2 - 8 \pi \sum_{i=1}^3 \Pi_i^2 - \fr{3}{2}
(\ps^2 \krr)^2 \right]^n_j \al^n_j  =  0 \,,
\eeq
\beq
 r_{_{\scr j-\ha}}\, \Drm \left(
\fr{\bt}{r}\right)^n_j = \murm \left[\fr{3}{2} \al \krr \right]^n_j \,.
\eeq

\noi
where  $r_{_{\scr j-\ha}} \eq (r_j + r_{j-1})/2$.  

The regularity conditions are
implemented as
\bea
  \ps\no &=& \fr{4 \ps^{^{\scr n}}_{_{\scr 2}} - \ps^{^{\scr n}}_{_{\scr
3}}}{3} \,, \\
 {\krr}\no &=& 0\,,\\
  \al\no &=& \fr{4 \al^{^{\scr n}}_{_{\scr 2}} - \al^{^{\scr n}}_{_{\scr
3}}}{3} \,,
\eea
\bea
 \mutp \left( (\ph_i)\no - \fr{4 (\ph_i)^{^{\scr n}}_{_{\scr 2}} - (\ph_i)^{^{\scr n}}_{_{\scr
3}}}{3}\right) &=& 0 \,, \\
 (\Ph_i)\no &=& 0\,,\\
 \mutp \left((\Pi_i)\no - \fr{4 (\Pi_i)^{^{\scr n}}_{_{\scr 2}} - (\Pi_i)^{^{\scr n}}_{_{\scr
3}}}{3}\right) &=& 0\,,
\eea
\noi
for all $i$ and $n$.  The outer boundary conditions are
\bea
\Dtp {\Ph_i}\nj  + \mutp \left( \Drzb {\Ph_i}\nj +
\fr{{\Ph_i}\nj}{r_j}\right) &=& 0\,, \\
\Dtp {\Pi_i}\nj  + \mutp \left( \Drzb {\Pi_i}\nj +
\fr{{\Pi_i}\nj}{r_j}\right) &=& 0\,.
\eea
\noi
We also adopt a scheme for 
numerical dissipation given by Kreiss and Oliger \cite{KO}.
In other words an additional term
\[
 \mutp \left( \Ddiss {\Ph_i}\nj \right)
\]

\noi
is added to the right hand side of (\ref{fd_klein_gordon_eq_2}), for $3 \le
j \le N_r-2$ (and
similarly for (\ref{fd_klein_gordon_eq_3}) for $\Pi_i$), where $\Ddiss$ is defined by

\beq \lab{KOdiss}
 \Ddiss u\nj = -\fr{\ep_d}{16 \De t} \left(  u\njpt - 4 u\njpo + 6 u\nj - 4
u\njmo + u\njmt \right) \,.
\eeq
Here, $\epsilon_d$ is an adjustable parameter satisfying $0\le\epsilon_d<1$, and is typically
chosen to be 0.5.  We note that the addition of Kreiss-Oliger dissipation changes 
the truncation error of the FDAs at $O(\Delta t^3,\Delta r^3)$  and thus does 
{\em not} effect the leading order error of a second order ($O(\Delta t^2,\Delta r^2)$) scheme.
The dissipation is useful for damping high frequency solution components that are 
often associated with numerical instability.

  \resetcounters

\chapter[Convergence Test of the Spherically Symmetric Code]
{} \label{ap6}
\vspace{-0.45in}
{\bf {\huge 
Convergence Test of the Spherically Symmetric Evolution Code
}}
\vspace{0.52in}

\noi
Here we present the results of a convergence test of the code that evolves boson stars in 
spherical symmetry. (See Chap.~\ref{bs1d} for results computed using this code,
and App.~\ref{ap4} for full details of the equations of motion and finite difference 
approximations used.)

\begin{figure}[h]
\begin{center}
\epsfxsize=12.0cm
\includegraphics[width=12.0cm,clip=true]{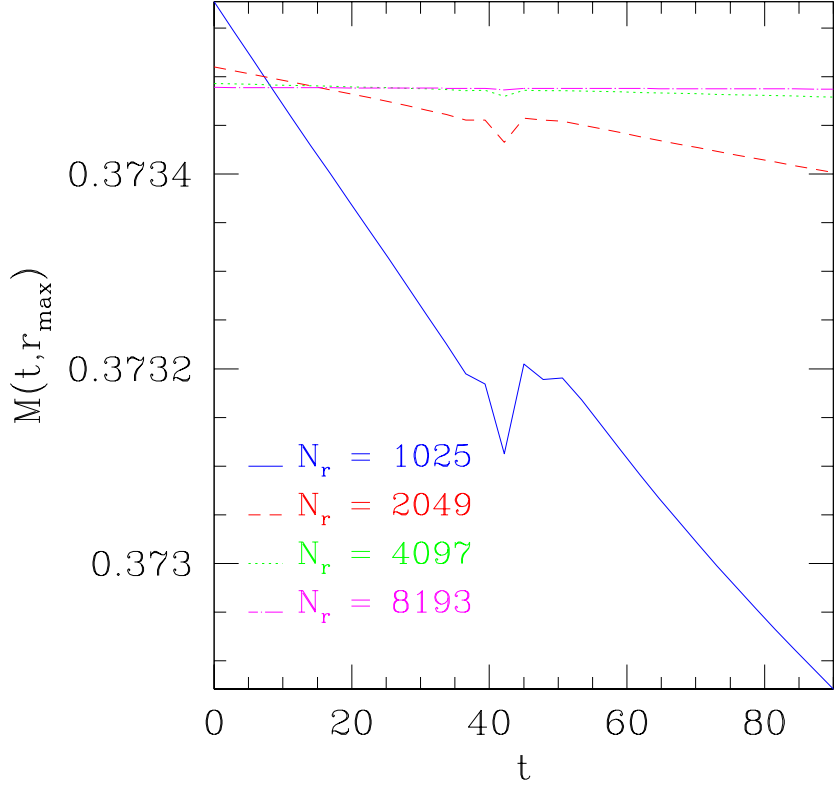}
    \caption[Convergence test of the spherically symmetric code]
{Convergence test of the spherically symmetric code.  The estimated
ADM mass, $M(t,r_{\rm max})$, is plotted against time, $t$, for four calculations 
using numbers of spatial grid points, $N_r$, of $1025, 2049, 4097$ and $8193$, so that 
the corresponding mesh spacings, $\Delta r$, are in a 8:4:2:1 ratio.
The initial data parameters for the
evolution are: $\ph_0 = 0.01$ for the complex field, and $A_3=0.001, r_0 = 40$ and $\si
= 3$ for the massless field (see~(\ref{ph3-gaussian})).  
The mass decreases with time in general, with a significant fluctuation 
at $40 \leq t \leq 50$, when the real scalar field is close to the origin
and strongly interacts with the boson star.  The change in mass tends to
vanish as we go to higher and higher resolutions.  Combining results from the 
four calculations we find strong evidence that the finite difference scheme is second
order accurate as expected.}
    \label{admm_0123outer}
\end{center}
\end{figure}

The ADM mass is one of the quantities which is most useful for diagnostic
purposes.  In Fig.~\ref{admm_0123outer} we plot the mass aspect function
at the outer boundary of the computational domain, $M(t,r_{\rm max})$,
as a function of time, and from four computations with grid spacings,
$\Delta r$, in a 8:4:2:1 ratio.  The simulation involves a pulse of 
massless scalar field imploding onto a stable boson star as in the calculations 
of the black hole threshold described in Sec.~\ref{critphenBS}.

The boson star has a central field value, $\ph_0 = 0.01$, while the incoming 
massless scalar field pulse is a gaussian of the form~(\ref{ph3-gaussian})
with $A_3 = 0.001, r_0 = 40$ and $\si = 3$.
The outer boundary is $r_{\mbox{max}} = 300$,
and $N_r = 1025, 2049, 4097, 8193$.  
During the time interval
$40 \leq t \leq 50$, the real scalar field
is concentrated near the the origin and interacts most strongly with the complex field. 
This results in a localized fluctuation of the ADM mass that is 
evident in the plots.  However, $M(t,r_{\rm max})$ clearly tends to a constant value 
as the resolution is increased. In addition, from the differences of $M(t,r_{\rm max})$
computed at different resolutions (e.g. $M^{\Delta r}(t,r_{\rm max}) - M^{2\Delta r}(t,r_{\rm max})$,
$M^{2\Delta r}(t,r_{\rm max}) - M^{4\Delta r}(t,r_{\rm max})$, etc.), we find strong
evidence that the overall difference scheme is converging in a second order fashion.
 
  \resetcounters

\chapter[Transformation from Areal to Isotropic Coordinates]
{}\label{ap3}
\vspace{-0.45in}
{\bf {\huge 
Transformation from Areal to Isotropic Spatial Coordinates
}}
\vspace{0.52in}

\noi
In this appendix we summarize the transformation from 
areal to isotropic spatial (radial) coordinates.  This 
transformation can be used to generate boson stars initial data 
in isotropic coordinates from data computed in areal coordinates.

Using areal coordinates, the {\em static}, spherically-symmetric metric is
\beq
ds^2 = - \al(R)^2 dt^2 + a(R)^2 dR^2 + R^2 d \Om^2 \,,
\eeq
\noi
while in isotropic coordinates it is
\beq
 ds^2 = - \al(r)^2 dt^2 + \ps(r)^4 \left( dr^2 + r^2 d \Om^2\right)\,.
\eeq
\noi
We thus 
seek the spatial coordinate transformation 
\beq
 r = r(R)\,,
\eeq
\noi
that leaves the $t$, $\te$ and $\vp$ coordinates unchanged.
Comparing coefficients of $d \Om^2$ and the two radial elements, we have
\bea
 R^2 &=& \ps^4 r^2\,,\\
 a^2 dR^2 &=& \ps^4 d r^2\,,
\eea
\noi
which immediately yields
\bea
 \ps &=& \sr{\fr{R}{r}} \,,\\ \lab{diffeqn}
 \fr{dr}{dR} &=& \pm\, a \fr{r}{R}\,.
\eea
\noi
Integrating the last equation, we have
\beq \lab{pa2mide}
 r = C \exp\left( \pm\int \fr{a}{R} dR\right)\,,
\eeq
\noi
where $C$ is some constant.
The sign and the integration constant can be fixed by choosing $R\to r$ when $R\to\infty$:
\beq \lab{pa2mibc}
\lim_{\Rtoi} \fr{r}{R} = 1\,.
\eeq
\noi
It is instructive to see the implication of the above by considering $R$ 
sufficiently large that the scalar field essentially vanishes.
In vacuum the Schwarzschild metric (areal coordinates) takes
the form
\beq
 a^2 = \left( 1-\fr{2M}{R}\right)^{-1}\,,
\eeq
\noi
where $M$ is the total mass inside the sphere of radius $R$.  
Expression~(\ref{diffeqn})
becomes
\beq
 \fr{dr}{r}= \fr{dR}{\sr{R^2-2MR}}\,.
\eeq
\noi
Integrating, we have
\beq
 R = \fr{C' r}{2} \left( 1+\fr{M}{C' r}\right)^2\,,
\eeq
\noi
where $C'$ is another integration constant. Applying the
boundary condition~(\ref{pa2mibc}) we have $C'=2$,
i.e.,
\beq \lab{rRrelation}
 R = r\left( 1+\fr{M}{2r}\right)^2\,.
\eeq
\noi
Thus, in a vacuum region, we have
\bea \nn
 \al^2 &=& 1 - \fr{2M}{R}\\
     &=& \left( \fr{1-M/2r}{1+M/2r}\right)^2\,,
\eea
\noi
and
\bea \nn
 \ps &=& \sr{\fr{R}{r}} \\
     &=& 1+\fr{M}{2r}\,.
\eea
\noi
Therefore, in vacuum, we can write the metric as 
\beq
 ds^2 = - \left( 1-\fr{2M}{R}\right)dt^2 + \left( 1-\fr{2M}{R}\right)^{-1} dR^2 + R^2
d \Om^2 \,,
\eeq
\noi
in Schwarzschild coordinates, and
\beq
 ds^2 = -\left( \fr{1-M/2r}{1+M/2r}\right)^2 dt^2 + \left(
1+\fr{M}{2r}\right)^4 \left( dr^2 + r^2 d \Om^2\right)\,.
\eeq
in isotropic coordinates.

From the above expression for the vacuum metric in isotropic coordinates we can
rewrite $M$ in terms of $\ps$ and hence find
\bea
 \nn \al &=& \fr{1-M/2r}{1+M/2r} \,\\
    &=& \fr{2}{\ps} -1 \,.
\eea
\noi
Therefore in isotropic coordinates,  the outer boundary condition for $\al$ is~(\ref{alouterBC})
\[
  \lim_{\rtoi}\al(r) = \fr{2}{\ps} - 1\,.
\]

If we have a numerical solution describing  a boson star in areal coordinates, we can 
transform
the solution to isotropic coordinates by numerically integrating
(\ref{pa2mide}) from the outer boundary inwards.  
In this case  we need to express $r$ in
terms of quantities known in the areal coordinate system.  
We may achieve this by noting that
from (\ref{rRrelation}) we have
\beq
\label{eq-a}
 r=\fr{R-M+\sr{R^2 -2MR}}{2}\,.
\eeq
\noi
On the other hand, in Schwarzschild coordinates we have
\beq
\label{eq-b}
 M=\fr{R}{2} \left( 1- \fr{1}{a}\right)\,.
\eeq
\noi
Substituting~(\ref{eq-b}) in~(\ref{eq-a}) we find
\beq
 r = \left( \fr{1+\sr{a}}{2}\right)^2 \fr{R}{a}\,.
\eeq
\noi
In other words, the coordinate transformation from areal to
isotropic coordinates is performed by solving

\bea  \nn
 \left. r\,\right|_{R=R_\mathrm{max}} &=& \left[\left(
\fr{1+\sr{a}}{2}\right)^2 \fr{R}{a}\right]_{R=R_\mathrm{max}}\,, \\
 \lab{ODE} \fr{dr}{dR} &=& \, a \fr{r}{R}\,.
\eea

On the other hand, under a coordinate transformation
\beq
 \tilde{x} = \tilde{x}(x)\,,
\eeq
\noi
we have
\beq
 {\tilde{g}_{\mu \nu}(\tilde{x})} = \fr{\pa {x}^{\al}}{\pa \tilde{x}^{\mu}}
\fr{\pa {x}^{\bt}}{\pa \tilde{x}^{\nu}} g_{\al \bt}(x)\,.
\eeq
\noi
Therefore, under the areal to isotropic coordinate
transformation, we have
\beq
 -\tilde{\al}(r)^2 = \left[ {\tilde{g}_{tt}}\right]_{\mathrm{I}} = \left[
g_{tt}\right]_{\mathrm{A}} = -\al(R)^2\,,
\eeq
\noi
where the subscripts I and A denote isotropic and areal
respectively, and where quantities with (without) a tilde
are functions of the isotropic (areal) radial coordinate.
The above expression implies 
\beq
 \tilde{\al}(r(R)) = \al(R)\,.
\eeq
\noi
In other words, the lapse function transforms as a scalar under the
specific transformation of areal to isotropic radial coordinates.  
If we now introduce a uniform grid in
$R$
\beq
 R_i = (i-1) \De R,\,\,\, i=1,\cdots, N+1
\eeq
\noi
where $\De R = R_\mathrm{max}/N$, then after solving
the initial value problem in areal coordinates, we have
\beq
 \al_i \eq \al(R_i) = \tilde{\al}(r(R_i)) \eq \tilde{\al}(r_i) \eq
\tilde{\al}_i \,,
\eeq
\noi
where we have defined $\al_i$, $r_i$ and $\tilde{\al}_i$ by the above
relations.  Note that $\al_i$ is simply equal to $\tilde{\al}_i$, but 
that the latter is defined on a non-uniform grid $r_i \eq r(R_i)$ (whose values
are obtained by solving the ODE (\ref{ODE})), and that therefore we need to do
some interpolation if we want to express the function $\tilde{\al}$ on a grid 
that is 
uniform in $r$.

We can make an identical argument for the scalar field, $\ph$, to
obtain the scalar field  on the isotropic coordinate grid.

For the conformal factor $\ps$ we have
\beq
 \tilde{\ps}(r)^4 = \left[ {\tilde{g}_{rr}}\right]_{\mathrm{I}} =
\left( \fr{dR}{dr}\right)^2 \left[
g_{RR}\right]_{\mathrm{A}} = \left( \fr{dR}{dr}\right)^2a^2 =
\left( \fr{R}{r}\right)^2\,,
\eeq 
\noi
and therefore (we drop the tilde since $\psi$ is obviously defined in the isotropic
coordinate system),
\beq
 \ps(r) = \sr{\fr{R}{r(R)}}\,,
\eeq
\noi
as it has to be.  Hence on the non-uniform grid, $r_i$, we have
\beq
 \ps_i \eq  \ps(r_i) = \sr{\fr{R_i}{r(R_i)}} = \sr{\fr{R_i}{r_i}}\,,
\eeq 
\noi
and, as before, we then need to do interpolation to obtain $\ps$ on a uniform
isotropic grid.

  \resetcounters

\chapter
[Numerical Determination of Spherical Boson Stars]
{
}\label{ap2}
\vspace{-0.45in}
{\bf {\huge 
Numerical Determination of Spherical Boson Stars Using a 
Compactified Isotropic Coordinate
}}
\vspace{0.52in}

\noi
As described in Chap.~\ref{bs1d}, the system of equations that 
determines spherically symmetric boson stars is an eigenvalue problem.
In this appendix we describe a numerical solution of this system that 
uses a compactified, isotropic radial coordinate.

Let $\ph(r)$, $\ps(r)$ and $\al(r)$ denote the modulus of the scalar field, 
the conformal factor,
and the lapse function respectively.  For notational simplicity let 
$\Ph \eq \ph'$, $\Ps \eq \ps'$, and $A \eq \al'$ denote the derivatives of 
the above functions with respect to $r$.
The relevant system of equations can then be written as

\bea
 \na^2 \ph + \left( \fr{A}{\al} + \fr{2 \Ps}{\ps}\right) \Ph - \ps^4
\left( m^2 - \fr{\om^2}{\al^2 }\right) \ph &=& 0 \,, \\
 \na^2 \ps + \pi \left[ \ps \Ph^2 + \ps^5 \left( \fr{\om^2}{\al^2} + m^2
\right) \ph^2\right] &=& 0 \,, \\
\na^2 \al + \fr{2 \Ps}{\ps} A - 4 \pi \ps^4 \al \left( \fr{2 \om^2}{\al^2 }
- m^2 \right) \ph^2  &=& 0 \,,
\eea

\noi
where 
$\na^2 \eq 3 \fr{d}{dr^3} (r^2 \fr{d}{dr})$. The regularity/boundary conditions
are:

\bea
 \Ph(0) &=& 0\,,\\
 \Ps(0) &=& 0\,,\\
 A(0) &=& 0\,,\\
 \lim_{\rtoi}\ph(r) &=& 0\,,\\
 \lim_{\rtoi}\ps(r) &=& 1\,,\\
 \lim_{\rtoi}\al(r) &=& 1\,.
\eea

We now perform a compactification of the spatial dimension by introducing a new 
coordinate, $\zeta$, such that

\beq
 \zeta = \fr{r}{1+r} \,,
\eeq

\noi
where $\zeta \in [0,1]$.  Under the above transformation we have

\bea
  r &=& \fr{\zeta}{1-\zeta} \,, \\
 \fr{d}{dr} &=& (1-\zeta)^2 \fr{d}{d \zeta} \,,\\
  \na^2 &=& (1-\zeta)^4 \na_\zeta^2 \,,
\eea

\noi
where 
$\na_\ze^2 \eq 3 \fr{d}{d\ze^3} (\ze^2 \fr{d}{d\ze})$. 
The previous system of equations can be rewritten as

\bea 
\na_\zeta^2 \ph + \left( \fr{A}{\al} + \fr{2 \Ps}{\ps}\right) \Ph -
\fr{\ps^4}{(1-\zeta)^4} \left( m^2 - \fr{\om^2}{\al^2}\right) \ph &=& 0
\,, \\
\na_\zeta^2 \ps + \pi \left[ \ps \Ph^2 + \fr{\ps^5}{(1-\zeta)^4} \left(
\fr{\om^2}{\al^2} + m^2 \right) \ph^2 \right] &=& 0 \,, \\
\na_\zeta^2 \al + \fr{2 \Ps}{\ps} A - 4 \pi \fr{\ps^4 \al}{(1-\zeta)^4}
 \left( \fr{2 \om^2}{\al^2 }
- m^2 \right) \ph^2  &=& 0 \,,
\eea

\noi
with similar regularity/boundary conditions except that the $r\to\infty$ limit
now becomes $\zeta\to 1$.
A second order finite difference form of the above equations is

\bea
3 \Dhrcz \lab{FDAph}
\left( r_{_{\scr j}}^2 \, \Dhrz \ph\nj \right) +
\left( \fr{\Drz \al\nj}{\al\nj} + \fr{2 \Drz \ps\nj}{\ps\nj}\right)
\Drz \ph\nj - \fr{{\ps\nj}^4}{(1-r_j)^4}
\left( m^2 - \fr{\om^2}{{\al\nj}^2 }\right) \ph\nj &=& 0\,,\;\;\;\; \\
3 \Dhrcz \lab{FDAps}
\left( r_{_{\scr j}}^2 \, \Dhrz \ps\nj \right) +
\pi \left[ \ps\nj \left( \Drz \ph\nj\right)^2 + \fr{{\ps\nj}^5}{(1-r_j)^4} 
\left( \fr{\om^2}{{\al\nj}^2  }+ m^2\right) {\ph\nj}^2 \right]
&=& 0\,,\;\;\;\; \\
3 \Dhrcz \lab{FDAal}
\left( r_{_{\scr j}}^2 \, \Dhrz \al\nj \right) +
\fr{2 \Drz{\ps\nj}}{\ps\nj} \Drz \al\nj - 4 \pi \fr{{\ps\nj}^4 \al\nj}{(1-r_j)^4}
\left( \fr{2 \om^2}{{\al\nj}^2} - m^2 \right) {\ph\nj}^2
&=& 0\,,\;\;\;\;
\eea
\noi
for $j = 2,\cdots, N_r-1$. 

The discrete regularity/boundary conditions are 
\bea
 \ph\no &=& \fr{4 \ph^{^{\scr n}}_{_{\scr 2}} - \ph^{^{\scr n}}_{_{\scr 
3}}}{3} \,, \\
\ph\nNr &=& 0\,,
\eea

\noi
with similar equations for $\ps\nj$ and $\al\nj$.

Equation 
(\ref{FDAph}) is linear in $\ph$ and is viewed as an eigenvalue equation with
eigenvalue $-\om^2$ and eigenvector $\ph$.  
More specifically, (\ref{FDAph}) can be written as a matrix equation
\beq  \lab{tridiag} 
\sum_j L_{i,j} \ph_j = - \om^2 \ph_j\,,
\eeq
\noi
where $L_{i,j}$ is a tridiagonal matrix.
The matrix $L_{i,j}$ is in general not symmetric.
Transformation to a symmetric tridiagonal form can be performed
using the {\tt FIGI} routine in the {\tt EISPACK}~\cite{eispack}
package,
and then (\ref{tridiag}) can be solved with the symmetric tridiagonal eigenvalue solver
{\tt RATQR} and {\tt TINVIT} from the same package.
Equation (\ref{FDAps})
is non-linear in $\ps_j$ and is solved by a global Newton method, with 
each Newton step involving a tridiagonal linear solve.
Finally, (\ref{FDAal}) can also be solved for $\al_j$ using
a tridiagonal solver.  The sequence of solving (\ref{FDAph}), 
(\ref{FDAps}) and (\ref{FDAal}) is repeated until the total change in the variables 
from iteration to iteration is within 
a prescribed tolerance, typically chosen to be $\ep = 10^{-10}$. 

  \resetcounters

\chapter[Summary of Equations for 2D Boson Stars]
{} \label{ap7}
\vspace{-0.45in}
{\bf {\huge 
Summary of Equations for Stationary, Axisymmetric Boson Stars and Associated Finite Difference Formulae
}}
\vspace{0.52in}

\noi
In this appendix we give a summary of the system of equations (System A of Chap.~\ref{bs2d}) that 
determines stationary, rotating, axisymmetric boson stars in quasi-isotropic coordinates, as well 
as our finite difference approximation of the system.

The spacetime metric is written as
\beq
 ds^2 = \left( -\al^2+\ps^4 r^2 \sin^2 \te e^{2 \si} \bt^2\right)dt^2
+\ps^4 \left( dr^2 + r^2 d \te^2 + r^2 \sin^2 \te e^{2 \si} d \vp^2\right)
+2 \ps^4  r^2 \sin^2 \te e^{2 \si} \bt dt d \vp
\eeq
\noi
and the equations for the five unknown functions $\al,\bt,\ps,\si$ and $\ph$ (which
depend on $r$ and $\te$ only) are

\beq
\begin{split}
\ph_{,rr}+\fr{2}{r} \ph_{,r} + \fr{1}{r^2}\ph_{,\te\te}+\fr{\cot\te}{r^2} \ph_{,\te}
- \fr{k^2}{r^2 \sin^2 \te e^{2\si}} \ph 
 + \left( \si_{,r}+\fr{\al_{,r}}{\al}+ \fr{2 \ps_{,r}}{\ps}\right) \ph_{,r} \\ +
\left( \si_{,\te}+\fr{\al_{,\te}}{\al}+ \fr{2 \ps_{,\te}}{\ps}\right) \fr{\ph_{,\te}}{r^2}
+ \left[ \left( \fr{\om-\bt k}{\al}\right)^2 -m^2 \right] \ps^4 \ph = 0\,,
\end{split}
\eeq

\bea \nn 
\ps_{,rr}+\fr{\ps_{,\te \te}}{r^2}+ \left(
\fr{1}{r}-\fr{\al_{,r}}{\al}-\fr{\ps_{,r}}{\ps}\right) \ps_{,r}
 - \left( \fr{\ps_{,\te}}{\ps}+\fr{\al_{,\te}}{\al}\right) \fr{\ps_{,\te}}{r^2}
-\fr{\ps^5 r^2\sin^2 \te e^{2 \si}}{8 \al^2}\left(\bt_{,r}\,^2+\fr{\bt_{,\te}\,^2}{r^2}
\right) \\ \nn
 -\fr{\ps}{2r} \left[ \left( 1+r \si_{,r}\right)\fr{\al_{,r}}{\al}+\fr{1}{r}\left( \cot \te +
\si_{,\te}\right) \fr{\al_{\te}}{\al}\right]
= \\ 
- 2 \pi \ps \left[ \left( \ph_{,r}\,^2+\fr{\ph_{,\te}\,^2}{r^2}\right)+ \ph^2 \left(
\fr{\ps^4}{\al^2}(w-\bt k)^2 - \fr{k^2}{r^2 \sin^2 \te e^{2 \si}}\right)\right]
\,,
\eea
\beq
 \bt_{,rr}+\fr{\bt_{,\te\te}}{r^2} + \left(\fr{4}{r}+3 \si_{,r}-\fr{\al_{,r}}{\al}+6
\fr{\ps_{,r}}{\ps}\right) \bt_{,r} +
\left(3 \cot \te+3 \si_{,\te}-\fr{\al_{,\te}}{\al}+6
\fr{\ps_{,\te}}{\ps}\right) \fr{\bt_{,\te}}{r^2} = -16 \pi \ph^2\fr{k(\om-\bt k)}{r^2\sin^2
\te e^{2 \si}}\,,
\eeq

\beq
\begin{split}
\al_{,rr}+\fr{2}{r} \al_{,r} +
\fr{1}{r^2}\al_{,\te\te}+\fr{\cot\te}{r^2} \al_{,\te} 
 + \left( \fr{2 \ps_{,r}}{\ps}+\si_{,r}\right) \al_{,r} + \left( 2 \fr{\ps_{,\te}}{\ps}
+ \si_{,\te}\right) \fr{\al_{,\te}}{r^2} \\
- \fr{\ps^4 r^2 \sin^2 \te e^{2 \si}}{2 \al}
\left( \bt_{,r}\,^2+\fr{\bt_{,\te}\,^2}{r^2}\right) + 
4 \pi \al \ps^4 \ph^2 \left( m^2 -
\fr{2 (\om - \bt k)^2}{\al^2}\right) = 0\,,
\end{split}
\eeq

\beq 
\begin{split}
\si_{,rr}+\fr{\si_{,\te\te}}{r^2} +
\left(\si_{,r} +4 \fr{\ps_{,r}}{\ps} + \fr{2
\al_{,r}}{\al} + \fr{3}{r}\right) \si_{,r} + 
\left(\si_{,\te} +4 \fr{\ps_{,\te}}{\ps} + \fr{2
\al_{,\te}}{\al} + 2\cot \te\right) \fr{\si_{,\te}}{r^2} \\ 
 +\fr{4}{\ps} \left[ \left( \fr{\ps_{,r}}{\ps}+\fr{\al_{,r}}{\al}+\fr{1}{r}\right) \ps_{,r} +
\left( \fr{\ps_{,\te}}{\ps}+\fr{\al_{,\te}}{\al}+\cot \te\right) \fr{\ps_{,\te}}{r^2}\right]
+\fr{2}{\al r} \left( \al_{,r}+\cot \te \fr{\al_{,\te}}{r}\right) \\ 
+\fr{3 \ps^4 r^2\sin^2 \te e^{2 \si}}{4 \al^2} \left(
\bt_{,r}\,^2+\fr{\bt_{,\te}\,^2}{r^2}\right) \\ 
+4 \pi \left[ \left( m^2+\fr{3k^2}{\ps^4 r^2\sin^2 \te e^{2 \si}}-\left(\fr{\om-\bt
k}{\al}\right)^2 \right) \ph^2 \ps^4 - \left( \ph_{,r}^2+\fr{\ph_{,\te}^2}{r^2}\right)\right]
= 0 \,.
\end{split}
\eeq

\noi
With the coordinate transformation
\bea
  \ze &=& \fr{r}{1+r} \,, \\
 s &=& \cos \theta \,,
\eea

\noi
the equations become
\bea \nn 
&&\left( \fr{1-\br}{\br}\right)^2 \left[ 3\br^2\left(1-\br\right)^2 \left(  \br^2
\ph_{,\br}\right)_{,\br^3}+\left((1-s^2)\ph_{,s} \right)_{,s} -
\fr{k^2}{\left( 1-s^2\right) e^{2 \si}} \ph\right] + \\ \nn
 &&\left( 1-\br\right)^4 
\left( \si_{,\br}+\fr{\al_{,\br}}{\al}+\fr{2 \ps_{,\br}}{\ps}\right) \ph_{,\br}
 + \left(\fr{1-\br}{\br}\right)^2 \left( 1-s^2\right)
\left(\si_{,s}+\fr{\al_{,s}}{\al} + \fr{2 \ps_{,s}}{\ps}\right) \ph_{,s}
\\  \lab{phcpteq}
&& \hspace{7cm} + \left[ \left( \fr{\om-\bt k}{\al}\right)^2 -m^2
\right] \ps^4 \ph = 0 \,,
\eea

\bea \nn
\left( \fr{1-\br}{\br}\right)^2 \left[ 3\br^2 \left(1-\br\right)^2 \left(  \br^2
\ps_{,\br}\right)_{,\br^3}+(1-s^2)\ps_{,ss} -s \ps_{,s}
\right] \\ \nn
- \left( 1-\br\right)^3 \left[ \fr{1}{\br}+
(1-\br)\left(\fr{\al_{,\br}}{\al} + \fr{\ps_{,\br}}{\ps}\right)\right] \ps_{,\br}\\
\nn
-(1-s^2)\bomrors \left( \fr{\ps_{,s}}{\ps}+\fr{\al_{,s}}{\al}\right) \ps_{,s}
\\ \nn
-\fr{\ps^5 \br^2(1-s^2)e^{2\si}}{8 \al^2}
\left[ (1-\br)^2 \bt_{,\br}\,^2+\fr{1}{\br^2} \omss \bt_{,s}\,^2\right] \\
\nn
-\fr{\ps}{2 \al} \left( \bomror\right)
\left[  (\bomr)^2 \left( 1+\br(\bomr) \si_{,\br}\right) \al_{,\br} +
\left( \bomror\right) \left( (1-s^2)\si_{,s} - s \right) \al_{,s}\right] \\
\nn
+ 2 \pi \ps \left[ (\bomr)^4 {\ph_{,\br}}^2 + \bomrors (1-s^2)
{\ph_{,s}}^2 + \right. \\ 
\left. \ph^2 \left( \fr{\ps^4}{\al^2} (\om - \bt k)^2 - \fr{k^2
(\bomr)^2}{\br^2 (1-s^2) e^{2 \si}}\right) \right] = 0\,,
\eea

\bea \nn
 \bomrors \left\{ 3 \br^2\left( \bomr\right)^2 \left( \br^2
\bt_{,\br}\right)_{,\br^3} + \left[ \left( 1-s^2\right) \bt_{,ss} - s
\bt_{,s} \right] \right\} \\ \nn
+ \left(\bomr\right)^3 \left[ \fr{2}{\br} + \left( \bomr\right) \left( 3 \si_{,\br} -
\fr{\al_{,\br}}{\al} + 6 \fr{\ps_{,\br}}{\ps}\right)\right]  \bt_{,\br} 
\\ \nn
+ \bomrors \left[ -3 s + \left( 1-s^2\right) \left( 3 \si_{,s} -
\fr{\al_{,s}}{\al} + 6 \fr{\ps_{,s}}{\ps}\right)\right] \bt_{,s}
\\ 
+ 16 \pi  \fr{k \left( \om - \bt k \right)}{ \left( 1-s^2\right) e^{2
\si}} \bomrors\ph^2 = 0\,,
\eea

\bea \nn
 \bomrors \left[ 3 \br^2 \left(\bomr\right)^2 \left( \br^2
\al_{,\br}\right)_{,\br^3} + \omss \al_{,ss} -s \al_{,s}\right]
+ \left(\bomr\right)^4 \left( \fr{2 \ps_{,\br}}{\ps}  + \si_{,\br}\right)
\al_{,\br} \\ \nn
+ \bomrors \left[ -s + \omss \left( \fr{2 \ps_{,s}}{\ps} +
\si_{,s}\right)\right] \al_{,s}  \\ \nn
- \ps^4 \fr{ \omss e^{2 \si}}{ 2 \al} \left[ \br^2 \left( \bomr\right)^2
{\bt_{,\br}}^2 + \omss {\bt_{,s}}^2 \right] +  \\
4 \pi \al \ps^4 \ph^2 \left[ m^2 - \fr{2 \left( \om - \bt k\right)^2}{
\al^2}\right] = 0\,,
\eea

\bea \nn
 \bomrors \left\{ 3 \br^2\left( \bomr\right)^2 \left( \br^2
\si_{,\br}\right)_{,\br^3} + \left[ \left( 1-s^2\right) \si_{,ss} - s
\si_{,s} \right] \right\} \\ \nn
+ \left(\bomr\right)^3\left[ \left( \bomr\right) \left( \si_{,\br} + 4 \fr{\ps_{,\br}}{\ps} + 2 
\fr{ \al_{,\br}}{\al}\right)+\fr{1}{\br}\right]  \si_{,\br} \\ \nn
+ \bomrors \left[ \omss \left( \si_{,s} + 4 \fr{\ps_{,s}}{\ps} + 2
\fr{\al_s}{\al} \right)- 2 s\right] \si_{,s}
\\ \nn
+ \fr{4}{\ps} \left\{ \left(\bomr\right)^3 \left[ (\bomr) \left( \fr{\ps_{,\br}}{\ps } +
\fr{\al_{,\br}}{\al}\right) + \fr{1}{\br}\right] \ps_{,\br} \right. \\ \nn
\left.
+ \bomrors \left[ \omss \left( \fr{\ps_{,s}}{\ps} + \fr{\al_{,s}}{\al}
\right) -s \right]\ps_{,s} 
\right\}
\\ \nn
+ \fr{2}{\al} \fr{(\bomr)^2}{\br} \left[ \left( \bomr\right) \al_{,\br} -
\fr{s}{\br} \al_{,s}\right]
+ \fr{3 \ps^4 \omss e^{2 \si}}{4 \al^2} \left[ \br^2 (\bomr)^2
{\bt_{,\br}}^2 + \omss {\bt_{,s}}^2 \right] \\ \nn
+ 4 \pi \left\{ \left[ m^2 + \fr{3 k^2}{ \ps^4 \omss e^{2 \si}} \bomrors -
\left( \fr{\om - \bt k}{\al}\right)^2 \right] \ph^2 \ps^4 - \right. \\
\lab{sicpteq}
\left. \left[
\left( \bomr\right)^4 {\ph_{,\br}}^2 + \bomrors \omss {\ph_{,s}}^2\right]
\right\} \,.
\eea

The regularity conditions for the system at $\ze = 0$ are given by
\bea
\ph(0,s) &=& 0\,, \\
\ps_{,\ze}(0,s) &=& 0\,, \\ 
\bt_{,\ze}(0,s) &=& 0 \,, \\ 
\al_{,\ze}(0,s) &=& 0 \,, \\ 
\si(0,s)  &=& 0 \,,
\eea

\noi
and boundary conditions at $\ze=1$ are
\bea
 \ph(1,s) &=& 0\,, \\
 \ps(1,s) &=& 1\,, \\
 \bt(1,s) &=& 0\,, \\
 \al(1,s) &=& 1\,, \\
 \si(1,s) &=& 0\,.
\eea

The boundary conditions at $s=0$ (the equatorial plane) are
\bea
\ph_{,s}(\zeta,0) &=& 0\,, \\
\ps_{,s}(\zeta,0) &=& 0\,, \\
\bt_{,s}(\zeta,0) &=& 0\,, \\
\al_{,s}(\zeta,0) &=& 0\,, \\
\si_{,s}(\zeta,0) &=& 0\,,
\eea

\noi
while at $s=1$ (the axis of symmetry), we have
\bea
\ph(\zeta,1) &=& 0\,, \\
\ps_{,s}(\zeta,1) &=& 0\,, \\
\bt_{,s}(\zeta,1) &=& 0\,, \\
\al_{,s}(\zeta,1) &=& 0\,, \\
\si_{,s}(\zeta,1) &=& 0\,.
\eea
For the finite difference formulae, since the expressions are lengthy and
messy, we simply note that we extend the difference operators defined in 
App.~\ref{ap5} to operators acting in two spatial dimensions in the obvious way. 
More explicitly, we will have

\begin{eqnarray*}
 \De^{^{\ze}}_{_{0}}  u_{_{{\scr i},{\scr j}}} &\eq& 
\fr{u_{_{{\scr i}+1,{\scr j}}} - u_{_{{\scr i}-1,{\scr j}}}}{2 \De \ze} \\
{\De^{^{\scr s}}_{_{\scr 0}}} u_{_{{\scr i},{\scr j}}} &\eq& 
\fr{u_{_{{\scr i},{\scr j}+1}} - u_{_{{\scr i},{\scr j}-1}}}{2 \De s} 
\end{eqnarray*}
\noi
etc..  

The difference equations are then obtained by performing the
following substitutions in (\ref{phcpteq})-(\ref{sicpteq}):

\begin{eqnarray*}
\fr{\pa}{\pa \br^3}&\longrightarrow & \De^{^{\scr \ze^3}}_{_{\scr 0}} \,,\\ 
\fr{\pa}{\pa \br} &\longrightarrow& \De^{^{\scr \ze}}_{_{\scr 0}}  \,,\\
\fr{\pa^2}{\pa s^2} &\longrightarrow& 
\De^{^{\scr s}}_{_{\scr +}}
\De^{^{\scr s}}_{_{\scr -}} \,,\\
\fr{\pa}{\pa s} &\longrightarrow& \De^{^{\scr s}}_{_{\scr 0}}\,.
\end{eqnarray*}

  \resetcounters

\chapter[Derivation of the PS system from the EKG system]
{} \label{ap8}
\vspace{-0.45in}
{\bf {\huge 
Weak Field Limit: Derivation of the Poisson-Schr\"{o}dinger system from
the Einstein-Klein-Gordon System
}}
\vspace{0.52in}

\noi
In this appendix we will consider the weak field limit (Newtonian limit) of the 
Einstein-Klein-Gordon (EKG) system---thereby deriving the Poisson-Schr\"{o}dinger (PS) system, which
represents {\em non-relativistic} Newtonian boson stars---with 
all approximations explicitly stated.  The main purpose of this derivation is to provide the
background for checking whether the stationary Newtonian solutions obtained by solving the
PS system provide good initial estimates for the algorithm used to determine 
the stationary general relativistic solutions~(see Sec.~\ref{numstr}).  
The assumptions made in deriving the Newtonian limit that need to be checked
on a solution-by-solution basis 
will be summarized at the end of the appendix.

Note that here we write the equations for the EKG system as:

\beq \lab{eineq}
 G_{\al \bt} = \ka T_{\al \bt} \,,\\
\eeq
\beq
 \na^\al \na_\al \ph - m^2 \ph = 0 \,,
\eeq
\noi
where $\ka$ is a constant which is conventionally chosen to be $8 \pi$.

\section{Poisson Equation}
The derivation of the Poisson equation for the Newtonian gravitational
potential from the Einstein field equation is quite standard \cite{schutz,wald}, 
although the right
hand side of the equation (matter coupling) is usually not worked out 
in detail, and the order of truncation is often not clearly
stated.
We suppose that spacetime is nearly flat.  The metric can then be written as

\beq
 g_{\al \bt} = \et_{\al \bt} + h_{\al \bt} \,,
\eeq

\noi
where
\beq \lab{appr1}
|h_{\al \bt}| = O(\ep) \ll 1\,,
\eeq
\noi
and $\et_{\al \bt}$ is the flat metric in Cartesian coordinates $\et =
\mathrm{diag}\left(-1,1,1,1\right)$.~\footnote{
This simply means that $h$ is a small number when compared to $\et$.  In the
context of numerical computation this has a well-defined meaning. Also note that
we restrict attention to Cartesian coordinates, since in other coordinate systems
the components, $g_{\al \bt}$, of the Minkowski metric can be arbitrarily small.  For instance, in
spherical coordinates, $g_{\te \te} = r^2\to 0$ as $r\to 0$.}
There are two classes of coordinate transformations
which preserve the above form: the \emph{background Lorentz
transformations} and \emph{gauge transformations} \cite{schutz}.
Gauge transformations can be written in the form

\beq
\label{gauge}
 x^{\bal}=x^{\al} + \xi^{\al}(x^{\bt})\,,
\eeq
where 
\beq \lab{appr2}
|\xi^\al\,_{,\bt}| = O(\ep)\,.
\eeq
\noi
The Jacobian matrix associated with~(\ref{gauge}) can be written as
\beq
\La^{\bal}\,_{\bt} = \de^{\al}\,_{\bt} + \xi^{\al}\,_{,\bt}\,,
\eeq
\noi
with an inverse given by 
\beq
\La^{\al}\,_{\bbt} = \de^{\al}\,_{\bt} - \xi^{\al}\,_{,\bt} + O(\ep^2)\,,
\eeq
\noi
as is established by the following:
\bea \nn
 \La^{\bal}\,_{\bt}\,\La^{\bt}\,_{\bga} &=& ( \de^{\al}\,_{\bt} + \xi^{\al}\,_{,\bt} )
 ( \de^{\bt}\,_{\ga} - \xi^{\bt}\,_{,\ga}) \\  \nn
&=& \de^{\al}\,_{\ga} + \xi^\al\,_{,\ga}- \xi^\al\,_{,\ga} + \xi^{\al}\,_{,\bt}\,
\xi^{\bt}\,_{,\ga} \\
&=& \de^{\al}\,_{\ga} + O(\ep^2)\,.
\eea
\noi
Now
\bea
\nn
g_{\bal\bbt} &=& \La^{\mu}\,_{\bal} \La^{\nu}\,_{\bbt} \,g_{\mu \nu} \\ \nn
&=&  \La^{\mu}\,_{\bal} \La^{\nu}\,_{\bbt}\, \et_{\mu \nu} + \La^{\mu}\,_{\bal}
\La^{\nu}\,_{\bbt} \,h_{\mu \nu} \\ \nn
&=& (\de^{\mu}\,_{\al} - \xi^{\mu}\,_{,\al})
    (\de^{\nu}\,_{\bt} - \xi^{\nu}\,_{,\bt}) \,\et_{\mu \nu} + 
    (\de^{\mu}\,_{\al} - \xi^{\mu}\,_{,\al})
    (\de^{\nu}\,_{\bt} - \xi^{\nu}\,_{,\bt}) \,h_{\mu \nu} +O(\ep^2)\\  \nn
&=& \et_{\al \bt} - \xi^{\mu}\,_{,\al}\,\et_{\mu \bt}
- \xi^{\nu}\,_{,\bt}\,\et_{\al \nu} + h_{\al \bt} + O(\ep^2) \\
&=& \et_{\al \bt} + h_{\al \bt} - \xi_{\al,\bt} - \xi_{\bt,\al} + O(\ep^2)\,,
\eea
\noi
or
\beq \lab{gauge_transf}
 h_{\bal\bbt} = h_{\al \bt} - \xi_{\al,\bt} - \xi_{\bt,\al} + O(\ep^2)\,,
\eeq
\noi
where in the last step we have adopted the {\em notational convention} that we
lower and raise tensor indices using $\et_{\al \bt}$.  That is, $h^{\al \bt}$ is,
{\em  by definition}, $\et^{\al \mu} \et^{\bt \nu} h_{\mu \nu}$. (Note that using this
convention we also have $g^{\al \bt} = \et^{\al \bt} - h^{\al \bt} + O(\ep^2)$.
In other words, for all quantities of order $O(\ep)$, raising and lowering of indices
using $\et$ rather than $g$ will lead to deviations of order $O(\ep^2)$. )

\noi
By definition, we have

\beq
-R^{\al}\,_{\bt \mu \nu} = \Ga^\al\,_{\bt \mu,\nu} - 
\Ga^\al\,_{\bt \nu,\mu} + 
\Ga^\al\,_{\si \nu}\,\Ga^{\si}\,_{\bt \mu}
-\Ga^\al\,_{\si \mu}\,\Ga^{\si}\,_{\bt \nu}\,,
\eeq

\noi
where

\beq
 \Ga^{\la}\,_{\mu \nu} = \ha \et^{\la \rh} 
\left( h_{\rh \nu,\mu} + h_{\mu \rh,\nu}- h_{\mu \nu,\rh}\right) + O(\ep^2)\,.
\eeq
Note that we have used the fact that we work in Cartesian coordinates 
(and hence that the derivatives of $\et$ vanish), and that we assume that derivatives of
small numbers are also small 
\beq \lab{appr3}
 h_{\al \bt, \ga} = O(\ep)\,
\eeq
(so that we can drop terms such as $h^{\al \bt} h_{\bt
\ga, \de}$).

\noi
Now we have
\beq
R^{\al}\,_{\bt \mu \nu} =
-\ha \et^{\al \rh} \left( h_{\rh\mu,\bt\nu} 
+ h_{\bt \rh, \mu \nu} - h_{\bt \mu, \rh \nu} \right)
-\left( \mu \leftrightarrow \nu \right)
+O(\ep^2)\,.
\eeq

\noi
The second terms within parentheses in the above cancel, so that we have

\beq
 R_{\al \bt \mu \nu} = \ha \left( 
h_{\al \nu, \bt \mu} - h_{\bt \nu,\al \mu}
-h_{\al \mu, \bt \nu} + h_{\bt \mu,\al \nu}
\right) + O(\ep^2)\,,
\eeq

\noi
and

\bea \nn
 R_{\bt\nu} &=& \ha \left( 
h^{\mu}\,_{\nu, \bt \mu} - h_{\bt \nu,\mu}\,^{\mu}
-h^{\mu}\,_{\mu, \bt \nu} + h_{\bt \mu,}\,^{\mu}\,_{\nu}
\right) + O(\ep^2)\\
&=& \ha \left( 
h^{\mu}\,_{\nu, \bt \mu} - \Box h_{\bt \nu}
-h\,_{,\bt \nu} + h_{\bt \mu,}\,^{\mu}\,_{\nu} 
\right)+ O(\ep^2)\,,
\eea
\noi
and
\beq
 R = \left( h_{\mu \nu,}\,^{\mu \nu} - \Box h  \right) + O(\ep^2)\,,
\eeq

\noi
where we have defined $h \eq h^\mu\,_{\mu}$ and $\Box h \eq h_{,\mu}\,^{\mu}$.

Constructing the Einstein tensor, we then have
\beq
 R_{\bt \nu} - \ha \et_{\bt \nu} R = 
 \ha \left( 
h^{\mu}\,_{\nu, \bt \mu} - \Box h_{\bt \nu}
-h\,_{,\bt \nu} + h_{\bt \mu,}\,^{\mu}\,_{\nu}
-\et_{\bt \nu} h_{\ga \de,}\,^{\ga \de} + \et_{\bt \nu}\Box h
\right) + O(\ep^2)\,.
\eeq

\noi
If we define
\beq
 \bar{h}^{\al \bt} \eq h^{\al \bt} - \ha \et^{\al \bt} h \,,
\eeq
\noi
then
\beq
 \bar{h}= -h\,,
\eeq
\noi
and
\beq
 h^{\al \bt} = \bar{h}^{\al \bt} - \ha \et^{\al \bt} \bar{h}\,.
\eeq
\noi
Now
\bea \nn
 G_{\bt \nu} &=&  
 \ha \left( 
\bh^{\mu}\,_{\nu,\bt \mu} - \ha \bh_{,\bt \nu}
-\Box \bh_{\bt \nu} + \ha \et_{\bt \nu} \Box \bh
+ \bh_{,\bt \nu}
+ \bh_{\bt \mu,}\,^{\mu}\,_{\nu} - \ha \bh_{,\bt \nu}
-\et_{\bt \nu} \bh_{\ga \de,}\,^{\ga \de} 
\right.\\ \nn
&& \left.
+ \ha \et_{\bt \nu } \Box \bh 
- \et_{\bt \nu}\Box \bh
\right) +O(\ep^2)\\
&=& \ha \left( - \Box \bh_{\bt \nu}
+ \bh^{\mu}\,_{\nu,\bt \mu}
+ \bh_{\bt \mu,}\,^{\mu}\,_{\nu} 
-\et_{\bt \nu} \bh_{\ga \de,}\,^{\ga \de} 
\right) + O(\ep^2)\,.
\eea

\noi

\noi
If we now impose the {\em Lorentz} (or harmonic) {\em gauge condition}
\beq
\bh^{\mu \nu}\,_{,\nu} = 0 \,,
\eeq
we have
\beq
 G_{\al \bt} = -\ha \Box \bh_{\al \bt} + O(\ep^2)\,,
\eeq
\noi
and hence
\beq \lab{lineq}
\Box \bh_{\al \bt} = -2 \ka T_{\al \bt} +O(\ep^2)\,.
\eeq

\noi
The existence of the Lorentz gauge can be shown as follows. From
(\ref{gauge_transf}) it follows that
\beq
 h^{\bal}\,_{\bal} = h - 2\xi^{\al},_{\al}+O(\ep^2)\,.
\eeq
\noi
Also, we have
\bea \nn
\bh_{\bal \bbt} &=& h_{\bal\bbt} - \ha \et_{\bal\bbt} h^{\bga}\,_{\bga} \\
&=& h_{\al \bt} - \xi_{\al,\bt} - \xi_{\bt,\al} - \ha \et_{\al \bt} (h - 2
\xi^{\ga}\,_{,\ga}) +O(\ep^2) \\
&=& \bh_{\al \bt}- \xi_{\al,\bt} - \xi_{\bt,\al} + \et_{\al \bt}
\xi^{\ga}\,_{,\ga} + O(\ep^2)\,.
\eea
\noi
Hence
\beq
 \bh^{\bal\bbt}\,_{,\bbt} = \bh^{\al \bt}\,_{,\bt} - \Box \xi^{\al} + O(\ep^2)\,,
\eeq
which leads to the inhomogeneous wave equation

\beq \lab{iwe}
 \Box \xi^{\al} = \bh^{\al \bt}\,_{,\bt}\,.
\eeq

\noi
Thus the existence of solutions to (\ref{iwe}) implies the viability of the 
Lorentz gauge condition for the ``barred" coordinate system, to order $O(\ep^2)$.
(Note that the solutions allow a homogeneous term.  Hence the Lorentz gauge is in
fact a class of gauges. )

Since we expect the Newtonian potential $V \sim h_{00}$,
\noi
we assume 
\beq \lab{appr8}
 |V| = O(\ep)\,.
\eeq
Dimensional analysis shows
$ v^2 \sim GM/r = V \Rightarrow |v| = O(\ep^{1/2})$.  We would expect $|T^{ij}| \sim
v |T^{0i}| \sim v^2 |T^{00}|$, or $|T^{00}| \gg |T^{0i}| \gg |T^{ij}|$.
This roughly gives $|\bh^{00}| \gg |\bh^{0i}| \gg |\bh^{ij}|$.  Therefore
we further assume

\bea \lab{appr4}
 \fr{T^{0i}}{T^{00}} &=& O(\ep^{1/2}) \,, \\  \lab{appr5}
 \fr{T^{ij}}{T^{00}} &=& O(\ep) \,, \\ \lab{appr6}
 \fr{\bh^{0i}}{\bh^{00}} &=& O(\ep^{1/2}) \,, \\ \lab{appr7}
 \fr{\bh^{ij}}{\bh^{00}} &=& O(\ep) \,.
\eea

\noi
Suppose we can ignore all components of $h^{\mu\nu}$ other than $h^{00}$;
then from (\ref{lineq}) we have

\beq
 \Box \bh^{00} = - 2 \ka \rh + O(\ep^2)\,,
\eeq

\noi
where we have defined $\rh \eq T^{00}$.  Since we are looking for
stationary solutions we have 

\beq \lab{appr9}
 \Box \bh^{\al \bt} = \na^2 \bh^{\al \bt} + O(\ep^2)\,.
\eeq
\noi
which will also be true for a slowly changing
gravitational field such that $h^{00}{}_{,00} /h^{00}{}_{,ij} = O(\epsilon)$.
Thus we have
\beq
\na^2 \bh^{00} = - 2 \ka \rh + O(\ep^2)\,.
\eeq
\noi
Comparing with Newton's law of gravity
\beq
 \na^2 V = \fr{\ka}{2} \rh\,,
\eeq
\noi
we get
\beq
 \bh^{00} = -4V + O(\ep^2)\,.
\eeq
\noi

\noi

\noi
Moreover,
\noi
\bea \nn
 h = - \bh &=& -\et_{00}\bh^{00} + O(\ep^{2}) \\ 
&=& \bh^{00}+ O(\ep^{2}) \,,
\eea
\bea \nn
 h^{00} &=& \bh^{00} + \ha \et^{00} h \\ \nn
&=& \bh^{00} - \ha \bh^{00}+ O(\ep^{2})  \\ \nn
&=& \ha \bh^{00}+ O(\ep^{2} )  \,,
\eea
\noi
and
\bea \nn
 \et_{00} h^{00}+\et_{ii} h^{ii} &=& h \\ 
 -\ha \bh^{00} + 3h^{11} &=& \bh^{00} + O(\ep^2)\,,
\eea
\noi
or
\beq
h^{11} = h^{22} = h^{33} = \ha \bh^{00}+ O(\ep^2)\,.
\eeq
\noi
That is
\beq
 h_{00} = h_{11} = h_{22} = h_{33} = -2V + O(\ep^2)\,,
\eeq
\noi
or
\beq
 h_{\mu \nu} = -2 \de_{\mu \nu} V + O(\ep^2)\,.
\eeq

\noi
The metric then becomes~\footnote{Note that if the original metric is written as
\[
ds^2=-\al^2 dt^2+\ps^4 (dx^2 + dy^2 + dz^2) \,,
\]
then
$\al^2=1+2V$ or $\al\approx 1+V$, and $\ps^4=1-2V$ or $\ps\approx1-V/2$.
}
\beq \lab{linmetric}
 ds^2 = -(1+2V) dt^2 + (1-2V)(dx^2+dy^2+dz^2)+ O(\ep^2) dx^\mu dx^\nu\,.
\eeq

\noi
The harmonic gauge implies

\beq
 \bh^{\mu \nu}\,_{,\nu} = h^{\mu \nu}\,_{,\nu} - \ha \et^{\mu \nu} h_{,\nu}
= 0 + O(\ep^2)\,,
\eeq

\noi
or
\beq
 -2 \de^{\mu \nu} V_{,\nu} = \ha \et^{\mu\nu} ( -4V_{,\nu}) + O(\ep^2)\,,
\eeq

\noi
i.e.,

\beq \lab{appr16}
 V_{,0} = 0 + O(\ep^2)\,,
\eeq

\noi
which is consistent with our assumption of a (near)-stationary solution.
In other words, the harmonic gauge condition will be satisfied as long as the
solutions are (near)-stationary.  Also note
that (\ref{appr3}) becomes

\beq \lab{appr17}
 V_{,\al} = O(\ep)\,.
\eeq
\noi
Moreover,
\beq
 \et_{00}\bh^{00} + \et_{ii}\bh^{ii} = \bh = -\bh^{00} + O(\ep^2)\,,
\eeq
\noi
which implies
\beq
 \bh^{ii} = O(\ep^2)\,,
\eeq
and justifies our assumption~(\ref{appr7}). (Equation (\ref{appr6}) is
satisfied for the diagonal metric (\ref{linmetric}), which implies
$\bh^{0i} = O(\ep^2)$. )
\noi

To evaluate the right hand side of the Einstein equation we consider

\beq 
T_{\mu \nu} = 
\ha \left[  \left( \na_{\mu} \ph \na_{\nu} \ph^\ast + \na_{\nu} \ph
\na_{\mu} \ph^\ast \right)
- g_{\mu \nu} \left( \ \na^\al \ph \na_\al \ph^\ast + m^2 |\ph|^2
\right)\right] \,.
\eeq
\noi
(Note that here $\na$ denotes the 4-dimensional covariant derivative operator. ) 
Since $\ph \sim e^{-i \om t}$, we have $\ph_{,0} \sim \om \ph \sim m
\ph$, where we have assumed
\beq \lab{appr10}
 \fr{\om}{m} = 1 + O(\ep^{1/2})\,.
\eeq  
\noi
Now
\bea \nn
 T^{00} &=& \ha g^{00} \left[2 g^{00} \ph^*_{,0}\ph_{,0}
- \left( g^{\al \bt} \ph^*_{,\al} \ph_{,\bt} +m^2 \ph^*\ph \right)
\right] + O(|\ep \ph|^2)\\ \nn
&=& \ha g^{00} \left[ g^{00}\ph^*_{,0} \ph_{,0} - \sum_{i=1}^3 g^{ii}
\ph^*_{,i} \ph_{,i}
-m^2 \ph^* \ph
  \right] + O(|\ep \ph|^2) \\ \nn
&=& \ha \left[ (1-2V)^2 \ph^*_{,0} \ph_{,0} +(1-2V)(1+2V) \sum_{i=1}^3 \ph^*_{,i} \ph_{,i} 
+m^2 (1-2V)\ph^* \ph \right] +O(|\ep \ph|^2)\\ \nn
&=& \ha \left[ (1-4V) \ph^*_{,0} \ph_{,0} +  \sum_{i=1}^3 \ph^*_{,i} \ph_{,i}
+m^2 (1-2V)\ph^* \ph \right] + O(|\ep \ph|^2)\,,
\eea
\noi
and

\bea \nn
T^{11} &=& 
\ha g^{11} \left[ 2g^{11} \ph^*_{,1}\ph_{,1}
- \left( g^{\al \bt} \ph^*_{,\al} \ph_{,\bt} +m^2 \ph^*\ph \right)
\right] +O(|\ep \ph|^2)\\ \nn
&=& \ha g^{11} \left[ g^{11} \ph^*_{,1}\ph_{,1}
- \sum_{\al=0,2,3} g^{\al \al} \ph^*_{,\al} \ph_{,\al} -m^2 \ph^*\ph
\right]+O(|\ep \ph|^2)\,.
\eea

\noi
Assumptions (\ref{appr4}), (\ref{appr5}) can be satisfied if
\beq \lab{appr18}
 \fr{\ph_{,i}}{ m \ph}  = O(\ep^{1/2})\,,
\eeq
\noi
or $\ph_{,i}/\ph_{,0} = O(\ep^{1/2})$, since $\ph_{,0} \sim m \ph$.
\noi
Then
\bea
 T^{00} &=& \ha \left( \om^2 + m^2\right) \ph^* \ph + O(\ep \ph^2)\\
        &=& m^2 \ph^{\ast} \ph + O(\ep \ph^2)\,,
\eea
\noi
and the Poisson equation becomes

\beq \lab{possion}
 \na^2 V = \fr{\ka}{2} m^2 \ph^* \ph + O(\ep \ph^2)\,.
\eeq
\noi
Also note that $T^{ii} = O(\ep \ph^2)$ and $T^{0i} = O(\ep^{1/2} \ph^2)$,
and hence the $00$-component of $T^{\mu\nu}$ is dominant.

\section{Schr\"{o}dinger Equation}
To derive the Schr\"{o}dinger equation for the complex scalar field, 
we first note that the Klein Gordon equation can be written as

\beq
 \fr{1}{\sr{-g}} \left( \sr{-g} g^{\mu \nu} \ph_{,\mu}\right)_{,\nu} - m^2
\ph = 0 \,.
\eeq

\noi
Now
\bea \nn
\sr{-g} &=& \sr{(1+2V)(1-2V)^3} + O(\ep^2)\\
 &=& 1-2V+O(\ep^2)\,.
\eea

\noi
The Klein Gordon Equation becomes:

\beq
\fr{1}{1-2V}\left[-\left( \fr{1-2V}{1+2V} \ph_{,0}\right)_{,0} +
\sum_{i=1}^3 \left( \fr{1-2V}{1-2V} \ph_{,i}\right)_{,i} 
\right]-m^2 \ph + O(\ep^2 \ph) = 0\,.
\eeq

\noi
Thus
\beq
 -\fr{1-4V}{1-2V}  \ph_{,00} - \fr{4V_{,0} \ph_{,0}}{1-2V}
+ (1+2V) \na^2 \ph - m^2 \ph + O(\ep^2 \ph)= 0\,,
\eeq

\noi
or
\beq
 -(1-2V) \ph_{,00}
+ (1+2V) \na^2 \ph - m^2 \ph = O(\ep^2 \ph)\,.
\eeq

\noi
Now let
\beq
 \ph(t,{\bf x}) \eq \Ph(t,{\bf x}) e^{-imt}\,,
\eeq
\noi
where $\Phi(t,{\bf x})$ is to be a slowly varying function of time.
Then

\beq
 \left[ -(1-2V) (\Ph_{tt} -m^2 \Ph -2im \Ph_{,t}) 
+(1+2V)\na^2 \Ph - m^2 \Ph 
\right] e^{-imt} = O(\ep^2 \Ph)\,,
\eeq

\noi
or 
\beq
-(1-2V) (\Ph_{tt} -2im \Ph_{,t}) 
+(1+2V)\na^2 \Ph - 2m^2 V \Ph 
= O(\ep^2 \Ph)\,.
\eeq

\noi
Let us assume
\bea \lab{appr11}
 \fr{\na^2 \Ph }{m^2 \Ph} &=& O(\ep)\,, \\ \lab{appr12}
 \fr{\Ph_{,t}}{m \Ph} &=& O(\ep)\,, \\ \lab{appr13}
 \fr{\Ph_{,tt}}{m^2 \Ph} &=& O(\ep^2)\,.
\eea
\noi
Then keeping track of terms up to $O(\epsilon)$, we have

\beq
 2im \Ph_{,t}+\na^2 \Ph - 2m^2 V \Ph = O(\ep^2)\,,
\eeq
\noi
or
\beq
 i \Ph_{,t} = -\fr{1}{2m} \na^2 \Ph + mV \Ph + O(\ep^2)\,.
\eeq

\noi
We now assume
\beq \lab{PheqPhe}
 \Ph(t,{\bf x}) = \Ph_0({\bf x}) e^{-iEt}\,,
\eeq

\noi
so that the full scalar field is

\beq
 \ph = \Ph e^{-imt} = \Ph_0 e^{-imt-iEt} = \Ph_0 e^{-iwt}\,,
\eeq

\noi
where $w=m+E$.
We then have

\beq
 E \Ph_0 = - \fr{1}{2m} \na^2 \Ph_0 + mV \Ph_0 + O(\ep^2 \Ph)\,.
\,
\eeq
\noi
Note that from (\ref{PheqPhe}) we have $\Ph_t \sim E \Ph$.
Therefore (\ref{appr12}) and (\ref{appr13}) will be satisfied if
\beq \lab{appr14}
 E = O(\ep)\,,
\eeq
\noi
or
\beq  
 \fr{\om}{m} = 1+O(\ep)\,.
\eeq
\noi
Note that (\ref{appr10}) is then automatically satisfied.  Also 
(\ref{appr11}) will be satisfied if
\beq \lab{appr15}
  \fr{\na^2 \Ph_0}{m^2 \Ph_0} = O(\ep)\,.
\eeq

\section{Summary}
The above derivation is lengthy and hence we present a summary of the 
development.
For our purposes stated at the beginning of this appendix, 
we first solve the PS system

\bea
 \na^2 V &=& \fr{\ka}{2} m^2 \Ph_0^{\ast} \Ph_0 \,,\\
 E \Ph_0 &=& - \fr{1}{2m} \na^2 \Ph_0 + mV \Ph_0 \,.
\eea
for $\Ph_0$ and $V$.

\noi
Then, ignoring $\Oeps$ terms, we have
\beq
\bh_{ \mu \nu} = 
\bh^{ \mu \nu} = 
 \left[ 
\ba{cccc}
-4V &  0& 0  &  0  \\
0& 0 & 0 & 0  \\
0        &  0     &  0 & 0 \\
0 & 0 & 0 & 0
\ea
 \right] \,,
\eeq
\noi
or
\beq
h_{ \mu \nu} = 
h^{ \mu \nu} = 
 \left[ 
\ba{cccc}
-2V &  0& 0  &  0  \\
0& -2V & 0 & 0  \\
0        &  0     &  -2V & 0 \\
0 & 0 & 0 & -2V
\ea
 \right] \,.
\eeq

\noi
The approximate metric for the EKG system is then
\beq 
 ds^2 = -(1+2V) dt^2 + (1-2V)(dx^2+dy^2+dz^2)\,,
\eeq
\noi
or
\bea
 \al  &=& 1+V \,, \\
  \ps &=& 1-\fr{V}{2} \,,\\
   \bt &=& 0\,, \\
   \si &=& 0\,.
\eea

\noi
For the scalar field we have
\bea
\ph_0 &=& \Ph_0 \,,\\
\om &=& m + E\,.
\eea
Note that we have assumed
(\ref{appr1}), (\ref{appr2}), (\ref{appr3}), (\ref{appr8}),
(\ref{appr4})-(\ref{appr7}),  (\ref{appr9}), (\ref{appr16}),
(\ref{appr17}), (\ref{appr10}), (\ref{appr18}),
(\ref{appr11})-(\ref{appr13}), (\ref{appr14}) and (\ref{appr15}).  
In
other words, solutions to the PS equations do {\em not} necessarily give good
initial guesses for the EKG system unless the approximations are satisfied.
Note that 
for our purpose of
finding {\em stationary} solutions, 
not all of the assumptions are independent. Specifically, 
the assumptions that we need to check on a case-by-case basis are 
(\ref{appr8}), (\ref{appr17}), (\ref{appr18}), (\ref{appr14}) and
(\ref{appr15}):
\bea
 V &=& O(\ep)\,, \\
 V_{,\al} &=& O(\ep)\,, \\
 E &=& O(\ep)\,,\\
 \fr{\Ph_{0,i}}{m \Ph_0}  &=& O(\ep^{1/2})\,, \\
  \fr{\na^2 \Ph_0}{m^2 \Ph_0} &=& O(\ep)\,.
\eea

\noi
Also
note that this check is, in general, nontrivial to carry out since in~(\ref{appr1}) the 
flat metric is expressed in Cartesian coordinates. Thus, we have to transform our 
solutions---which are computed in polar coordinates---to Cartesian coordinates in 
order to verify the validity of the above approximations.

  \resetcounters

\def\graxii{{\tt graxi}$_{\tt 0}$}
\def\Test{{Unigrid Static Boson Star Code Test}}
\def\test#1{{Unigrid static boson star code test: {#1}}}

\chapter[\graxi\ Unigrid Static Boson Star Code Test]
{} \label{ap9}
\vspace{-0.45in}
{\bf {\huge 
\graxi\ Unigrid Static Boson Star Code Test
}}
\vspace{0.52in}

\noi
In this appendix we present the results of a test of the 
code, \graxi, that we use to evolve the massive, complex scalar field and 
massless real scalar field in axisymmetry~(see Chap.~\ref{bs2d}).  

\section{Background}
{\graxi} is based on a unigrid (i.e.\ no AMR) code described in~\cite{graxi:2003} that 
solves the Einstein and massless scalar wave equations in axisymmetry. 
AMR was incorporated into that code by Pretorius~\cite{fransp:phd,graxi:2003b}, who later
extended the model to include a self-interacting complex scalar field~\cite{fransp:notes-complex-03}.
In original work for this thesis, we extended that code so that it could be initialized with interpolated 
boson star initial data as described in Sec.~\ref{bincol}; this produced the final
version of the code that we refer to as \graxi\ in the thesis.

Note that the version of \graxi\ described in~\cite{fransp:phd,graxi:2003b} has been
thoroughly convergence tested, so we restrict attention to a test that verifies the 
boson star initial data interface (our current work) as well
as the modifications for the complex scalar field {\em per se} (Pretorius' work). 

\section{Test Definition}
The test involves evolution of static boson star initial data.
Although this problem may sound trivial, it is not. 
In particular, for the complex scalar field it is anything but a time-independent calculation,
as a glance at the ansatz~(\ref{bs1dansatz}) 
\[
\ph(t,r) = \ph_0(r)\,e^{-i \om t}\,,
\]
will confirm (recall that we evolve the real and imaginary parts of the complex 
field separately).  In addition, the initial data, $\ph_0(r_i)$, that results from
solving a discrete form of the spherically symmetric Einstein-Klein-Gordon system with the 
above ansatz, must always be interpolated to the cylindrical grid $(\rho_j,z_k)$ and 
this naturally introduces a certain level of error (including departure from 
strict spherical symmetry) in the initial conditions $\phi(0,\rho_j,z_k)$ and $\Pi(0,\rho_j,z_k)$.
This the finite difference solutions $\ph^h$ are not expected to be precisely
time-independent.

The parameters for the test are as follows.  We initialize the complex 
scalar field to a single boson star with $\ph_0(0) = 0.02$, centered 
at the origin $(\rho,z) = (0,0)$, on the domain $0 \le \rho \le 48$, 
$-48 \le z \le 48$. 
We use a Courant factor $\De t/\De \rh = \De t/\De z = 0.3$ and a 
dissipation coefficient $\ep_d = 0.5$.

AMR is not enabled for this test.  We note that we have also tested the
code with AMR enabled, but due to our choice of problem parameters 
(in particular the truncation error threshold that controls the overall
regridding algorithm) no refinement is triggered and the results
are thus essentially identical to those presented here.

In order to perform convergence tests, we use the above initial data and code 
parameters and then perform calculations 
with $N =64, 128, 256$ and $512$ ($N_\rh = N+1, N_z=2N+1$).

\section{Test Results}
\subsection{Time-independence of $\vert\phi(t,\rho,z)\vert$ and $\psi(t,\rho,z)$}
\begin{figure}
\begin{center}
\epsfxsize=15.0cm
\ifthenelse{\equal{\highQ}{true}} {
\includegraphics[width=15.0cm,clip=true]{eps/graxi_csf/bs.1s.N/modph4to1.eps}
}{
\includegraphics[width=15.0cm,clip=true]{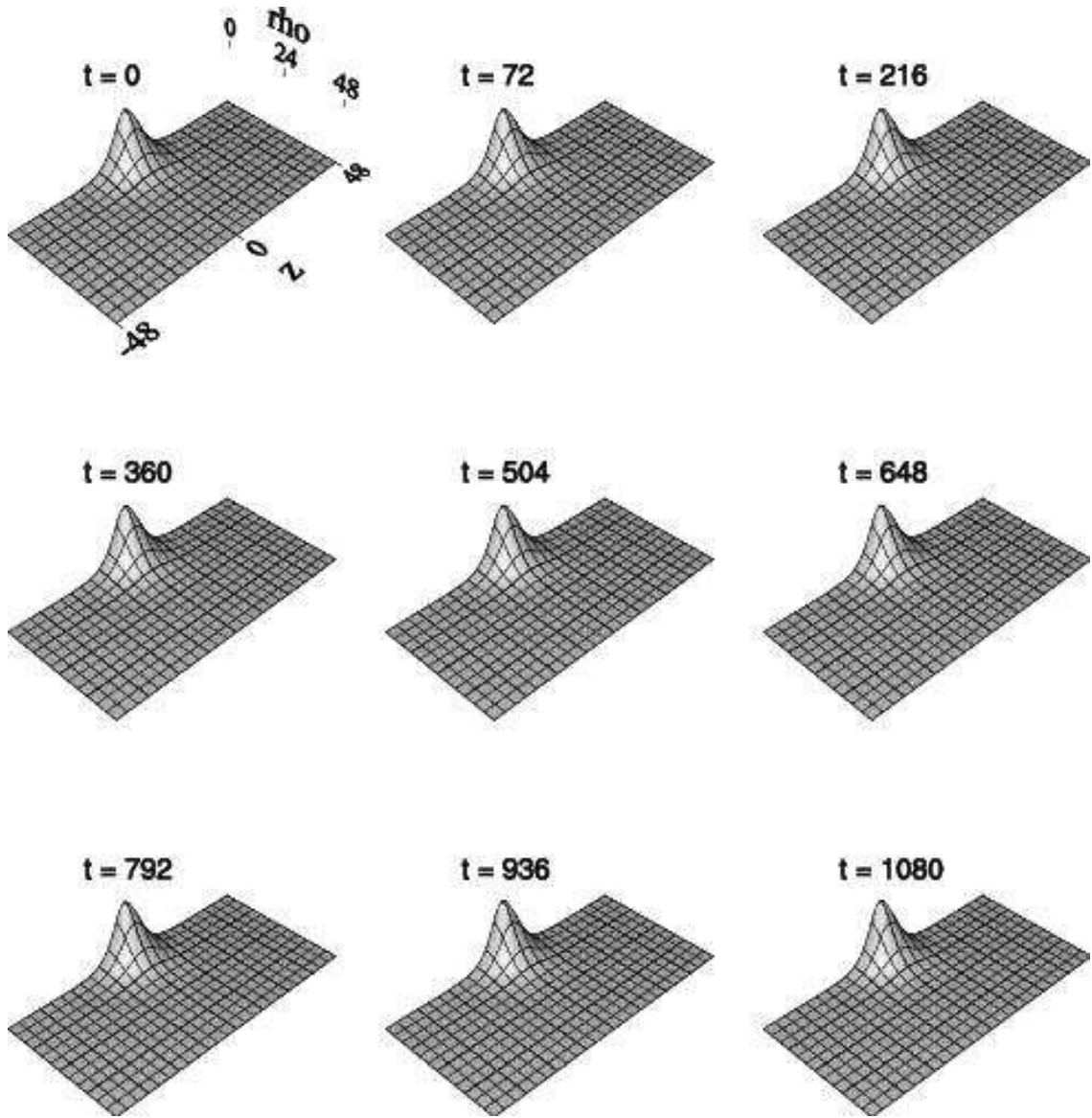}
}
\caption
[\test{$\vert\phi(t,\rho,z)\vert$}]
{\test{$\vert\phi(t,\rho,z)\vert$}. 
The figure shows $N=128$ calculation.
The initial central value of the scalar field
is $\ph_0(0) = 0.02$.  The period of evolution 
shown corresponds to 
roughly 20-crossing times and 160 oscillations of $\phi_1$ and $\phi_2$.
To ``eye-ball'' accuracy, 
$|\ph(t,\rh,z)|$ is time-independent.
$0.0 \le |\ph(t,\rh,z)| \le 0.02\times \fr{1}{\sr{2}}$.
}
\label{phcvtest}
\end{center}
\end{figure}

\begin{figure}
\begin{center}
\epsfxsize=15.0cm
\ifthenelse{\equal{\highQ}{true}} {
\includegraphics[width=15.0cm,clip=true]{eps/graxi_csf/bs.1s.N/psi4to1.eps}
}{
\includegraphics[width=15.0cm,clip=true]{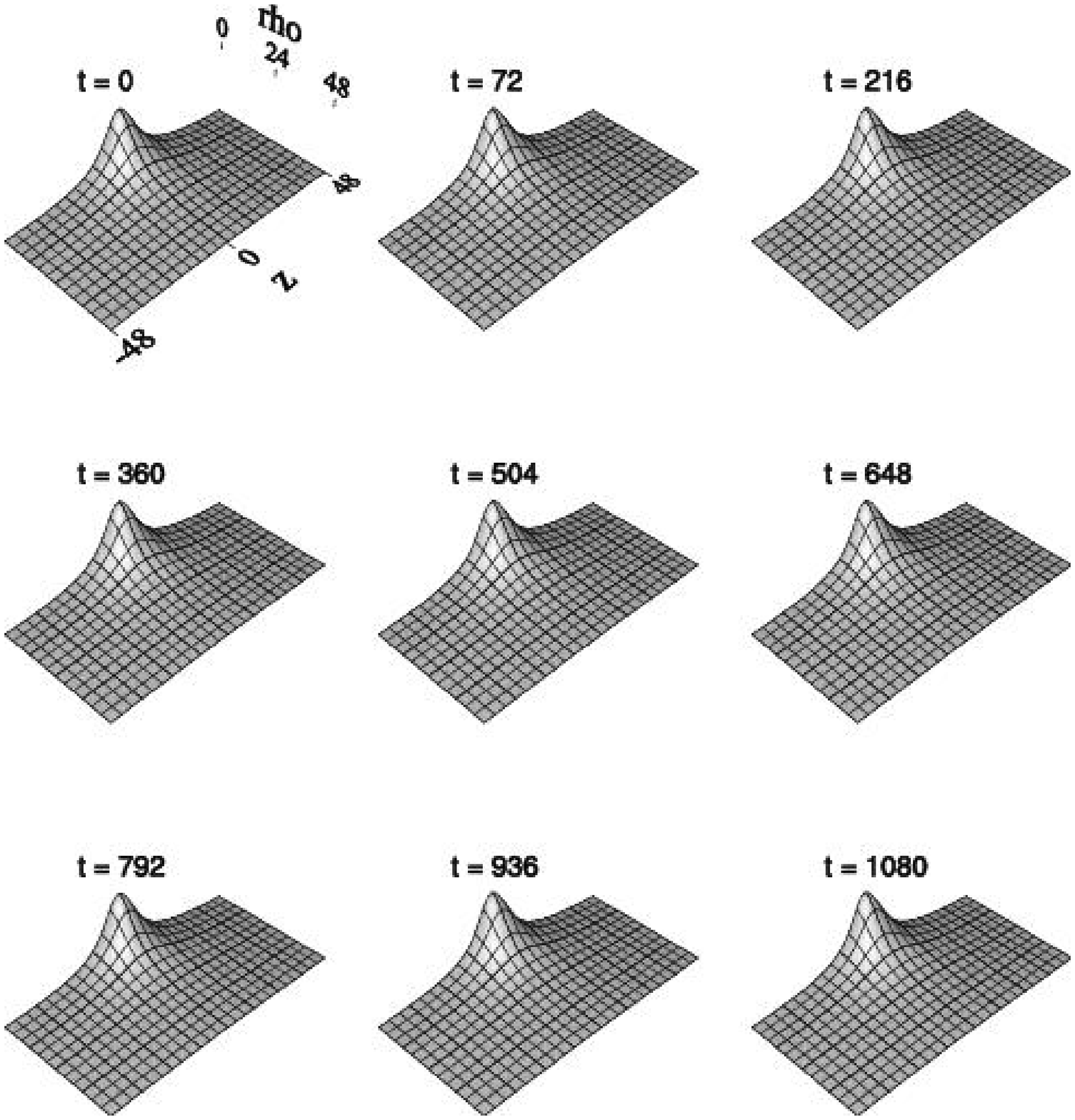}
}
\caption
[\test{$\psi(t,\rho,z)$}]
{\test{$\psi(t,\rho,z)$}. 
The figure shows $N=128$ calculation.
The initial central value of the scalar field
is $\ph_0(0) = 0.02$.  
As is the case for $\vert\phi(t,\rho,z)\vert$ (Fig.~\ref{phcvtest}), 
$\psi(t,\rh,z)$ is time-independent to eye-ball accuracy.
$1.004 \le \ps(t,\rh,z) \le 1.048$.
}
\label{psicvtest}
\end{center}
\end{figure}

Figs.~\ref{phcvtest} and \ref{psicvtest} are 
`eye-ball'' tests in which we display plots of $\vert\phi(t,\rho,z)\vert$ 
and $\psi(t,\rho,z)$ that should be static, and which {\em do} exhibit 
time-independence to ``eye-ball'' accuracy.
The data shown here came from the $N = 128$ 
calculation.  The length of time spanned by the plots corresponds to slightly
more than 20 light-crossing times, and about 160 oscillations of each of the 
component fields $\phi_1$ and $\phi_2$.

\subsection{Time-independence of $\max_{\rh,z}|\ph(t,\rh,z)|$}
\begin{figure}
\begin{center}
\epsfxsize=10.5cm
\includegraphics[width=10.5cm,clip=true]{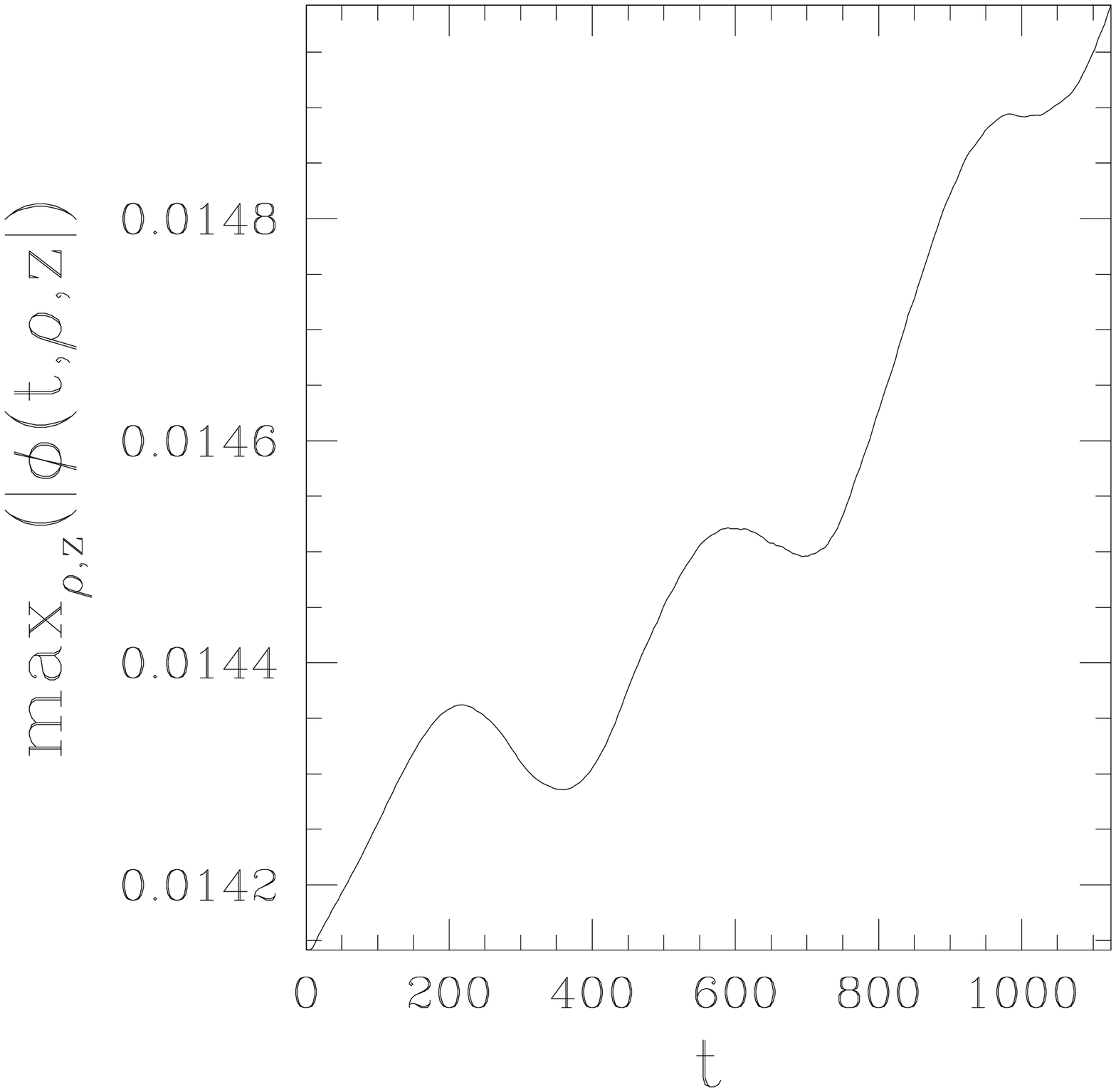}
\caption
[\test{Oscillation of $\max_{\rh,z}|\ph(t,\rh,z)|$ as a measure of
perturbation.}]
{
\test{Oscillation of $\max_{\rh,z}|\ph(t,\rh,z)|$ as a measure of
perturbation}.
Results shown here are from the $N=128$ calculation.
The initial central value of the scalar field
is $\ph_0(0) = 0.02$.  
The observed oscillation frequency is roughly
$\si^2 \approx 0.00032$, and is in good agreement with 
the fundamental mode frequency computed from perturbation analysis, $\si^2 \approx 0.00035$.
}
\label{maxmodph}
\end{center}
\end{figure}

Fig.~\ref{maxmodph} plots the maximum value of the modulus of the scalar field,
$\max_{\rh,z}|\ph(t,\rh,z)|$, {\em vs} $t$ from the $N=128$ calculation.
The figure shows that the maximum value exhibits periodic oscillations as well as 
a drift that is due to the accumulation of solution error.  We fully expect 
to see such oscillations since numerical errors act as a 
perturbation to the ``exact'' stationary solution.~\footnote{
Note that the notion of ``exact'' in this case is relative to the discretization
used by \graxi\, and not to the continuum.  Thus, at any resolution,
the ``exact'' stationary solution,
should one exist, would be the solution of the difference equations for {\graxi}
at that resolution, (including all boundary and regularity conditions) with all 
discrete time 
derivatives set to zero.
}

It is an instructive exercise to compare the observed period of oscillation
with that computed from perturbation analysis of the 
initial boson star  as  described in Sec.~(\ref{perttheory}).
Assuming harmonic time-dependence for the perturbed scalar field and metric
component (\ref{harmonicdep1}), (\ref{harmonicdep2}), and assuming that the oscillation is 
in the fundamental mode, we
obtain a theoretical value of $\si^2 \approx 0.00035$.  
(Time averaging $\ph_0(t,0)$ gives $\langle \ph_0(t,0) \rangle \approx 0.02
\times \sr{4 \pi} = 0.071$. ) 
On the other hand,
Fig.~\ref{maxmodph} shows a period of oscillation $T\approx 382$, corresponding
to a frequency $\si \eq 2 \pi /T = 0.0164$.  
The average of the lapse function is $\langle \al(t,0) \rangle = 0.993$, therefore
$\si^2/\al^2 \approx 0.00032$, in good agreement with the perturbation analysis value
$0.00035$.   
 
\subsection{Time-independence of ADM Mass}
\begin{figure}
\begin{center}
\epsfxsize=12.0cm
\includegraphics[width=12.0cm,clip=true]{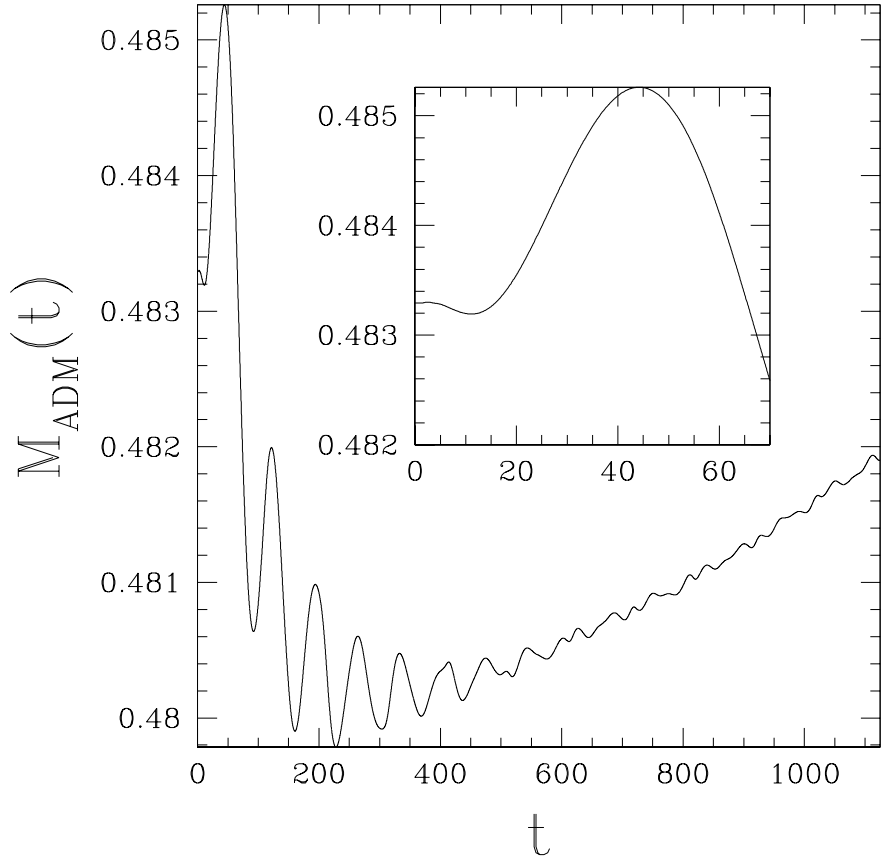}
\caption
[\test{$M_{\rm ADM}(t)$} for $N=128$ calculation]
{\test{$M_{\rm ADM}(t)$ for $N=128$ calculation}.
The initial central value of the scalar field
is $\ph_0(0) = 0.02$.  
The general increase in mass starting at $t\approx 450$ is indicative of
an instability, and implies
that the evolution will generically break down for a sufficiently long evolution.
In the inset, we show a detail view of the initial fluctuations in $M_{\rm ADM}(t)$.
}
\label{massvst}
\end{center}
\end{figure}

Fig.~\ref{massvst} shows the ADM mass $M_{\rm ADM}(t)$ {\em vs} time $t$ for the $N=128$ simulation.
The general increase in mass starting at $t\approx 450$ is indicative of 
an instability, and implies
that the evolution will generically break down for a sufficiently long evolution.

\subsection{Convergence tests of $\max_{\rh,z}(|\ph(t,\rh,z)|)$ and $M_{\rm ADM}(t)$}
\begin{figure}
\begin{center}
\epsfxsize=8.5cm
\includegraphics[width=8.5cm,clip=true]{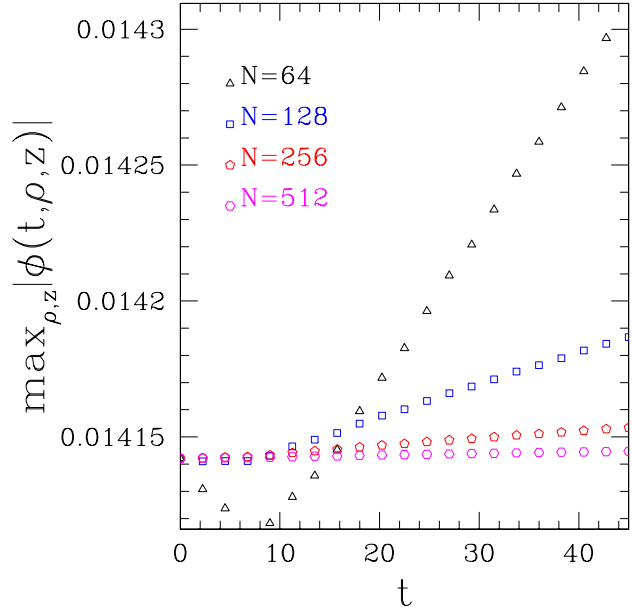}
\caption
[\test{4-level convergence test of $\max_{\rh,z}|\ph(t,\rh,z)|$}]
{
\test{4-level convergence test of $\max_{\rh,z}|\ph(t,\rh,z)|$}.
The figure shows that $\max_{\rh,z}|\ph(t,\rh,z)|$ converges to a
constant as $\Delta \rho, \Delta z \to 0$.
}
\label{cvph4level}
\end{center}
\end{figure}

\begin{figure}
\begin{center}
\epsfxsize=8.5cm
\includegraphics[width=8.5cm,clip=true]{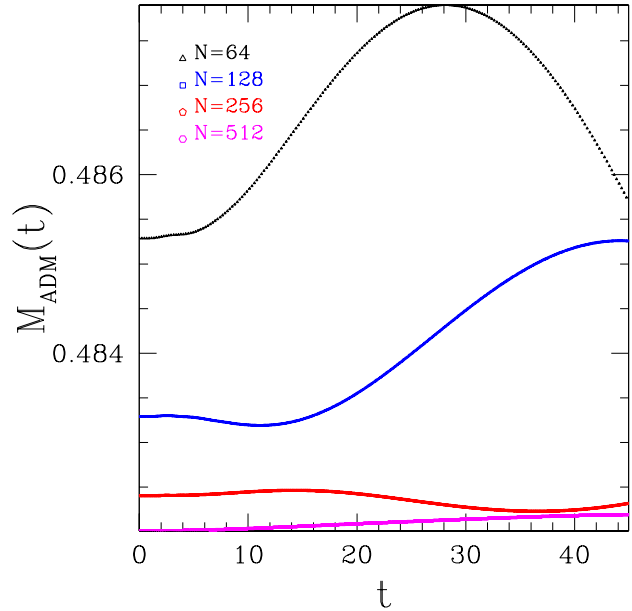}
\caption
[\test{4-level convergence test of ADM mass $M_{\rm ADM}(t)$}]
{
\test{4-level convergence test of ADM mass $M_{\rm ADM}(t)$}.
The figure shows that $M_{\rm ADM}(t)$ generally converges to a
constant as $\Delta \rho, \Delta z \to 0$, although there 
are clearly indications of some problems at later times in this plot which are 
almost certainly a result of imperfect boundary conditions.
}
\label{cvM4level}
\end{center}
\end{figure}

\begin{figure}
\begin{center}
\epsfxsize=8.5cm
\includegraphics[width=8.5cm,clip=true]{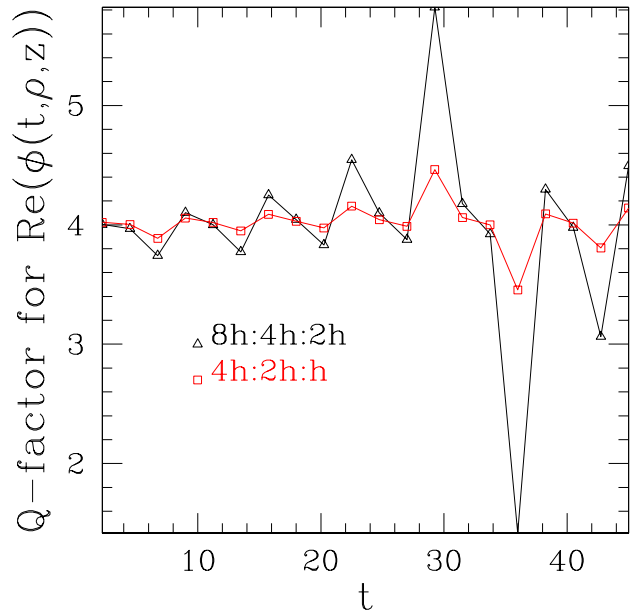}
\caption
[\test{Convergence rate of ${\rm Re}(\phi(t,\rho,z))\equiv\phi_1(t,\rho,z)$}]
{
\test{Convergence rate of ${\rm Re}(\phi(t,\rho,z))\equiv\phi_1(t,\rho,z)$}.
The theoretical value for a 
second-order convergent scheme is $Q=4$. See text for further explanation.
}
\label{convfactor}
\end{center}
\end{figure}

Figs.~\ref{cvph4level} shows the maximum value of the modulus of the scalar field,
$\max_{\rh,z}(|\ph(t,\rh,z)|)$, {\em vs} $t$ from the calculations at the four different
finite difference resolutions, $N=64,128,256$ and $512$. 
Fig.~\ref{cvM4level} similarly shows the ADM mass as a function of time for the four calculations.  
The two figures provide strong evidence that both quantities are converging.
Finally, 
Fig.~\ref{convfactor} shows the convergence $Q$-factor for ${\rm
Re}(\phi(t,\rho,z))\equiv\phi_1(t,\rho,z)$,
where the $Q$-factor for a general grid function $u^h$ is defined by
\beq \lab{q_fact}
Q \eq \fr{\|u^{4h}-u^{2h}\|_2}{\|u^{2h} - u^{h}\|_2}\,,
\eeq
and which has a theoretical value $Q=4$ for a second order scheme.  Here $\|\cdot\|_2$
denotes the $l_2$ norm, and $h=\Delta \rho=\Delta z$.

\end{appendix}
\end{document}